\documentclass[aps,prx,reprint,superscriptaddress,nofootinbib,floatfix]{revtex4-2}

\usepackage{amsmath,amssymb}
\usepackage{mathrsfs}
\usepackage{booktabs}
\usepackage{array}[=2016-10-06]
\usepackage{multirow}
\newcolumntype{C}[1]{>{\centering\arraybackslash}p{#1}}
\usepackage{cancel}
\usepackage{graphicx}
\usepackage{placeins}
\usepackage{bm}
\usepackage{color}
\usepackage[colorlinks=true,citecolor=blue,linkcolor=blue,urlcolor=blue]{hyperref}
\usepackage[ruled,linesnumbered,longend]{algorithm2e}
\usepackage{amsthm}

\definecolor{RED}{rgb}{1,0,0}

\newif\ifshowrev \showrevfalse % set \showrevfalse to render all revision markup in normal color
\ifshowrev
  \newcommand{\rev}[1]{\textcolor{red}{#1}}
  \newcommand{\revcolor}{\color{red}}
\else
  \newcommand{\rev}[1]{#1}
  \newcommand{\revcolor}{\normalcolor}
\fi

\begin{document}

\title{Resolving Quantum Criticality in the Honeycomb Hubbard Model}
\author{Fo-Hong Wang}
\affiliation{Tsung-Dao Lee Institute, Key Laboratory of Artificial Structures and Quantum Control (Ministry of Education), School of Physics and Astronomy, Shanghai Jiao Tong University, Shanghai 200240, China}
\author{Jakob V\"{a}th}
\affiliation{Institut f\"{u}r Theoretische Physik und Astrophysik and W\"{u}rzburg-Dresden Cluster of Excellence ctd.qmat, Universit\"{a}t W\"{u}rzburg, 97074 W\"{u}rzburg, Germany}
\author{Fanjie Sun}
\affiliation{Tsung-Dao Lee Institute, Key Laboratory of Artificial Structures and Quantum Control (Ministry of Education), School of Physics and Astronomy, Shanghai Jiao Tong University, Shanghai 200240, China}
\author{Cheng-Hao He}
\affiliation{Tsung-Dao Lee Institute, Key Laboratory of Artificial Structures and Quantum Control (Ministry of Education), School of Physics and Astronomy, Shanghai Jiao Tong University, Shanghai 200240, China}
\author{Fakher F. Assaad}
\email{fakher.assaad@physik.uni-wuerzburg.de}
\affiliation{Institut f\"{u}r Theoretische Physik und Astrophysik and W\"{u}rzburg-Dresden Cluster of Excellence ctd.qmat, Universit\"{a}t W\"{u}rzburg, 97074 W\"{u}rzburg, Germany}
\author{Xiao Yan Xu}
\email{xiaoyanxu@sjtu.edu.cn}
\affiliation{Tsung-Dao Lee Institute, Key Laboratory of Artificial Structures and Quantum Control (Ministry of Education), School of Physics and Astronomy, Shanghai Jiao Tong University, Shanghai 200240, China}
\date{\today}

%%%%%%%%%%%%%%%%%%%%%%%%%%%%%%%%%%%%%%%%%%%%%%%%%%%%%
\begin{abstract}
Quantum phase transitions driven by electronic correlations are central to understanding the physics of graphene and related two-dimensional materials.
A paradigmatic example is the semimetal-to-Mott-insulator transition on the honeycomb lattice, governed by the Gross-Neveu-Heisenberg universality class, yet consensus on its critical exponents has remained elusive for over a decade due to severe finite-size effects and the absence of rigorous conformal bootstrap benchmarks.
Here we resolve this long-standing controversy by performing projector determinant quantum Monte Carlo simulations on lattices of unprecedented size, reaching 10,368 sites.
By developing a novel projected submatrix update algorithm, we achieve a significant algorithmic speedup \rev{that brings such large system sizes within reach}.
\rev{We find that the critical point and boson anomalous dimension extracted from the equal-time spin structure factor continue to drift even at the largest sizes due to a slowly decaying short-range background.
Suppressing this background at the observable level brings all critical parameters onto converged plateaus without empirical extrapolation, reconciling the honeycomb Hubbard exponents with field-theoretic predictions.}
To validate our analysis, we perform parallel large-scale simulations of the spinless $t$-$V$ model on the honeycomb lattice, which belongs to the Gross-Neveu-Ising class.
Our results for the $t$-$V$ model show agreement with conformal bootstrap predictions, thereby corroborating the robustness of our methodology.
Our work provides state-of-the-art critical exponents for the honeycomb Hubbard model and establishes a systematic finite-size scaling workflow applicable to a broad class of strongly correlated quantum systems, paving the way for resolving other challenging fermionic quantum critical phenomena.
\end{abstract}

%%%%%%%%%%%%%%%%%%%%%%%%%%%%%%%%%%%%%%%%%%%%%%%%%%%%%

\maketitle

%%%%%%%%%%%%%%%%%%%%%%%%%%%%%%%%%%%%%%%%%%%%%%%%%%%%%
\section{Introduction}
%%%%%%%%%%%%%%%%%%%%%%%%%%%%%%%%%%%%%%%%%%%%%%%%%%%%%
Understanding how quantum materials transition between distinct phases of matter is a central challenge in condensed matter physics.
When massless Dirac fermions---the relativistic quasiparticles underlying graphene's electronic properties and emerging in diverse quantum materials from $d$-wave superconductors to surface of topological insulators~\cite{Annu.Rev.Condens.MatterPhys.2014Vafek,Adv.Phys.2014Wehling}---interact strongly, they can undergo quantum phase transitions that lie beyond conventional theoretical paradigms.
A prototypical example is the interaction-driven semimetal-to-Mott-insulator transition in the honeycomb Hubbard model~\cite{Europhys.Lett.1992Sorella,Phys.Rev.B2005Paiva,Phys.Rev.Lett.2006Herbut,Nature2010Meng,SciRep2012Sorella,Phys.Rev.X2013Assaad}, where increasing electron-electron repulsion spontaneously generates mass and magnetic order.
This relativistic Mott transition has recently been observed experimentally in an artificial graphene realized by twisted double bilayer WSe$_2$~\cite{Nat.Mater.2025L.Ma,ArXiv2509B.Hawashin,ArXiv2509J.Biedermann}, in which the twist angle tunes the interaction strength.
The associated quantum critical point belongs to the Gross-Neveu-Heisenberg (GNH) universality class~\cite{Phys.Rev.D2017Zerf,Phys.Rev.B2023Ladovrechis,Phys.Rev.D2018Gracey,PRB2014L.Janssen,Phys.Rev.B2018Knorr,ArXiv2609H.Du,PRL2026YK.Yu,Nat.Commun.2025Zeng,Phys.Rev.B2021Ostmeyer,Phys.Rev.B2019Buividovich,Phys.Rev.X2016Otsuka,Phys.Rev.B2015ParisenToldin,ArXiv2611J.Vaeth,Phys.Rev.B2026Hands,Phys.Rev.B2025Lang,Phys.Rev.Lett.2021Xu,Phys.Rev.B2021Liu,Phys.Rev.B2020Otsuka,Science2018Tang,Phys.Rev.B2018Buividovich} and connects fundamental questions in quantum materials to high-energy physics concepts such as chiral symmetry breaking and dynamical mass generation.

\begin{figure*}[htbp]
    \centering
    \includegraphics[width=\textwidth]{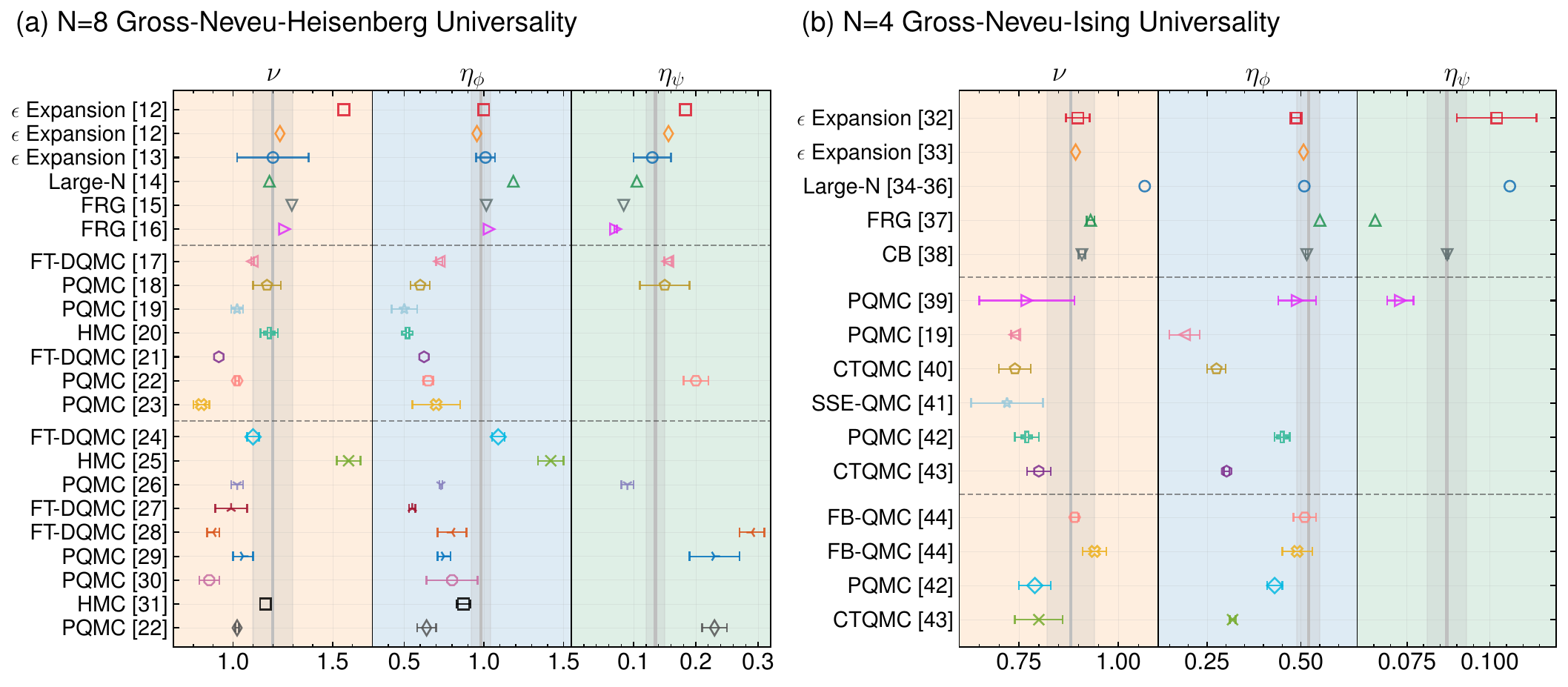}
    \caption{Summary of critical exponents $\nu$, $\eta_\phi$, and $\eta_\psi$ reported in the literature and obtained in this work.
    (a) $N=8$ Gross-Neveu-Heisenberg universality class~\cite{Phys.Rev.D2017Zerf,Phys.Rev.B2023Ladovrechis,Phys.Rev.D2018Gracey,PRB2014L.Janssen,Phys.Rev.B2018Knorr,ArXiv2609H.Du,PRL2026YK.Yu,Nat.Commun.2025Zeng,Phys.Rev.B2021Ostmeyer,Phys.Rev.B2019Buividovich,Phys.Rev.X2016Otsuka,Phys.Rev.B2015ParisenToldin,ArXiv2611J.Vaeth,Phys.Rev.B2026Hands,Phys.Rev.B2025Lang,Phys.Rev.Lett.2021Xu,Phys.Rev.B2021Liu,Phys.Rev.B2020Otsuka,Science2018Tang,Phys.Rev.B2018Buividovich}.
    (b) $N=4$ Gross-Neveu-Ising universality class~\cite{Phys.Rev.B2018Ihrig,PRD2025JA.Gracey,Phys.Lett.B1992Gracey,Int.J.Mod.Phys.A1994Graceya,Int.J.Mod.Phys.A1994Gracey,Phys.Rev.B2016Knorr,J.HighEnerg.Phys.2023Erramilli,Sci.Adv.2026YK.Yu,Nat.Commun.2025Zeng,Phys.Rev.B2016Hesselmann,Phys.Rev.B2016Wang,NewJ.Phys.2015Li,NewJ.Phys.2014Wang,Phys.Rev.D2020Huffman}.
    The gray horizontal dashed lines divide the plot into three blocks: analytical results (top), QMC results for the same model as in this work (middle), and QMC results for other models in the same universality class (bottom).
    CB denotes conformal bootstrap~\cite{J.HighEnerg.Phys.2023Erramilli}.
    See Tables~\ref{tab:gn_heis} and~\ref{tab:gn_is} in Appendix~\ref{app:Gross-Neveu} for the numerical values and methodological details.}
    \label{fig:critical_exponents_literature}
\end{figure*}

Despite decades of effort, consensus on the precise critical exponents governing this transition remains elusive, as illustrated in Fig.~\ref{fig:critical_exponents_literature}.
Analytical approaches---including $\epsilon$ expansions, large-$N$ expansions, and functional renormalization group (FRG) calculations---yield estimates that are only partially compatible, with a particularly pronounced spread in the fermion anomalous dimension $\eta_\psi$~\cite{Phys.Rev.D2017Zerf,Phys.Rev.B2023Ladovrechis,Phys.Rev.D2018Gracey,Phys.Rev.B2018Knorr}.
Quantum Monte Carlo (QMC) results are even less uniform: for the honeycomb Hubbard model, most studies report a boson anomalous dimension $\eta_\phi \lesssim 0.65$~\cite{PRL2026YK.Yu,Nat.Commun.2025Zeng,Phys.Rev.B2021Ostmeyer,Phys.Rev.B2019Buividovich,Phys.Rev.X2016Otsuka}, \rev{well below analytical predictions near unity}~\cite{Phys.Rev.D2017Zerf,Phys.Rev.B2023Ladovrechis,Phys.Rev.D2018Gracey,Phys.Rev.B2018Knorr}, while different lattice realizations in the same universality class yield noticeably larger values.
In contrast, for the $N=4$ Gross-Neveu-Ising (GNI) universality class~\cite{Phys.Rev.B2018Ihrig,PRD2025JA.Gracey,Phys.Lett.B1992Gracey,Int.J.Mod.Phys.A1994Graceya,Int.J.Mod.Phys.A1994Gracey,Phys.Rev.B2016Knorr,J.HighEnerg.Phys.2023Erramilli,Sci.Adv.2026YK.Yu,Nat.Commun.2025Zeng,Phys.Rev.B2016Hesselmann,Phys.Rev.B2016Wang,NewJ.Phys.2015Li,NewJ.Phys.2014Wang,Phys.Rev.D2020Huffman}, large-scale QMC simulations~\cite{Phys.Rev.D2020Huffman} have produced critical exponents consistent with conformal bootstrap predictions~\cite{J.HighEnerg.Phys.2023Erramilli}, highlighting the achievable precision when finite-size effects are properly controlled.

The root cause of the GNH controversy lies in severe finite-size effects that render extracted exponents highly sensitive to the choice of system sizes and fitting procedures.
\rev{Irrelevant operators from competing interaction channels yield a small correction exponent of about $0.3$~\cite{Phys.Rev.B2023Ladovrechis}, while the large $\eta_\phi$ induces a strong, slowly decaying analytic background~\cite{Phys.Rev.B2015ParisenToldin}.}
\rev{The large $\eta_\phi$ also leads to large statistical fluctuations in susceptibilities sampled from QMC~\cite{Phys.Rev.B2021Liu}.}
Alternative simulation setups have been proposed to mitigate these effects, including extended-hopping models that enlarge the linear dispersion region~\cite{Phys.Rev.Lett.2019Lang,Phys.Rev.Lett.2022Tabatabaei,Phys.Rev.B2020Liu,Phys.Rev.B2025Lang} and elective-momentum ultrasize QMC methods~\cite{Phys.Rev.B2023Wanga}.
Remarkably, SLAC-fermion formulations produce high-precision exponents and suggest that honeycomb estimates drift toward SLAC values once the smallest system sizes are excluded~\cite{Phys.Rev.B2025Lang}, though a noticeable discrepancy with analytical predictions persists.
However, the honeycomb Hubbard model remains the minimal model for the GNH transition with direct experimental relevance to graphene and related two-dimensional materials, making it essential to resolve the controversy within this original setting.
While hybrid Monte Carlo (HMC) simulations~\cite{Phys.Rev.B2021Ostmeyer} have achieved lattices with $L>100$, ergodicity issues restrict reliable results to high temperatures, preventing direct access to ground-state properties. Their extracted correlation length exponent $\nu$ falls within the range spanned by various analytical estimates, yet the boson anomalous dimension $\eta_\phi$ remains notably smaller than most field-theoretic predictions, and $\eta_\psi$ was not determined~\cite{Phys.Rev.B2021Ostmeyer}.

In this work, we revisit this long-standing problem with the numerically exact projector determinant quantum Monte Carlo (PQMC)~\cite{Ann.Phys.1986Sugiyama,Europhys.Lett.1989Sorella,Phys.Rev.B1989White} method on lattices up to linear size $L=72$ ($10{,}368$ sites), twice that of previous PQMC studies~\cite{Phys.Rev.X2016Otsuka}.
This unprecedented scale is enabled by a new submatrix update algorithm that optimizes cache utilization in CPU operations, achieving state-of-the-art performance without introducing any additional approximations.
Through a straightforward yet systematic finite-size scaling (FSS) analysis, \rev{we show how a slowly decaying short-range background biases conventional structure-factor estimates even at $L=72$, and overcome it via background-suppressed equal-time observables (cross-checked by finite-temperature spin susceptibility~\cite{ArXiv2611J.Vaeth}) to obtain plateau-converged critical exponents consistent with field-theoretic predictions}.
Parallel simulations of the spinless $t$-$V$ model belonging to $N=4$ GNI universality class validate our approach and demonstrate that scaling up to such large sizes reveals the characteristic finite-size behaviors essential for resolving fermionic quantum criticality.

\rev{The remainder of this paper is organized as follows.
In Sec.~\ref{sec:hubbard_model}, we introduce the honeycomb Hubbard model, the PQMC setup, and the correlation-ratio crossing-point analysis.
A sliding-window data-collapse analysis of conventional equal-time observables is then carried out in Sec.~\ref{sec:gnh_criticality} alongside the spinless $t$-$V$ benchmark, revealing persistent finite-size drift.
To overcome this bottleneck, Sec.~\ref{sec:bg_suppressed} introduces background-suppressed equal-time observables that establish plateau convergence in both models, together with finite-temperature spin-susceptibility cross-checks for the Hubbard model.
Finally, we summarize our critical-exponent estimates, reconcile prior QMC discrepancies, and discuss broader implications in Sec.~\ref{sec:discussion}, while leaving algorithmic details, the FSS workflow, and supporting analyses to Appendices~\ref{app:methods}--\ref{app:Gross-Neveu}.}

%%%%%%%%%%%%%%%%%%%%%%%%%%%%%%%%%%%%%%%%%%%%%%%%%%%%%
\section{Semimetal-Mott-insulator transition in the honeycomb Hubbard model}\label{sec:hubbard_model}
\begin{figure}[t]
    \centering
    \includegraphics[width=\columnwidth]{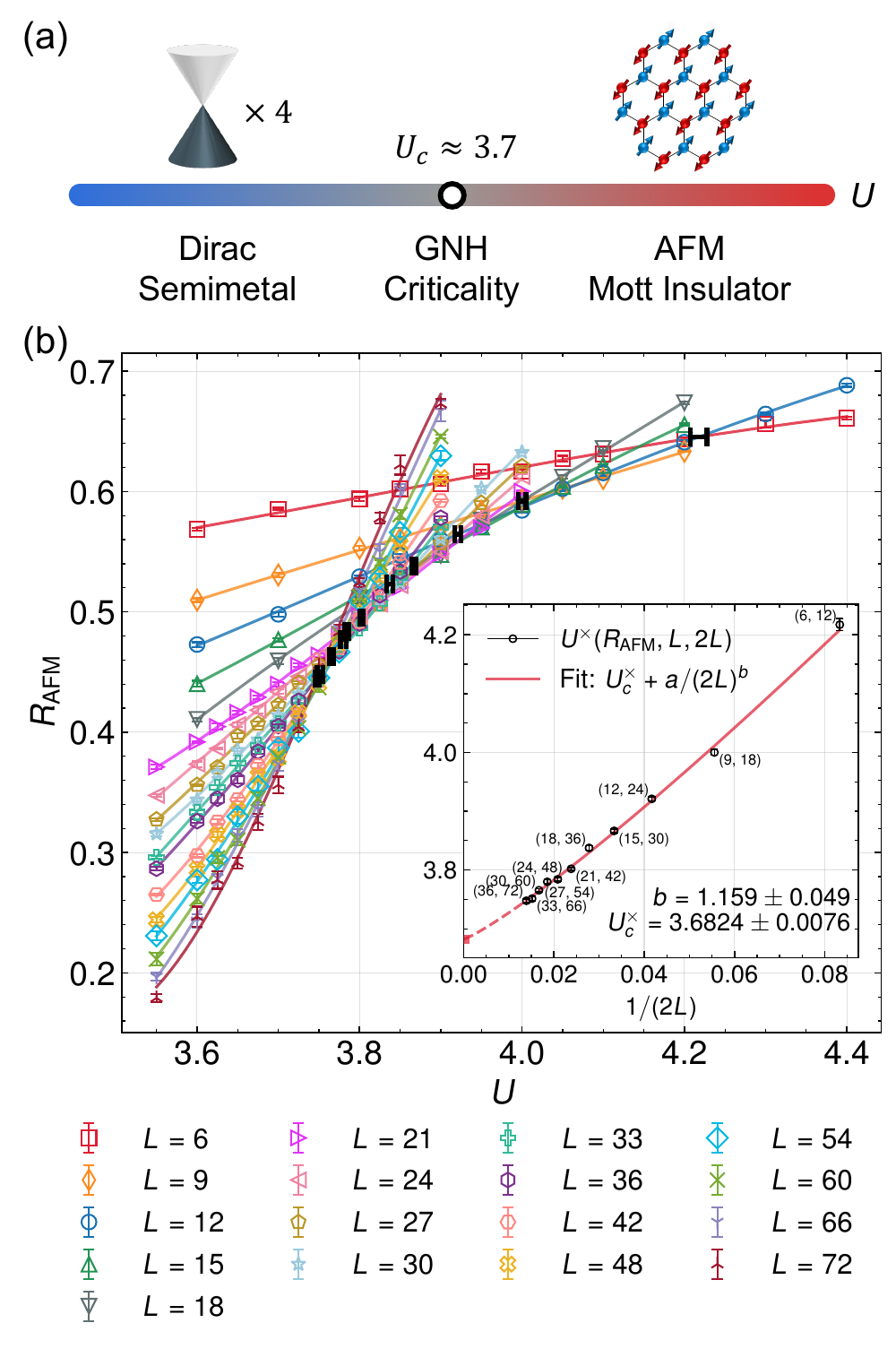}
    \caption{Phase diagram and crossing-point analysis of the correlation ratio.
    (a) Schematic phase diagram of the interaction-driven semimetal-Mott insulator transition with a Gross-Neveu-Heisenberg critical point.
    (b) Correlation ratio \rev{$R_{\mathrm{AFM}}$} as a function of $U$ for various system sizes $L$. Horizontal black error bars mark the crossing points $U^\times$ of size pairs $(L,2L)$, determined by \rev{cubic} interpolation of the data for each size. Inset: power-law extrapolation of the crossing points.}
    \label{fig:correlation_ratio_crossing}
\end{figure}

%%%%%%%%%%%%%%%%%%%%%%%%%%%%%%%%%%%%%%%%%%%%%%%%%%%%%
We study the half-filled Hubbard model on the honeycomb lattice.
The Hamiltonian is given by $\hat{H}=-t\sum_{\langle i, j\rangle,\sigma}(c_{i \sigma}^{\dagger}c_{j \sigma}+\mathrm{H.c.})+U \sum_{i}\left(\hat{n}_{i \uparrow}-\frac{1}{2}\right)\left(\hat{n}_{i \downarrow}-\frac{1}{2}\right)$, where $t$ is the nearest-neighbor hopping amplitude and $U$ is the Hubbard interaction strength.
We impose periodic boundary conditions on an $L\times L$ honeycomb lattice with two sublattices, yielding $N_{\mathrm{s}}=2L^2$ lattice sites.
At half filling, the low-energy excitations are spin-1/2 interacting Dirac fermions at the Dirac points $\mathrm{K}$ and $\mathrm{K'}$.
Increasing $U$ drives a continuous Mott transition belonging to the $N=8$ GNH universality class, from a semimetal to an antiferromagnetic (AFM) insulator with spontaneous spin-rotation symmetry breaking~\cite{Europhys.Lett.1992Sorella,Phys.Rev.Lett.2006Herbut,Nature2010Meng,SciRep2012Sorella,Phys.Rev.X2013Assaad,PRL2026YK.Yu,Nat.Commun.2025Zeng,Phys.Rev.B2021Ostmeyer,Phys.Rev.B2019Buividovich,Phys.Rev.X2016Otsuka,Phys.Rev.B2015ParisenToldin,Phys.Rev.B2025Lang}, as illustrated in Fig.~\ref{fig:correlation_ratio_crossing}(a).
We employ the sign-problem-free PQMC~\cite{Ann.Phys.1986Sugiyama,Europhys.Lett.1989Sorella,Phys.Rev.B1989White} to obtain the ground-state properties of this model, with the electron number fixed to $N_{\mathrm{s}}$ to enforce half filling.
The formulation, optimization, implementation, and parameter choices of the algorithm are provided in Appendices~\ref{app:methods}--\ref{app:convergence_test}.

To detect AFM long-range order on finite lattices, we define the correlation function in momentum space:
\begin{equation}\label{equ:C_k_def}
    \rev{C_{\mathrm{AFM}}(\mathbf{k})=\frac{1}{N_{\mathrm{s}}^{2}}\sum_{i,j}\epsilon_{i}\epsilon_{j}\,e^{-\mathrm{i}\mathbf{k}\cdot(\mathbf{r}_{i}-\mathbf{r}_{j})}\left\langle \mathbf{\hat{S}}_{i}\cdot\mathbf{\hat{S}}_{j}\right\rangle} ,
\end{equation}
\rev{where $i$ and $j$ run over all lattice sites, $\mathbf{r}_{i}$ is the position of site $i$, and $\epsilon_{i}=+1$ ($-1$) for $i$ on sublattice $A$ ($B$).}
\rev{The local spin operator is $\mathbf{\hat{S}}_{i}=\frac{1}{2}c_{i\mu}^{\dagger}\boldsymbol{\sigma}_{\mu\nu}c_{i\nu}$, where $\boldsymbol{\sigma}=(\sigma_x,\sigma_y,\sigma_z)$ and $\mu=\uparrow,\downarrow$.}
\rev{Notably, $C_{\mathrm{AFM}}(\mathbf{k})$ respects the full $C_{6v}$ point-group symmetry of the lattice.}
At $\mathbf{k}=0$, the correlation function provides a finite-size proxy of the order parameter,
\begin{equation}
    m_{\mathrm{AFM}}^{2}\equiv\frac{1}{N_{\mathrm{s}}^{2}}\left\langle \left(\sum_{i\in A}\mathbf{\hat{S}}_{i}-\sum_{i\in B}\mathbf{\hat{S}}_{i}\right)^{2}\right\rangle =\rev{C_{\mathrm{AFM}}(\mathbf{k}=0)}.
\end{equation}
A finite squared staggered magnetization $m^2_{\mathrm{AFM}}$ in the thermodynamic limit (TDL) indicates long-range AFM order, whereas it vanishes in the disordered phase.

Around the critical point, we employ the correlation ratio, an RG-invariant quantity constructed from the spin-spin correlation function, to sensitively probe the phase transition~\cite{Phys.Rev.B2015ParisenToldin,Phys.Rev.Lett.2022Liu,Phys.Rev.B2025Lang}:
\begin{equation}\label{equ:R_AFM_def}
\rev{R_{\mathrm{AFM}}=1-\frac{C_{\mathrm{AFM}}(\delta\mathbf{k}_{1})}{C_{\mathrm{AFM}}(0)}},
\end{equation}
\rev{where $\delta\mathbf{k}_{1}$ denotes a nearest-neighbor momentum around the $\Gamma$ point with $|\delta\mathbf{k}_{1}|=|\mathbf{b}_{1,2}|/L$ ($\mathbf{b}_1$ and $\mathbf{b}_2$ being the reciprocal lattice vectors), and $C_{\mathrm{AFM}}(\delta\mathbf{k}_{1})$ is averaged over the six $C_{6v}$-equivalent momenta in this first shell.}
In the TDL, $R_{\mathrm{AFM}}$ vanishes in the disordered phase and approaches unity in the ordered phase.
At the critical point, $R_{\mathrm{AFM}}$ converges to a universal constant $R_{\mathrm{AFM}}^*$, signaled by the crossing of $R_{\mathrm{AFM}}$-$U$ curves for various $L$.

\begin{figure*}[t]
    \centering
    \includegraphics[width=0.9\textwidth]{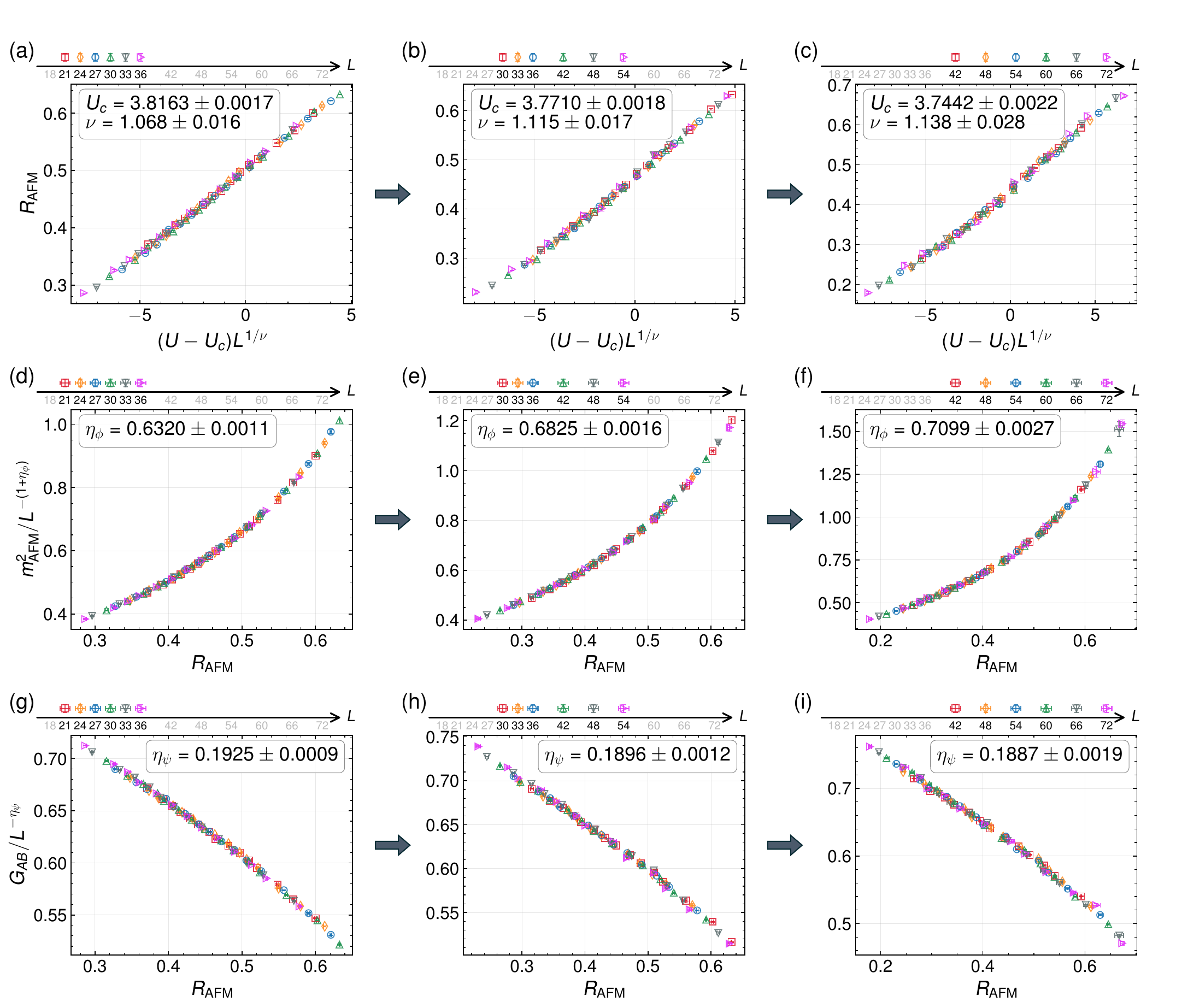}
    \caption{Sliding-window data-collapse analysis for the Hubbard model.
    Columns show three representative fit windows with $L_{\mathrm{max}}=36$, 54, and 72 (see Appendix~\ref{app:FSS_Hubbard} for all windows).
    The $L$-axis above each panel indicates the sizes included in the fit, with greyed-out sizes excluded.
    Rows: \rev{$R_{\mathrm{AFM}}$} collapse yielding $(U_c,\nu)$ (top), $m^2_{\mathrm{AFM}}$ collapse yielding $\eta_\phi$ (middle), and \rev{$G_{AB}$} collapse yielding $\eta_\psi$ (bottom).
    \rev{Each panel is drawn with all fitting parameters fixed at their bootstrap means.}}
    \label{fig:data_collapse_hubbard_main}
\end{figure*}

For finite systems, the crossing point between a pair of system sizes exhibits a drift due to scaling corrections~\cite{Phys.Rev.B2014Campostrini,Phys.Rev.B2015ParisenToldin}.
As shown in Fig.~\ref{fig:correlation_ratio_crossing}(b), only for $L\gtrsim 36$ do the $R_{\mathrm{AFM}}$-$U$ curves \rev{visually cluster near $U\approx 3.75$}, indicating strong finite-size effects in the honeycomb Hubbard model.
To quantify this drift, we denote the crossing point between sizes $L$ and $2L$ as $U^{\times}(L,2L)$ [horizontal bars in Fig.~\ref{fig:correlation_ratio_crossing}(b)] and fit $U^{\times}(L,2L)=U_c^{\times}+a(2L)^{-b}$.
The inset of Fig.~\ref{fig:correlation_ratio_crossing}(b) shows that this power-law extrapolation yields \rev{$b_{\mathrm{fit}}=1.159(49)$}.
\rev{For the correlation ratio, finite-size scaling predicts the asymptotic drift exponent $b=1/\nu+\min(\omega_{\mathrm{irr}},2-z-\eta_\phi)$~\cite{Phys.Rev.B2014Campostrini}, where $\omega_{\mathrm{irr}}$ is the leading irrelevant exponent and $2-z-\eta_\phi$ is the correction exponent associated with the analytic background.}
\rev{With $z=1$, the data-collapse estimates from the same observables in Sec.~\ref{sec:gnh_criticality} give $1-\eta_\phi\approx0.22$, below the one-loop $\epsilon$-expansion value $\omega_{\mathrm{irr}}\sim0.3$ for the $N=8$ GNH fixed point~\cite{Phys.Rev.B2023Ladovrechis}.}
\rev{Under this background-dominated assumption, we obtain $b_{\mathrm{collapse}}=1.101(34)$, consistent with $b_{\mathrm{fit}}$ within error bars.}
\rev{Because $b_{\mathrm{fit}}$ depends on the fitted size range~\cite{Science2016Shao} and crossing pairs~\cite{Phys.Rev.B2020Otsuka} while both analyses reflect effective scaling over the accessible sizes, this agreement serves as an internal consistency check on these observables.}

%%%%%%%%%%%%%%%%%%%%%%%%%%%%%%%%%%%%%%%%%%%%%%%%%%%%%
\section{\rev{Gross-Neveu criticality and persistent finite-size drift}}\label{sec:gnh_criticality}
%%%%%%%%%%%%%%%%%%%%%%%%%%%%%%%%%%%%%%%%%%%%%%%%%%%%%

\rev{Having located $U_c$ from pairwise crossings, we now extract the full set of critical parameters} via a systematic FSS analysis.
The critical point $U_c$ and exponent $\nu$ are obtained from the ansatz for the RG-invariant correlation ratio:
\begin{equation}\label{equ:fss_R_AF}
    \begin{aligned}
        \rev{R_{\mathrm{AFM}}(U,L)}&=f^{R}\left(uL^{1/\nu},L^{-\omega}\right)\\&\approx f_{0}^{R}\left(uL^{1/\nu}\right)+L^{-\omega}f_{1}^{R}\left(uL^{1/\nu}\right),
    \end{aligned}
\end{equation}
where $u=U-U_c$ is the distance from criticality, and $\omega$ denotes the leading correction-to-scaling exponent arising from irrelevant operators~\cite{Phys.Rev.B2023Ladovrechis} or non-universal analytic contributions~\cite{Phys.Rev.B2015ParisenToldin} (see also the preceding discussion).
The second line expands $f^{R}$ to first order in $L^{-\omega}$.
In \rev{the main-text} analysis, only $f_0^R$ is fitted while the correction term is neglected---a procedure termed \textit{data collapse}, as data at different $U$ and $L$ collapse onto a single curve when $U_c$ and $\nu$ are properly chosen.
\rev{Explicitly fitting the $L^{-\omega}$ term proves unstable because the data do not constrain $\omega$ and competing fit branches (multiple local minima) emerge, yielding broadly consistent thermodynamic-limit estimates but with much larger uncertainties (Appendix~\ref{app:correction_initialization}).}

The anomalous dimensions are extracted from the correlation functions.
For the boson anomalous dimension $\eta_\phi$, the squared staggered magnetization obeys the FSS ansatz
\begin{equation}\label{equ:fss_m2}
    \rev{m^2_{\mathrm{AFM}}\left(R_{\mathrm{AFM}},L\right)=L^{-1-\eta_{\phi}}f_{0}^{m}\left(R_{\mathrm{AFM}}\right).}
\end{equation}
Instead of fitting against $uL^{1/\nu}$, the scaling is analyzed with respect to \rev{$R_{\mathrm{AFM}}$}~\cite{Phys.Rev.B2015ParisenToldin,Phys.Rev.Lett.2022Liu,Phys.Rev.B2025Lang}, which bypasses prior knowledge of $U_c$ and $\nu$ and yields a more stable single-parameter fit.
The fermion anomalous dimension $\eta_\psi$ is extracted from the off-diagonal Green's function in momentum space, $G_{AB}(\mathbf{k})=\sum_{\mathbf{r}}e^{-\mathrm{i}\mathbf{k}\cdot\mathbf{r}}\langle c_{\mathbf{r}A\mu}^{\dagger}c_{0B\mu}\rangle$, where $\mu=\uparrow$ or $\downarrow$.
The averaged modulus $\left\langle \left|G_{AB}(\mathrm{K}+\delta\mathbf{k})\right|\right\rangle _{\delta\mathbf{k}}$ over the six momenta nearest to the Dirac point $\mathrm{K}$\rev{, denoted simply by $G_{AB}$ hereafter (see Appendix~\ref{app:GAB-free-fermion} for the modulus prescription),} follows the FSS ansatz~\cite{Phys.Rev.Lett.2019Lang,Phys.Rev.Lett.2022Tabatabaei,Phys.Rev.B2025Lang,PRL2026YK.Yu,Phys.Rev.X2016Otsuka}
\begin{equation}\label{equ:fss_G_AB}
    \rev{G_{AB}\left(R_{\mathrm{AFM}},L\right)=L^{-\eta_{\psi}}f_{0}^{G}\left(R_{\mathrm{AFM}}\right).}
\end{equation}

\begin{figure}[t]
    \centering
    \includegraphics[width=\columnwidth]{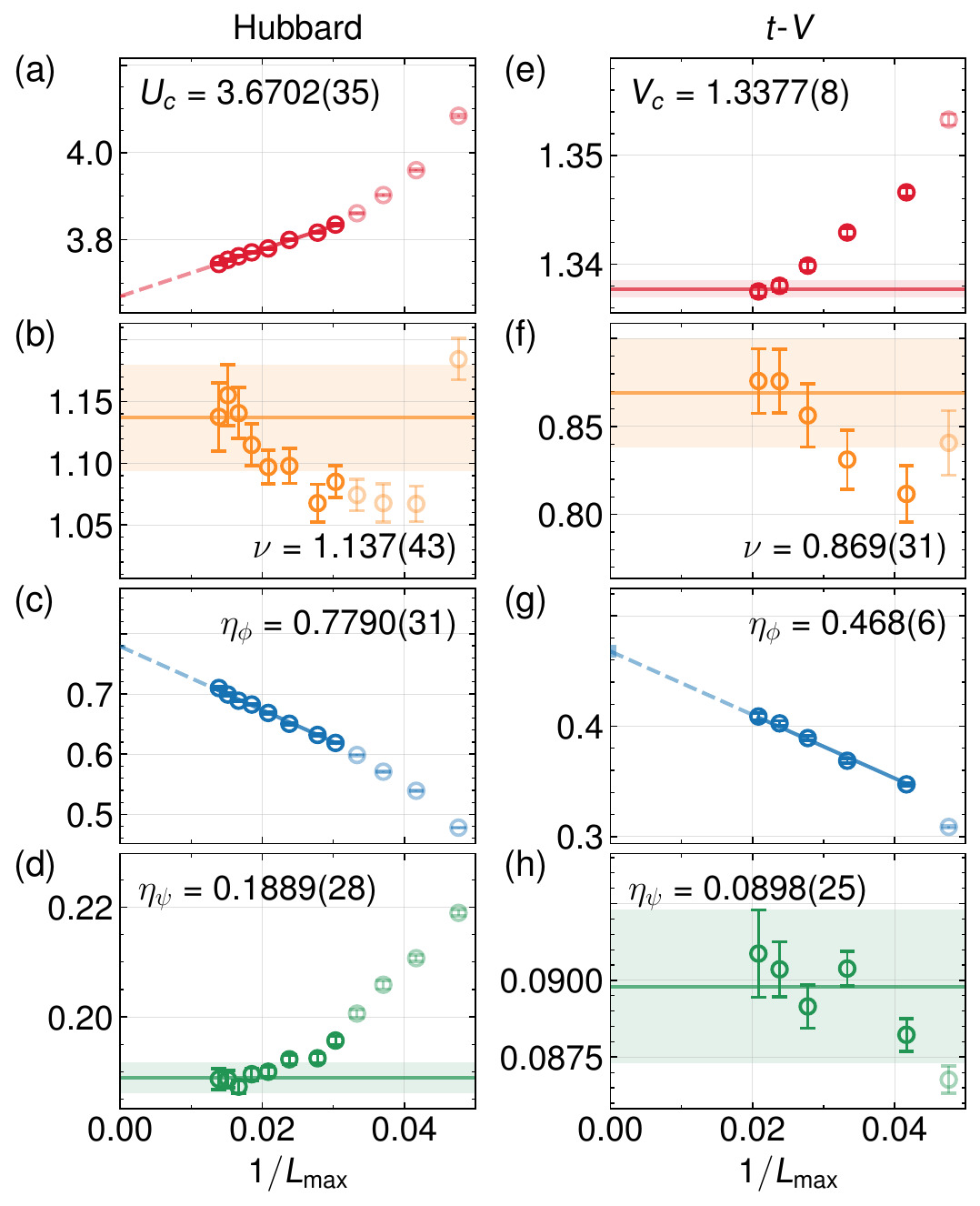}
    \caption{Fitting-window dependence of critical parameters for the Hubbard and spinless $t$-$V$ models.
    (a)--(d) Hubbard model; (e)--(h) $t$-$V$ model.
    Each point represents the fitted value of $U_c$ (or $V_c$), $\nu$, $\eta_\phi$, or $\eta_\psi$ obtained from a fitting window of six consecutive sizes, with $L_{\mathrm{max}}$ denoting the largest size in each window.
    \rev{Horizontal lines and shaded bands indicate envelope estimates and their uncertainties; solid lines show linear fits in $1/L_{\mathrm{max}}$, and dashed lines extend them to $1/L_{\mathrm{max}}\to 0$.}
    \rev{Faded symbols mark excluded windows ($L_{\min}<18$ for Hubbard, $L_{\min}<9$ for $t$-$V$).}}
    \label{fig:fit_extrapolation_bothmodels}
\end{figure}

\rev{When $L^{-\omega}$ is omitted in Eqs.~\eqref{equ:fss_R_AF}--\eqref{equ:fss_G_AB}, any remaining scaling corrections may be absorbed into size-dependent effective parameters, so a single fit cannot tell \emph{a priori} whether the TDL has been reached.}
\rev{Assessing convergence toward the TDL therefore requires both a sequence of estimates that advances to larger $L$ and enough sizes per fit to suppress large-$L$ statistical noise (see Table~\ref{tab:analysis_strategies} in Appendix~\ref{app:conventional_analysis} for a detailed comparison of strategies).}
\rev{Pairwise crossing-point analysis~\cite{Phys.Rev.Lett.2022Liu} advances $L$ directly via $(L,2L)$ pairs, but relying on two sizes per point amplifies large-$L$ noise.}
\rev{Conversely, in the conventional \textit{shrinking-window} strategy of discarding smaller sizes until $\chi^2_{\mathrm{red}}$ becomes acceptable~\cite{Phys.Rev.B2015ParisenToldin,Phys.Rev.B2020Otsuka,Phys.Rev.Lett.2022Liu,Phys.Rev.B2025Lang} (where Ref.~\cite{Phys.Rev.B2025Lang} noted an upward drift of $\eta_\phi$ toward the SLAC-fermion value), $L_{\max}$ is fixed and $L_{\min}$ is thereby bounded from above, so it serves only as a consistency check within the current $L_{\max}$ while narrowing the window leaves fewer sizes in the fit and increases sensitivity to the noisier large-$L$ data (Appendix~\ref{app:shrinking_window}).}

\rev{Our extended size range instead enables a \textit{sliding-window} analysis, stepping a fixed-width window $[L_{\min},L_{\max}]$ upward in $L$ to compare on an equal footing how the fitted parameters evolve as $L_{\max}$ grows toward the TDL.}
\rev{A two-dimensional scan over $(L_{\min},L_{\max})$ shows that overly narrow windows lose constraints and inflate uncertainties, whereas overly wide windows accumulate scaling corrections and raise $\chi^2_{\mathrm{red}}$~\cite{Phys.Rev.B2015ParisenToldin,Phys.Rev.Lett.2022Liu}; we therefore fix the width at six consecutive sizes, with widths five, seven, and eight yielding consistent results (Appendix~\ref{app:fitting_window_scan}).}
Within each window, we fit Eqs.~\eqref{equ:fss_R_AF}--\eqref{equ:fss_G_AB} following Appendix~\ref{app:data_collapse_analysis} and label the window by $L_{\mathrm{max}}$.
\rev{Figure~\ref{fig:data_collapse_hubbard_main} illustrates this progression for the Hubbard model: at $L_{\mathrm{max}}=36$ (left column), the fits closely match earlier $L\le 36$ PQMC studies~\cite{Phys.Rev.X2016Otsuka}, whereas at $L_{\mathrm{max}}=54$ and $72$, $\nu$ and $\eta_\psi$ level off while $U_c$ and $\eta_\phi$ continue to shift substantially, with $\eta_\phi$ rising from about $0.63$ to $0.71$.}
We also perform the same FSS analysis for the spinless $t$-$V$ model on the honeycomb lattice, whose critical point belongs to the $N=4$ GNI universality class, with system sizes up to $L=48$ (see Appendix~\ref{app:FSS_tV} for details).

\rev{To trace this size evolution toward the TDL}, we plot \rev{the fitted parameters} against $1/L_{\mathrm{max}}$ for both models in Fig.~\ref{fig:fit_extrapolation_bothmodels}.
\rev{Small-size windows whose $\chi^2_{\mathrm{red}}$ jumps well above that of larger windows ($L_{\min}<18$ for the Hubbard model and $L_{\min}<9$ for the $t$-$V$ model; Appendix~\ref{app:data_collapse_analysis}), where adding an $L^{-\omega}$ correction markedly improves the fit (Appendix~\ref{app:correction_initialization}), are shown faded and excluded when extracting the thermodynamic-limit estimates.}
Three distinct patterns emerge as $1/L_{\mathrm{max}}$ decreases: \rev{convergence to an apparent plateau} ($V_c$ in the $t$-$V$ model, \rev{$\nu$ in both models,} and $\eta_\psi$ in the Hubbard model), stable oscillations ($\eta_\psi$ in the $t$-$V$ model), and persistent monotonic drift ($U_c$ in the Hubbard model, $\eta_\phi$ in both models).
For the first two patterns, \rev{we take envelope averages over the stabilized windows}.
For the last kind of pattern, \rev{where the window estimates evolve monotonically toward the analytical predictions, we perform an empirical linear extrapolation to $1/L_{\mathrm{max}}\to 0$ (Appendices~\ref{app:data_collapse_analysis} and~\ref{app:covariance_extrapolation})}.
\rev{The resulting estimates, annotated in Fig.~\ref{fig:fit_extrapolation_bothmodels}, are consistent with $(L,2L)$ crossing-point extrapolations (Appendix~\ref{app:crossing_point}).}
\rev{In the $t$-$V$ model, $\nu$ and $\eta_\psi$ agree with conformal bootstrap predictions within $1.3\sigma$, whereas even the extrapolated $\eta_\phi=0.468(6)$ remains well below the bootstrap value $0.5155(30)$~\cite{J.HighEnerg.Phys.2023Erramilli}, pointing to an unresolved finite-size bias in these observables.}

%%%%%%%%%%%%%%%%%%%%%%%%%%%%%%%%%%%%%%%%%%%%%%%%%%%%%
\section{\rev{Background-suppressed equal-time observables and spin susceptibility}}\label{sec:bg_suppressed}
%%%%%%%%%%%%%%%%%%%%%%%%%%%%%%%%%%%%%%%%%%%%%%%%%%%%%
\begingroup
\revcolor

Even at $L=72$, the fitted $U_c$ and $\eta_\phi$ from the conventional equal-time observables $R_{\mathrm{AFM}}$ and $m^2_{\mathrm{AFM}}$ continue to drift monotonically, in contrast to $\nu$ and $\eta_\psi$, which level off at large sizes [Figs.~\ref{fig:fit_extrapolation_bothmodels}(a)--\ref{fig:fit_extrapolation_bothmodels}(d)]; $\eta_\phi$ in the $t$-$V$ model similarly drifts up to $L=48$ [Fig.~\ref{fig:fit_extrapolation_bothmodels}(g)].
This persistent drift stems from the large boson anomalous dimension $\eta_\phi$, which causes the critical spin correlations $\epsilon_{i}\epsilon_{j}\langle \mathbf{\hat{S}}_{i}\cdot\mathbf{\hat{S}}_{j}\rangle\sim |\mathbf{r}_{i}-\mathbf{r}_{j}|^{-(1+\eta_\phi)}$ to decay rapidly with distance and leaves a substantial analytic background $B(U)$ from nonuniversal short-range correlations in the structure factor $S_{\mathrm{AFM}}\equiv N_{\mathrm{s}}m^{2}_{\mathrm{AFM}}$~\cite{Phys.Rev.B2014Campostrini,Phys.Rev.B2015ParisenToldin,Phys.Rev.B2021Liu} (and analogously in $S_{\mathrm{CDW}}$ for the $t$-$V$ model).
At the critical point, $S_{\mathrm{AFM}}=L^{1-\eta_{\phi}}\left[A+B(U_c)L^{-\omega_{\eta}}\right]$, where the analytic background acts as a subleading correction with exponent $\omega_\eta\equiv 1-\eta_{\phi}$ (given $z=1$).
Since the relative suppression factor $L^{-\omega_\eta}=e^{-(1-\eta_\phi)\ln L}$ depends only on $(1-\eta_\phi)\ln L$, suppressing the background requires $(1-\eta_\phi)\ln L$ of order unity, so the crossover system size grows exponentially in $1/(1-\eta_\phi)$ as $\eta_\phi\to 1$.

Before reaching this scale, leading-order collapses yield only slowly drifting effective parameters.
Extrapolating that drift to $1/L_{\max}\to 0$ is uncontrolled because the window trajectory obeys no known asymptotic functional form, as reflected in the $t$-$V$ linear extrapolation undershooting the conformal bootstrap benchmark and in $\chi^2_{\mathrm{red}}>1$ for the Hubbard linear fits (Appendices~\ref{app:covariance_extrapolation} and~\ref{app:strategy_tradeoffs}).
Explicitly adding a subleading $L^{-\omega}$ term to the collapse ansatz does not cure the bias either, since near degeneracy with the leading term at small $\omega$ makes the fit ill-conditioned and leaves $\eta_\phi$ similarly underestimated (Appendix~\ref{app:correction_initialization}).

A more direct solution is to suppress or eliminate this short-range background at the level of the observables themselves.
Within equal-time measurements, we replace $m^2_{\mathrm{AFM}}$ by two background-suppressed order-parameter proxies: the real-space correlation function at maximum lattice separation~\cite{Phys.Rev.B1997Sandvik,JStatMech2005Laflorencie,Phys.Rev.B2009Varney,AIPConfProc2010Sandvik,SciRep2012Sorella,Phys.Rev.Lett.2013Pujari},
\begin{equation}\label{equ:C_rmax_def}
    C^{s}_{AA}(\mathbf{r}_{\max})\equiv\frac{1}{L^2}\sum_{i,j\in A}\delta_{\mathbf{r}_{i}-\mathbf{r}_{j},\,\mathbf{r}_{\max}}\left\langle\mathbf{\hat{S}}_{i}\cdot\mathbf{\hat{S}}_{j}\right\rangle,
\end{equation}
with $\mathbf{r}_{\max}=\lfloor L/2\rfloor(\mathbf{a}_{2}-\mathbf{a}_{1})$ ($\mathbf{a}_{1,2}$ being the primitive lattice vectors), which avoids short-range correlations altogether, and a momentum-subtracted order parameter,
\begin{equation}\label{equ:Delta_m2_def}
    \Delta m^2_{\mathrm{AFM}}\equiv C_{\mathrm{AFM}}(0)-C_{\mathrm{AFM}}(\delta\mathbf{k}_{1})=R_{\mathrm{AFM}}\,m^2_{\mathrm{AFM}}.
\end{equation}
Since inversion symmetry reduces $e^{-\mathrm{i}\delta\mathbf{k}_{1}\cdot(\mathbf{r}_{i}-\mathbf{r}_{j})}$ to $\cos[\delta\mathbf{k}_{1}\cdot(\mathbf{r}_{i}-\mathbf{r}_{j})]$ in Eq.~\eqref{equ:C_k_def}, subtracting $C_{\mathrm{AFM}}(\delta\mathbf{k}_{1})$ weights real-space correlations by $1-\cos[\delta\mathbf{k}_{1}\cdot(\mathbf{r}_{i}-\mathbf{r}_{j})]\propto(|\mathbf{r}_{i}-\mathbf{r}_{j}|/L)^2$ at $|\mathbf{r}_{i}-\mathbf{r}_{j}|\ll L$.
This $(|\mathbf{r}_{i}-\mathbf{r}_{j}|/L)^2$ weighting makes $\Delta m^2_{\mathrm{AFM}}$ the momentum-space counterpart to the numerator of the real-space correlation ratio in Ref.~\cite{Phys.Rev.B2015ParisenToldin}, smoothly filtering out short-range contributions without the sharp real-space cutoff of Ref.~\cite{Phys.Rev.B2009Varney}.
Meanwhile, in both $R_{\mathrm{AFM}}=\Delta m^2_{\mathrm{AFM}}/C_{\mathrm{AFM}}(0)$ and that real-space ratio~\cite{Phys.Rev.B2015ParisenToldin}, this short-range background remains in the denominator $C_{\mathrm{AFM}}(0)$.
To eliminate it from the denominator as well while avoiding the lattice-commensurability constraints of real-space ratios~\cite{Phys.Rev.B2002Tomita,Phys.Rev.B2015ParisenToldin}, we construct a background-subtracted correlation ratio from the first and second momentum shells $\delta\mathbf{k}_{1}$ and $\delta\mathbf{k}_{2}$ around $\Gamma$ [with $|\delta\mathbf{k}_{2}|=|\mathbf{b}_{1}+\mathbf{b}_{2}|/L$ and $C_{\mathrm{AFM}}(\delta\mathbf{k}_{2})$ likewise averaged over the six $C_{6v}$-equivalent momenta],
\begin{equation}\label{equ:R_Delta_def}
    R^{\Delta}_{\mathrm{AFM}}\equiv\frac{C_{\mathrm{AFM}}(0)-C_{\mathrm{AFM}}(\delta\mathbf{k}_{1})}{C_{\mathrm{AFM}}(0)-C_{\mathrm{AFM}}(\delta\mathbf{k}_{2})},
\end{equation}
which is RG-invariant and obeys Eq.~\eqref{equ:fss_R_AF}.
Paired with $R^{\Delta}_{\mathrm{AFM}}$, both $C^{s}_{AA}(\mathbf{r}_{\max})$ and $\Delta m^2_{\mathrm{AFM}}$ obey the same FSS form as $m^2_{\mathrm{AFM}}$ in Eq.~\eqref{equ:fss_m2}.

We now apply the same sliding-window analysis as in Sec.~\ref{sec:gnh_criticality} to these background-suppressed equal-time observables and their CDW counterparts in the $t$-$V$ model, with $R^{\Delta}$ replacing $R$ in Eqs.~\eqref{equ:fss_R_AF}--\eqref{equ:fss_G_AB} and $\Delta m^2$ or $C_{AA}(\mathbf{r}_{\max})$ replacing $m^2$.
In contrast to the conventional observables, the separate leading-form collapses remain acceptable down to the smallest windows ($\chi^2_{\mathrm{red}}\lesssim 6$), so no small-size windows need to be excluded; adding an explicit $L^{-\omega}$ correction leaves the resulting estimates unchanged within uncertainties (Appendix~\ref{app:correction_bgsup}).

Figure~\ref{fig:fit_extrapolation_bgsup} shows that suppressing the short-range background arrests the persistent drifts of Fig.~\ref{fig:fit_extrapolation_bothmodels}, allowing most critical parameters to converge onto plateaus at large sizes without empirical extrapolation.
Two quantities, $\eta_\psi$ in the Hubbard model [panel~(d)] and $V_c$ in the $t$-$V$ model [panel~(e)], already fluctuate around a flat baseline right from the smallest window and are thus averaged over the full range.
Five others retain a modest small-size drift before leveling off, at $L_{\max}\ge 54$ in the Hubbard model [$U_c$, $\nu$, and $\eta_\phi$, with $\Delta m^2_{\mathrm{AFM}}$ and $C^{s}_{AA}(\mathbf{r}_{\max})$ giving consistent $\eta_\phi$] and at $L_{\max}\ge 36$ in the $t$-$V$ model [$\nu$ and $\eta_\phi$].
In particular, the $t$-$V$ $\eta_\phi$ plateau at $0.512(19)$ resolves the conventional bias in Fig.~\ref{fig:fit_extrapolation_bothmodels}(g) and agrees directly with the conformal bootstrap benchmark $0.5155(30)$~\cite{J.HighEnerg.Phys.2023Erramilli}.
Only $t$-$V$ $\eta_\psi$ [panel~(h)] continues climbing across all accessible windows without a clear linear trend; we therefore record its largest-window ($L_{\max}=48$) value, $0.0833(18)$, as a finite-size frontier estimate that serves as a lower bound (Appendix~\ref{app:FSS_tV}).

\begin{figure}[t]
    \centering
    \includegraphics[width=\columnwidth]{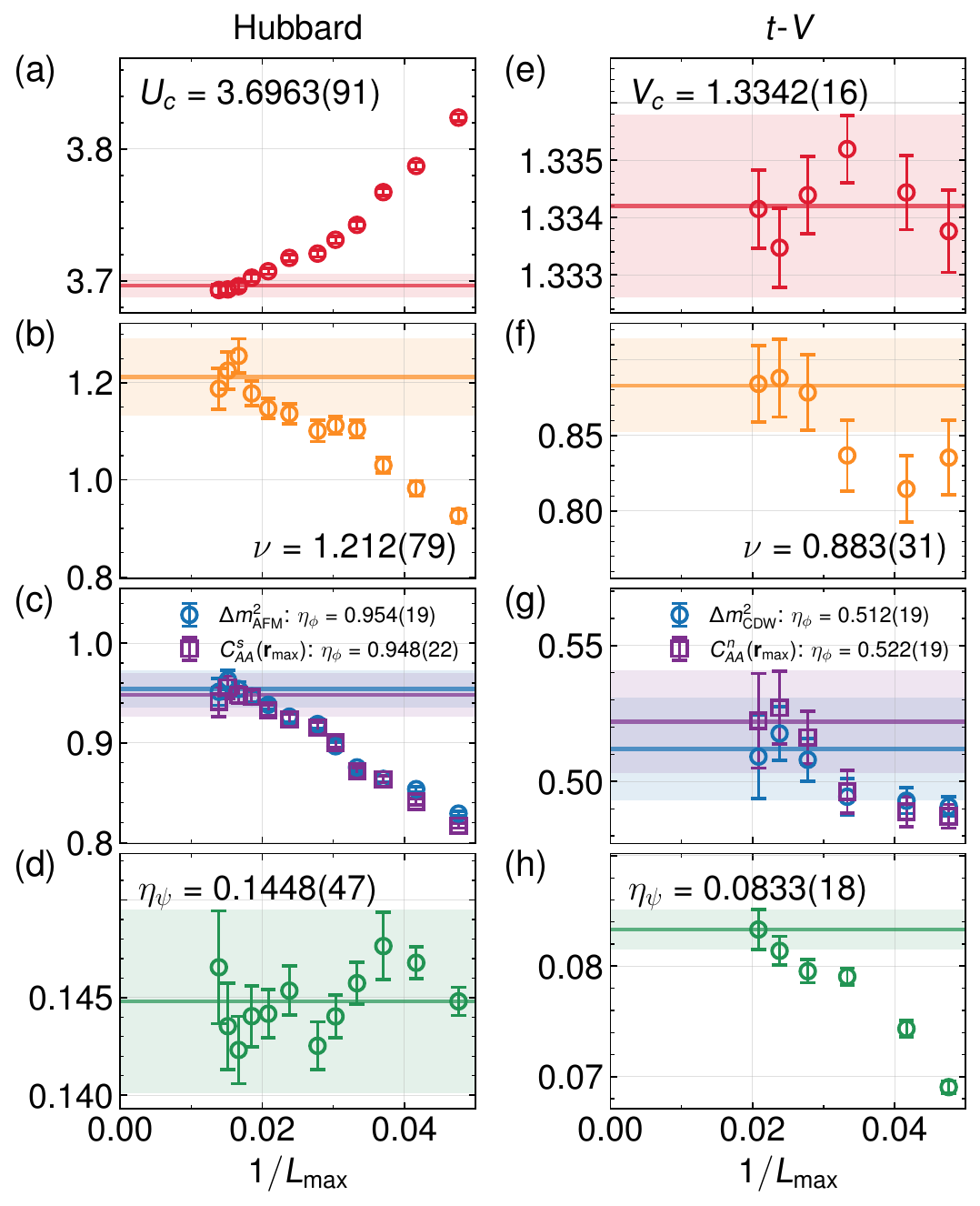}
    \caption{\rev{Fitting-window dependence of critical parameters from the background-suppressed equal-time observables, in the same layout as Fig.~\ref{fig:fit_extrapolation_bothmodels}.
    (a)--(d) Hubbard model; (e)--(h) $t$-$V$ model.
    $U_c$ (or $V_c$) and $\nu$ are obtained from the collapse of $R^{\Delta}$, $\eta_\phi$ from the collapses of $\Delta m^2$ and $C_{AA}(\mathbf{r}_{\max})$ against $R^{\Delta}$, and $\eta_\psi$ from the collapse of $G_{AB}$ against $R^{\Delta}$.
    Horizontal lines and shaded bands indicate plateau envelope estimates [or, in panel~(h), the largest-window estimate] and their uncertainties.}}
    \label{fig:fit_extrapolation_bgsup}
\end{figure}

\begin{figure*}[t]
    \centering
    \includegraphics[width=0.9\textwidth]{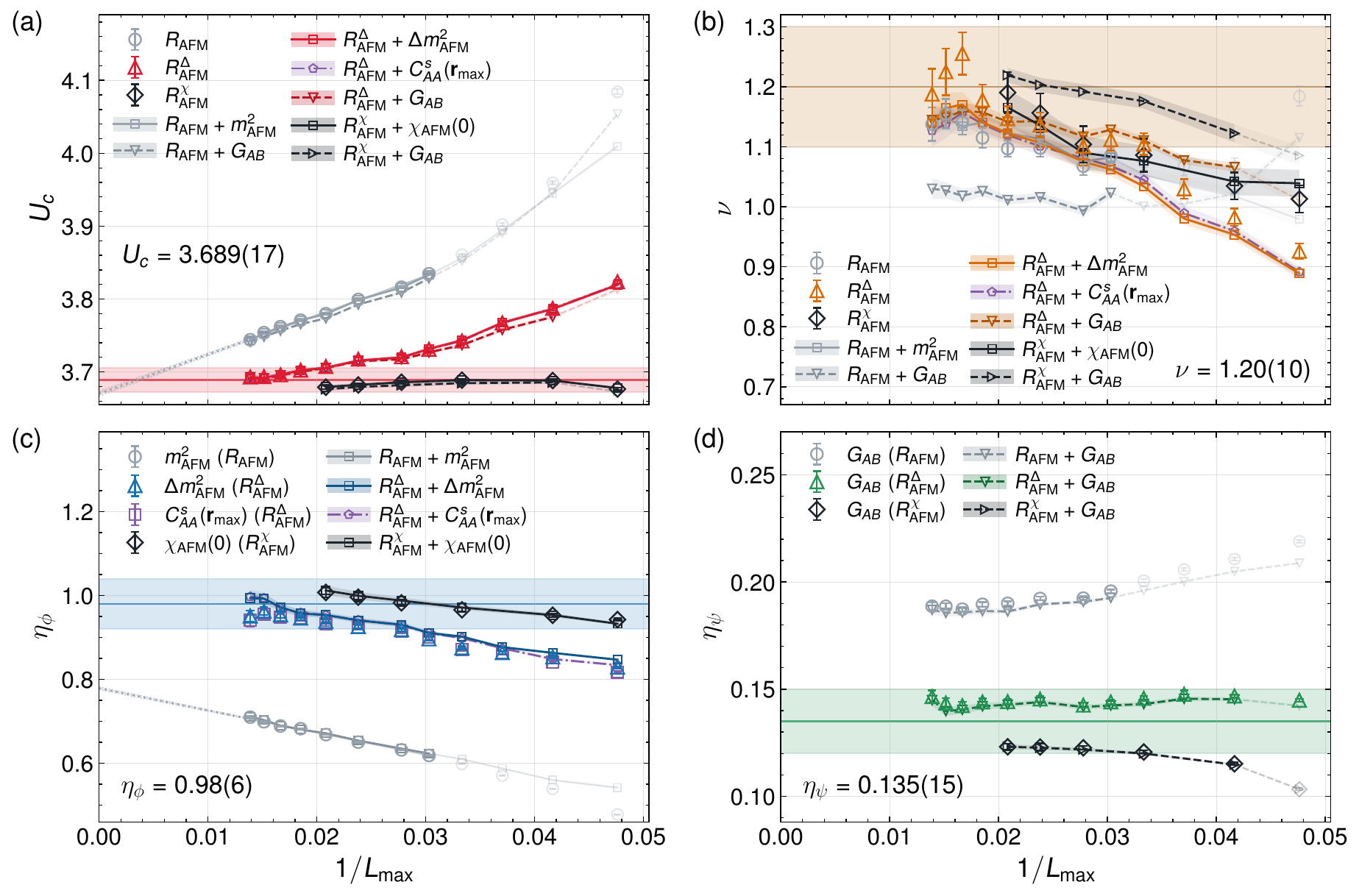}
    \caption{\rev{Critical quantities of the Hubbard model across observable families and data-collapse schemes.
    Width-six sliding-window estimates of (a) $U_c$, (b) $\nu$, (c) $\eta_\phi$, and (d) $\eta_\psi$ versus $1/L_{\mathrm{max}}$ are compared across the conventional equal-time observables, background-suppressed equal-time observables, and $\beta=L$ spin susceptibility ($L\le 48$).
    Unconnected symbols with error bars denote the separate collapses of Secs.~\ref{sec:gnh_criticality} and~\ref{sec:bg_suppressed} [Eqs.~\eqref{equ:fss_R_AF}--\eqref{equ:fss_G_AB}], whereas connected curves with shaded $1\sigma$ uncertainty bands denote joint collapses against $U$.
    Dotted lines with shaded $1\sigma$ fans show the linear extrapolations of the conventional separate-collapse $U_c$ and $\eta_\phi$ trajectories, faded symbols and curves mark excluded small-size windows (Appendix~\ref{app:data_collapse_analysis}), and horizontal lines and shaded bands indicate our combined final Hubbard estimates in Eq.~\eqref{equ:final_estimates}.}}
    \label{fig:fit_extrapolation_hubbard_comparison}
\end{figure*}

To see how background suppression alters the extracted Hubbard parameters, Fig.~\ref{fig:fit_extrapolation_hubbard_comparison} overlays the background-suppressed trajectories directly onto their conventional counterparts from Fig.~\ref{fig:fit_extrapolation_bothmodels}.
While the two families yield consistent $\nu$ at large sizes [panel~(b)], substantial shifts emerge in $U_c$, $\eta_\phi$, and $\eta_\psi$.
In panel~(a), removing the short-range background lowers the finite-size window estimates of $U_c$ by $\Delta U_c\approx 0.05$ across intermediate and large sizes while arresting the downward drift at $3.6963(91)$, slightly above the conventional linear intercept $3.6702(35)$ (dotted line).
In panel~(c), $\eta_\phi$ from both $\Delta m^2_{\mathrm{AFM}}$ and $C^{s}_{AA}(\mathbf{r}_{\max})$ starts above $0.8$ already at $L_{\max}=21$, exceeding even the conventional linear extrapolation $0.7790(31)$ (dotted line) before climbing to the $0.954(19)$ plateau.
Most notably, in panel~(d), although the fermionic observable $G_{AB}$ itself is unchanged, replacing $R_{\mathrm{AFM}}$ with $R^{\Delta}_{\mathrm{AFM}}$ in Eq.~\eqref{equ:fss_G_AB} eliminates the small-size drift and shifts the $\eta_\psi$ plateau downward from $0.1889(28)$ to $0.1448(47)$.

These shifts in $U_c$, $\eta_\phi$, and $\eta_\psi$ are mutually connected.
Although Eqs.~\eqref{equ:fss_m2} and~\eqref{equ:fss_G_AB} do not fit $U_c$ explicitly, collapsing against $R$ still probes scaling around the $R$-crossing coupling.
In the conventional analysis, the short-range background in $R_{\mathrm{AFM}}$ overestimates this finite-size transition scale by $\Delta U_c\approx 0.05$ [Fig.~\ref{fig:fit_extrapolation_hubbard_comparison}(a)], pushing it slightly into the Mott regime where emerging spin order flattens the size decay of $C^{s}_{AA}(\mathbf{r}_{\max})$ while vanishing quasiparticle weight steepens that of $G_{AB}$~\cite{Phys.Rev.B2019Seki}.
Specifically, between $U=3.70$ and $3.75$, the effective power-law exponents of their large-$L$ decay shift from about $0.97$ down to $0.7$ for $\eta_\phi$ and from about $0.15$ up to $0.19$ for $\eta_\psi$ (Figs.~\ref{fig:ss_decay_Hubbard} and~\ref{fig:gab_decay_Hubbard} in Appendix~\ref{app:FSS_Hubbard}).
This coupling offset compounds the direct background in $m^2_{\mathrm{AFM}}$ to depress $\eta_\phi$ [Fig.~\ref{fig:fit_extrapolation_hubbard_comparison}(c)] and alone accounts for the upward bias of $\eta_\psi$ in the $G_{AB}(R_{\mathrm{AFM}})$ collapse [Fig.~\ref{fig:fit_extrapolation_hubbard_comparison}(d)], whose apparent plateau merely reflects the slow $L^{-\omega_\eta}$ decay of the $R_{\mathrm{AFM}}$ background.
An analogous fixed-$V$ analysis for the $t$-$V$ model likewise confirms the mutual consistency of $V_c$, $\eta_\phi$, and $\eta_\psi$ (Figs.~\ref{fig:decay_tV} and~\ref{fig:gab_decay_tV} in Appendix~\ref{app:FSS_tV}).

As a cross-check, we replace $O(R,L)=L^{-y_O}f_0^{O}(R)$ with $O(U,L)=L^{-y_O}\tilde{f}_0^{O}(uL^{1/\nu})$ to fit each observable directly against $U$.
In a standalone three-parameter fit of $O(U,L)$, shifts in $U_c$ and in $y_O$ can compensate each other while maintaining good collapse quality over a wide range, leaving the parameters poorly constrained.
We therefore perform a \textit{joint collapse} of each observable with its corresponding correlation ratio [Eq.~\eqref{equ:fss_R_AF}] sharing $(U_c,\nu)$.
As shown by the shaded bands in Fig.~\ref{fig:fit_extrapolation_hubbard_comparison}, the joint and separate collapses agree closely within each observable family; for the background-suppressed observables, jointly constraining $(U_c,\nu)$ also smooths out the large-$L$ fluctuations of $R^{\Delta}_{\mathrm{AFM}}$ and nearly halves the $\nu$ uncertainty [panel~(b)].
The sole exception is the conventional $R_{\mathrm{AFM}}+G_{AB}$ fit, where $U_c$ tracks $R_{\mathrm{AFM}}$ and $\eta_\psi$ stays near $0.19$, whereas $\nu$ drops to $1.021(27)$ [panel~(b)], departing from all other fits.
Forcing the background-free $G_{AB}$ to share the overestimated $U_c$ of $R_{\mathrm{AFM}}$ distorts $\nu$, a mismatch that disappears once $R_{\mathrm{AFM}}$ is replaced by $R^{\Delta}_{\mathrm{AFM}}$ and independently rules out $\eta_\psi\approx 0.19$.
Joint collapses of the $t$-$V$ model likewise agree closely with the separate-collapse values across both equal-time families (Figs.~\ref{fig:fit_extrapolation_tv_comparison} and~\ref{fig:tv_joint_collapse}, Table~\ref{tab:fss_summary}).

Finally, as an independent consistency check beyond equal-time observables, we perform finite-temperature DQMC simulations of the Hubbard model at $\beta=L$ ($L\le 48$) with Trotter step $\Delta_\tau t=0.05$ (Appendix~\ref{app:convergence_test}) to measure the zero-frequency spin susceptibility~\cite{Phys.Rev.B2015ParisenToldin,Phys.Rev.B2021Liu},
\begin{equation}\label{equ:chi_AFM_def}
    \chi_{\mathrm{AFM}}(\mathbf{k})=\frac{1}{N_{\mathrm{s}}}\int_{0}^{\beta}\mathrm{d}\tau\sum_{i,j}\epsilon_{i}\epsilon_{j}\,e^{-\mathrm{i}\mathbf{k}\cdot(\mathbf{r}_{i}-\mathbf{r}_{j})}\left\langle\mathbf{\hat{S}}_{i}(\tau)\cdot\mathbf{\hat{S}}_{j}(0)\right\rangle,
\end{equation}
and the associated correlation ratio $R^{\chi}_{\mathrm{AFM}}=1-\chi_{\mathrm{AFM}}(\delta\mathbf{k}_{1})/\chi_{\mathrm{AFM}}(0)$, where $\chi_{\mathrm{AFM}}(\delta\mathbf{k}_{1})$ is likewise averaged over the six first-shell momenta.
Integrating over $\tau\in[0,\beta]$ enhances the singular part by $\beta=L$ relative to $S_{\mathrm{AFM}}$, so $L^{-3}\chi_{\mathrm{AFM}}(0)\sim L^{-(1+\eta_\phi)}$ obeys Eq.~\eqref{equ:fss_m2} with the analytic-background exponent increased from $1-\eta_\phi$ to $2-\eta_\phi$.
Because direct DQMC measurements of $\chi_{\mathrm{AFM}}$ suffer from severe fat-tailed distributions~\cite{Phys.Rev.E2016Shi,Phys.Rev.B2021Liu,Phys.Rev.E2022Ulybyshev}, we adopt a newly developed technique~\cite{ArXiv2611J.Vaeth} to measure $\chi_{\mathrm{AFM}}(\mathbf{k})$ and $R^{\chi}_{\mathrm{AFM}}$ robustly, while measuring $G_{AB}$ in additional standard DQMC runs at the same $\beta$ and $\Delta_\tau$.
Analyzed with both separate and joint collapses in Fig.~\ref{fig:fit_extrapolation_hubbard_comparison}(a), these data yield a critical coupling that varies by only about $0.01$ across all windows, far less than the conventional and background-suppressed drifts.
The separate collapse of $R^{\chi}_{\mathrm{AFM}}$ gives the envelope $U_c=3.6834(87)$, the joint collapses agree with it within $0.002$, and all are consistent with the $R^{\Delta}_{\mathrm{AFM}}$ plateau.
In panels~(b)--(d), both sets of trajectories show $\nu$ rising into the equal-time plateau band and $\eta_\phi\approx 0.93$--$1.01$ lying near unity across all windows, independently corroborating the equal-time results, while $\eta_\psi$ levels off at $0.1229(18)$ for $L_{\max}\ge 36$ (see also Appendix~\ref{app:crossing_bgsup}).
Relative to the background-suppressed equal-time estimates, the slightly larger $\eta_\phi$ and smaller $\eta_\psi$ are consistent with the coupling sensitivity above and likely stem from the lower finite-size $U_c$ of $R^{\chi}_{\mathrm{AFM}}$ compared with $R^{\Delta}_{\mathrm{AFM}}$, making this $\eta_\psi$ plateau a conservative lower bound.
Because $\nu$ and $\eta_\phi$ still climb over the accessible sizes, larger lattices will be needed to extract their thermodynamic-limit values and to test whether these differences close.
Illustrative linear extrapolations of the separate and joint trajectories over all windows give $\nu\approx 1.23$--$1.30$ and $\eta_\phi\approx 1.04$--$1.07$.

\endgroup

%%%%%%%%%%%%%%%%%%%%%%%%%%%%%%%%%%%%%%%%%%%%%%%%%%%%%
\section{Discussion}\label{sec:discussion}
%%%%%%%%%%%%%%%%%%%%%%%%%%%%%%%%%%%%%%%%%%%%%%%%%%%%%
\rev{Bringing together the observables and collapse schemes introduced above, we consolidate our results into combined final estimates for both models [shown as horizontal bands in Fig.~\ref{fig:fit_extrapolation_hubbard_comparison} for the Hubbard model, with the parallel comparison for the $t$-$V$ model given in Fig.~\ref{fig:fit_extrapolation_tv_comparison} of Appendix~\ref{app:FSS_tV}],
\begin{equation}\label{equ:final_estimates}
    \begin{aligned}
        \text{Hubbard}\quad & U_c = 3.689(17), && \nu = 1.20(10), \\
                            & \eta_\phi = 0.98(6), && \eta_\psi = 0.135(15); \\[0.3em]
        t\text{-}V\quad & V_c = 1.3343(19), && \nu = 0.88(6), \\
                        & \eta_\phi = 0.520(31), && \eta_\psi = 0.087(6).
    \end{aligned}
\end{equation}
Each interval is taken as the conservatively rounded union of the $1\sigma$ ranges of the width-six background-suppressed estimates (separate and joint collapses), supplemented by the $\beta=L$ spin-susceptibility $U_c$ and $\eta_\psi$ plateaus and largest-window $\nu$ and $\eta_\phi$ values in the Hubbard model and by the conventional $G_{AB}$ plateaus for $\eta_\psi$ in the $t$-$V$ model.
All values remain consistent within uncertainties with the corresponding alternative window widths and cross-checks in Appendices~\ref{app:fitting_window_scan}--\ref{app:fss_summary}.}

Placed in the broader context of prior studies, our results are compared with representative analytical calculations and large-scale QMC simulations in Fig.~\ref{fig:critical_exponents_literature}.
For the $N=4$ GNI transition realized in the spinless $t$-$V$ model, our $\nu$ and $\eta_\phi$ agree with both conformal bootstrap (CB) predictions~\cite{J.HighEnerg.Phys.2023Erramilli} and fermion-bag QMC simulations~\cite{Phys.Rev.D2020Huffman}.
Most notably, our estimate of $\eta_\psi$ in this model presents the first QMC determination that is consistent with the CB prediction within uncertainties.
For the $N=8$ GNH transition in the honeycomb Hubbard model, the wide confidence interval of our $\nu$ encompasses most prior QMC and analytical values.
\rev{With the short-range background removed, our $\eta_\phi$ rises above earlier QMC estimates into direct agreement with four-loop $\epsilon$-expansion and FRG predictions~\cite{PRB2014L.Janssen,Phys.Rev.D2017Zerf,Phys.Rev.B2018Knorr,Phys.Rev.B2023Ladovrechis} and is also consistent with recent finite-temperature DQMC results for the $\pi$-flux Hubbard model in the same GNH universality class~\cite{ArXiv2611J.Vaeth}.
Meanwhile, our $\eta_\psi$ shifts downward from most earlier QMC values of about $0.2$--$0.3$~\cite{Phys.Rev.X2016Otsuka,Phys.Rev.B2020Otsuka,Phys.Rev.B2021Liu} to align with four-loop $\epsilon$-expansion results~\cite{Phys.Rev.D2017Zerf,Phys.Rev.B2023Ladovrechis}, although spread among analytical approaches themselves remains.}

\rev{The sliding-window trajectories in Fig.~\ref{fig:fit_extrapolation_hubbard_comparison} thus connect earlier QMC studies directly to our final estimates within a single dataset.
Restricted to $L_{\max}=36$, the conventional observables reproduce prior honeycomb Hubbard PQMC results~\cite{Phys.Rev.X2016Otsuka,Phys.Rev.B2025Lang} (Fig.~\ref{fig:data_collapse_hubbard_main}, left column).
Advancing the window to $L_{\max}=72$ drives $\eta_\phi$ upward toward field-theoretic predictions, yet the extrapolated $\eta_\phi$ and the apparent $\eta_\psi$ plateau near $0.19$ still miss them because of the slowly decaying $L^{-(1-\eta_\phi)}$ background and the associated $\Delta U_c\approx 0.05$ shift.
Suppressing this background in the same QMC data closes the remaining gap and reconciles the honeycomb Hubbard exponents with field theory.}

In summary, \rev{by developing a projected submatrix update algorithm, we have pushed ground-state PQMC simulations of the honeycomb Hubbard model to $10{,}368$ sites ($L=72$), alongside benchmark simulations of the spinless $t$-$V$ model.
Tracking the finite-size evolution across sliding windows shows that, for conventional equal-time structure-factor observables, even $L=72$ remains far from the asymptotic regime because of the slowly decaying $L^{-(1-\eta_\phi)}$ short-range background at large $\eta_\phi$.
Suppressing this background, whether within equal-time measurements or via the imaginary-time-integrated spin susceptibility, removes this dominant bias.
Crucially, reaching these sizes allows the background-suppressed equal-time estimates to settle onto explicit plateaus ($L_{\max}\ge 54$) that bound the critical exponents directly, without extrapolating drifting small-size windows of unknown functional form.}

Looking ahead, \rev{pairing background-suppressed observables with sliding-window plateau diagnostics offers a practical route for other quantum phase transitions with severe finite-size effects, especially when only equal-time data are available or dynamical measurements are noisy.}
With both the computational power to reach larger sizes and \rev{this systematic scaling workflow}, the path toward resolving other outstanding problems in strongly correlated quantum systems---particularly those involving competing orders, emergent gauge structures, or unconventional quantum criticality---is now significantly clearer.

Further progress in the GNH criticality will also benefit from independent benchmarks from experiments and from CB.
In particular, measurements of critical exponents in artificial graphene realized by moir\'e materials~\cite{Nat.Mater.2025L.Ma} may become possible in the future.
On the theory side, extending CB constraints available for the $O(N)$ Gross-Neveu-Yukawa model~\cite{J.HighEnerg.Phys.2023Erramilli} to the GNH fixed point would offer a complementary nonperturbative reference.

%%%%%%%%%%%%%%%%%%%%%%%%%%%%%%%%%%%%%%%%%%%%%%%%%%%%%
% \section{Acknowledgments}
%%%%%%%%%%%%%%%%%%%%%%%%%%%%%%%%%%%%%%%%%%%%%%%%%%%%%
\begin{acknowledgments}
    We thank J. A. Gracey, T. C. Lang, M. Scherer, L. Thremer, and F. Parisen Toldin for helpful discussions.
    This work was supported by the National Key R\&D Program of China (Grant No. 2022YFA1402702, No. 2021YFA1401400), the National Natural Science Foundation of China (Grants No. 12447103, No. 12674187 and No. 125B2077), Yangyang Development Fund, and Shanghai Jiao Tong University 2030 Initiative.
    \mbox{F.-H.W.} is supported by T.D. Lee scholarship.
    F.F.A. and J.V. acknowledge financial support by the Deutsche Forschungsgemeinschaft (DFG, German Research Foundation) through the W\"urzburg-Dresden Cluster of Excellence ctd.qmat -- Complexity, Topology and Dynamics in Quantum Matter (EXC 2147, Project No. 390858490).
    The computations in this paper were partially run on the Siyuan-1 cluster supported by the Center for High Performance Computing at Shanghai Jiao Tong University.
\end{acknowledgments}

\appendix

%%%%%%%%%%%%%%%%%%%%%%%%%%%%%%%%%%%%%%%%%%%%%%%%%%%%%
\section{Methods summary}\label{app:methods}
%%%%%%%%%%%%%%%%%%%%%%%%%%%%%%%%%%%%%%%%%%%%%%%%%%%%%
Here we outline the key computational details and parameter choices underlying our results.                           
A more comprehensive account of the PQMC algorithm and the update techniques is provided in the following Appendices.

%%%%%%%%%%%%%%%%%%%%%%%%%%%%%%%%%%%%%%%%%%%%%%%%%%%%%
\subsection{Projector determinant quantum Monte Carlo}
%%%%%%%%%%%%%%%%%%%%%%%%%%%%%%%%%%%%%%%%%%%%%%%%%%%%%
The projector determinant quantum Monte Carlo (PQMC) algorithm evaluates ground-state properties by imaginary-time projection of a trial Slater determinant $\left|\Psi_{T}\right\rangle$~\cite{Ann.Phys.1986Sugiyama,Europhys.Lett.1989Sorella,Phys.Rev.B1989White}:
\begin{equation}
    \langle{\hat{O}}\rangle=\frac{\left\langle \Psi_{T}\left|e^{-\Theta\hat{H}}\hat{O}e^{-\Theta\hat{H}}\right|\Psi_{T}\right\rangle }{\left\langle \Psi_{T}\left|e^{-2\Theta\hat{H}}\right|\Psi_{T}\right\rangle }.
\end{equation}
We use the symmetric Trotter decomposition to discretize the projection length $\beta\equiv2\Theta=L_{\tau}\Delta_{\tau}$: $e^{-\Delta_{\tau}\hat{H}}\approx e^{-\Delta_{\tau}\hat{H}_{0}/2}e^{-\Delta_{\tau}\hat{H}_{I}}e^{-\Delta_{\tau}\hat{H}_{0}/2}$.
The trial wave function $|\Psi_T\rangle$ is chosen as the ground state of the non-interacting Hamiltonian $\hat{H}_0$, with small random hoppings added on each nearest-neighbor bond to lift energy level degeneracies at Dirac points.
The interaction term $\hat{H}_{I}$ is decoupled via the Hubbard-Stratonovich (HS) transformation.
For the repulsive Hubbard model, we employ the discrete HS transformation in the density channel~\cite{Phys.Rev.B1983Hirsch}:
\begin{equation}
e^{-\Delta_{\tau}U\left(\hat{n}_{i\uparrow}-\frac{1}{2}\right)\left(\hat{n}_{i\downarrow}-\frac{1}{2}\right)}={\gamma}\sum_{s_i=\pm1}e^{\mathrm{i}{\alpha}s_i\left(\hat{n}_{i\uparrow}+\hat{n}_{i\downarrow}-1\right)},
\end{equation}
where ${\gamma} = \frac{1}{2}e^{\Delta_\tau U/4}$ and $\cos{\alpha} = e^{-\Delta_\tau U/2}$.
For the spinless $t$-$V$ model, we decouple the nearest-neighbor repulsion to the hopping channel~\cite{Phys.Rev.B2015Li,Phys.Rev.Lett.2015Wang}:
\begin{equation}\label{equ:S2_HStV}
e^{-\Delta_\tau V\left(\hat{n}_{j}-\frac{1}{2}\right)\left(\hat{n}_{k}-\frac{1}{2}\right)}=\zeta\sum_{s_{jk}=\pm1 }e^{\lambda s_{jk}\left(c_{j}^{\dagger}c_{k}+c_{k}^{\dagger}c_{j}\right)},
\end{equation}
where $\zeta=\frac{1}{2}e^{-\frac{V\Delta_\tau}{4}}$ and $\cosh \lambda=e^{\frac{V \Delta_\tau}{2}}$.
In the Hubbard model, the auxiliary fields are defined on lattice sites, whereas in the $t$-$V$ model, they reside in nearest-neighbor bonds. We denote by $\mathbf{s}$ the complete set of auxiliary fields across all spatial locations and imaginary-time slices.

Systematic errors in PQMC arise primarily from the Trotter discretization $\Delta_\tau$ and the finite projection length $\beta$.
We control these errors by performing convergence tests for both models (see Appendix~\ref{app:convergence_test} for details).
For the Hubbard model, we adopt $\Delta_\tau t = 0.1$~\cite{Phys.Rev.B2015ParisenToldin,Phys.Rev.X2016Otsuka,Phys.Rev.B2025Lang} \rev{(with $\Delta_\tau t = 0.05$ in both finite-temperature $\beta=L$ DQMC cross-checks of Sec.~\ref{sec:bg_suppressed})}, whereas for the $t$-$V$ model, a smaller time step $\Delta_\tau t = 0.05$ is required~\cite{Nat.Commun.2025Zeng}.
The projection length is scaled with the linear system size as $\beta t = L + 12$ for both models, which is sufficient to render systematic errors smaller than statistical fluctuations.

%%%%%%%%%%%%%%%%%%%%%%%%%%%%%%%%%%%%%%%%%%%%%%%%%%%%%
\subsection{Submatrix update algorithm for PQMC}\label{app:submatrix_T_summary}
%%%%%%%%%%%%%%%%%%%%%%%%%%%%%%%%%%%%%%%%%%%%%%%%%%%%%
We first briefly review the traditional fast update algorithm for $T\equiv (LR)^{-1}$. More details can be found in Appendix~\ref{app:PQMC}.
After tracing out the fermionic degrees of freedom, the Monte Carlo weight for configuration $\mathbf{s}$ is proportional to a fermionic determinant $\det[L(\tau)R(\tau)]$, where $R(\tau)$ (size $N \times N_p$) and $L(\tau)$ (size $N_p \times N$) are rectangular matrices encoding the imaginary-time propagation from the trial state to time slice $\tau$ from $0$ and $\beta$, respectively.
Here $N$ typically denotes the total number of lattice sites and $N_p$ is the number of particles per spin block.

In PQMC, the most computationally intensive part is updating $T$ upon changes in auxiliary fields~\cite{Phys.Rev.B2024Sun}.
A local update of the auxiliary field at specific spatial position $x$ and time slice $\tau$ modifies $R$ by a rank-$k$ correction, $R'=(I+\Delta)R$, and the Metropolis acceptance ratio involves the determinant ratio $r = \det[LR']/\det[LR]$.
The Sherman-Morrison-Woodbury formula allows for a fast update of $T$:
\begin{equation}\label{equ:fast_update_T}
T' = T - T U (I_k + V U)^{-1} V,
\end{equation}
where $U = L \Delta P_{N\times k}$ and $V=P_{k\times N}R T$ are small rectangular matrices.
Here, $\Delta$ is a sparse diagonal matrix with only $k$ non-zero entries at the specific lattice sites being updated ($x=\{x_1,\ldots,x_k\}$), and we introduce the \textit{index matrices}
\begin{equation}
P_{N\times k}=\left[e_{x_{1}}| e_{x_{2}}|\cdots| e_{x_{k}}\right]\text{ and }P_{k\times N}=P_{N\times k}^{T},
\end{equation}
where $e_{x_j}=[0,\ldots,0,1,0,\ldots,0]^T$ is the $x_j$-th unit vector of length $N$ (several properties of the index matrices are discussed in Appendix~\ref{app:PQMC}).
The acceptance ratio is given by $r=\det[I_k+V U]$.
The per-move cost is $\mathcal{O}(N_{p}^{2})$ and the per-time-slice cost is $\mathcal{O}(N_{p}^{2} N)$.

While this fast update formula enables efficient per-move computation, it relies heavily on rank-$k$ matrix-vector operations (BLAS-1\&2) that are not cache-friendly.
To overcome this bottleneck, \textit{delayed update} algorithms, accumulating multiple local updates before performing a single, cache-efficient, higher-rank batch operation (BLAS-3), were proposed~\cite{Phys.Rev.B2024Sun,SciPostPhys.2025Sun,Phys.Rev.B2025Du}.
In Appendix~\ref{app:delayed_update_algorithms}, we provide a self-contained review of delay-G~\cite{Phys.Rev.B2024Sun}, submatrix-G~\cite{SciPostPhys.2025Sun}, and delay-T~\cite{Phys.Rev.B2025Du} algorithms.

\begin{figure}[tbp]
    \centering
    \includegraphics[width=0.48\textwidth]{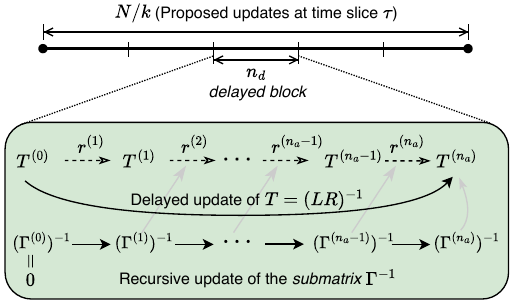}
    \caption{Schematic illustration of the submatrix-T update algorithm.
    At a specific time slice, there are $N/k$ proposed local updates in total, which is divided into several \textit{delayed blocks} of size $n_d$.
    Within each delayed block, we do not update $T$ (indicated by dashed arrows) but instead perform a recursive update of the \textit{submatrix} $\Gamma^{-1}$ (indicated by solid arrows in black), which helps in calculating not only the intermediate acceptance ratios $r^{(i)}$ but also the final $T^{(n_a)}$ (indicated by solid arrows in grey), where $n_a$ denotes the total accepted moves in the current delayed block.}
    \label{fig:submatrix_LR_scheme}
\end{figure}

In this work, we introduce a new ``submatrix-T'' update algorithm which outperforms all existing delayed update algorithms for PQMC simulations.
As illustrated in Fig.~\ref{fig:submatrix_LR_scheme}, a series of accepted local updates are accumulated within a block (dubbed ``delayed block'') before performing a full update.
The key formulas are the recursive update of the acceptance ratio $r$, the recursive update of the submatrix $\Gamma^{-1}$, and the full update of $T$ after each delayed block.
For implementation details including complete derivations, workflow flowchart, computational complexity analysis, hardware specifications, comprehensive $n_d$ optimization results, \rev{and numerical-stability tests,} see Appendix~\ref{app:submatrix_T}.
\rev{Figure~\ref{fig:efficacy_benchmark} benchmarks the submatrix-T algorithm against the fast update and the other delayed-update algorithms in the production regime of this work.}
\rev{The speedup of submatrix-T over the fast update grows with system size to more than two orders of magnitude at the largest lattices, whether accounted by the update time alone [panel (a)] or by the total sweep time that additionally includes imaginary-time propagation, numerical stabilization, and measurements [panel (b)].}
The superior performance stems from replacing cache-inefficient BLAS-1\&2 operations with batched BLAS-3 routines, which not only improves cache utilization but also provides better amenability to parallel acceleration on multi-threaded (e.g., OpenMP) and GPU architectures.
\rev{Against delay-T, the speedup in total sweep time is about $1.36$ at the largest sizes, a saving of about $25\%$--$30\%$ in computational cost, and this advantage still grows overall with system size.}
\rev{The benchmark data underlying Fig.~\ref{fig:efficacy_benchmark} can be explored through an \href{https://wangfh5.github.io/honeycomb-dqmc-benchmark/}{interactive online report}~\cite{Web-Wang-HoneycombDQMCBenchmark}.}

\begin{figure}[!t]
    \centering
    \includegraphics[width=0.46\textwidth]{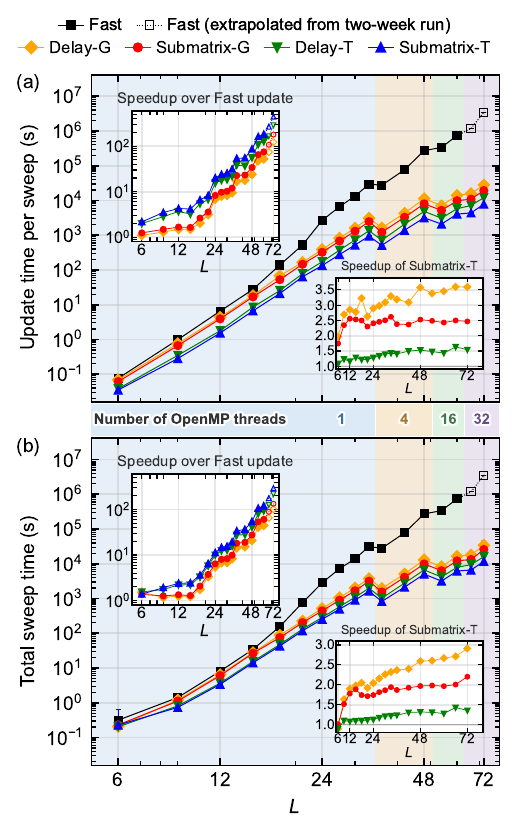}
    \caption{\rev{Production-regime benchmark of the update algorithms in PQMC simulations of the honeycomb Hubbard model ($U/t=3.7$, $\beta t=L+12$, $\Delta_\tau t=0.1$, stabilization every $10$ time slices).
    (a) Update time per Monte Carlo sweep and (b) total sweep time, for the fast update and the four delayed-update algorithms.
    The total sweep time additionally includes imaginary-time propagation, numerical stabilization, and measurements.
    Upper-left insets: speedup of each delayed-update algorithm over the fast update.
    Lower-right insets: speedup of submatrix-T over the other three delayed-update algorithms.
    All delayed-update algorithms use their respective optimal $n_d$ values.
    Pastel bands mark the number of OpenMP threads for one Markov chain.
    For delayed-update algorithms, symbols are medians over at least five independent fixed-seed runs; bars span the observed range (often smaller than symbols).
    For the fast update, solid symbols denote completed single runs, and open symbols with dotted extensions show per-sweep times at $L=66$ and $72$ extrapolated from stable per-slice costs recorded over two-week runs.}}
    \label{fig:efficacy_benchmark}
\end{figure}

As a low-level optimization, our submatrix-T algorithm can be broadly applied to accelerate fermionic computations within the PQMC framework, including entanglement measures~\cite{Phys.Rev.Lett.2013Grover,Phys.Rev.B2014Assaad,Nat.Commun.2025Wanga,ArXiv2025Wang-UntwistedTwistedRenyi}, stabilizer R\'{e}nyi entropy~\cite{ArXiv2026Fang-TwoPointStabilizerRenyi}, and finite-temperature schemes evaluating Fock-state expectations~\cite{Commun.Phys.2025Ding,ArXiv2026Hong-LinearCanonicalEnsembleQuantum}.

%%%%%%%%%%%%%%%%%%%%%%%%%%%%%%%%%%%%%%%%%%%%%%%%%%%%%
\subsection{Data collapse analysis}\label{app:data_collapse_analysis}
%%%%%%%%%%%%%%%%%%%%%%%%%%%%%%%%%%%%%%%%%%%%%%%%%%%%%
Finite-size scaling (FSS) characterizes critical behavior when the linear system size is comparable to the correlation length and thus serves as the most relevant length scale, i.e., $\xi\sim L$.
\rev{One expects a singular quantity to scale as $Q(u,L)\sim L^{y_Q} g(uL^{1/\nu})$ with $u$ measuring the distance of coupling from criticality and $y_Q$ a universal exponent associated with $Q$.}
Within a fixed fitting window of system sizes and couplings around the transition, we perform a one-parameter data-collapse fit using the scaling ansatz
\begin{equation}\label{equ:scaling_form}
Y(X,L)=L^{c_2}F\left[(X-X_c)L^{c_1}\right],
\end{equation}
where the scaling function $F$ is \emph{a priori} unknown but shared by all $L$.
A successful \textit{data collapse} is achieved when $Y/L^{c_2}$ for different $L$ falls onto a single curve when plotted against the scaling variable $(X-X_c)L^{c_1}$.

The scaling form in Eq.~\eqref{equ:scaling_form} is applied to the correlation ratio $R$, the squared order parameter $m^2_{\mathrm{AFM}}$, and the averaged modulus $G_{AB}$ of the off-diagonal Green's function.
For the dimensionless $R$ [Eq.~\eqref{equ:fss_R_AF}], we set $X$ to the tuning parameter ($U$ for the Hubbard and $V$ for the $t$-$V$ model) and fix $c_2=0$ to extract $X_c$ and $c_1=1/\nu$.
For $m^2_{\mathrm{AFM}}$ and $G_{AB}$ [Eqs.~\eqref{equ:fss_m2} and~\eqref{equ:fss_G_AB}], we use $X=R$ and fix $X_c=c_1=0$, yielding $Y(R,L)=L^{c_2}F(R)$.
Here $c_2$ encodes the anomalous dimensions: $c_2=-(1+\eta_{\phi})$ for $m^2_{\mathrm{AFM}}$ and $c_2=-\eta_{\psi}$ for $G_{AB}$.
\rev{The observables of Sec.~\ref{sec:bg_suppressed} enter the same scaling forms, with $R^{\Delta}$ (or $R^{\chi}_{\mathrm{AFM}}$) in place of $R$ and with $\Delta m^2$ or $C_{AA}(\mathbf{r}_{\max})$ [or $L^{-3}\chi_{\mathrm{AFM}}(0)$] in place of $m^2_{\mathrm{AFM}}$.
In the joint collapses, the correlation ratio ($c_2=0$) and the paired observable instead both set $X$ to the tuning parameter in Eq.~\eqref{equ:scaling_form} and share $(X_c,c_1)$.}

To avoid choosing an explicit ansatz for $F$, we adopt the Bayesian scaling analysis (BSA) based on Gaussian-process regression~\cite{Phys.Rev.E2011Harada}.
The scaling function $F$ is modeled as a Gaussian process defined by its kernel hyperparameters\rev{, with independent Gaussian processes for the two observables in a joint collapse}.
The physical scaling parameters ($X_c$, $c_1$, $c_2$) and the hyperparameters are fitted jointly; in what follows, we refer to all of them collectively as the \textit{fitting parameters}.
We use the reduced chi-squared $\chi^2_{\mathrm{red}}=\chi^2/(N_{\mathrm{data}}-N_{\mathrm{free}})$ to evaluate the goodness of fit, where $N_{\mathrm{data}}$ is the number of data points in the fitting window and $N_{\mathrm{free}}$ is the number of unfixed fitting parameters.
The chi-squared is computed as $\chi^{2}=\sum_{i}\left[F(X_{i})-Y_{i}\right]^2/E_{i}^{2}$, where 
\begin{equation}\label{equ:effective_variance}
E_{i}^{2}=\sigma_{Y_{i}}^{2}+\left[F^{\prime}(X_{i})\right]^{2}\sigma_{X_{i}}^{2}-2F^{\prime}(X_{i})\sigma_{X_{i}Y_{i}}
\end{equation}
is the effective variance that accounts for uncertainties in both $X$ and $Y$ data, as well as their covariance $\sigma_{X_{i}Y_{i}}$~\cite{Am.J.Phys.1982Orear,ChemicalGeology2024Daeron}.
\rev{The variances $\sigma_{X_{i}}^{2}$ and $\sigma_{Y_{i}}^{2}$ and the covariance $\sigma_{X_{i}Y_{i}}$ follow by error propagation from the covariance matrix of the correlation functions measured at each simulation point, introduced below.}
\rev{When $X$ is the tuning parameter (in the correlation-ratio and joint collapses), only $\sigma_{Y_{i}}$ contributes.}

\begin{figure}[t]
    \centering
    \includegraphics[width=\columnwidth]{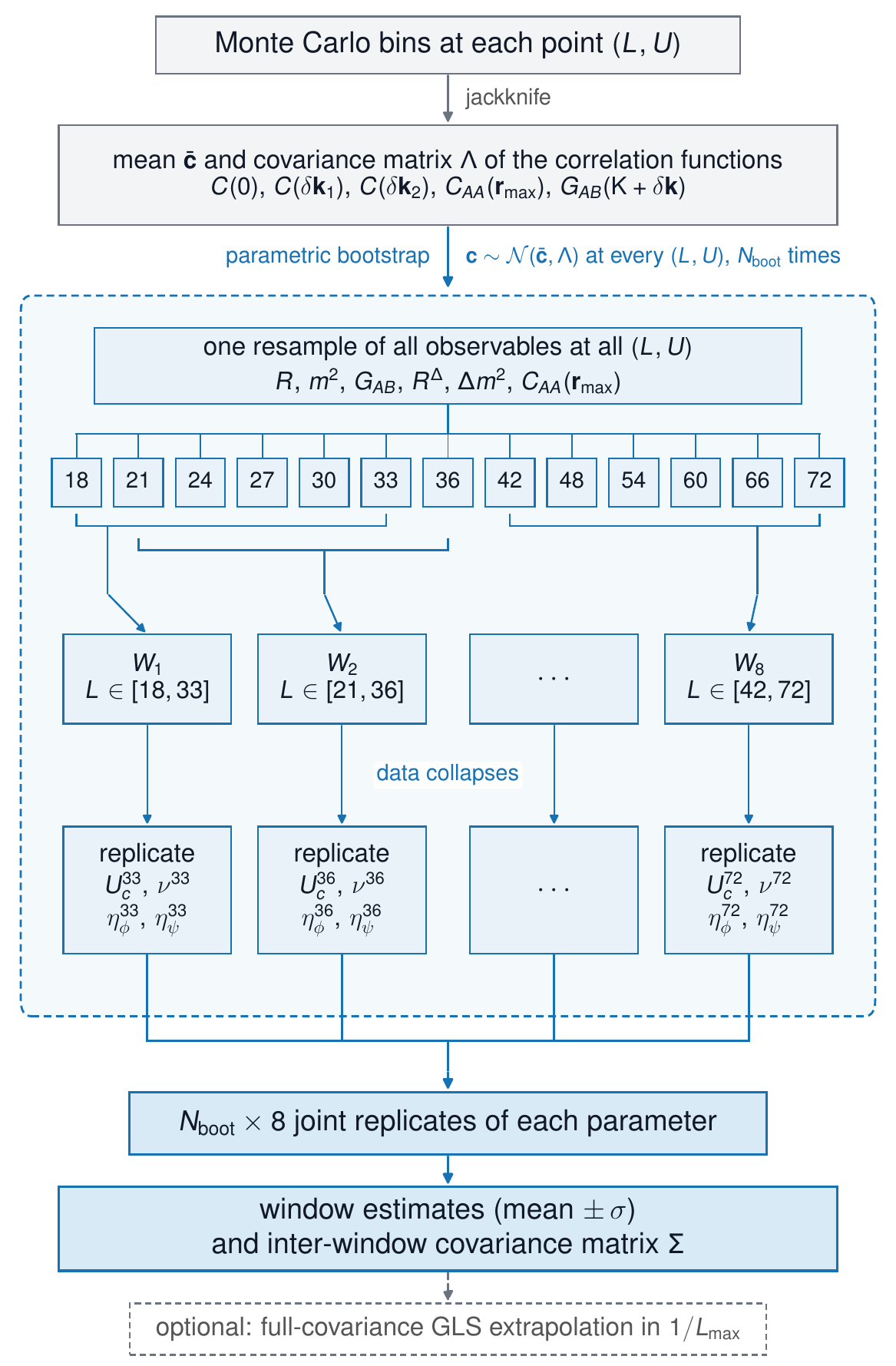}
    \caption{\rev{Joint-bootstrap workflow of the sliding-window data-collapse analysis (steps 1--4 in the text), illustrated with the Hubbard windows of six consecutive sizes, $W_1,\ldots,W_8$.
    The dashed frame encloses one of the $N_{\mathrm{boot}}$ resamples, in which every window is fitted to the same resampled data.}}
    \label{fig:joint_bootstrap_workflow}
\end{figure}

\rev{We estimate the fitting parameters and their uncertainties with a \textit{joint bootstrap} over all simulation points (Fig.~\ref{fig:joint_bootstrap_workflow}), in which a single set of resamples serves every fitting window and every scaling form.}
\rev{Appendix~\ref{app:covariance_extrapolation} examines the correlations among observables and among windows that are thereby carried through to the fitted critical parameters.}
\rev{The individual fits are run through our open-source Julia package at \url{https://github.com/wangfh5/BSAHelper}, a downstream utility for the BSA program.}
\begingroup
\revcolor
\begin{enumerate}
    \item \textbf{Statistics at each simulation point.} All observables at a simulation point, $(L,U)$ for the Hubbard model or $(L,V)$ for the $t$-$V$ model, are functions of a vector $\mathbf{c}$ collecting the measured correlation functions $C(0)$, $C(\delta\mathbf{k}_{1})$, $C(\delta\mathbf{k}_{2})$, $C_{AA}(\mathbf{r}_{\max})$, and $G_{AB}(\mathrm{K}+\delta\mathbf{k})$ at the six momenta nearest to $\mathrm{K}$ (AFM/spin correlations for the Hubbard model and CDW/density correlations for the $t$-$V$ model; we omit these channel labels below).
    A jackknife over the Monte Carlo bins after warm-up gives its mean $\bar{\mathbf{c}}$ and intra-run covariance matrix $\Lambda$, which contains the covariances between different quantities measured in the same simulation run.
    For any two observables $X(\mathbf{c})$ and $Y(\mathbf{c})$ at that point, linear error propagation around $\bar{\mathbf{c}}$ gives $\sigma_{XY}=(\nabla_{\mathbf{c}}X)^{\mathsf{T}}\Lambda\,\nabla_{\mathbf{c}}Y$, with $\sigma_X^2=\sigma_{XX}$ and $\sigma_Y^2=\sigma_{YY}$.
    \item \textbf{Joint resampling.} At every simulation point, draw a resampled vector $\mathbf{c}$ from the multivariate Gaussian distribution $\mathcal{N}(\bar{\mathbf{c}},\Lambda)$, independently between points, which come from separate simulations.
    Evaluate $R$, $m^2$, $G_{AB}$, and the background-suppressed observables from the drawn vector.
    This parametric bootstrap preserves the intra-run covariance, so that $R$ and $m^2$, which share $C(0)$, fluctuate together in each resample.
    \item \textbf{Window fits.} For every fitting window and every scaling form, run the BSA program once on the resampled data and record the best-fit values of the fitting parameters.
    Each data point $i$ is weighted by a frozen effective standard error $E_i$ [Eq.~\eqref{equ:effective_variance}], evaluated once using $\sigma_{X_i}^2$, $\sigma_{Y_i}^2$, and $\sigma_{X_iY_i}$ from step~1, with $F^{\prime}(X_i)$ estimated from fixed-$L$ cubic fits to the mean data.
    We turn off the program's internal uncertainty estimation for the fitting parameters, so $E_i$ enters only as the Gaussian-process noise weight without double counting the statistical fluctuations injected in step~2.
    \item \textbf{Replicate statistics.} Repeat steps 2--3 for $N_{\mathrm{boot}}$ resamples and discard the rare failed fits.
    For a given parameter, let $p_i^{(b)}$ denote its replicate in window $i$ from resample $b$.
    The replicates give the window means and the inter-window covariance matrix
    \begin{equation}\label{equ:window_covariance}
    \begin{aligned}
        \bar p_i&=\frac{1}{N_{\mathrm{boot}}}\sum_{b=1}^{N_{\mathrm{boot}}}p_i^{(b)},\\
        \Sigma_{ij}&=\frac{1}{N_{\mathrm{boot}}-1}\sum_{b=1}^{N_{\mathrm{boot}}}(p_i^{(b)}-\bar p_i)(p_j^{(b)}-\bar p_j).
    \end{aligned}
    \end{equation}
    The window estimate is $\bar p_i$ with uncertainty $\sqrt{\Sigma_{ii}}$, and the off-diagonal elements of $\Sigma$ quantify the correlations between windows that share system sizes.
    \item \textbf{Reconstruction and chi-squared test.} Using the original Monte-Carlo-averaged data, rerun the BSA program with all fitting parameters fixed to their bootstrap means to reconstruct the scaling function and evaluate $\chi^2_{\mathrm{red}}$ via Eq.~\eqref{equ:effective_variance}, using $\sigma_{X_i}^2$, $\sigma_{Y_i}^2$, and $\sigma_{X_iY_i}$ from step~1 and taking $F^{\prime}(X_i)$ from the reconstructed scaling function.
    All data-collapse figures in this work show this reconstruction.
    In both separate and joint collapses, the smallest windows can show a $\chi^2_{\mathrm{red}}$ far above that of larger windows, dropping abruptly at some $L_{\min}$; this jump signals a breakdown of the leading-form collapse, which an $L^{-\omega}$ correction largely repairs (Appendix~\ref{app:correction_initialization}).
    Such windows, and those collapsed against an excluded correlation ratio, are shown faded and excluded when extracting any thermodynamic-limit estimate.
\end{enumerate}
\endgroup

\rev{All fits start from the same initial values, shared by all resamples and all windows.}
For the correlation-ratio collapse ($Y=R$), we initialize $U_{c,0}=3.75$ and $c_{1,0}=0.9$ for the Hubbard model, and $V_{c,0}=1.35$ and $c_{1,0}=1.1$ for the $t$-$V$ model.
For the analysis of $m^2_{\mathrm{AFM}}$ ($G_{AB}$) versus $R$, \rev{the exponent $c_2$ is initialized at} $c_{2,0}=-1.7$ ($-0.15$) for the Hubbard model and $c_{2,0}=-1.5$ ($-0.08$) for the $t$-$V$ model.
\rev{The collapses of the background-suppressed observables start from the same initial values as the corresponding collapses of $R$ and $m^2$, except that $R^{\Delta}_{\mathrm{CDW}}$ in the $t$-$V$ model is initialized near its crossing ($V_{c,0}=1.334$) to avoid an unphysical $V_c<0$ branch in the $L_{\min}=6$ windows.}
\rev{We use $N_{\mathrm{boot}}=1000$ joint resamples} and confirm that increasing $N_{\mathrm{boot}}$ does not significantly change the results.
\rev{Throughout this work, bootstrap uncertainties are rounded up to the precision supported by their error of error, $\sigma/\sqrt{2(N_{\mathrm{boot}}-1)}$.}

\rev{When the window estimates of a parameter drift monotonically, we select the larger-size windows over which the drift is approximately linear in $x_i=1/L_{\max,i}$ and extrapolate them by a full-covariance generalized least-squares (GLS) fit using the inter-window covariance matrix $\Sigma$ (see also Appendix~\ref{app:covariance_extrapolation}).}
\rev{With $f_i(p_\infty,a)=p_\infty+a x_i$, we minimize}
\begin{equation}\label{equ:window_gls}
    \rev{\chi^2_{\mathrm{GLS}}=(\bar{\mathbf p}-\mathbf f)^{T}\Sigma^{-1}(\bar{\mathbf p}-\mathbf f),}
\end{equation}
\rev{where $\bar{\mathbf p}$ and $\mathbf f$ collect $\bar p_i$ and $f_i$ over the selected windows, and report the fitted intercept $\hat p_\infty$.}
\rev{Its uncertainty is the standard deviation of the intercepts obtained by refitting each replicate vector $\mathbf p^{(b)}$ with the same $\Sigma$, plus the absolute difference between their mean and $\hat p_\infty$.}
\rev{When the window estimates oscillate or settle on a plateau, we instead report an envelope average, whose central value is the mean of the selected window estimates and whose uncertainty is the smallest symmetric interval that covers all their error bars.}

%%%%%%%%%%%%%%%%%%%%%%%%%%%%%%%%%%%%%%%%%%%%%%%%%%%%%
\section{Projector determinant quantum Monte Carlo}\label{app:PQMC}
%%%%%%%%%%%%%%%%%%%%%%%%%%%%%%%%%%%%%%%%%%%%%%%%%%%%%

\begin{figure*}[htbp]
    \centering
    \includegraphics[width=0.9\textwidth]{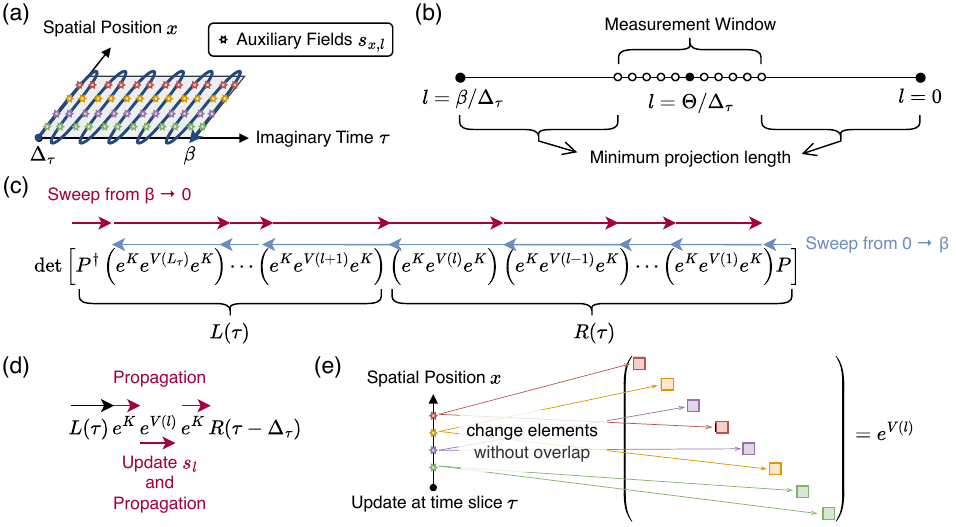}
    \caption{Illustrations of the sweep in PQMC. (a) The sequential local update scheme (only show the $0\rightarrow \beta$ half of the sweep). (b) Measurements of equal-time and time-displaced observables are performed within the measurement window around $\Theta$. (c) The sweep process from the perspective of weight/ratio calculation. (d) The partial propagation and update at time slice $\tau$. (e) At time slice $\tau$, updating a single auxiliary field proposes changes in $k$ (we take 2 for illustration) elements of $e^{V(l)}$.  
    $\beta=2\Theta=L_{\tau}\Delta_{\tau}$ and $\tau=l\Delta_{\tau}$.}
    \label{fig:PQMCsweep}
\end{figure*}

The projector determinant quantum Monte Carlo (PQMC) method evaluates ground-state properties by imaginary-time projection of a trial Slater determinant~\cite{Ann.Phys.1986Sugiyama,Europhys.Lett.1989Sorella,Phys.Rev.B1989White}. 
In this section, we briefly review the basic formalism of PQMC following Ref.~\cite{Assaad2008WorldlineDeterminantalQuantum}.
In the language of second quantization, the trial state reads $\left|\Psi_{T}\right\rangle=\prod_{n=1}^{N_{p}} (\mathbf{c}^{\dagger} P)_{n} \left|0\right\rangle$, where $\mathbf{c}^\dagger=(c_{1\uparrow}^{\dagger},\cdots,c_{N_{s}\uparrow}^{\dagger},c_{1\downarrow}^{\dagger},\cdots,c_{N_{s}\downarrow}^{\dagger})$ collects all the local fermionic creation operators on a lattice with $N_{s}$ sites, $N_p$ is the number of particles in the trial state and $P$ is a $N(=2N_{s})\times N_p$ matrix that defines the Slater determinant.
Provided $\langle \Psi_{T}|\Psi_{0}\rangle\neq 0$, the ground state is given by $\left|\Psi_{0}\right\rangle=\lim_{\Theta\rightarrow \infty}e^{-\Theta \hat{H}}\left|\Psi_{T}\right\rangle$ with $\hat{H}=\hat{H}_{0}+\hat{H}_{I}$ being the full Hamiltonian.
Ground-state expectation value of any observable $\hat{O}$ is then computed as
\begin{equation}\label{equ:PQMC_O_general}
    \langle{\hat{O}}\rangle=\frac{\left\langle \Psi_{0}|\hat{O}|\Psi_{0}\right\rangle }{\left\langle \Psi_{0}\mid\Psi_{0}\right\rangle }=\frac{\left\langle \Psi_{T}\left|e^{-\Theta\hat{H}}\hat{O}e^{-\Theta\hat{H}}\right|\Psi_{T}\right\rangle }{\left\langle \Psi_{T}\left|e^{-2\Theta\hat{H}}\right|\Psi_{T}\right\rangle }
\end{equation}
To evaluate the above expectation value, we first discretize the projection length, $\beta=2\Theta=L_{\tau}\Delta_{\tau}$, and employ the symmetric Trotter decomposition 
\begin{equation}\label{equ:Trotter_decomposition}
    e^{-2\Theta H}=\left(e^{\hat{K}}e^{-\Delta_{\tau}\hat{H}_{I}}e^{\hat{K}}\right)^{L_{\tau}}+\mathcal{O}(\Delta_{\tau}^{3}).
\end{equation}
Here we have defined $\hat{K}\equiv-\Delta_{\tau}\hat{H}_{0}/2\equiv \mathbf{c}^{\dagger}K\mathbf{c}$ which is a fermion bilinear operator.
Next, we decouple $\hat{H}_{I}$ on each time slice via a Hubbard--Stratonovich (HS) transformation, 
\begin{equation}\label{equ:HS_general}
    e^{-\Delta_{\tau}\hat{H}_{I}}=\sum_{s}\alpha\left[s\right]e^{\hat{V}\left[s\right]},
\end{equation}
where $s=\{s_x\}$ is a set of auxiliary fields at different \textit{spatial positions} $x$ (which may be lattice sites, bonds, or others), $\hat{V}\left[s\right]=\mathbf{c}^{\dagger}V[s]\mathbf{c}$ is a fermion bilinear operator and $\alpha\left[s\right]$ is a coupling coefficient. 
Doing HS transformation at all time slices introduces space-time-dependent auxiliary fields denoted as $\mathbf{s}=\{s_{l}\}=\{s_{x,l}\}$ (illustrated in Fig.~\ref{fig:PQMCsweep}(a)) and yields factorized weight and expectation
\begin{gather}
    \left\langle \Psi_{T}\left|e^{-2\Theta
    \hat{H}}\right|\Psi_{T}\right\rangle \approx\sum_{\mathbf{s}}w_{\mathbf{s}}=\sum_{\mathbf{s}}
    w_{b}[\mathbf{s}]w_{f}[\mathbf{s}],\\
    \langle{\hat{O}}\rangle\approx\sum_{\mathbf{s}}{p}_{\mathbf{s}}\langle{\hat{O}}\rangle_{\mathbf{s}}
    , p_{\mathbf{s}}=\frac{w_{\mathbf{s}}}{\sum_{\mathbf{s}}w_{\mathbf{s}}}=\frac{w_{b}[\mathbf{s}]w_{f}[\mathbf{s}]}{\sum_{\mathcal{\mathbf{s}}}w_{b}[\mathbf{s}]w_{f}[\mathbf{s}]}.
    \label{equ:PQMC_O}
\end{gather}
Here we use symbol ``$\approx$'' since the Trotter decomposition introduces an error of order $\mathcal{O}(\Delta_{\tau}^{3})$.
The scalar weight is $w_{b}[\mathbf{s}]=\prod_{l}\alpha\left[s_{l}\right]$ and the fermionic weight is given by
\begin{equation}\label{equ:PQMC_w_f}
    w_{f}[\mathbf{s}] = \det\left[ P^{\dagger} B_{\mathbf{s}}(2\Theta,0)P \right],
\end{equation}
where the single-particle propagator is a ordered product of exponential matrices over time slices,
\begin{equation}
    B_{\mathbf{s}}(\tau_{2}=l_2\Delta_{\tau},\tau_{1}=l_1\Delta_{\tau})= \prod_{l=l_{2}}^{l_{1}+1} \big( e^{K} \, e^{V(l)} \, e^{K} \big), 
\end{equation}
where $V(l)=V[s_{l}]$ denotes the decoupled interaction matrix at time slice $l$. We also define the operator version as
\begin{equation}
    \hat{B}_{\mathbf{s}}(\tau_{2}=l_2\Delta_{\tau},\tau_{1}=l_1\Delta_{\tau})= \prod_{l=l_{2}}^{l_{1}+1} \big( e^{\hat{K}} \, e^{\hat{V}(l)} \, e^{\hat{K}} \big).
\end{equation}

The weighted sum in Eq.~\eqref{equ:PQMC_O} is done by Monte Carlo sampling of the auxiliary-field configurations. 
We adopt a common used local update scheme for PQMC, i.e., sequentially updating the auxiliary fields at each spatial position and then at each time slice. A \textit{sweep} includes updates of all the auxiliary fields, from $\tau=0$ to $\tau=\beta$ and then back to $\tau=0$, and from the first spatial position to the last one within each time slice, as illustrated in Fig.~\ref{fig:PQMCsweep}(a). 
An equal-time observable is a generalization of Eq.~\eqref{equ:PQMC_O_general}, defined at each time slice:
\begin{gather}
    \langle\hat{O}(\tau)\rangle\equiv\frac{\left\langle \Psi_{T}\left|e^{-\left(2\Theta-\tau\right)\hat{H}}Oe^{-\tau\hat{H}}\right|\Psi_{T}\right\rangle }{\left\langle \Psi_{T}\left|e^{-2\Theta\hat{H}}\right|\Psi_{T}\right\rangle }=\sum_{\mathbf{s}}p_{\mathbf{s}}\langle \hat{O}\left(\tau\right)\rangle _{\mathbf{s}},\\
    \langle \hat{O}\left(\tau\right)\rangle_{\mathbf{s}}=\frac{\left\langle
    \Psi_{T}\left|\hat{B}_{\mathbf{s}}\left(2\Theta,\tau\right)\hat{O}\hat{B}_{\mathbf{s}}\left(\tau,0\right)\right|\Psi_{T}\right\rangle
    }{\left\langle \Psi_{T}\left|e^{-2\Theta \hat{H}}\right|\Psi_{T}\right\rangle }.
\end{gather}
In practice, we perform measurements of equal-time (and also time-displaced) observables within the measurement window around $\Theta$, as illustrated in Fig.~\ref{fig:PQMCsweep}(b). 
All the left and right projection times $2\Theta-\tau$ and $\tau$ inside the measurement window are considered long enough to project out the excited states. 
In particular, the equal-time Green's function at slice $\tau$ (whose elements are defined as $G_{ij}(\tau,\tau)\equiv \langle c_{i}(\tau) c_{j}^{\dagger}(\tau)\rangle$) is given by~\cite{Assaad2008WorldlineDeterminantalQuantum}
\begin{equation}\label{equ:PQMC_G}
    G(\tau,\tau)= I - R(\tau) [L(\tau) R(\tau)]^{-1}L(\tau), 
\end{equation}
where $R(\tau)=B_{\mathbf{s}}(\tau,0) P$ (size $N\times N_{p}$) and $L(\tau)= P^{\dagger} B_{\mathbf{s}}(2\Theta,\tau)$ (size $N_{p}\times N$) are factorized from the matrix appearing in Eq.~\eqref{equ:PQMC_w_f}, i.e., $P^\dagger B_{\mathbf{s}}(\beta,0) P=L(\tau)R(\tau)$. 
Figure~\ref{fig:PQMCsweep}(c) shows different factorizations during the whole sweep process. 

Consider a local update of the auxiliary field at spatial position $x$ at time slice $l$, $s_{x,l}\rightarrow s_{x,l}'$, which induces a change of matrix $e^{V(l)}\rightarrow e^{V'(l)}\equiv(I+\Delta)e^{V(l)}$. The acceptance ratio is given by $\min\{1,r_{\rm tot}\}$ with
\begin{equation}
    r_{{\rm tot}}=\frac{w_{\mathbf{s}'}}{w_{\mathbf{s}}}=\frac{\alpha[s_{l}']}{\alpha[s_{l}]}r, 
\end{equation}
where $r$ is the ratio of the fermionic weight, given by
\begin{equation}
    \begin{aligned}
        r&=\frac{\det\left[P^{\dagger}B_{\mathbf{s}^{\prime}}\left(\beta,0\right)P\right]}{\det\left[P^{\dagger}B_{\mathbf{s}}\left(\beta,0\right)P\right]}\\&=\frac{\det\left[\left(L\left(\tau\right)e^{K}\right)\left(I+\Delta\right)\left(e^{V\left(l\right)}e^{K}R\left(\tau-\Delta_{\tau}\right)\right)\right]}{\det\left[\left(L\left(\tau\right)e^{K}\right)\left(e^{V\left(l\right)}e^{K}R\left(\tau-\Delta_{\tau}\right)\right)\right]}
    \end{aligned}
\end{equation}
Due to the symmetric Trotter decomposition, we have to partially propagate $L(\tau)$ and $R(\tau)$ to $\tilde{L}(\tau)\equiv L(\tau)e^{K}$ and $\tilde{R}(\tau)\equiv e^{-K}R(\tau)$ respectively before the update (see Fig.~\ref{fig:PQMCsweep}(d) for illustration). 
For simplicity, we still use the notation $L(\tau)$ ($R(\tau)$) to label $\tilde{L}(\tau)$ ($\tilde{R}(\tau)$) in the following update formulas, and write
\begin{equation}\label{equ:PQMC_r_LRtau}
    r=\frac{\det\left[L\left(\tau\right)\left(I+\Delta\right)R\left(\tau\right)\right]}{\det\left[L\left(\tau\right)R\left(\tau\right)\right]}.
\end{equation}

The efficacy of the local update scheme relies on the fact that changing a single local auxiliary field only affects a few elements of matrix $V(l)$ and, more importantly, these elements being changed do not overlap with those being changed by updates at other spatial positions. 
This property not only makes the update order of the spatial positions irrelevant but also allows us to develop cache-efficient delayed update algorithms~\cite{Phys.Rev.B2024Sun,SciPostPhys.2025Sun,Phys.Rev.B2025Du}. 
In the following, we only consider the cases where $e^{V(l)}$, as well as $\Delta$, are diagonal matrices with rank $k\ll N$. 
In Fig.~\ref{fig:PQMCsweep}(e), we illustrate an example of $k=2$ case. 
This sparse structure of $\Delta$ makes the calculation of ratio $r$ much more efficient since using $\Delta$ to left (or right) multiply a matrix simply scales $k$ rows (or columns) of the matrix while making other elements zero. 
To better play with the sparsity of $\Delta$, we denote the locations of the $k$ non-zero entries of $\Delta$ as $x=\{x_1,x_2,\cdots,x_k\}$ and introduce the \textit{index matrices}
\begin{equation}
    P_{N\times k}\left[x\right]=\left[e_{x_{1}}| e_{x_{2}}|\cdots| e_{x_{k}}\right]\text{ and }P_{k\times N}=P_{N\times k}^{T},
\end{equation}
where $e_{x_j}$ is the $x_j$-th unit vector of length $N$. 
$P_{N\times k}$ (or $P_{k\times N}$) can be viewed as being cropped from the identity matrix $I_{N}$ by keeping only the columns (or rows) indexed by $x$. 
They are uniquely determined by $\Delta$, so when unambiguous we omit the argument $x$ of $P_{N\times k}$ and $P_{k\times N}$ to simplify the notation. 
Below we list some evident properties of the index matrices. 
First, for arbitrary $N\times N$ matrix $A$, $AP_{N\times k}$ extracts the $k$ columns of $A$ indexed by $x$, while $P_{k\times N}A$ extracts the corresponding $k$ rows.
Second, $P_{k\times N}\,\Delta\,P_{N\times k}=\mathrm{diag}(\Delta_{x_1 x_1},\dots,\Delta_{x_k x_k})$ is a full-rank diagonal matrix and $P_{k\times N}P_{N\times k}=I_k$ is a $k\times k$ identity matrix.
Third, although $P_{N\times k}P_{k\times N}\neq I_{N}$, one has identity
\begin{equation}\label{equ:Delta_P_P}
    \Delta=\Delta P_{N\times k}P_{k\times N}=P_{N\times k}P_{k\times N}\Delta,
\end{equation}
so whenever $\Delta$ appears in a product one may safely insert $P_{N\times k}P_{k\times N}$ on either side.
Furthermore, this identity is invariant under any permutation of the columns of $P_{N\times k}$ (and corresponding rows of $P_{k\times N}$). Formally, for any $k$-th order permutation matrix $\mathbf{S}_k$,
\begin{equation}\label{equ:Delta_PS_SP}
\Delta=\Delta\left(P_{N\times k}\mathbf{S}_k\right)\left(\mathbf{S}_k^{T}P_{k\times N}\right)=\left(P_{N\times k}\mathbf{S}_k\right)\left(\mathbf{S}_k^{T}P_{k\times N}\right)\Delta.
\end{equation}
This property underpins the construction of accumulated update matrices in delayed update algorithms.

Before introducing the update algorithms, we present two concrete model examples with different update ranks $k$.
For the repulsive Hubbard model, we apply the discrete HS transformation in the density channel~\cite{Phys.Rev.B1983Hirsch}:
\begin{equation}
\begin{aligned}
&e^{-\Delta_{\tau}U\sum_{i=1}^{N_\mathrm{s}}\left(\hat{n}_{i \uparrow}-\frac{1}{2}\right)\left(\hat{n}_{i \downarrow}-\frac{1}{2}\right)}\\
&= \gamma^{N_\mathrm{s}}\sum_{\{s_i=\pm1\}}e^{\mathrm{i}{\alpha}\sum_{i=1}^{N_\mathrm{s}}s_i\left(\hat{n}_{i\uparrow}+\hat{n}_{i\downarrow}-1\right)},
\end{aligned}
\end{equation}
with $\gamma=\frac{1}{2}e^{\Delta_\tau U/4}$ and $\cos{\alpha} = e^{-\Delta_\tau U/2}$. The auxiliary fields $s_i$ are defined on lattice sites, introducing a diagonal interaction matrix $V(l)$ that implies a rank $k=1$ local update.
The spatial positions that auxiliary fields dwell in are simply all the lattice sites ($x=i\in \{1,\ldots,N_\mathrm{s}\}$). 
Compare with the general form in Eq.~\eqref{equ:HS_general}, this transformation introduces a diagonal matrix $V[s] = V_\uparrow[s] \oplus V_\downarrow[s]$ with $V_\uparrow=V_\downarrow=\mathrm{i}{\alpha}\,\mathrm{diag}(s_1,\ldots,s_{N_\mathrm{s}})$, and the coupling coefficient $\alpha[s] = {\gamma}^{N_\mathrm{s}}e^{-\mathrm{i}\alpha\sum_{i=1}^{N_\mathrm{s}}s_i}$.
Since the Hamiltonian contains no spin-flip terms, and furthermore the spin blocks are degenerate, allowing us to track only one spin block. Therefore, the rank of each local update is $k=1$.

For the spinless $t$-$V$ model, we decouple the interaction to the hopping channel~\cite{Phys.Rev.B2015Li,Phys.Rev.Lett.2015Wang}:
\begin{equation}\label{equ:S2_HStV_SM}
\begin{aligned}
&e^{-\Delta_\tau V \sum_{\langle j,k\rangle}\left(\hat{n}_{j}-\frac{1}{2}\right)\left(\hat{n}_{k}-\frac{1}{2}\right)}\\
&= \zeta^{N_c} \sum_{\{s_{jk}=\pm1\} }e^{\lambda \sum_{\langle j,k\rangle} s_{jk}\left(c_{j}^{\dagger}c_{k}+c_{k}^{\dagger}c_{j}\right)},
\end{aligned}
\end{equation}
where $\zeta=\frac{1}{2}e^{-\frac{V\Delta_\tau}{4}}$, $\cosh \lambda=e^{\frac{V \Delta_\tau}{2}}$ and $N_c$ denotes the total number of nearest-neighbor bonds. 
The auxiliary fields $s_{jk}$ reside on bonds, and updating a single field changes only the hopping between sites $j$ and $k$, suggesting a rank $k=2$ local update. 
To make this precise, we further Trotter-decompose the bond term into $N_{\rm fml}$ disjoint ``bond families'' (the so-called ``checkerboard decomposition''), 
$e^{\lambda\sum_{\left\langle jk\right\rangle }s_{jk}\left(c_{j}^{\dagger}c_{k}+c_{k}^{\dagger}c_{j}\right)}\approx\prod_{n=1}^{N_{\rm fml}}e^{\lambda\sum_{b\in\text{fml }n}s_{b}\left(c_{b_{j}}^{\dagger}c_{b_{k}}+c_{b_{k}}^{\dagger}c_{b_{j}}\right)}$. 
For example, $N_{\rm fml}=4$ for the square lattice and $N_{\rm fml}=3$ for the honeycomb lattice. 
Each factor can then be diagonalized by a unitary transformation $U_n$ consisting of a permutation and several $\frac{1}{\sqrt{2}}\left(\begin{array}{cc} 1 & 1 \\ 1 & -1 \end{array}\right)$ blocks, i.e., $e^{V(l)}\approx\prod_{n=1}^{N_{\rm fml}}e^{V_{n}(l)}=\prod_{n=1}^{N_{\rm fml}}U_{n}e^{\bar{V}_{n}(l)}U_{n}^{\dagger}$. 
To work with the diagonal exponential $e^{\bar{V}_{n}(l)}$, one additionally performs a partial propagation by $U_n$ or $U_n^{\dagger}$. 
As illustrated in Fig.~\ref{fig:PQMCsweep}(d), we write $LR=L(\tau)e^K \prod_{n=1}^{N_{\rm fml}} (U_n e^{\bar{V}_{n}(l)}U_{n}^{\dagger}) e^K R(\tau-\Delta_\tau)$ and sweep through the factors sequentially: propagate by $e^K$, apply $U_1$, update and apply $e^{\bar{V}_{1}(l)}$, apply $U_1^{\dagger}$, then proceed to next bond family $n=2$, and so on. 
When updating a bond $b$ within a given family, only the two diagonal entries associated with its endpoints $b_{j}$ and $b_{k}$ are modified (Fig.~\ref{fig:PQMCsweep}(e), illustrated here for $k=2$), hence the corresponding $\Delta$ has rank $k=2$. 

To calculate the acceptance ratio, two different update objects can be chosen: $T\equiv (LR)^{-1}$ and $G$, both of which have \textit{fast update} formulas.

\subsection{Local update formulas for $T\equiv(LR)^{-1}$}

Following Eq.~\eqref{equ:PQMC_r_LRtau}, the most natural choice is to update $(LR)^{-1}$: 
\begin{equation}\label{equ:fast_LR_update}
    \begin{aligned}
        \left[\left(LR\right)^{-1}\right]^{\prime}&=(LR+L\left(\Delta P_{k\times N}P_{N\times k}\right)R)^{-1}\\&=(LR)^{-1}\left(I_{N_{p}}+UV\right)^{-1}\\&=(LR)^{-1}-(LR)^{-1}U(I_{k}+VU)^{-1}V,
    \end{aligned}
\end{equation}
where we have used Eq.~\eqref{equ:Delta_P_P} and the Woodbury identity. $U\equiv L\Delta P_{N\times k}$ and $V\equiv P_{k\times N}R(LR)^{-1}$ are $N_p\times k$ and $k\times N_p$ matrices respectively.
The fermionic weight ratio is obtained from the second line of Eq.~\eqref{equ:fast_LR_update} and Sylvester's determinant identity
\begin{equation}\label{equ:fast_LR_ratio}
    r=\det\left[I_{N_{p}}+L\Delta R\left(LR\right)^{-1}\right]=\det\left[I_{k}+VU\right].
\end{equation}
The per-move cost is $\mathcal{O}(N_{p}^{2})$ and the per-time-slice cost is $\mathcal{O}(N_{p}^{2} N)$.

\subsection{Local update formulas for $G$}

Alternatively, one may update the equal-time Green's function defined in Eq.~\eqref{equ:PQMC_G} (note that partial propagation is also needed before the update).
Using $VU = P_{k\times N} (I - G) \Delta P_{N\times k}$, the same ratio becomes
\begin{equation}\label{equ:fast_G_ratio}
    r=\det\left[I_{N}+\Delta(I-G)\right]=\det\left[I_{k}+\tilde{V}\tilde{U}\right], 
\end{equation}
where we define $\tilde{U}\equiv \Delta P_{N\times k}$ and $\tilde{V}\equiv P_{k\times N}(I-G)$. A fast update formula for $G$ follows from Woodbury identity
\begin{equation}\label{equ:fast_G_update}
    G' = G  \left[ I + \Delta (I - G) \right]^{-1} = G - G \tilde{U}  \left(I_{k} + \tilde{V}\tilde{U}\right)^{-1} \tilde{V}.
\end{equation}
The per-move cost is $\mathcal{O}(N^{2})$ and the per-time-slice cost is $\mathcal{O}(N^{3})$, which is more expensive than the $(LR)^{-1}$ update.

%%%%%%%%%%%%%%%%%%%%%%%%%%%%%%%%%%%%%%%%%%%%%%%%%%%%%
\section{Brief review of former delayed update algorithms}\label{app:delayed_update_algorithms}
%%%%%%%%%%%%%%%%%%%%%%%%%%%%%%%%%%%%%%%%%%%%%%%%%%%%%

For the simulations on CPU, one of the performance bottlenecks comes from the cache usage. 
This bottleneck is evidently seen in the worst case $k=1$, where the update of $T$ and $G$ is full of vector-vector multiplications, which are BLAS-1 operations that are not cache-friendly. 
To boost the performance, the \textit{delay update} and \textit{submatrix update} algorithms have been proposed~\cite{Phys.Rev.B2024Sun,SciPostPhys.2025Sun,Phys.Rev.B2025Du}. 
Both of them introduce a parameter $n_d$ to control the delayed steps before fully updating the update objects, including $T$ and $G$, using BLAS-3 matrix-matrix multiplications, thus improving the performance even with higher computational complexity. 
In this work, all these kind of algorithms are called \textit{delayed update} algorithms uniformly and different specific realizations are denoted by ``delay-$\mathrm{O}$'' and ``submatrix-$\mathrm{O}$'', where $\mathrm{O}\in\{\mathrm{T},\mathrm{G}\}$ specifies the chosen update object. 
For PQMC, the delay-G~\cite{Phys.Rev.B2024Sun} and submatrix-G~\cite{SciPostPhys.2025Sun} algorithms only improve performance for large system sizes since they update the Green's function $G$, which increases the computational complexity from \rev{$\mathcal{O}(NN_p^2)$} to $\mathcal{O}(N^3)$.
Recently, the delay-T algorithm has been proposed, which designed a delayed update strategy for matrix $T$~\cite{Phys.Rev.B2025Du}. 

In this section, we review three established delayed-update algorithms (delay-G, delay-T, and submatrix-G) that provide the foundation for our work.
These approaches enhance computational performance by accumulating multiple rank-$k$ updates and executing them collectively as a single, cache-efficient, higher-rank batch operation.

\subsection{Delay-G~\cite{Phys.Rev.B2024Sun}}\label{app:delay_G}

Let $G^{(i)}$ denote the Green's function after $i$ accepted updates within a delayed block, with $G^{(0)}$ representing the initial matrix.
For the $(i+1)$-th proposed update on sites $x^{(i+1)}=\{x_1^{(i+1)}, \ldots, x_k^{(i+1)}\}$, we define the index matrices $P_{N\times k}^{(i+1)}$ and $P_{k\times N}^{(i+1)}$, along with $\tilde{U}^{(i+1)} = \Delta^{(i+1)} P_{N\times k}^{(i+1)}$ and $\tilde{V}^{(i+1)} = P_{k\times N}^{(i+1)}(I - G^{(i)})$.
The fermionic acceptance ratio is given by
\begin{equation}
    r^{(i+1)} = \det\left[I_{k} + \tilde{V}^{(i+1)} \tilde{U}^{(i+1)}\right] = \det \tilde{S}^{(i+1)},
\end{equation}
where
\begin{equation}
\begin{aligned}
    &\tilde{S}^{(i+1)} = I_{k} + \tilde{V}^{(i+1)} \tilde{U}^{(i+1)} \\
                    &= I_{k} + \left[P_{k\times N}^{(i+1)} \left(I - G^{(i)}\right) P_{N\times k}^{(i+1)} \right] \left( P_{k\times N}^{(i+1)} \Delta^{(i+1)} P_{N\times k}^{(i+1)}\right).
\end{aligned}
\end{equation}
The Green's function evolves according to the recursive relation
\begin{equation}
\begin{aligned}
    G^{(i)} &= G^{(0)} - \sum_{m=1}^{i} \tilde{X}_{N\times k}^{(m)} \tilde{Y}_{k\times N}^{(m)} \\
              &= G^{(0)} - \tilde{X}_{N\times ik}^{(i)} \tilde{Y}_{ik\times N}^{(i)} \,,
\end{aligned}
\end{equation}
where $\tilde{X}_{N\times ik}^{(i)}$ and $\tilde{Y}_{ik\times N}^{(i)}$ represent the accumulated matrices from $i$ accepted updates.
Upon acceptance of the $(i+1)$-th move, the corresponding single-update matrices are computed as
\begin{equation}
\begin{aligned}
    \tilde{X}_{N\times k}^{(i+1)} &= G^{(i)} \tilde{U}^{(i+1)} \big(\tilde{S}^{(i+1)}\big)^{-1} \\
                                 &= G_{:,\{x_j^{(i+1)}\}}^{(i)} \left(P_{k\times N}^{(i+1)} \Delta^{(i+1)} P_{N\times k}^{(i+1)}\right) \big(\tilde{S}^{(i+1)}\big)^{-1} \,, \\
    \tilde{Y}_{k\times N}^{(i+1)} &= \tilde{V}^{(i+1)} = P_{k\times N}^{(i+1)} - G_{\{x_j^{(i+1)}\},:}^{(i)} \,.
\end{aligned}
\end{equation}
These matrices are subsequently appended to construct the accumulated matrices:
\begin{equation}
\begin{aligned}
    \tilde{X}_{N\times (i+1)k}^{(i+1)} &= \big[\tilde{X}_{N\times ik}^{(i)} \mid \tilde{X}_{N\times k}^{(i+1)}\big] \,, \\
    \tilde{Y}_{(i+1)k\times N}^{(i+1)} &= \begin{bmatrix} \tilde{Y}_{ik\times N}^{(i)} \\ \tilde{Y}_{k\times N}^{(i+1)} \end{bmatrix} \,.
\end{aligned}
\end{equation}

\subsection{Submatrix-G~\cite{SciPostPhys.2025Sun}}\label{app:submatrix_G}

Let $F^{(i)} \equiv I - G^{(i)}$ denote the update object after $i$ accepted updates within a delayed block, with $F^{(0)} = R^{(0)}T^{(0)}L$ representing the initial matrix\footnote{To reduce the symbols, we define a slightly different update object from Ref.~\cite{SciPostPhys.2025Sun}, where $F_{\text{SciPost}}^{(i)}\equiv R^{(0)}T^{(i)}L=(I+\Delta^{(i)}_{\mathrm{all}})^{-1}F^{(i)}$. These two matrices have very similar update formulas since they are connected by a simply diagonal matrix. }.
For the $(i+1)$-th proposed update on sites $x^{(i+1)}=\{x_1^{(i+1)}, \ldots, x_k^{(i+1)}\}$, we define $\tilde{U}^{(i+1)} = \Delta^{(i+1)} P_{N\times k}^{(i+1)}$ and $\tilde{V}^{(i+1)} = P_{k\times N}^{(i+1)}F^{(i)}$ as in Appendix~\ref{app:delay_G}. 
The acceptance ratio is expressed as
\begin{equation}
r^{(i+1)} = \det\left[I_{k} + \tilde{V}^{(i+1)} \tilde{U}^{(i+1)}\right] = \det \tilde{S}^{(i+1)},
\end{equation}
where
\begin{equation}
\begin{aligned}
\tilde{S}^{(i+1)} &= I_{k} + \tilde{V}^{(i+1)} \tilde{U}^{(i+1)} = I_{k} + \tilde{\mathcal{V}}_{k\times k}^{(i+1)}D_{k\times k}^{(i+1)},
\end{aligned}
\end{equation}
with $\tilde{\mathcal{V}}_{k\times k}^{(i+1)} = P_{k\times N}^{(i+1)}F^{(i)}P_{N\times k}^{(i+1)}$ and $D_{k\times k}^{(i+1)} = P_{k\times N}^{(i+1)}\Delta^{(i+1)}P_{N\times k}^{(i+1)}$.
The essential insight lies in expressing $\tilde{\mathcal{V}}^{(i+1)}$ through a general term formula that circumvents the explicit computation of the matrix $F^{(i)}$.
The update object evolves according to
\begin{equation}
\begin{aligned}
F^{(i)} &= \big(I+\Delta^{(i)}_{\mathrm{all}}\big)F^{(0)} \\
&\quad \times \Big[I-\tilde{U}_{N\times ik}^{(i)}\big(I+\tilde{V}_{ik\times N}^{(i)}\tilde{U}_{N\times ik}^{(i)}\big)^{-1}\tilde{V}_{ik\times N}^{(i)}\Big] \\
&= \big(I+\Delta^{(i)}_{\mathrm{all}}\big) \\
&\quad \times \Big[F^{(0)}-F^{(0)}P_{N\times ik}^{(i)}\big(\Gamma^{(i)}\big)^{-1}P_{ik\times N}^{(i)}F^{(0)}\Big],
\end{aligned}
\end{equation}
where $\tilde{U}_{N\times ik}^{(i)} = \Delta^{(i)}_{\mathrm{all}}P_{N\times ik}^{(i)}$ and $\tilde{V}_{ik\times N}^{(i)} = P_{ik\times N}^{(i)}F^{(0)}$ denote the accumulated intermediate matrices.
This formulation naturally introduces the submatrix $\Gamma^{(i)}$, which shares an identical structure with that in the submatrix-T algorithm (see Appendix~\ref{app:submatrix_T}):
\begin{equation}
\begin{aligned}
\Gamma^{(i)} &= \big(P_{ik\times N}^{(i)}\Delta^{(i)}_{\mathrm{all}}P_{N\times ik}^{(i)}\big)^{-1} \\
&\quad + P_{ik\times N}^{(i)}F^{(0)}P_{N\times ik}^{(i)}.
\end{aligned}
\end{equation}
This yields the explicit expression
\begin{equation}
\begin{aligned}
\tilde{\mathcal{V}}_{k\times k}^{(i+1)} &= P_{k\times N}^{(i+1)}F^{(0)}P_{N\times k}^{(i+1)} \\
&\quad - P_{k\times N}^{(i+1)}F^{(0)}P_{N\times ik}^{(i)}\big(\Gamma^{(i)}\big)^{-1}P_{ik\times N}^{(i)}F^{(0)}P_{N\times k}^{(i+1)},
\end{aligned}
\end{equation}
which enables efficient evaluation of the acceptance ratio. 

Upon acceptance of the $(i+1)$-th move, the submatrix $\Gamma^{(i+1)}$ must be expanded from its current $ik \times ik$ dimensions to accommodate the new $(i+1)k \times (i+1)k$ structure.
This ``growth'' of submatrix is achieved through the same block matrix inversion technique employed in submatrix-T (see Eq.~\eqref{equ:Gammainv_update}), enabling efficient computation of $\big(\Gamma^{(i+1)}\big)^{-1}$ from the existing $\big(\Gamma^{(i)}\big)^{-1}$.

\subsection{Delay-T~\cite{Phys.Rev.B2025Du}}\label{app:delay_LR}

Let $T^{(i)} \equiv (LR^{(i)})^{-1}$ denote the update object after $i$ accepted updates within a delayed block, with $T^{(0)}$ representing the initial matrix.
While the left factor $L$ remains unchanged, the right factor evolves according to $R^{(i)} = (I + \sum_{m=1}^{i} \Delta^{(m)}) R^{(0)}$ as proposed updates are accepted.
For the $(i+1)$-th proposal, we define $U^{(i+1)} = L \Delta^{(i+1)} P_{N\times k}^{(i+1)} $ and $V^{(i+1)} = P_{k\times N}^{(i+1)} R^{(i)} T^{(i)}$. 
Since different updates operate on non-overlapping sites, we have $P_{N\times k}^{(j)}R^{(i)}=P_{N\times k}^{(j)}R^{(0)}$ for $j>i$, allowing the acceptance ratio to be computed as
\begin{equation}\label{equ:delay_LR_ratio}
    r^{(i+1)}=\det\left[I_{k}+V^{(i+1)}U^{(i+1)}\right]=\det S^{(i+1)},
\end{equation}
where
\begin{equation}\label{equ:delay_LR_S}
    \begin{aligned}
        S^{(i+1)}&=I_{k}+\big(P_{k\times N}^{(i+1)}R^{(0)}\big)\big(T^{(0)}L\Delta^{(i+1)}P_{N\times k}^{(i+1)}\big)\\&\quad-\sum_{n=1}^{i}\big(P_{k\times N}^{(i+1)}R^{(0)}X_{N_{p}\times k}^{(n)}\big)\big(Y_{k\times N_{p}}^{(n)}L\Delta^{(i+1)}P_{N\times k}^{(i+1)}\big).
    \end{aligned}
\end{equation}
This expression utilizes the recursive expression of the $T$ matrix:
\begin{equation}\label{equ:delay_LR_T_recursive}
\begin{aligned}
    T^{(i)} &= T^{(0)} - \sum_{m=1}^{i} X_{N_{p}\times k}^{(m)} Y_{k\times N_{p}}^{(m)} \\
              &= T^{(0)} - X_{N_{p}\times ik}^{(i)} Y_{ik\times N_{p}}^{(i)} \,,
\end{aligned}
\end{equation}
where $X_{N_{p}\times ik}^{(i)}$ and $Y_{ik\times N_{p}}^{(i)}$ represent the accumulated matrices from $i$ accepted updates.
Upon acceptance of the $(i+1)$-th move, the corresponding single-update matrices $X_{N_{p}\times k}^{(i+1)}$ and $Y_{k\times N_{p}}^{(i+1)}$ are evaluated as
\begin{equation}\label{equ:delay_LR_X_update}
    \begin{aligned}
        &X_{N_{p}\times k}^{(i+1)} = T^{(i)} U^{(i+1)} \left(S^{(i+1)}\right)^{-1} \\
        &= \left[T^{(0)} L \Delta^{(i+1)} P_{N\times k}^{(i+1)} - \sum_{n=1}^{i} X_{N_{p}\times k}^{(n)} \big(Y_{k\times N_{p}}^{(n)} L \Delta^{(i+1)} P_{N\times k}^{(i+1)}\big)\right]\\
        &\quad\times \left(S^{(i+1)}\right)^{-1},
    \end{aligned}
\end{equation}
\begin{equation}\label{equ:delay_LR_Y_update}
    \begin{aligned}
        &Y_{k\times N_{p}}^{(i+1)} = P_{k\times N}^{(i+1)} R^{(i)} T^{(i)} \\
        &= P_{k\times N}^{(i+1)} R^{(0)} T^{(0)} - \sum_{n=1}^{i} \big(P_{k\times N}^{(i+1)} R^{(0)} X_{N_{p}\times k}^{(n)}\big) Y_{k\times N_{p}}^{(n)}.
    \end{aligned}
\end{equation}
These matrices are then concatenated to form the accumulated matrices:
\begin{equation}
\begin{aligned}
    X_{N_{p}\times (i+1)k}^{(i+1)} &= \big[X_{N_{p}\times ik}^{(i)} \mid X_{N_{p}\times k}^{(i+1)}\big] \,, \\
    Y_{(i+1)k\times N_{p}}^{(i+1)} &= \begin{bmatrix} Y_{ik\times N_{p}}^{(i)} \\ Y_{k\times N_{p}}^{(i+1)} \end{bmatrix} \,.
\end{aligned}
\end{equation}

Following the approach in Ref.~\cite{Phys.Rev.B2025Du}, the delay-T algorithm pre-computes selected rows of $R^{(0)}T^{(0)}$ and columns of $T^{(0)}L^{(0)}$ at the beginning of each delayed block to reduce BLAS-2 operations during ratio calculations and matrix updates.
Table~\ref{tab:delay_LR_complexity} provides a detailed breakdown of the delay-T algorithm's computational complexity, categorized by operation type and BLAS level.

\begin{table}[htbp]
    \centering
    \caption{Detailed computational complexity breakdown for the delay-T algorithm. The complexities are given per time slice and the BLAS levels are based on $k=1$ case. $p_a$ denotes the acceptance probability of local updates (see Appendix~\ref{app:submatrix_T} for details).}
    \label{tab:delay_LR_complexity}
    \begin{tabular}{p{4.0cm}p{2.5cm}c}
    \toprule
    Operation & Complexity & BLAS Level \\
    \midrule
    Full update \newline (Eq.~\eqref{equ:delay_LR_T_recursive}) & $\mathcal{O}(p_aNN_{p}^{2})$ & BLAS-3 \\
    Pre-computation \newline ($R^{(0)}T^{(0)}$~and~$T^{(0)}L^{(0)}$) & $\mathcal{O}(2NN_{p}^{2})$ & BLAS-3 \\
    Ratio calculation \newline (Eq.~\eqref{equ:delay_LR_S}) & $\mathcal{O}(n_{d}NN_{p})$ & BLAS-2 \\
    Update matrices \newline (Eqs.~\eqref{equ:delay_LR_X_update}, \eqref{equ:delay_LR_Y_update}) & $\mathcal{O}(p_{a}^{2}n_{d}NN_{p})$ & BLAS-2 \\
    \bottomrule
    \end{tabular}
\end{table}

%%%%%%%%%%%%%%%%%%%%%%%%%%%%%%%%%%%%%%%%%%%%%%%%%%%%%
\section{Submatrix-T update algorithm for PQMC}\label{app:submatrix_T}
%%%%%%%%%%%%%%%%%%%%%%%%%%%%%%%%%%%%%%%%%%%%%%%%%%%%%
In this section, we provide complete derivations, implementation details, computational complexity analysis, efficacy benchmarks, \rev{and numerical-stability tests,} for the submatrix-T algorithm introduced in Appendix~\ref{app:methods}.

\subsection{Complete derivations}
We consider how to write down the update formula (as in Eq.~\eqref{equ:fast_LR_update}) and ratio formula using the original $T$ without any updating.
To express these formulas, it is convenient to introduce a superscript $[\cdot]^{(i)}$ which labels quantities after $i$ accepted moves or quantities at the proposed move after $i-1$ accepted moves in the current delayed block.
The lattice sites being updated or may be updated is denoted by $x^{(i)} = \{x_1^{(i)}, \ldots, x_k^{(i)}\}$, which further determines the difference $\Delta^{(i)}$ and the index matrix $P_{N\times k}^{(i)}$.
Other examples include the update object $T^{(i)}=(LR^{(i)})^{-1}$ and intermediate matrices like $U^{(i)}\equiv L^{(i-1)}\Delta^{(i)}P_{N\times k}^{(i)}$ and $V^{(i)}\equiv P_{k\times N}^{(i)}R^{(i-1)}T^{(i-1)}$.

The most essential insight for designing submatrix-T update is that $L$, $R$ and $T$ can find their general term formulas after arbitrary number of accepted local updates.
Within our sweep strategy, at a given time slice, $L$ does not update so $L^{(i)}=L^{(0)}=L$ (c.f. Eq.~\eqref{equ:PQMC_r_LRtau}).
Since different local updates operate on disjoint lattice sites, the matrix $R$ can be updated by successive multiplication:
\begin{equation}\label{equ:R_general_term}
R^{(i)} = \prod_{m=1}^{i}(I+\Delta^{(m)})R^{(0)} = \left(I+\Delta^{(i)}_{\mathrm{all}}\right)R^{(0)},
\end{equation}
where $\Delta^{(i)}_{\mathrm{all}} = \sum_{m=1}^{i}\Delta^{(m)}$ is the accumulated diagonal matrix containing $ik$ non-zero elements.
This operation is computationally inexpensive, requiring only element-wise scaling of the affected rows of $R^{(0)}$.
Another result is that at next proposed move, since $P_{N\times k}^{(i+1)}$ do not overlap with $\Delta^{(i)}_{\mathrm{all}}$, we have $P_{N\times k}^{(i+1)}R^{(i)}=P_{N\times k}^{(i+1)}R^{(0)}$, or more generally,
\begin{equation}\label{equ:PjRi}
    P_{N\times k}^{(j)}R^{(i)}=P_{N\times k}^{(j)}R^{(0)}\text{ provided that }j>i.
\end{equation}
The general term formula for $T$ directly follows from Eq.~\eqref{equ:R_general_term}. After utilizing the Woodbury matrix identity, we obtain:
\begin{equation}\label{equ:T_general_term}
\begin{aligned}
T^{(i)} &= \big(L(I+\Delta^{(i)}_{\mathrm{all}})R^{(0)}\big)^{-1} \\
&= T^{(0)} - T^{(0)}U_{N_p\times ik}^{(i)} \\
&\quad \times \big(I_{ik}+V_{ik\times N_p}^{(i)}U_{N_p\times ik}^{(i)}\big)^{-1}V_{ik\times N_p}^{(i)} \,,
\end{aligned}
\end{equation}
where the accumulated intermediate matrices are defined as $U_{N_p\times ik}^{(i)} = L\Delta^{(i)}_{\mathrm{all}}P_{N\times ik}^{(i)}$ and $V_{ik\times N_p}^{(i)} = P_{ik\times N}^{(i)}R^{(0)}T^{(0)}$.
Here, $P_{N\times ik}^{(i)}$ and $P_{ik\times N}^{(i)}$ are the accumulated index matrices which are constructed by concatenating the index matrices from each accepted update:
\begin{equation}
\begin{aligned}
	P_{N\times ik}^{(i)} & =\left[P_{N\times k}^{(1)}|P_{N\times k}^{(2)}|\cdots|P_{N\times k}^{(i)}\right] \\
	                     & =\left[e_{x_{1}^{(1)}}|\cdots|e_{x_{k}^{(1)}}|e_{x_{1}^{(2)}}|\cdots|e_{x_{k}^{(2)}}|\cdots|e_{x_{1}^{(i)}}|\cdots|e_{x_{k}^{(i)}}\right],\\
    P_{ik\times N}^{(i)}&=(P_{N\times ik}^{(i)})^T.
\end{aligned}
\end{equation}
Crucially, the order of columns in $P_{N\times ik}^{(i)}$ is determined by the sequence of accepted moves, and does not necessarily correspond to the ordered indices of the non-zero elements in $\Delta^{(i)}_{\mathrm{all}}$. This is permissible due to the permutation invariance: for any $k\times k$ permutation matrix $\mathbf{S}_k$, we have $\Delta=\Delta\left(P_{N\times k}\mathbf{S}_k\right)\left(\mathbf{S}_k^{T}P_{k\times N}\right)$.

The next step of designing the submatrix-T update algorithm is to figure out how to compute the acceptance ratio for next proposed move recursively.
We start with the fast update ratio formula $r=\det[I_k+VU]$:
\begin{equation}\label{equ:submatrix_LR_ratio}
\begin{aligned}
r^{(i+1)} &= \frac{\det[L R^{(i+1)}]}{\det[L R^{(i)}]} = \det[I_k+V_{k\times N_p}^{(i+1)}U_{N_p\times k}^{(i+1)}] \\
&= \det[I_k+\mathcal{V}_{k\times k}^{(i+1)}D_{k\times k}^{(i+1)}] \,,
\end{aligned}
\end{equation}
where we further factorize the $k \times k$ determinant with matrices $\mathcal{V}_{k\times k}^{(i+1)}\equiv P_{k\times N}^{(i+1)}F^{(i)}P_{N\times k}^{(i+1)}$ and $D_{k\times k}^{(i+1)}\equiv P_{k\times N}^{(i+1)}\Delta^{(i+1)}P_{N\times k}^{(i+1)}$.
The $N\times N$ matrix $F^{(i)}\equiv R^{(i)}T^{(i)}L$ has a similar role as the Green's function $G^{(i)}$ (which equals to $(I-F^{(i)})$), and it is hired as the update object in the submatrix-G algorithm.
Its general term formula follows from Eqs.~\eqref{equ:R_general_term} and \eqref{equ:T_general_term}:
\begin{equation}
    \begin{aligned}
        F^{(i)}=&\left(I+\Delta_{\mathrm{all}}^{(i)}\right)F^{(0)}\Big[I-\Delta_{\mathrm{all}}^{(i)}P_{N\times ik}^{(i)}\\&\times\left(I_{ik}+P_{ik\times N}^{(i)}F^{(0)}\Delta_{\mathrm{all}}^{(i)}P_{N\times ik}^{(i)}\right)^{-1}P_{ik\times N}^{(i)}F^{(0)}\Big].
    \end{aligned}
\end{equation}
Substituting this into $\mathcal{V}^{(i+1)}$ and utilizing Eq.~\eqref{equ:PjRi} yields
\begin{equation}\label{equ:Vcal_general_term}
\begin{aligned}
\mathcal{V}_{k\times k}^{(i+1)} &= P_{k\times N}^{(i+1)}F^{(i)}P_{N\times k}^{(i+1)} \\
&= P_{k\times N}^{(i+1)}F^{(0)}P_{N\times k}^{(i+1)} \\
&\quad - P_{k\times N}^{(i+1)}F^{(0)}P_{N\times ik}^{(i)}\big(\Gamma^{(i)}\big)^{-1}P_{ik\times N}^{(i)}F^{(0)}P_{N\times k}^{(i+1)}.
\end{aligned}
\end{equation}
where we identify the $ik \times ik$ submatrix $\Gamma^{(i)}$ defined as:
\begin{equation}\label{equ:Gamma}
    \begin{aligned}
        \Gamma^{(i)}&=\left(P_{ik\times N}^{(i)}\Delta_{\mathrm{all}}^{(i)}P_{N\times ik}^{(i)}\right)^{-1}+P_{ik\times N}^{(i)}F^{(0)}P_{N\times ik}^{(i)}\\&=\left(\begin{array}{ccc}
        \left(D_{k\times k}^{(1)}\right)^{-1}\\
            & \ddots\\
            &  & \left(D_{k\times k}^{(i)}\right)^{-1}
        \end{array}\right)+P_{ik\times N}^{(i)}F^{(0)}P_{N\times ik}^{(i)}.
    \end{aligned}
\end{equation}
This submatrix $\Gamma^{(i)}$ is the same as that in the submatrix-G algorithm~\cite{SciPostPhys.2025Sun}.
In addition to calculating acceptance ratios, it also aids in performing full update of $T$. Following Eq.~\eqref{equ:T_general_term} we have
\begin{equation}\label{equ:T_general_term2}
\begin{aligned}
T^{(i)} = T^{(0)} - T^{(0)}\big(LP_{N\times ik}^{(i)}\big)\big(\Gamma^{(i)}\big)^{-1}\big(P_{ik\times N}^{(i)}R^{(0)}T^{(0)}\big).
\end{aligned}
\end{equation}

The first term of $\Gamma^{(i)}$ is a block-diagonal matrix determined by the difference $\Delta$ of each accepted move, while the second term simply comprises selected elements of $F^{(0)}$.
As $i$ increases, $\Gamma^{(i)}$ simply expands and satisfies the following recursive relation:
\begin{equation}
    \Gamma^{(i+1)}=\begin{pmatrix}\Gamma^{(i)} & F_{ik,k}^{(i+1)}\\
        F_{k,ik}^{(i+1)} & (D_{k\times k}^{(i+1)})^{-1}+F_{k,k}^{(i+1)}
        \end{pmatrix},
\end{equation}
with short-hand notation for blocks of $F^{(0)}$:
\begin{equation}\label{equ:F_elements}
    \begin{aligned}
        F_{ik,k}^{(i+1)}&\equiv P_{ik\times N}^{(i)}F^{(0)}P_{N\times k}^{(i+1)},\\F_{k,ik}^{(i+1)}&\equiv P_{k\times N}^{(i+1)}F^{(0)}P_{N\times ik}^{(i)},\\F_{k,k}^{(i+1)}&\equiv P_{k\times N}^{(i+1)}F^{(0)}P_{N\times k}^{(i+1)}.
    \end{aligned}
\end{equation}
Using the block matrix inversion identity, $(\Gamma^{(i)})^{-1}$ can also be updated efficiently~\cite{SciPostPhys.2025Sun}:
\begin{equation}\label{equ:Gammainv_update}
    \left(\Gamma^{(i+1)}\right)^{-1}=\begin{pmatrix}\left(\Gamma^{(i)}\right)^{-1}\Upsilon & -\left(\Gamma^{(i)}\right)^{-1}F_{ik,k}^{(i+1)}\Sigma\\
        -\Sigma F_{k,ik}^{(i+1)}\left(\Gamma^{(i)}\right)^{-1} & \Sigma
        \end{pmatrix},
\end{equation}
where
\begin{equation}
    \begin{aligned}
        \Upsilon&\equiv\left(I+F_{ik,k}^{(i+1)}\Sigma F_{k,ik}^{(i+1)}\left(\Gamma^{(i)}\right)^{-1}\right),\\\Sigma&\equiv\left(\left(D_{k\times k}^{(i+1)}\right)^{-1}+F_{k,k}^{(i+1)}-F_{k,ik}^{(i+1)}\left(\Gamma^{(i)}\right)^{-1}F_{ik,k}^{(i+1)}\right)^{-1}\\&=D_{k\times k}^{(i+1)}\left(I_{k}+\mathcal{V}_{k\times k}^{(i+1)}D_{k\times k}^{(i+1)}\right)^{-1}.
    \end{aligned}
\end{equation}
We note that the matrix $(I_k+\mathcal{V}_{k\times k}^{(i+1)}D_{k\times k}^{(i+1)})$ also appears in the acceptance ratio in Eq.~\eqref{equ:submatrix_LR_ratio}.

We use Eqs.~\eqref{equ:submatrix_LR_ratio} and \eqref{equ:Vcal_general_term} to compute acceptance ratios.
We keep track of $(\Gamma^{(i)})^{-1}$ and use Eq.~\eqref{equ:Gammainv_update} to update it after each accepted move.
Finally, we use Eqs.~\eqref{equ:R_general_term} and \eqref{equ:T_general_term2} to perform full update of $R$ and $T$ respectively after a full delayed block.
These formulas form the core of the submatrix-T algorithm illustrated in Fig.~\ref{fig:submatrix_LR_scheme}.

\subsection{Practical implementation}
Unlike Refs.~\cite{Phys.Rev.B2024Sun} and \cite{SciPostPhys.2025Sun}, we interpret $n_d$ as the number of proposed updates, rather than accepted updates, within a delayed block. 
The number of accepted updates within a delayed block is denoted as $n_a$ (see Fig.~\ref{fig:submatrix_LR_scheme}).
This strategy was discussed and also utilized in the delay-T algorithm~\cite{Phys.Rev.B2025Du}. 

\begin{figure}[htbp]
    \centering
    \includegraphics[width=0.48\textwidth]{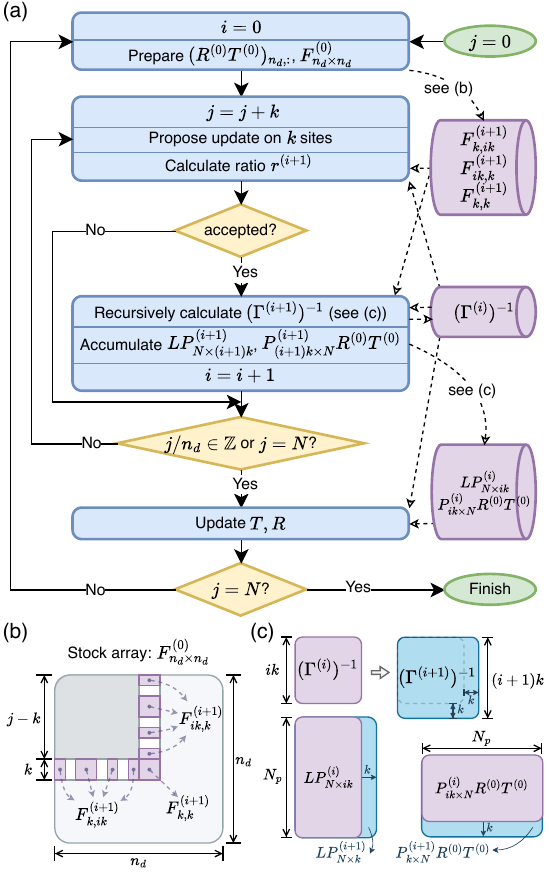}
    \caption{(a) Flowchart of the submatrix-T algorithm implementation. Solid arrows indicate the control flow, while dashed arrows indicate the data flow.
    (b) When proposing a local update after $i$ accepted moves, we extract $F^{(i+1)}_{k,ik}$, $F^{(i+1)}_{ik,k}$, and $F^{(i+1)}_{k,k}$ from the pre-computed array $F^{(0)}_{n_d\times n_d}$. 
    White rectangles between the extracted purple blocks represent redundant calculations arising from rejected updates. 
    (c) Upon accepting a local update, we recursively update the submatrix $(\Gamma^{(i)})^{-1}$ and append columns and rows to the accumulated matrices $LP^{(i)}_{N\times ik}$ and $P^{(i)}_{ik\times N}R^{(0)}T^{(0)}$, respectively. }
    \label{fig:submatrix_LR_flowchart}
\end{figure}

The complete workflow is shown in Fig.~\ref{fig:submatrix_LR_flowchart}(a).
Our implementation employs three types of storage arrays, represented by the purple cylinders in the flowchart, to optimize performance. 
We pre-compute $n_d$ columns of $R^{(0)}T^{(0)}$ and a corresponding $n_d\times n_d$ slice of $F^{(0)}=(R^{(0)}T^{(0)})L$ at the beginning of each delayed block. 
The specific locations of these columns and slices are determined by the sequence of proposed updates. 
These pre-computed data enables efficient calculation of acceptance ratios and updates to $\Gamma^{-1}$, providing the matrix elements of $F^{(0)}$ required in Eqs.~\eqref{equ:Vcal_general_term} and \eqref{equ:Gammainv_update}, as illustrated in Fig.~\ref{fig:submatrix_LR_flowchart}(b). 
Pre-computation avoids numerous low-rank matrix-vector multiplications that would otherwise occur when computing these elements on-the-fly from rows of $R^{(0)}T^{(0)}$ and columns of $L$, following the same motivation as all delayed update algorithms. 
However, rejected updates lead to redundant calculations, making this approach most effective for models with high acceptance probability $p_a=kn_a/n_d$~\cite{Phys.Rev.B2025Du}. 
Figure~\ref{fig:submatrix_LR_flowchart}(c) illustrates the updating process for the latter two kinds of storage arrays.
The second storage array maintains the submatrix $(\Gamma^{(i)})^{-1}$, which facilitates both acceptance ratio calculations and updates to $T$. 
This submatrix is updated via Eq.~\eqref{equ:Gammainv_update} following each accepted local update. 
The third kind of storage arrays accumulate the matrices $LP^{(i)}_{N\times ik}$ and $P^{(i)}_{ik\times N}R^{(0)}T^{(0)}$ after each accepted move, aiding in full updates of $T$ at the end of each delayed block through Eq.~\eqref{equ:T_general_term2}. 

\subsection{Computational complexity}

In Table~\ref{tab:submatrix_LR_complexity}, we provide a detailed breakdown of the submatrix-T algorithm's computational complexity, categorized by operation type and BLAS level.
The full update operations use Eq.~\eqref{equ:T_general_term2} to update $T$ after each delayed block.
The complexity of this formula is $\mathcal{O}[2N_p^2kn_a + N_p(kn_a)^2]$, and since $N/n_d$ such full updates are performed, the total complexity per time slice becomes $\mathcal{O}(2p_aNN_p^2 + p_a^2n_dNN_p)$.
Other computational complexity comes from three main contributions. 
First, pre-computing selected rows of $R^{(0)}T^{(0)}$ and elements of $F^{(0)}$ costs $\mathcal{O}(n_dN_p^2+n_d^2N_p)$ per delayed block and $ \mathcal{O}(N_p^2N+n_dNN_p)$ per time slice.
Second, using Eq.~\eqref{equ:Vcal_general_term} to compute acceptance ratios costs $ \mathcal{O}(n_d^2 N)$ per time slice. 
Third, updating the submatrix $(\Gamma^{(i)})^{-1}$ using Eq.~\eqref{equ:Gammainv_update} after each accepted move also contributes $\mathcal{O}(n_d^2N)$ per time slice, where we assume that $n_a$ is comparable to $n_d$ and $n_d\gg1$. 

We now compare the theoretical performance of different delayed update algorithms.
Compared to delay-G~\cite{Phys.Rev.B2024Sun} and submatrix-G~\cite{SciPostPhys.2025Sun} algorithms, submatrix-T benefits from updating the $(LR)^{-1}$ object instead of $G$, reducing the update scaling from $\mathcal{O}(N^3)$ to $\mathcal{O}(NN_p^2)$.
Table~\ref{tab:complexity_comparison} focuses on the more subtle comparison with delay-T~\cite{Phys.Rev.B2025Du}, which shares the same update object.
Submatrix-T offers two advantages: the prefactor of the leading scaling $\mathcal{O}(NN_p^2)$ reduces from $(2+p_a)$ to $(1+2p_a)$, and more importantly, the subleading scaling $\mathcal{O}(n_dNN_p)$ convert from BLAS-2 to BLAS-3.
The trade-off is additional $\mathcal{O}(n_d^2N)$ BLAS-2 overhead from submatrix updates.
Since this overhead scales less sensitively with $n_d$ than $\mathcal{O}(n_d NN_p)$, submatrix-update algorithms can achieve larger optimal $n_d$ values, more effectively exploiting the efficiency of BLAS-3 operations~\cite{SciPostPhys.2025Sun}.

\begin{table}[htbp]
\centering
\caption{Computational complexity breakdown for the submatrix-T algorithm. The complexities are given per time slice and the BLAS levels are based on $k=1$ case. }
\label{tab:submatrix_LR_complexity}
\begin{tabular}{p{3.0cm}cc}
\toprule
Operation & Complexity & BLAS Level \\
\midrule
Full update \newline (Eq.~\eqref{equ:T_general_term2}) & $\mathcal{O}(2p_{a}NN_{p}^{2}+p_{a}^{2}n_{d}NN_{p})$ & BLAS-3 \\
Pre-computation \newline ($R^{(0)}T^{(0)}$ and $F^{(0)}$) & $\mathcal{O}(NN_{p}^{2}+n_{d}NN_{p})$ & BLAS-3 \\
Ratio calculation \newline (Eq.~\eqref{equ:Vcal_general_term}) & $\mathcal{O}(n_{d}^{2}N)$ & BLAS-2\&1 \\
Submatrix update \newline (Eq.~\eqref{equ:Gammainv_update}) & $\mathcal{O}(n_{d}^{2}N)$ & BLAS-2\&1 \\
\bottomrule
\end{tabular}
\end{table}

\begin{table}[htbp]
\centering
\caption{Computational complexity comparison of fast, delay-T and submatrix-T update algorithms. The complexities are given per time slice and categorized by BLAS level when $k=1$.}
\label{tab:complexity_comparison}
\begin{tabular}{p{2.8cm}p{3.0cm}c}
\toprule
Algorithm & BLAS-3 & BLAS-1\&2 \\
\midrule
Fast update & $\mathcal{O}(0)$ & $\mathcal{O}[(1+2p_{a})NN_{p}^{2}]$ \\
delay-T & $\mathcal{O}[(2+p_{a})NN_{p}^{2}]$ & $\mathcal{O}(n_{d}NN_{p})$ \\
submatrix-T & $\mathcal{O}[(1+2p_{a})NN_{p}^{2} ]$ \newline $+\mathcal{O}(n_{d}NN_{p})$ & $\mathcal{O}(n_{d}^{2}N)$ \\
\bottomrule
\end{tabular}
\end{table}

\subsection{Efficacy benchmarks for submatrix-T algorithm}\label{app:optimal_nd}

We conducted tests on a computing node with CPU [$2\times$ Intel Xeon ICX Platinum 8358 (2.6GHz, 32 cores, 48MB Intel Cache)] and memory [$16\times 32$ GB TruDDR4 3200 MHz (2R $\times$ 8 1.2V) RDIMM] to determine optimal $n_d$ values and benchmark different update algorithms for the Hubbard model on honeycomb lattices of different sizes.
The simulation parameters were fixed at $U/t=3.5$, $\beta t=4$, and $\Delta_\tau t=0.1$. We test $n_d$ from $4$ to $544$.
We used exclusive node allocation with MPI parallelization across 64 processes, where each process runs an independent Markov chain.
To ensure fair comparison, all test groups employed the same initial random seed, yielding identical simulation process and outcomes.
Performance measurements were obtained by running 10 sweeps and averaging the per-sweep update time.

The optimal $n_d$ is hardware-dependent, primarily constrained by cache size.
Our benchmark over various system sizes shows that the empirical formula $n_d = 2^{\lambda}$ with $\lambda = \max\{\lceil\log_2 \frac{N_p}{14}\rceil, 5\}$ yields good performance.
Table~\ref{tab:optimal_nd_comparison} presents the optimal $n_d$ values and corresponding per-sweep times for different update algorithms across various honeycomb lattice sizes. 
One can see that the optimal $n_d$ of submatrix-T is generally larger than those of delay-T, which is consistent with our complexity analysis in Table~\ref{tab:complexity_comparison}.

\begin{table}[htbp]
\centering
\caption{Optimal $n_d$ values and per-sweep update times for different algorithms on honeycomb lattices. System sizes are characterized by $L \times L$ unit cells with $N_s = 2L^2$ sites.}
\label{tab:optimal_nd_comparison}
\begin{tabular}{cccccc}
\toprule
$L$ & Fast update & \multicolumn{2}{c}{Delay-T} & \multicolumn{2}{c}{Submatrix-T} \\
\cmidrule(lr){3-4} \cmidrule(lr){5-6}
& Time (s) & Optimal $n_d$ & Time (s) & Optimal $n_d$ & Time (s) \\
\midrule
6  & 0.0114 & 32  & 0.0145 & 32  & 0.0134 \\
9  & 0.1336 & 16  & 0.0766 & 32  & 0.0606 \\
12 & 0.6557 & 16  & 0.3165 & 32  & 0.2590 \\
15 & 2.7332 & 32  & 1.1697 & 32  & 0.9531 \\
18 & 17.4867 & 32  & 3.6227 & 32  & 3.1229 \\
21 & 57.2817 & 64  & 9.2109 & 96  & 7.8123 \\
24 & 130.8077 & 64  & 20.4088 & 192 & 15.4356 \\
27 & 526.7491 & 64  & 38.8925 & 192 & 29.1392 \\
30 & 1151.7527 & 96  & 71.5170 & 192 & 51.2235 \\
33 & 2065.6961 & 96  & 119.7856 & 192 & 86.0148 \\
36 & 3628.5876 & 96  & 198.9442 & 192 & 141.4112 \\
39 & -- & 96  & 309.3411 & 192 & 215.6543 \\
42 & -- & 96  & 473.0774 & 192 & 337.0433 \\
\bottomrule
\end{tabular}
\end{table}

\begin{figure}[htbp]
    \centering
    \includegraphics[width=0.48\textwidth]{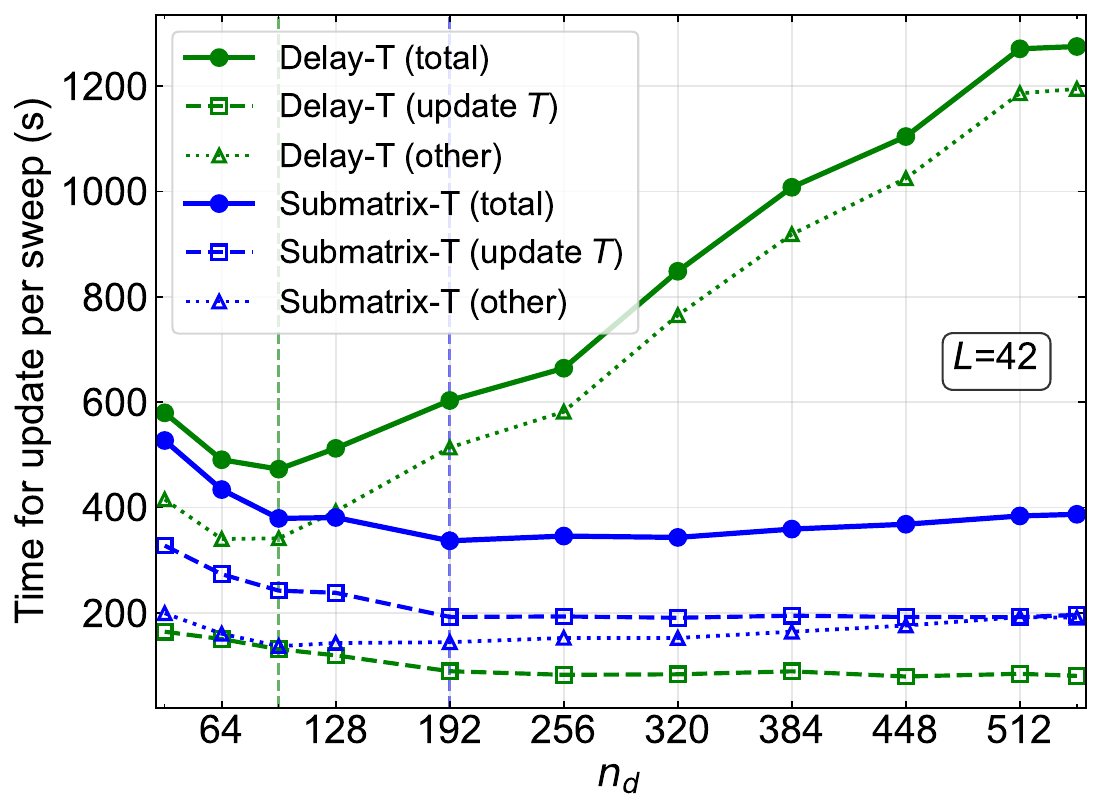}
    \caption{Per-sweep update time breakdown for delay-T and submatrix-T algorithms as a function of $n_d$ on a honeycomb lattice with $L=42$. 
    The total time (solid circles and lines) is decomposed into update-$T$ (hollow squares and dashed lines) and other operations (hollow triangles and dot-dashed lines). 
    Dashed vertical lines indicate the optimal $n_d$ values for each algorithm. }
    \label{fig:optimal_nd_example}
\end{figure}

For clearer illustration, we take $L=42$ as an example to show the update time breakdown for different $n_d$ values in Fig.~\ref{fig:optimal_nd_example}, where the total per-sweep time is decomposed into two components: update-$T$ and other operations (ratio calculations, submatrix updates, etc.).
As expected, the update-$T$ time of submatrix-T is consistently larger than that of delay-T.
However, as $n_d$ increases, the other operations of submatrix-T become significantly more efficient than those of delay-T, leading to better overall performance.
The optimal $n_d$ values (marked by dashed vertical lines) demonstrate that submatrix-T can effectively utilize larger $n_d$ values ($n_d=192$) compared to delay-T ($n_d=96$).
Furthermore, a key advantage of the submatrix-T algorithm is its robustness: performance remains nearly constant across a wide range of large $n_d$ values~\cite{SciPostPhys.2025Sun}.
This robustness obviates the need for fine-tuning; identifying a broad optimal range for $n_d$ is sufficient for practical implementation.

\begin{figure}[tbp]
    \centering
    \includegraphics[width=0.48\textwidth]{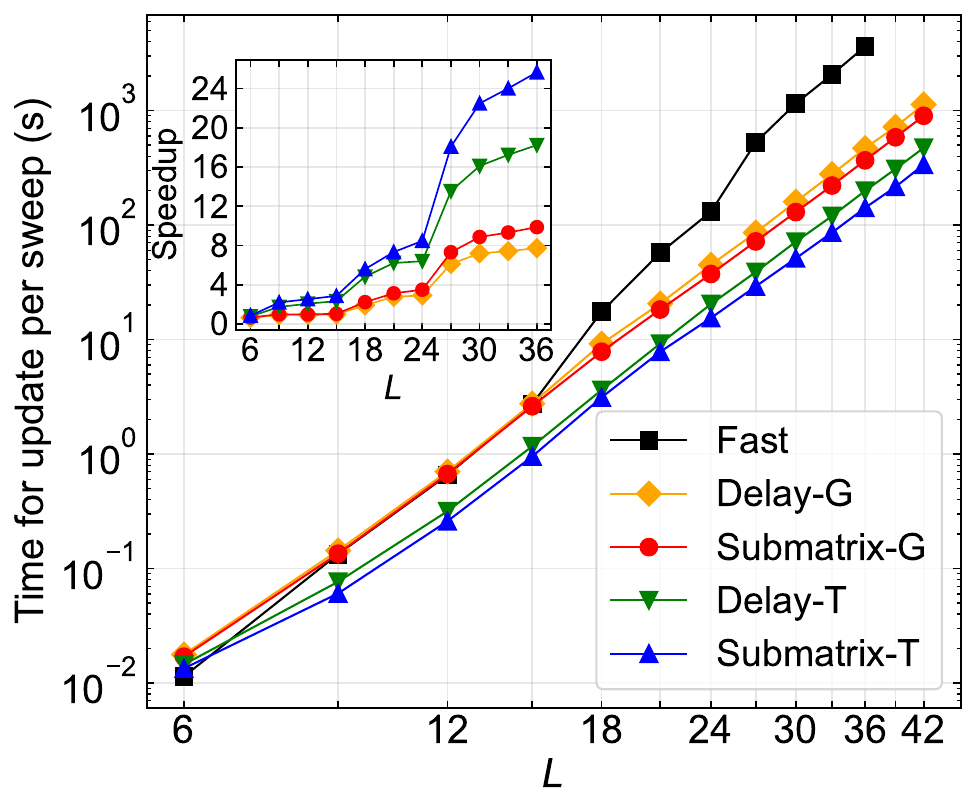}
    \caption{Per-sweep update time comparison for different update algorithms in PQMC simulations of the Hubbard model on honeycomb lattices \rev{($U/t=3.5$, $\beta t=4$, $\Delta_\tau t=0.1$)}.
    All delayed-update algorithms of various system sizes use their respective optimal $n_d$ values, and the inset illustrates the speedups achieved by the delayed-update algorithms compared to the fast update.}
    \label{fig:update_time_benchmark}
\end{figure}

\rev{With each delayed-update algorithm at its respective optimal $n_d$, Fig.~\ref{fig:update_time_benchmark} benchmarks the update time of the submatrix-T algorithm against the fast update and the other delayed-update algorithms.}
Submatrix-T consistently outperforms all alternatives.
\rev{For the $L=36$ honeycomb lattice in single-threaded execution, submatrix-T achieves more than $24\times$ speedup over the fast update, and this factor increases with system size.}

%%%%%%%%%%%%%%%%%%%%%%%%%%%%%%%%%%%%%%%%%%%%%%%%%%%%%
\subsection{\rev{Numerical stability of the submatrix-T update}}\label{app:numerical_stability}
%%%%%%%%%%%%%%%%%%%%%%%%%%%%%%%%%%%%%%%%%%%%%%%%%%%%%
\rev{This subsection summarizes the numerical-stability tests of the submatrix-T update, from fixed-seed regression tests to production runs.
During the development stage, the submatrix-T backend was validated against the fast update by fixed-seed regression tests on small systems of both the Hubbard and the $t$-$V$ model.
The two backends produce byte-identical auxiliary-field trajectories, and every intermediate quantity in the output---the per-proposal acceptance ratios, the intermediate matrices, the stabilization diagnostics---is identical to all printed digits, as are all binned observables.}

\rev{At large system sizes, where outputting every intermediate quantity becomes impractical, we quantify numerical stability directly with two diagnostics.
In our code, numerical stabilization is performed every $n_{\mathrm{wrap}}=10$ time slices by a fresh QR factorization of the accumulated matrix product.
At each stabilization point, the propagated Green's function and weight phase are compared with their freshly recomputed counterparts, and the maximal differences define the wrap error $\epsilon_{\mathrm{wrap}}=\max_{ij}|G^{\mathrm{prop}}_{ij}-G^{\mathrm{stab}}_{ij}|$ and the weight-phase drift $\epsilon_{\mathrm{phase}}=\max|\mathrm{phase}^{\mathrm{prop}}-\mathrm{phase}^{\mathrm{stab}}|$.
Each diagnostic thus covers the error accumulated over $n_{\mathrm{wrap}}$ time slices, far beyond a single delayed block, which spans only part of one time slice.
We have benchmarked the submatrix-T update against the fast update on the spinless $t$-$V$ model, where stabilization is most delicate, at fixed Trotter step, over several medium system sizes ($L\leq12$) and interaction strengths up to $V/t=4$.
In these benchmarks, the submatrix-T update reproduces the fast update to all printed digits in both diagnostics, and both diagnostics are independent of the delayed-block size, unchanged from $n_d=4$ up to $n_d=N_{\mathrm{s}}$.}

\rev{To demonstrate that the submatrix-T update remains unbiased and numerically stable in the production regime, we have performed additional control runs of the fast update.
They cover six interaction strengths around the critical point for the Hubbard model at $L=24$ and for the spinless $t$-$V$ model at $L=18$, using the same projection length, Trotter step, stabilization frequency, and binning as the production simulations.}

\begin{figure}[htbp]
    \centering
    \includegraphics[width=\columnwidth]{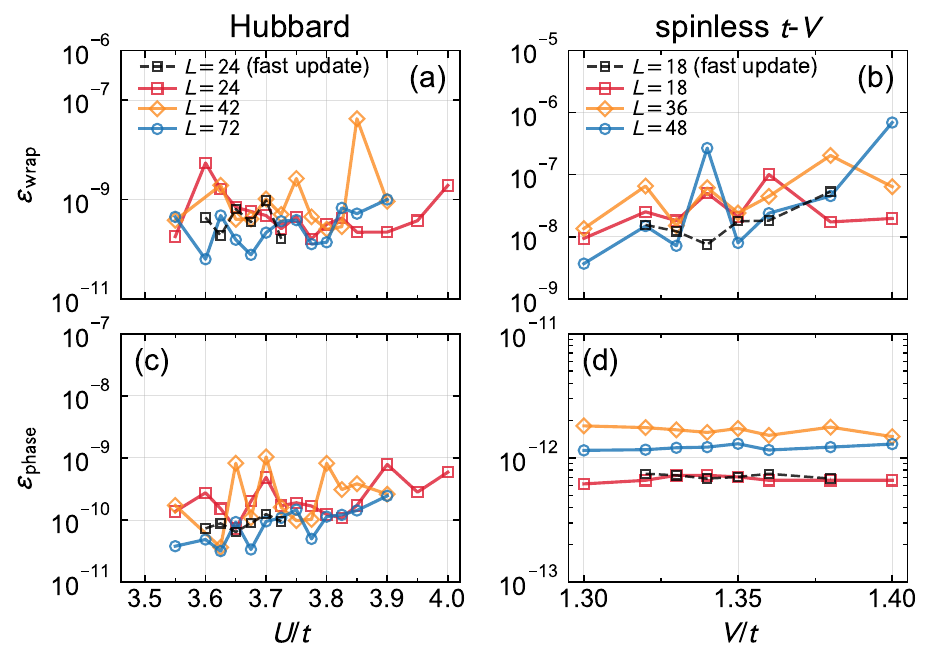}
    \caption{\rev{Stabilization diagnostics of representative production runs behind this work (submatrix-T update; e.g., the data shown in Figs.~\ref{fig:correlation_ratio_crossing}--\ref{fig:fit_extrapolation_hubbard_comparison} of the main text).
    Panels (a) and (c) refer to the Hubbard model, (b) and (d) to the spinless $t$-$V$ model; the top row gives the maximal wrap error $\epsilon_{\mathrm{wrap}}$ and the bottom row the maximal weight-phase drift $\epsilon_{\mathrm{phase}}$, versus interaction strength in the critical windows.
    Colored curves trace the worst value over each run at one specific interaction strength, for three representative sizes.
    Dashed curves show the same diagnostics for the fast-update control runs ($L=24$, Hubbard; $L=18$, $t$-$V$).}}
    \label{fig:stability_production}
\end{figure}

\begin{figure}[htbp]
    \centering
    \includegraphics[width=\columnwidth]{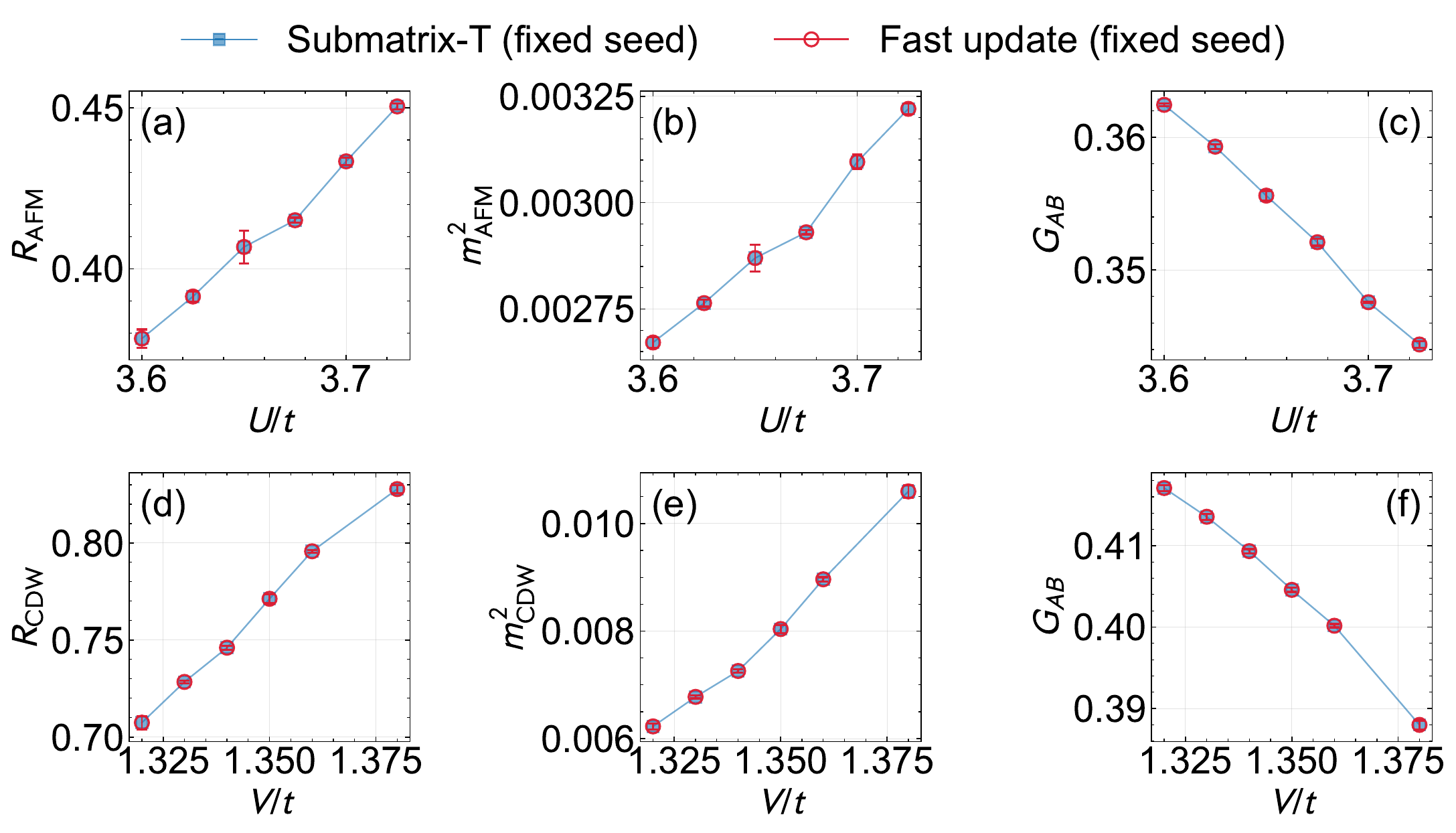}
    \caption{\rev{Observable-level comparison of the fast update (open circles) with the submatrix-T update (filled squares) in paired production-regime control runs with identical seeds.
    In every panel the two data sets coincide point by point to all printed digits, error bars included.
    The top row refers to the Hubbard model at $L=24$, the bottom row to the spinless $t$-$V$ model at $L=18$; the columns show, from left to right, the correlation ratio $R$, the squared order parameter $m^{2}$, and the averaged modulus $G_{AB}$.}}
    \label{fig:app_correctness_benchmark}
\end{figure}

\rev{Figure~\ref{fig:stability_production} shows the two stabilization diagnostics for representative production runs, which use the submatrix-T update throughout.
The fast-update control runs are overlaid as dashed curves for both models.
The wrap error stays below $10^{-6}$ in both models [Figs.~\ref{fig:stability_production}(a) and (b)], and the weight-phase drift below $10^{-7}$ [Figs.~\ref{fig:stability_production}(c) and (d)].
Both errors are orders of magnitude below the precision required by our observables, and the sign-free nature of the two models is preserved.
In each model, the fast-update control runs and the production curves are of comparable magnitude, with no sign that the submatrix-T update degrades numerical stability.
We have also re-run the submatrix-T update on the same random-number sequence as these fast-update control runs.
The two updates produce the same final auxiliary-field configurations, overall acceptance ratios, and physical observables, identical digit by digit (Fig.~\ref{fig:app_correctness_benchmark}).
In conclusion, relative to the fast update, the submatrix-T update remains unbiased and equally stable at the observable level.}

%%%%%%%%%%%%%%%%%%%%%%%%%%%%%%%%%%%%%%%%%%%%%%%%%%%%%
\section{Convergence test of Trotter error and projection length}\label{app:convergence_test}
%%%%%%%%%%%%%%%%%%%%%%%%%%%%%%%%%%%%%%%%%%%%%%%%%%%%%

There are two main sources of systematic error in PQMC: Trotter error and finite projection length. 
The Trotter decomposition in Eq.~\eqref{equ:Trotter_decomposition} introduces an error of order $\mathcal{O}(\Delta_{\tau}^{2})$ for a complete sweep over all time slices, i.e., $\langle O\rangle = \langle O\rangle_{\mathrm{exact}} + \mathcal{O}(\Delta_{\tau}^{2})$. 
Thus, a sufficiently small $\Delta_{\tau}$ is required to ensure that the Trotter error remains smaller than the statistical error.
The trial wave function $|\Psi_{T}\rangle$ is chosen as the ground state of a trial Hamiltonian $\hat{H}_{T}$, which is essentially the non-interacting Hamiltonian $\hat{H}_{0}$ with small random hoppings added on each nearest-neighbor bond to lift energy level degeneracies at Dirac points. 
To reach the true ground state, a sufficiently large projection length $\beta$ is required to project out all excited states. Since the first excited state typically has a gap of order $1/L$ in our cases~\cite{SciRep2012Sorella}, a sufficiently large projection length usually scales as $\beta \sim L$. 

\begin{figure}[htbp]
    \centering
    \includegraphics[width=0.45\textwidth]{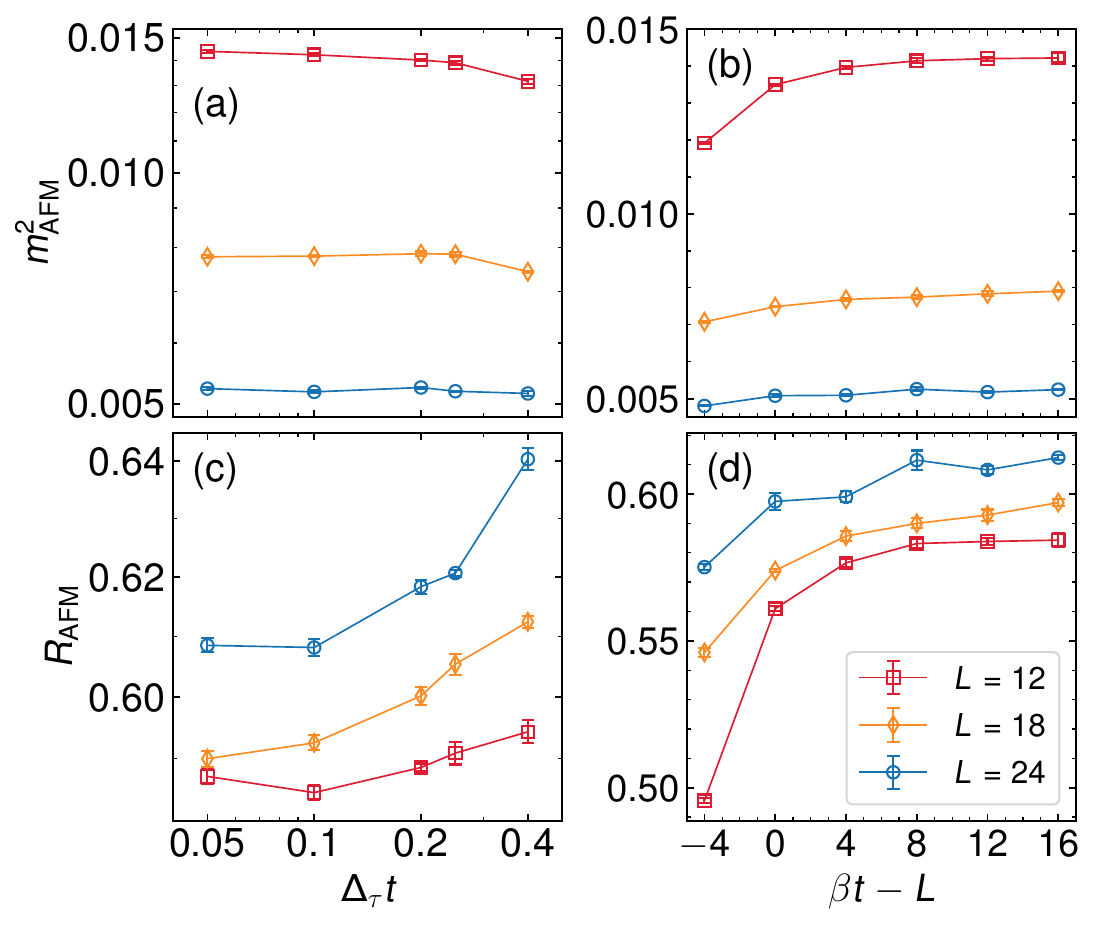}
    \caption{Convergence tests for the Trotter error and projection length in the Hubbard model ($U/t=4.0$) on honeycomb lattices. 
    Panels (a) and (b) show the squared staggered magnetization $m^2_{\mathrm{AFM}}=C_{\mathrm{AFM}}(\mathbf{k}=0)$, while panels (c) and (d) show the correlation ratio \rev{$R_{\mathrm{AFM}}$}. 
    Panels (a) and (c) are displayed on a log--log scale. In (a) and (c), $\beta t= 36$ is fixed; in (b) and (d), $\Delta_\tau t=0.1$ is fixed.
    }
    \label{fig:convergence_test}
\end{figure}

\begin{figure}[htbp]
    \centering
    \includegraphics[width=0.45\textwidth]{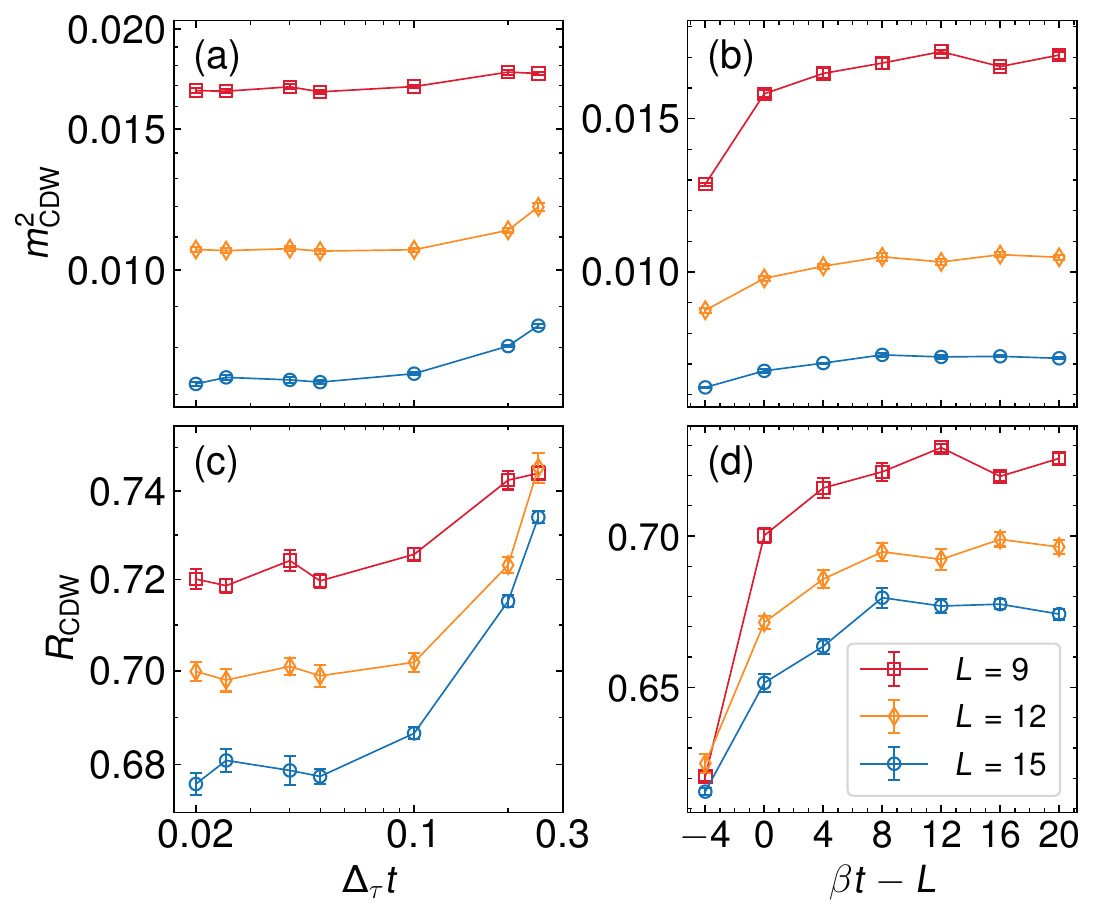}
    \caption{Convergence tests for the Trotter error and projection length in the spinless $t$-$V$ model on honeycomb lattices. 
    Panels (a) and (b) show the CDW correlation \rev{$m^2_{\mathrm{CDW}}=C_{\mathrm{CDW}}(\mathbf{k}=0)$}, while panels (c) and (d) show the correlation ratio \rev{$R_{\mathrm{CDW}}$}. 
    Panels (a) and (c) are displayed on a log--log scale. In (a) and (c), $\beta t=L+16$ is fixed; in (b) and (d), $\Delta_\tau t=0.05$ is fixed.
    }
    \label{fig:convergence_test_tv}
\end{figure}

\begin{figure}[htbp]
    \centering
    \includegraphics[width=0.45\textwidth]{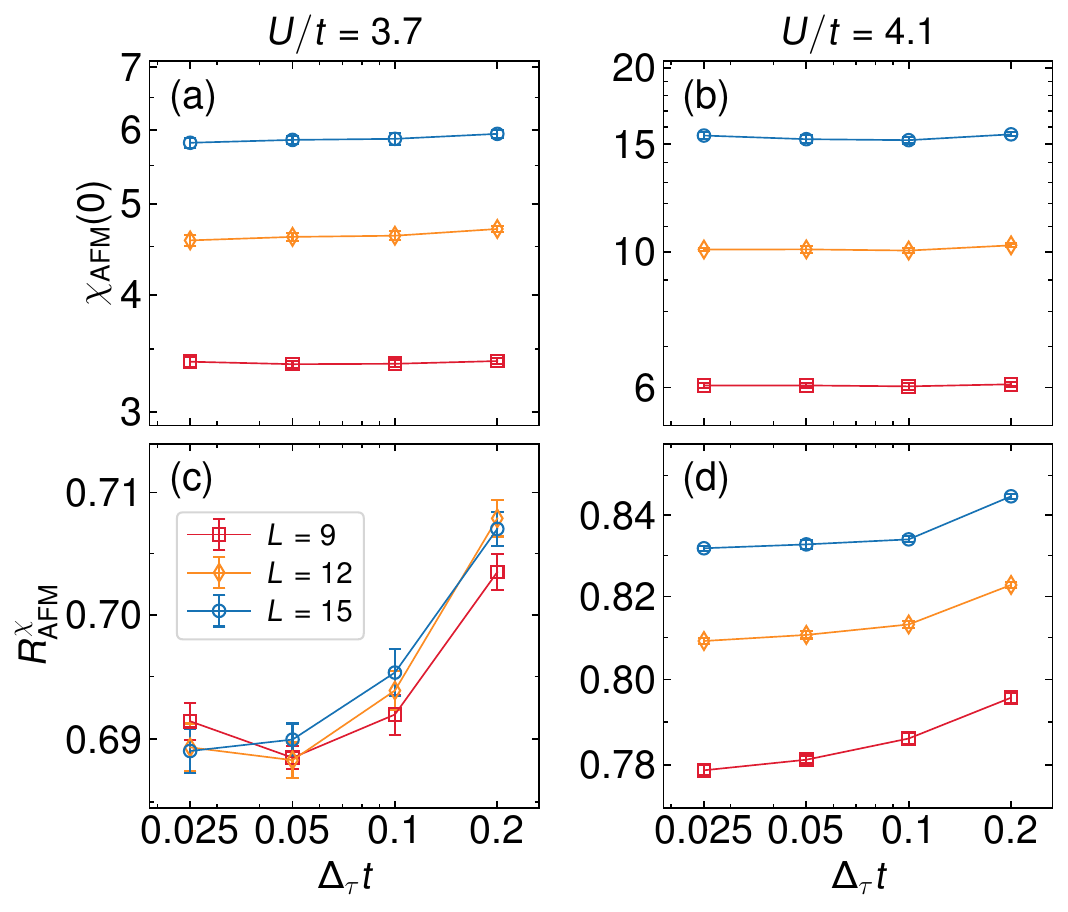}
    \caption{\rev{Trotter-step convergence tests for the finite-temperature spin-susceptibility simulations~\cite{ArXiv2611J.Vaeth} of the honeycomb Hubbard model at $\beta t=L$ for $U/t=3.7$ [panels~(a) and~(c)] and $U/t=4.1$ [panels~(b) and~(d)].
    Panels~(a) and~(b) show the zero-frequency staggered spin susceptibility $\chi_{\mathrm{AFM}}(0)$, while panels~(c) and~(d) show the correlation ratio $R^{\chi}_{\mathrm{AFM}}$, all on a log--log scale.}}
    \label{fig:convergence_test_susc}
\end{figure}

To quantify the systematic errors associated with the Trotter discretization $\Delta_\tau$ and the finite projection length $\beta$, we benchmarked both the Hubbard and the spinless $t$-$V$ models on honeycomb lattices. 
Figures~\ref{fig:convergence_test} and \ref{fig:convergence_test_tv} display the squared staggered magnetization $m^2_{\mathrm{AFM}}$ and correlation ratio \rev{$R_{\mathrm{AFM}}$}, and the CDW correlation $m^2_{\mathrm{CDW}}$ and correlation ratio \rev{$R_{\mathrm{CDW}}$}, respectively, as functions of the Trotter time slice $\Delta_\tau t$ and the projection length $\beta t$. 
For the Hubbard model, the observables become insensitive to further decreasing $\Delta_\tau t$ at $\Delta_\tau t=0.1$~\cite{Phys.Rev.B2015ParisenToldin,Phys.Rev.X2016Otsuka,Phys.Rev.B2025Lang}, whereas in the spinless $t$-$V$ model they converge at $\Delta_\tau t=0.05$~\cite{Nat.Commun.2025Zeng}; in both cases $\beta t=L+12$ is sufficient to render systematic errors negligible.
\rev{For the finite-temperature spin-susceptibility cross-checks of the Hubbard model at $\beta t=L$ (Sec.~\ref{sec:bg_suppressed}~\cite{ArXiv2611J.Vaeth}), we test the Trotter error at $U/t=3.7$ and $U/t=4.1$ (Fig.~\ref{fig:convergence_test_susc}); both $\chi_{\mathrm{AFM}}(0)$ and $R^{\chi}_{\mathrm{AFM}}$ converge at $\Delta_\tau t=0.05$, agreeing with the $\Delta_\tau t=0.025$ values within statistical uncertainties.}

%%%%%%%%%%%%%%%%%%%%%%%%%%%%%%%%%%%%%%%%%%%%%%%%%%%%%
\section{Off-diagonal Green's function on honeycomb lattices}\label{app:GAB-free-fermion}
%%%%%%%%%%%%%%%%%%%%%%%%%%%%%%%%%%%%%%%%%%%%%%%%%%%%%
The off-diagonal Green's function $G_{AB}(\mathbf{k})\equiv\langle c_{\mathbf{k}A}^{\dagger}c_{\mathbf{k}B}\rangle$, defined via the Fourier transform $c_{\mathbf{k}\alpha}^{\dagger}=\frac{1}{L}\sum_{\mathbf{r}}e^{-\mathrm{i}\mathbf{k}\cdot\mathbf{r}}c_{\mathbf{r}\alpha}^{\dagger}$ (with spin indices suppressed), has been employed to extract the fermionic anomalous dimension $\eta_\psi$ in SLAC-fermion models hosting chiral-Heisenberg~\cite{Phys.Rev.Lett.2019Lang,Phys.Rev.B2025Lang} and chiral-Ising~\cite{Phys.Rev.Lett.2022Tabatabaei} critical points.
In these square-lattice models, the gapless Dirac points are located at $\Gamma$; consequently, $G_{AB}(\mathbf{k})$ is analyzed at momenta $\delta\mathbf{k}$ near $\Gamma$ to capture low-energy excitations.
Notably, $G_{AB}(\mathbf{k})$ is proportional to the quasiparticle weight $Z$~\cite{Phys.Rev.Lett.2019Lang}, validating its utility in extracting $\eta_\psi$~\cite{Phys.Rev.X2016Otsuka,Phys.Rev.B2019Seki,Phys.Rev.B2020Otsuka,Phys.Rev.B2021Liu,Phys.Rev.Lett.2022Liu}.
In this work, we extend this approach to honeycomb lattices, where the relevant low-energy physics occurs near the Dirac points $\pm\mathrm{K}$.
Specifically, we demonstrate that the modulus $|G_{AB}(\mathrm{K}+\delta\mathbf{k})|$ serves as a robust observable for determining $\eta_\psi$.

To establish a baseline for interacting systems, we first examine the finite-size scaling behavior of $|G_{AB}|$ in the free-fermion limit. 
In momentum space, the tight-binding Hamiltonian reads $\hat{H}_0=\sum_{\mathbf{k}}\mathbf{c}_{\mathbf{k}}^{\dagger}H_0\left(\mathbf{k}\right)\mathbf{c}_{\mathbf{k}}$, where $\mathbf{c}_{\mathbf{k}}=\left(c_{\mathbf{k}A},c_{\mathbf{k}B}\right)^T$ and
\begin{equation}
    H_{0}\left({\bf k}\right)=\left(\begin{matrix}0 & -tf(\mathbf{k})\\
        -tf^{*}(\mathbf{k}) & 0
        \end{matrix}\right),f(\mathbf{k})=1+e^{-\mathrm{i}\mathbf{k}\cdot\mathbf{a}_{1}}+e^{\mathrm{i}\mathbf{k}\cdot\mathbf{a}_{2}}.
\end{equation}
In our calculation, we define the primitive lattice vectors as $\mathbf{a}_{1}=(\frac{\sqrt{3}}{2},-\frac{3}{2})$ and $\mathbf{a}_{2}=(\frac{\sqrt{3}}{2},\frac{3}{2})$ in units of $d$, the side length of the hexagonal plaquettes. 
Diagonalizing $H_0\left(\mathbf{k}\right)$ yields the zero-temperature off-diagonal Green's function
\begin{equation}
    G_{AB}\left(\mathbf{k}\right)=\frac{1}{2}\frac{f^{*}({\mathbf{k}})}{f({\mathbf{k}})},\quad \mathbf{k}\neq\pm\mathrm{K}=\pm\frac{1}{3}(\mathbf{b}_{1}+\mathbf{b}_{2}).
\end{equation}
At the two inequivalent Dirac points $\pm\mathrm{K}$, $G_{AB}\left(\mathbf{k}\right)$ vanishes.
For momenta near $\pm\mathrm{K}$, we expand $f(\mathbf{k})$ as $f\left(\pm\mathrm{K}+\mathbf{q}\right)=\frac{3}{2}\left(\mp q_{x}-\mathrm{i}q_{y}\right)$, where $\mathbf{q}=(q_x,q_y)$ (in units of $d^{-1}$) denotes a small deviation from $\pm\mathrm{K}$. 
This gives the low-energy Green's function
\begin{equation}
    G_{AB}\left(\pm\mathrm{K}+\mathbf{q}\right)\approx\frac{\mp q_{x}+\mathrm{i}q_{y}}{q}. 
\end{equation}
Fourier transforming to real space, we obtain the long-distance behavior of $G_{AB}(\mathbf{r})$~\cite{Phys.Rev.B2019Seki}:
\begin{equation}\label{equ:GAB_r_free_fermion}
    \begin{aligned}
        G_{AB}\left(\mathbf{r}\right)&=\frac{1}{L^{2}}\sum_{\mathbf{k}\neq\pm\mathrm{K}}\frac{1}{2}\frac{f^{*}\left(\mathbf{k}\right)}{\left|f\left(\mathbf{k}\right)\right|}e^{\mathrm{i}\mathbf{k}\cdot\mathbf{r}}\\&\approx-\frac{S_{\mathrm{U}}}{4\pi}\left[e^{\mathrm{i}\mathrm{K}\cdot\mathbf{r}}\frac{y+\mathrm{i}x}{r^{3}}+e^{-\mathrm{i}\mathrm{K}\cdot\mathbf{r}}\frac{y-\mathrm{i}x}{r^{3}}\right],
    \end{aligned}
\end{equation}
where $S_{\mathrm{U}}={3\sqrt{3}d^{2}}/{2}$ is the area of the unit cell. 
The two terms in the square bracket arise from the two Dirac points $\pm\mathrm{K}$, respectively. 
Each carries a phase factor $e^{\pm\mathrm{i}\mathrm{K}\cdot\mathbf{r}}$, and their coherent interference contrasts with that of SLAC fermions on square lattices~\cite{Phys.Rev.Lett.2019Lang}, where the only Dirac point lies at $\Gamma$.

Based on the low-energy results in free-fermion systems, we conjecture that the real-space Green's function has the following scaling form near the critical point~\cite{Phys.Rev.B2021Liu}: 
\begin{equation}
    G(\mathbf{r},\xi,L) =\sum_{\sigma=\pm}\frac{e^{\sigma\mathrm{i}\mathrm{K}\cdot\mathbf{r}}e^{\sigma\mathrm{i}\alpha\theta_{\mathbf{r}}}}{\left|\mathbf{r}\right|^{2+\eta_\psi}}f\left(\frac{|\mathbf{r}|}{L},\frac{\xi}{L},L^{-\omega}\right),
\end{equation}
where $\theta_{\mathbf{r}}=\arctan(y/x)$ denotes the direction of $\mathbf{r}$, $\alpha$ is a constant, and setting $\alpha=-1$ recovers Eq.~\eqref{equ:GAB_r_free_fermion}. 
The two Dirac points ($\sigma=\pm1$) contribute to the real-space Green's function with different phases $e^{\sigma\mathrm{i}(\mathrm{K}\cdot\mathbf{r}+\alpha\theta_{\mathbf{r}})}$. 
The function $f$ describes the corrections from short-range and finite-size effects, where $\xi\sim |U-U_c|^{-1/\nu}$ is the correlation length and $\omega$ is the leading-correction-to-scaling exponent.
Transforming to momentum space, we have the following scaling form for $G_{AB}(\mathrm{K}+\mathbf{q})$ in the continuum limit:
\begin{equation}\label{equ:GAB_K_q_scaling}
    \begin{aligned}
        &\quad G_{AB}\left(\mathrm{K}+\mathbf{q}\right)=\int\mathrm{d}^{2}\mathbf{r}e^{-\mathrm{i}\left(\mathrm{K}+\mathbf{q}\right)\cdot\mathbf{r}}G_{AB}(\mathbf{r})\\&\approx\int\mathrm{d}^{2}\mathbf{r}e^{-\mathrm{i}\mathbf{q}\cdot\mathbf{r}}\frac{e^{\mathrm{i}\alpha\theta_{\mathbf{r}}}}{\left|\mathbf{r}\right|^{2+\eta_{\psi}}}f\left(\frac{|\mathbf{r}|}{L},\frac{\xi}{L},L^{-\omega}\right)\\&=L^{2-\left(2+\eta_{\psi}\right)}\int\mathrm{d}^{2}\mathbf{r}^{\prime}e^{-\mathrm{i}\left(L\mathbf{q}\right)\cdot\mathbf{r}^{\prime}}\frac{e^{\mathrm{i}\alpha\theta_{\mathbf{r}^{\prime}}}}{\left|\mathbf{r}^{\prime}\right|^{2+\eta_{\psi}}}f\left(\left|\mathbf{r}^{\prime}\right|,\frac{\xi}{L},L^{-\omega}\right)\\&=L^{-\eta_{\psi}}e^{\mathrm{i}\alpha\phi\left[\theta_{\mathbf{q}}\right]}f^\prime\left(L\left|\mathbf{q}\right|,\frac{\xi}{L},L^{-\omega}\right).
    \end{aligned}    
\end{equation}
In the second line we drop the contribution from the other Dirac point ($-\mathrm{K}$) since it is far away from $\mathrm{K}+\mathbf{q}$. More rigorously, this contribution vanishes due to the rapidly oscillating factor $e^{-2\mathrm{i}\mathrm{K}\cdot\mathbf{r}}$. 
In the third line, we perform a change of variable $\mathbf{r}^\prime=\mathbf{r}/L$, and the resulting integral defines the function $f^\prime$, which in general can still be complex but, importantly, depends only on the magnitude of $\mathbf{q}$. 
The entire angular dependence on $\theta_{\mathbf{q}}=\arctan(q_y/q_x)$ is contained in the phase prefactor $e^{\mathrm{i}\alpha\phi\left[\theta_{\mathbf{q}}\right]}$, where $\phi\left[\theta_{\mathbf{q}}\right]$ is a function of $\theta_{\mathbf{q}}$. 
The last line follows from comparing $G_{AB}(\mathrm{K}+\mathbf{q})$ and $G_{AB}(\mathrm{K}+R_\phi\mathbf{q})$, where $R_\phi$ is a rotation matrix that rotates the vector $\mathbf{q}$ by an angle $\phi$ anticlockwise. 
Changing the integral variable from $\mathbf{r}^\prime$ to $\mathbf{r}^{\prime\prime}=R_\phi\mathbf{r}^\prime$ renders the two integrals identical, up to a \textit{space-independent} phase factor $e^{\mathrm{i}\alpha (\theta_{\mathbf{r}^\prime}-\theta_{\mathbf{r}^{\prime\prime}})}=e^{-\mathrm{i}\alpha\phi}$. 

Since $G_{AB}(\mathrm{K}+\mathbf{q})$ lacks rotational symmetry and is gauge variant, we consider instead its modulus, 
\begin{equation}
    \left|G_{AB}\left(\mathrm{K}+\mathbf{q}\right)\right|=L^{-\eta_{\psi}}F^\prime\left(L\left|\mathbf{q}\right|,\frac{\xi}{L},L^{-\omega}\right),
\end{equation}
which is real, gauge invariant, and possesses $C_6$ rotational symmetry in reciprocal space. 
For free fermions, $|G_{AB}(\mathrm{K}+\mathbf{q})|=1$ for sufficiently small $\left|\mathbf{q}\right|$, consistent with $\eta_\psi=0$.
For interacting fermions, we have verified numerically that the six nearest-neighbor momenta to the Dirac point $\mathrm{K}$, namely $\mathbf{q}=\pm\mathbf{b}_1/L,\pm\mathbf{b}_2/L,\pm(\mathbf{b}_1-\mathbf{b}_2)/L$, yield identical values of $|G_{AB}(\mathrm{K}+\mathbf{q})|$. 
Averaging over these six momenta improves the statistical robustness. 
Since these momenta scale as $1/L$, the scaling form of $|G_{AB}(\mathrm{K}+\mathbf{q})|$ reduces to
\begin{equation}
    \left\langle \left|G_{AB}(\mathrm{K}+\delta\mathbf{k})\right|\right\rangle _{\delta\mathbf{k}}=L^{-\eta_{\psi}}F\left(\xi/L,L^{-\omega}\right). 
\end{equation}

Finally, we note that $|G_{AB}(\mathrm{K}+\delta\mathbf{k})|$ is equivalent to the occupation jump used in Ref.~\cite{Phys.Rev.X2016Otsuka} to extract the fermion anomalous dimension.
In that approach, one considers the bare eigenstates $\psi_{\mathbf{k},\pm}=\left(c_{\mathbf{k}A}\pm c_{\mathbf{k}B}f(\mathbf{k})/|f(\mathbf{k})|\right)/\sqrt{2}$ with bare energies $\varepsilon_{\mathbf{k},\pm}=\pm t|f(\mathbf{k})|$, and their occupation numbers are $n(\varepsilon_{\mathbf{k},\pm})=\langle\psi_{\mathbf{k},\pm}^{\dagger}\psi_{\mathbf{k},\pm}\rangle$.
The quantity $\Delta n$ is the difference between the occupation of the lower band ($\varepsilon_{\mathrm{K}+\delta\mathbf{k},-}<0$) and that of the upper band ($\varepsilon_{\mathrm{K}+\delta\mathbf{k},+}>0$) at the same momentum $\mathrm{K}+\delta\mathbf{k}$, which is one of the closest points to the Fermi level.
Using the relation $n(\varepsilon_{\mathbf{k},\pm})=\frac{1}{2}\mp\frac{1}{2}(\frac{f(\mathbf{k})}{\left|f(\mathbf{k})\right|}G_{AB}\left(\mathbf{k}\right)+\mathrm{H.c.})$, we obtain
\begin{equation}
    \begin{aligned}
        \Delta n&=n\left(\varepsilon_{\mathrm{K}+\delta\mathbf{k},-}\right)-n\left(\varepsilon_{\mathrm{K}+\delta\mathbf{k},+}\right)\\&=\frac{f\left(\mathrm{K}+\delta\mathbf{k}\right)}{\left|f\left(\mathrm{K}+\delta\mathbf{k}\right)\right|}G_{AB}\left(\mathrm{K}+\delta\mathbf{k}\right)+\mathrm{H.c.},
    \end{aligned}
\end{equation}
where the prefactor $f(\mathrm{K}+\delta\mathbf{k})/|f(\mathrm{K}+\delta\mathbf{k})|$ is a pure phase depending on the orientation of $\delta\mathbf{k}$.
More rigorously, one should diagonalize the $2\times2$ one-body density matrix $(\rho_{\mathbf{k}})_{\alpha\beta}=\langle c_{\mathbf{k}\alpha}^{\dagger}c_{\mathbf{k}\beta}\rangle$ ($\alpha,\beta\in\{A,B\}$).
This gives the natural orbitals $\bar{\psi}_{\mathbf{k},\pm}=(c_{\mathbf{k}A}\pm c_{\mathbf{k}B}G_{AB}^{*}(\mathbf{k})/|G_{AB}(\mathbf{k})|)/\sqrt{2}$ and the natural occupation numbers $\bar{n}(\mathbf{k},\pm)=\frac{1}{2}\pm|G_{AB}(\mathbf{k})|$.
The corresponding natural occupation jump (here $+$ labels the larger eigenvalue),
\begin{equation}
    \Delta\bar{n}=\bar{n}(\mathrm{K}+\delta\mathbf{k},+)-\bar{n}(\mathrm{K}+\delta\mathbf{k},-)=2\left|G_{AB}\left(\mathrm{K}+\delta\mathbf{k}\right)\right|,
\end{equation}
is directly proportional to $|G_{AB}(\mathrm{K}+\delta\mathbf{k})|$, and hence to the quasiparticle weight $Z$ and the fermionic anomalous dimension $\eta_\psi$~\cite{Phys.Rev.Lett.2019Lang}.
It was further verified that the bare and natural orbital distributions are indistinguishable even in the interacting case~\cite{Phys.Rev.X2016Otsuka}.
In evaluating $n(\varepsilon_{\mathbf{k},\pm})$ for $\Delta n$, Ref.~\cite{Phys.Rev.X2016Otsuka} averages over equivalent momenta sharing the same bare energy, analogous to our definition of $\left\langle \left|G_{AB}(\mathrm{K}+\delta\mathbf{k})\right|\right\rangle _{\delta\mathbf{k}}$.

%%%%%%%%%%%%%%%%%%%%%%%%%%%%%%%%%%%%%%%%%%%%%%%%%%%%%
\section{Finite-size scaling analysis for Hubbard model}\label{app:FSS_Hubbard}
%%%%%%%%%%%%%%%%%%%%%%%%%%%%%%%%%%%%%%%%%%%%%%%%%%%%%

\begin{figure*}[htbp]
    \centering
    \includegraphics[width=\textwidth]{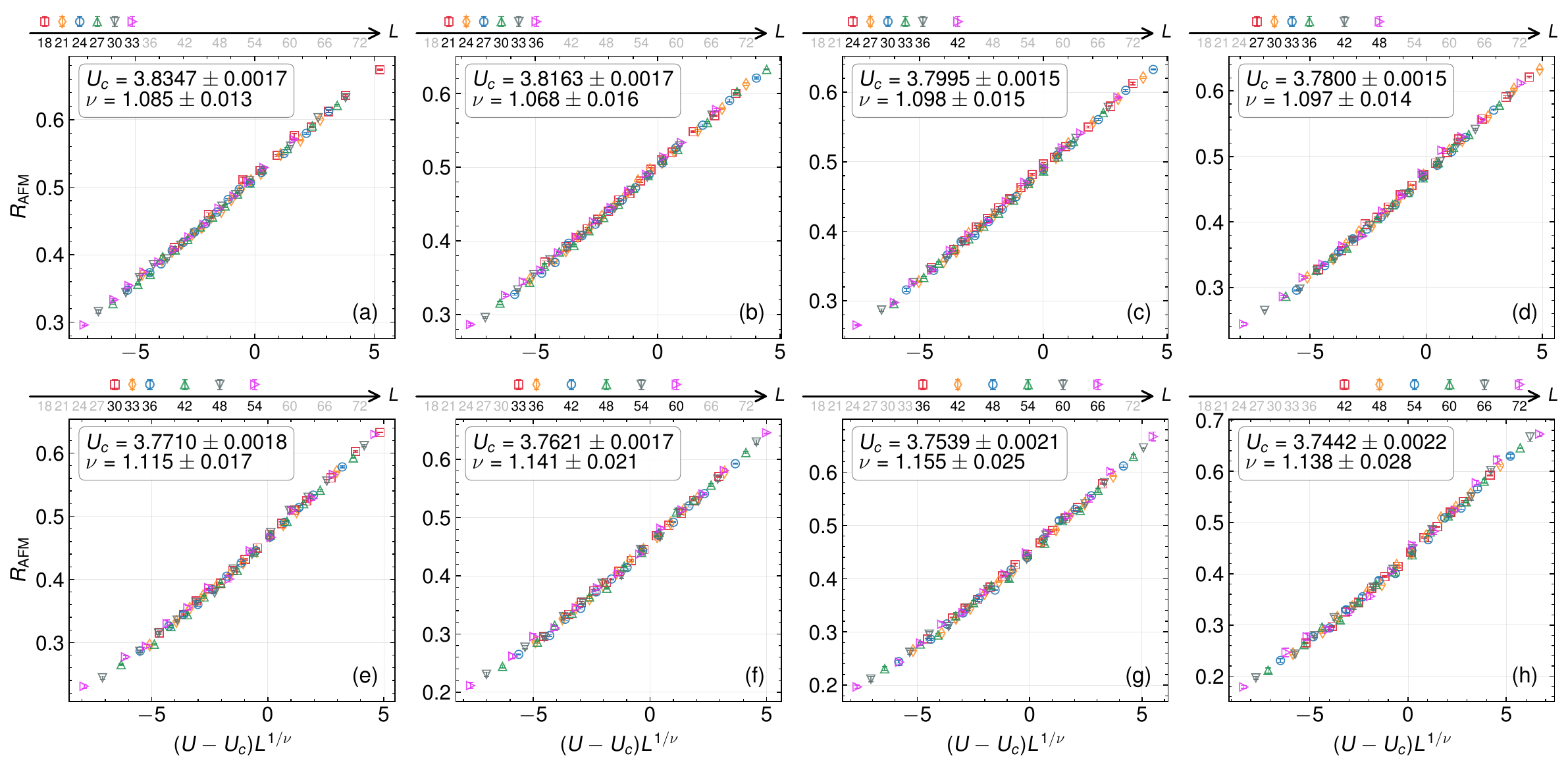}
    \caption{Sliding-window data-collapse analysis of the correlation ratio \rev{$R_{\mathrm{AFM}}$} for the Hubbard model. 
    The data collapse is performed over a window of system sizes with a fixed width (covering 6 consecutive sizes) that shifts across the \rev{windows with $L_{\min}\geq18$ used in the main text}. 
    The finite-size scaling ansatz \rev{$R_{\mathrm{AFM}}(U,L)=f^{R}_0(uL^{1/\nu})$} is employed to extract the critical point $U_c$ and critical exponent $\nu$. 
    Panels show the collapsed curves and the extracted parameters $U_c$ and $\nu$ for each window.
    }
    \label{fig:data_collapse_CR}
\end{figure*}

\begin{figure*}[htbp]
    \centering
    \includegraphics[width=\textwidth]{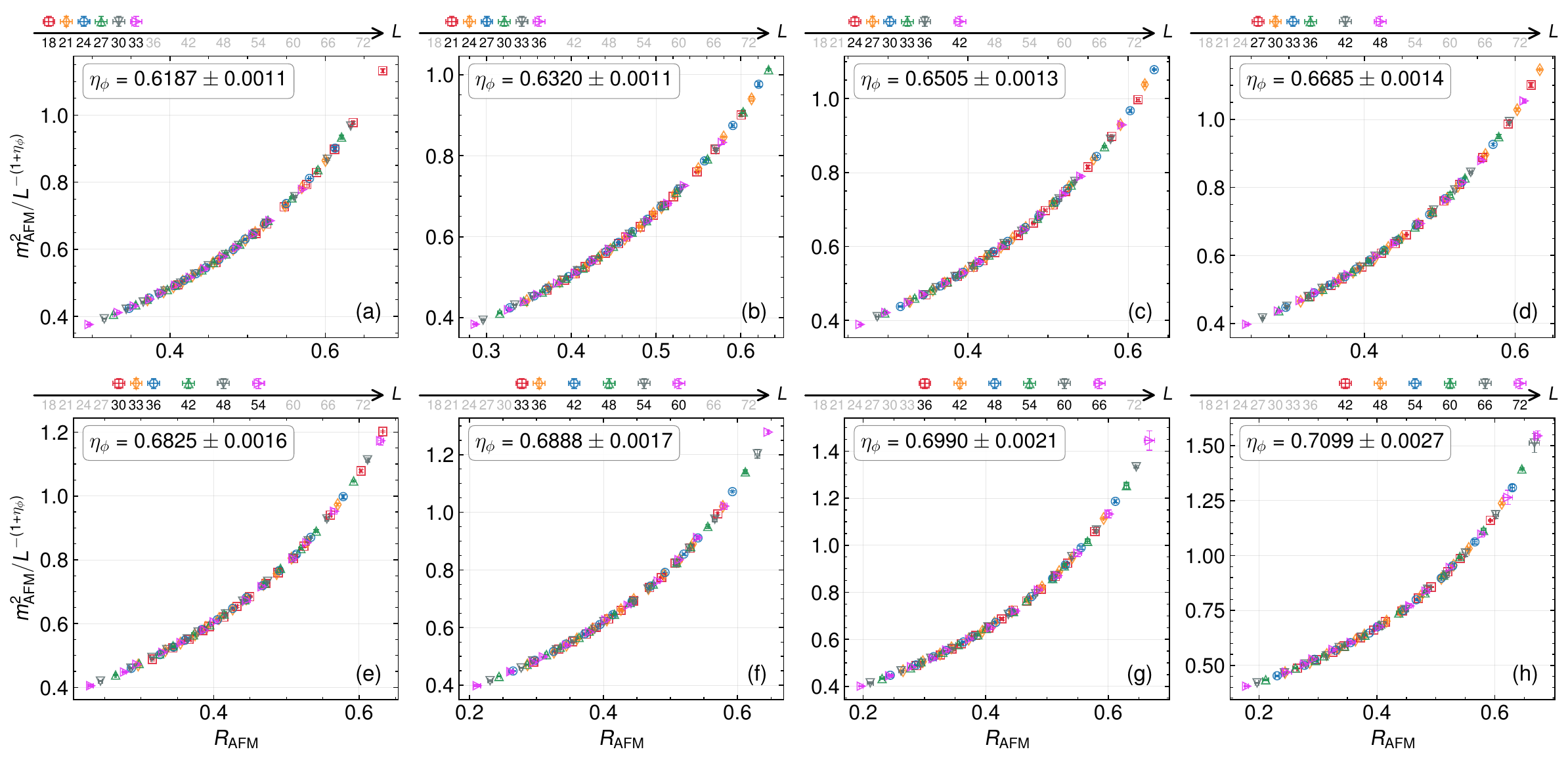}
    \caption{Sliding-window data-collapse analysis of the squared staggered magnetization $m_{\mathrm{AFM}}^2$ for the Hubbard model. 
    The data collapse is performed over a window of system sizes with a fixed width (covering 6 consecutive sizes) that shifts across the \rev{windows with $L_{\min}\geq18$ used in the main text}. 
    The finite-size scaling ansatz $m_{\mathrm{AFM}}^{2}(R_{\mathrm{AFM}},L)=L^{-1-\eta_{\phi}}f_{0}^{m}(R_{\mathrm{AFM}})$ is employed to extract the bosonic anomalous dimension $\eta_\phi$. 
    Panels show the collapsed curves and the extracted parameter $\eta_\phi$ for each window.
    }
    \label{fig:data_collapse_m2}
\end{figure*}

\begin{figure*}[htbp]
    \centering
    \includegraphics[width=\textwidth]{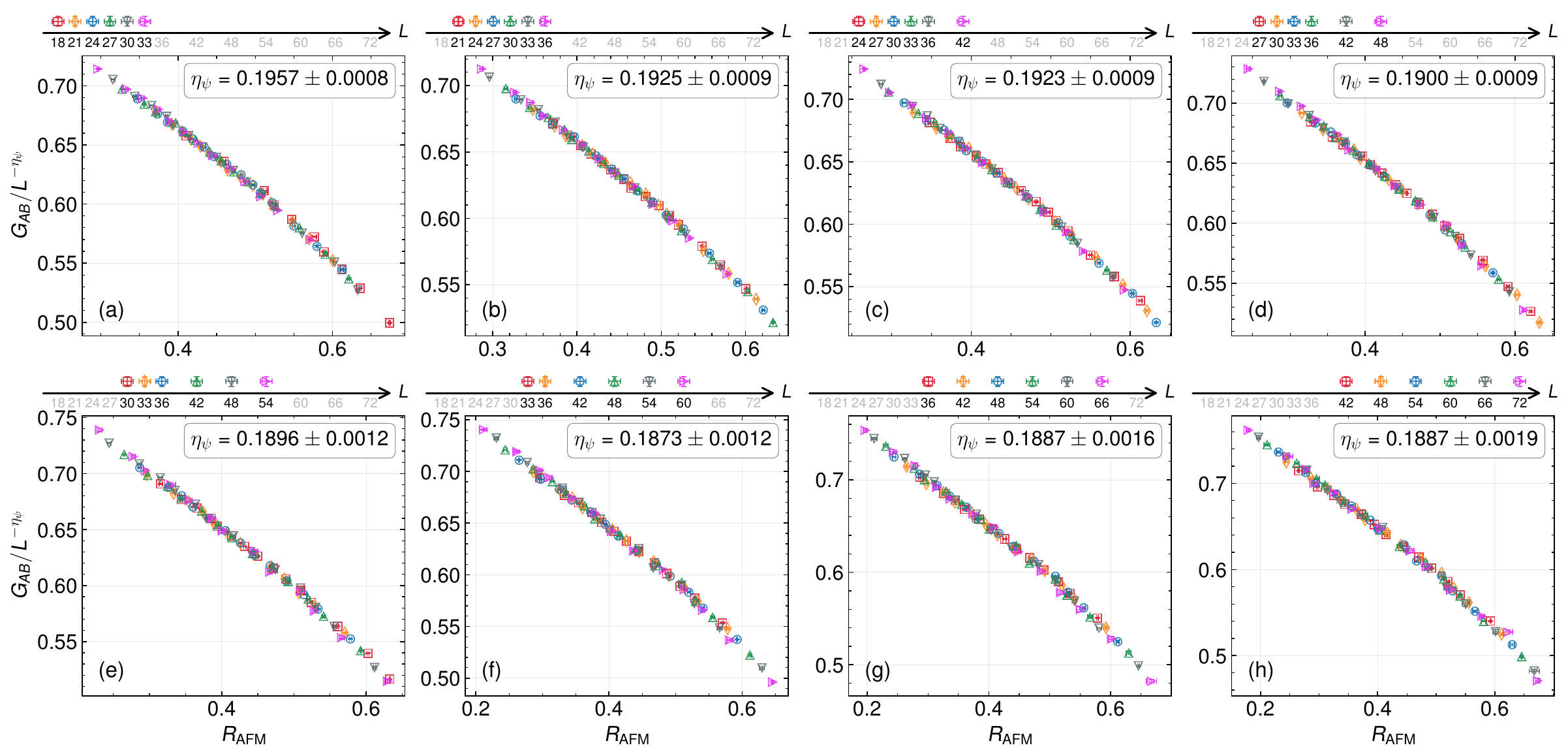}
    \caption{Sliding-window data-collapse analysis of the averaged modulus of the off-diagonal Green's function \rev{$G_{AB}$} for the Hubbard model. 
    The data collapse is performed over a window of system sizes with a fixed width (covering 6 consecutive sizes) that shifts across the \rev{windows with $L_{\min}\geq18$ used in the main text}. 
    The finite-size scaling ansatz \rev{$G_{AB}(R_{\mathrm{AFM}},L)=L^{-\eta_{\psi}}f_{0}^{G}(R_{\mathrm{AFM}})$} is employed to extract the fermionic anomalous dimension $\eta_\psi$. 
    Panels show the collapsed curves and the extracted parameter $\eta_\psi$ for each window.
    }
    \label{fig:data_collapse_Gab}
\end{figure*}

In this section, we present the detailed finite-size scaling analysis for the Hubbard model. 
We show the sliding-window data collapse results for the correlation ratio (Fig.~\ref{fig:data_collapse_CR}), the squared staggered magnetization (Fig.~\ref{fig:data_collapse_m2}), and the off-diagonal Green's function (Fig.~\ref{fig:data_collapse_Gab}). 

\begin{figure*}[htbp]
    \centering
    \includegraphics[width=0.85\textwidth]{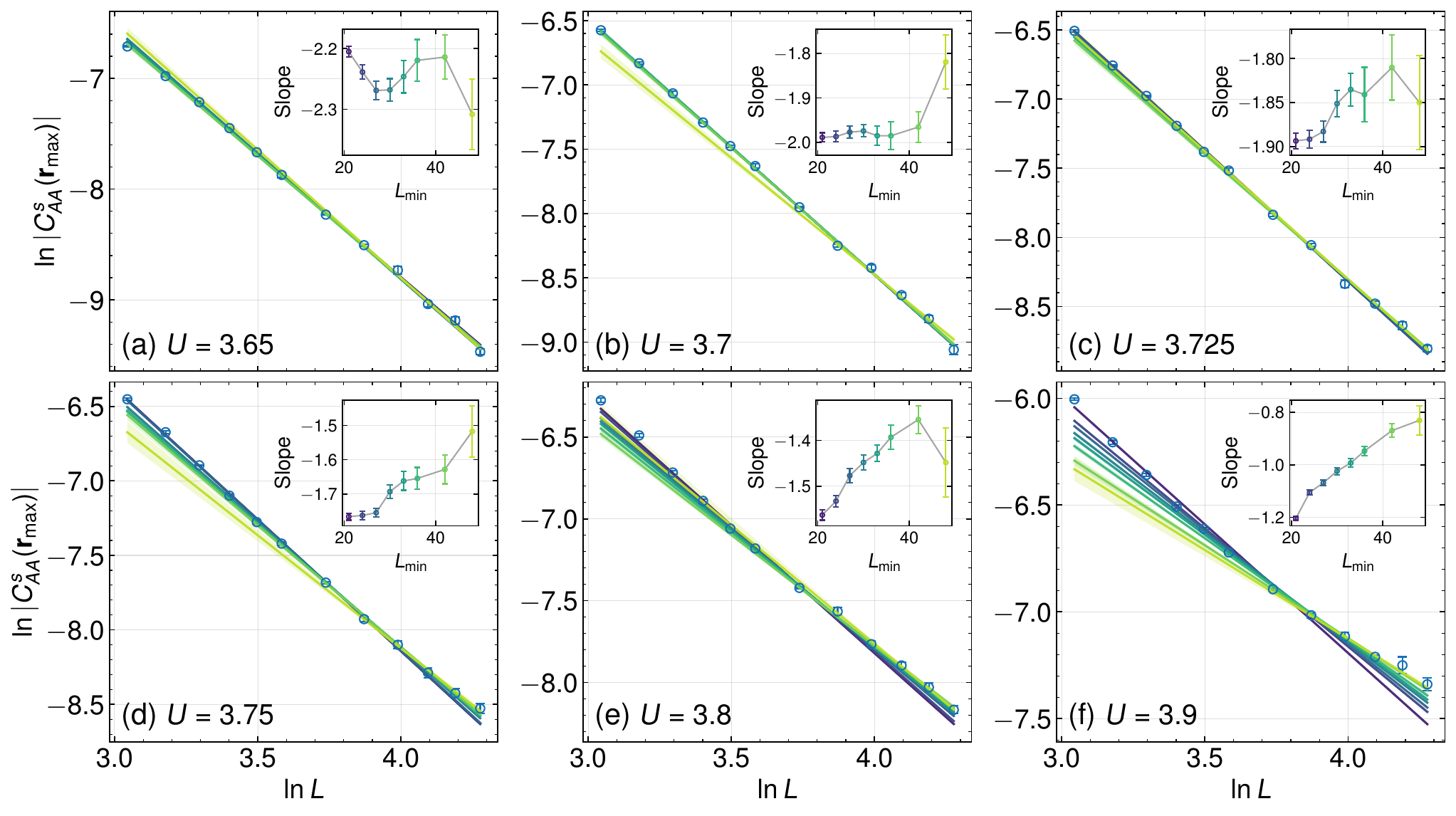}
    \caption{Finite-size scaling of the spin-spin correlation function \rev{$C^{s}_{AA}(\mathbf{r}_{\max})$ [Eq.~\eqref{equ:C_rmax_def}]} at the maximum distance $\mathbf{r}_{\max} = \lfloor L/2 \rfloor (\mathbf{a}_2-\mathbf{a}_1)$ for the Hubbard model; at the critical point, we expect a scaling behavior of $\sim 1/L^{1+\eta_\phi}$. 
    The panels show the double-logarithmic scaling of the correlation magnitude versus system size $L$ for various interaction strengths $U$. 
    Solid lines represent linear fits performed over the window $[L_{\min}, 72]$. 
    Insets display the fitted slope as a function of the fitting window lower bound $L_{\min}$.
    }
    \label{fig:ss_decay_Hubbard}
\end{figure*}

\begin{figure*}[htbp]
    \centering
    \includegraphics[width=0.85\textwidth]{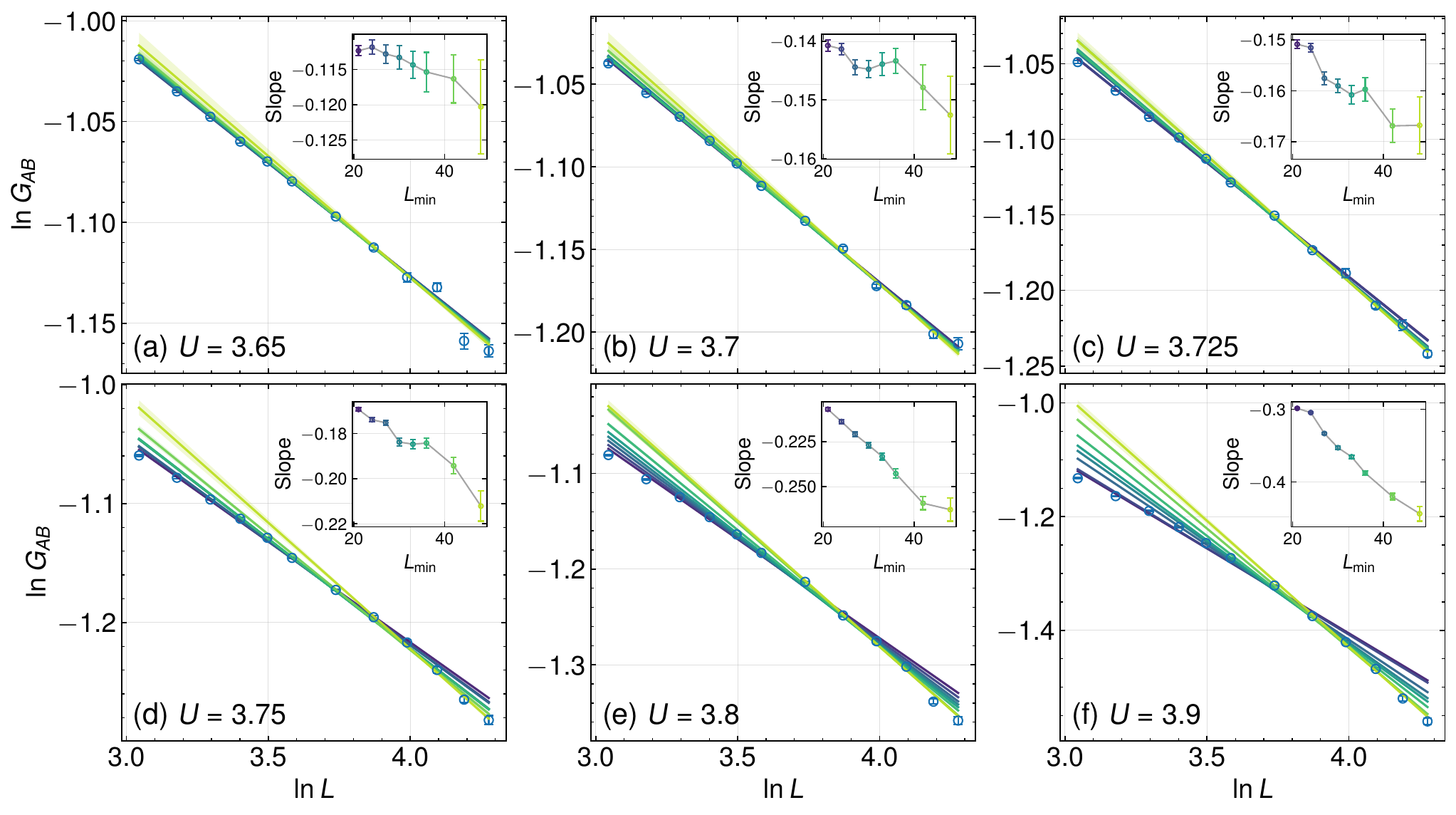}
    \caption{\rev{Finite-size scaling of the averaged off-diagonal Green's function modulus $G_{AB}$ [Eq.~\eqref{equ:fss_G_AB}] at fixed interaction strengths $U$ for the Hubbard model, in the same layout as Fig.~\ref{fig:ss_decay_Hubbard}; at the critical point, $G_{AB}\sim L^{-\eta_\psi}$.
    Solid lines and shaded bands show linear fits of $\ln G_{AB}$ versus $\ln L$ over $[L_{\min}, 72]$ and their $1\sigma$ uncertainties, and insets display the fitted slope as a function of $L_{\min}$.}}
    \label{fig:gab_decay_Hubbard}
\end{figure*}

\rev{Additionally, Figs.~\ref{fig:ss_decay_Hubbard} and~\ref{fig:gab_decay_Hubbard} show the double-logarithmic size dependence at fixed $U$ for $C^{s}_{AA}(\mathbf{r}_{\max})$ in Eq.~\eqref{equ:C_rmax_def} and $G_{AB}$ in Eq.~\eqref{equ:fss_G_AB}, where each panel fits linear slopes from a lower cutoff $L_{\min}$ up to $L=72$ and tracks the slope versus $L_{\min}$ in its inset.
At criticality, algebraic decay produces a straight line whose slope is insensitive to $L_{\min}$, equal to $-(1+\eta_\phi)$ for $C^{s}_{AA}(\mathbf{r}_{\max})$ and $-\eta_\psi$ for $G_{AB}$.
Once the system enters the gapped AFM Mott phase, however, algebraic scaling gives way to exponential decay with opposite consequences for the two observables, as long-range spin order provides a nonzero constant floor that flattens $\ln |C^{s}_{AA}(\mathbf{r}_{\max})|$ at large $L$ and pushes its fitted slope above $-1$ in the inset of Fig.~\ref{fig:ss_decay_Hubbard}(f) at $U=3.9$, whereas the quasiparticle weight has no constant floor and decays exponentially toward zero~\cite{Phys.Rev.B2019Seki}, bending $\ln G_{AB}$ downward so that its slope plunges continuously with $L_{\min}$ in the inset of Fig.~\ref{fig:gab_decay_Hubbard}(f).
In panels~(b) of Figs.~\ref{fig:ss_decay_Hubbard} and~\ref{fig:gab_decay_Hubbard} at $U=3.7$, matching the background-suppressed plateau of $U_c$ in Fig.~\ref{fig:fit_extrapolation_bgsup}(a), both insets indeed stay nearly flat across $L_{\min}$, with large-size fits from $L_{\min}=42$ giving slopes of $-1.97(3)$ and $-0.148(4)$ that agree directly with the plateaus of $\eta_\phi$ and $\eta_\psi$ in Figs.~\ref{fig:fit_extrapolation_bgsup}(c) and~\ref{fig:fit_extrapolation_bgsup}(d).
By contrast, in panels~(d) at $U=3.75$, near the conventional large-size crossing in Fig.~\ref{fig:correlation_ratio_crossing}(b) and the $L=42$--$72$ data-collapse result in Fig.~\ref{fig:data_collapse_hubbard_main}(c), the insets of Figs.~\ref{fig:ss_decay_Hubbard}(d) and~\ref{fig:gab_decay_Hubbard}(d) already drift upward and downward with $L_{\min}$, and their local slopes yield a smaller $\eta_\phi$ around $0.7$ together with a steeper $G_{AB}$ slope of $-0.194(4)$ from $L_{\min}=42$, reproducing the biased conventional data-collapse estimates in Figs.~\ref{fig:data_collapse_hubbard_main}(f) and~\ref{fig:data_collapse_hubbard_main}(i).}

%%%%%%%%%%%%%%%%%%%%%%%%%%%%%%%%%%%%%%%%%%%%%%%%%%%%%
\section{Finite-size scaling analysis for \texorpdfstring{$t$-$V$}{t-V} model}\label{app:FSS_tV}
%%%%%%%%%%%%%%%%%%%%%%%%%%%%%%%%%%%%%%%%%%%%%%%%%%%%%
In this section, we present the parallel finite-size scaling analysis for the half-filled spinless $t$-$V$ model on the honeycomb lattice, described by the Hamiltonian~\cite{Phys.Rev.B2015Li,Phys.Rev.Lett.2015Wang}: $\hat{H}=-t\sum_{\langle i,j\rangle}(c_{i}^{\dagger}c_{j}+\mathrm{H.c.})+V\sum_{\langle i,j\rangle}(\hat{n}_{i}-\frac{1}{2})(\hat{n}_{j}-\frac{1}{2})$, where $t$ is the nearest-neighbor hopping amplitude and $V$ is the nearest-neighbor repulsion.
Increasing $V$ drives a continuous quantum phase transition from a semimetal to a charge-density-wave (CDW) insulator, which is found to belong to the $N=4$ Gross-Neveu-Ising universality class~\cite{Phys.Rev.B2018Ihrig,PRD2025JA.Gracey,Phys.Lett.B1992Gracey,Int.J.Mod.Phys.A1994Graceya,Int.J.Mod.Phys.A1994Gracey,Phys.Rev.B2016Knorr,J.HighEnerg.Phys.2023Erramilli,Sci.Adv.2026YK.Yu,Nat.Commun.2025Zeng,Phys.Rev.B2016Hesselmann,Phys.Rev.B2016Wang,NewJ.Phys.2015Li,NewJ.Phys.2014Wang,Phys.Rev.D2020Huffman}.

To detect CDW long-range order, we define the CDW correlation function in momentum space:
\begin{equation}
    C_{\mathrm{CDW}}(\mathbf{k})=\rev{\frac{1}{N_{\mathrm{s}}^{2}}\sum_{i,j}\epsilon_{i}\epsilon_{j}\,e^{-\mathrm{i}\mathbf{k}\cdot(\mathbf{r}_{i}-\mathbf{r}_{j})}\left\langle \left(\hat{n}_{i}-\tfrac{1}{2}\right)\left(\hat{n}_{j}-\tfrac{1}{2}\right)\right\rangle} ,
\end{equation}
\rev{with $\hat{n}_{i}=c_{i}^{\dagger}c_{i}$ and the same site positions $\mathbf{r}_{i}$ and sublattice signs $\epsilon_{i}$ as in Eq.~\eqref{equ:C_k_def}.}
The squared CDW order parameter is given by $m_{\mathrm{CDW}}^2(L)=C_{\mathrm{CDW}}(\mathbf{k}=0)$.
To sensitively probe the phase transition, we employ the RG-invariant correlation ratio:
\begin{equation}
\rev{R_{\mathrm{CDW}}=1-\frac{C_{\mathrm{CDW}}(\delta\mathbf{k}_{1})}{C_{\mathrm{CDW}}(0)}},
\end{equation}
\rev{with the first-shell momentum $\delta\mathbf{k}_{1}$ defined as in Eq.~\eqref{equ:R_AFM_def} and $C_{\mathrm{CDW}}(\delta\mathbf{k}_{1})$ averaged over the six $C_{6v}$-equivalent momenta.}
The definition of the off-diagonal Green's function $G_{AB}(\mathbf{k})$ and the finite-size scaling ansatzes for all three quantities follow those of the Hubbard model in Sec.~\ref{sec:gnh_criticality} and Appendix~\ref{app:data_collapse_analysis}.

\begin{figure*}[htbp]
    \centering
    \includegraphics[width=0.9\textwidth]{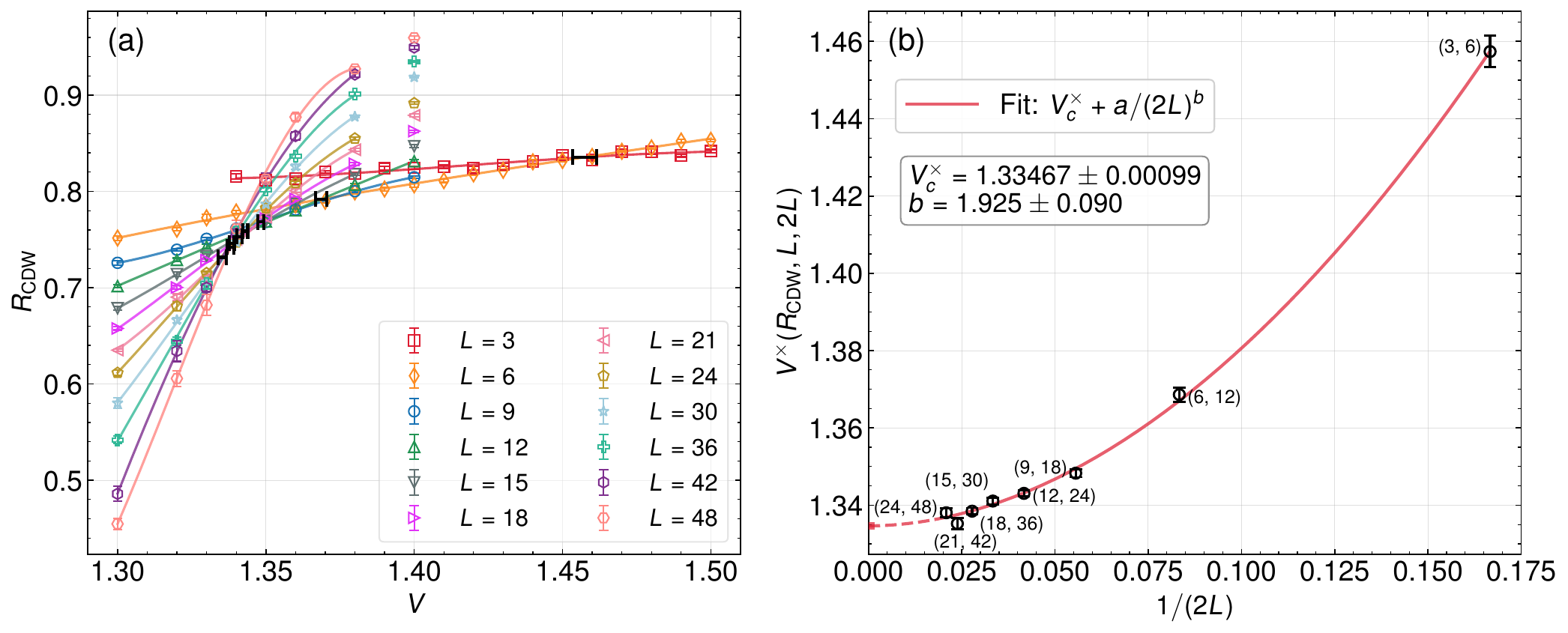}
    \caption{Determination of the critical interaction strength $V_c$ for the $t$-$V$ model using the crossing-point analysis of the correlation ratio \rev{$R_{\mathrm{CDW}}$}. 
    (a) $V$-dependence of \rev{$R_{\mathrm{CDW}}$} for various system sizes $L$. 
    The crossing points $V^{\times}(L, 2L)$ between system sizes $L$ and $2L$ are indicated by the horizontal error bars in black, determined by \rev{cubic} interpolation of the data for each size \rev{(restricted to $V\le 1.38$ for $L\ge 15$, with the saturated $V=1.40$ points retained as symbols for reference)}.
    (b) Extrapolation of the crossing points $V^{\times}(L, 2L)$ as a function of $1/(2L)$. 
    The solid curve represents a power-law fit $V_c^{\times} + a(2L)^{-b}$, yielding the critical point \rev{$V_c^{\times} = 1.33467(99)$} in the thermodynamic limit.
    }
    \label{fig:crossing_tV}
\end{figure*}

As shown in Fig.~\ref{fig:crossing_tV}(a), the $R_{\mathrm{CDW}}$-$V$ curves for $L\geq 12$ almost cross at a common point near $V\approx 1.34$, indicating that finite-size effects are relatively weak in the $t$-$V$ model compared to the honeycomb Hubbard model.
This can be attributed to two factors: (i) the GNI universality class lacks competing four-fermion interaction channels~\cite{Phys.Rev.B2023Ladovrechis}, leading to a larger \rev{leading irrelevant exponent $\omega_{\mathrm{irr}}$}; (ii) the smaller bosonic anomalous dimension $\eta_\phi$ reduces the contribution from the analytic part of the free energy.
\rev{Because the CDW phase breaks discrete $Z_2$ symmetry and is fully gapped, $R_{\mathrm{CDW}}$ saturates rapidly toward unity above $V_c$, reaching $\approx 0.96$ (and $R^{\Delta}_{\mathrm{CDW}}\approx 0.992$) at $V=1.40$ for $L=48$.}
\rev{To stay within the critical scaling regime and avoid saturation-induced curvature in cubic interpolations and data collapses, we restrict $L\ge 15$ to $V\le 1.38$ throughout all $t$-$V$ crossing-point and data-collapse analyses, while retaining $V\le 1.40$ for $L=9,12$ (and $V\le 1.50$ for $L=3,6$ in the crossing-point analysis to bracket the $(3,6)$ crossing, trimmed to $V\le 1.40$ for $L=6$ in the data collapses).}
To quantify the residual drift, we denote the crossing point between sizes $L$ and $2L$ as $V^{\times}(L,2L)$ [horizontal bars in panel (a)] and fit $V^{\times}(L,2L)=V_c^{\times}+a(2L)^{-b}$.
The inset of Fig.~\ref{fig:crossing_tV}(b) shows that this power-law extrapolation yields \rev{$V_c^{\times}=1.33467(99)$ and $b_{\mathrm{fit}}=1.925(90)$}.
\rev{The asymptotic crossing relation is $b=1/\nu+\min(\omega_{\mathrm{irr}},2-z-\eta_\phi)$ with $z=1$~\cite{Phys.Rev.B2014Campostrini}.}
\rev{The data-collapse estimates $\nu=0.869(31)$ and $\eta_\phi=0.468(6)$ from the same observables (Sec.~\ref{sec:gnh_criticality}) already give $1-\eta_\phi\approx0.53$, below the four-loop Pad\'e estimates $\omega_{\mathrm{irr}}\approx0.76$--$0.78$ for the GNI universality class~\cite{Phys.Rev.D2017Zerf}.}
\rev{Assuming that the analytic background controls the asymptotic drift, these estimates yield $b_{\mathrm{collapse}}=1.682(42)$, consistent with $b_{\mathrm{fit}}$ within $3\sigma$ as an internal consistency check over the accessible sizes (Sec.~\ref{sec:hubbard_model}).}

\begin{figure*}[htbp]
    \centering
    \includegraphics[width=\textwidth]{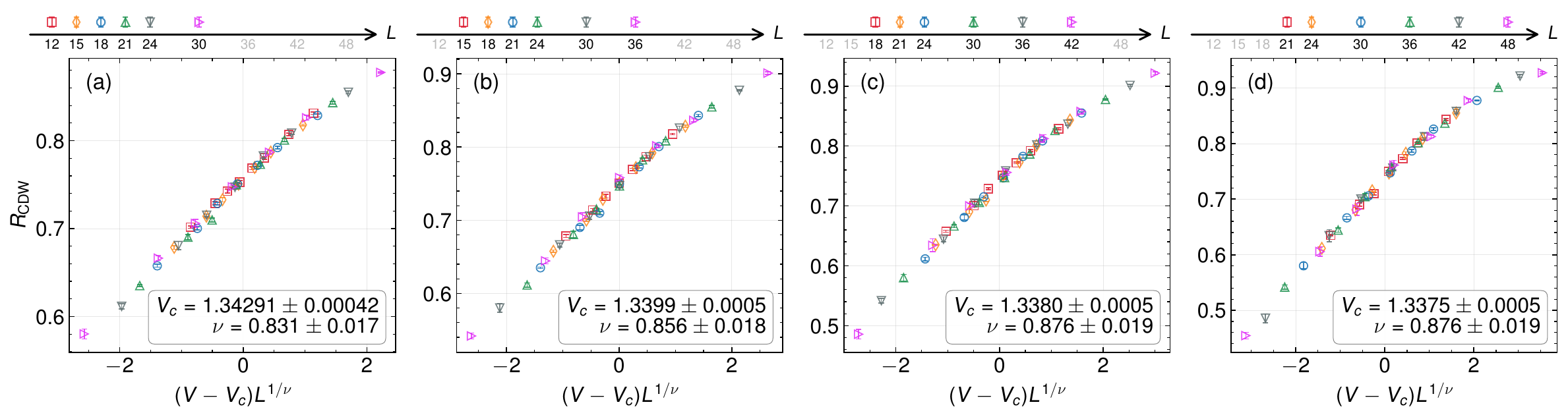}
    \caption{Sliding-window data-collapse analysis of the correlation ratio \rev{$R_{\mathrm{CDW}}$} for the $t$-$V$ model. 
    The data collapse is performed over a window of system sizes with a fixed width (covering 6 consecutive sizes)\rev{; four representative windows with $L_{\max}=30$--$48$ are shown}. 
    The finite-size scaling ansatz \rev{$R_{\mathrm{CDW}}(V,L)=f^{R}_0(uL^{1/\nu})$} is employed to extract the critical point $V_c$ and critical exponent $\nu$. 
    Panels show the collapsed curves and the extracted parameters $V_c$ and $\nu$ for each window.
    }
    \label{fig:data_collapse_CR_tV}
\end{figure*}

\begin{figure*}[htbp]
    \centering
    \includegraphics[width=\textwidth]{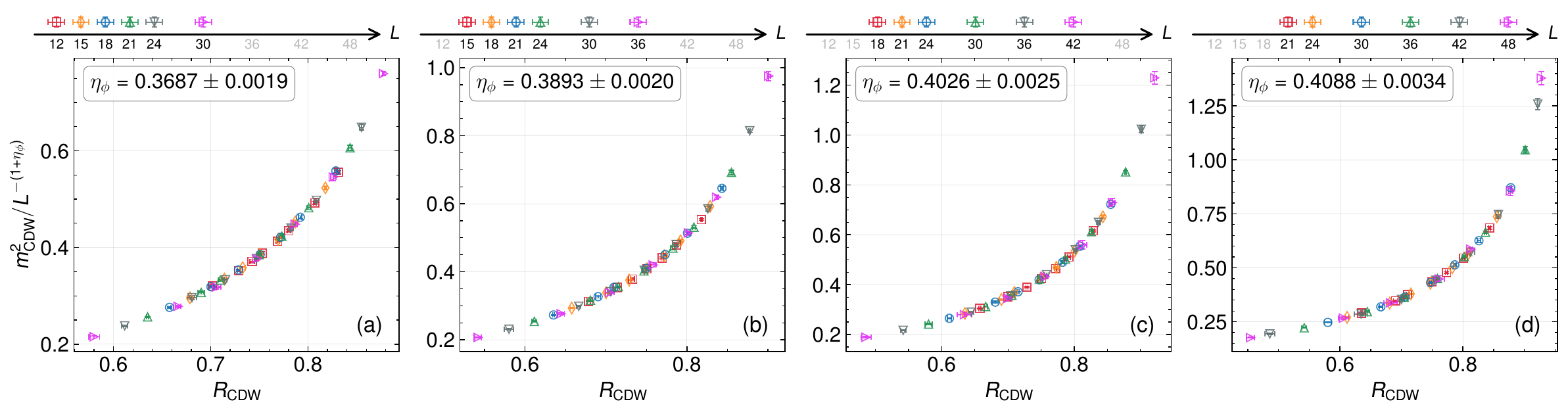}
    \caption{Sliding-window data-collapse analysis of the squared charge-density-wave order parameter $m_{\mathrm{CDW}}^2$ for the $t$-$V$ model. 
    The data collapse is performed over a window of system sizes with a fixed width (covering 6 consecutive sizes)\rev{; four representative windows with $L_{\max}=30$--$48$ are shown}. 
    The finite-size scaling ansatz $m_{\mathrm{CDW}}^{2}(R_{\mathrm{CDW}},L)=L^{-1-\eta_{\phi}}f_{0}^{m}(R_{\mathrm{CDW}})$ is employed to extract the bosonic anomalous dimension $\eta_\phi$. 
    Panels show the collapsed curves and the extracted parameter $\eta_\phi$ for each window.
    }
    \label{fig:data_collapse_m2_tV}
\end{figure*}

\begin{figure*}[htbp]
    \centering
    \includegraphics[width=\textwidth]{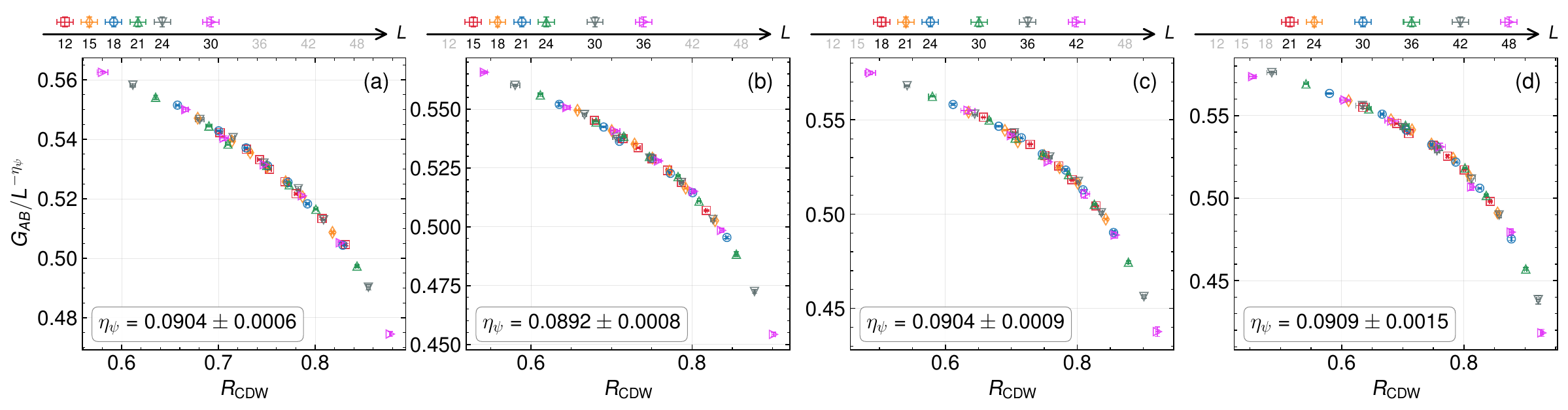}
    \caption{Sliding-window data-collapse analysis of the averaged modulus of the off-diagonal Green's function \rev{$G_{AB}$} for the $t$-$V$ model. 
    The data collapse is performed over a window of system sizes with a fixed width (covering 6 consecutive sizes)\rev{; four representative windows with $L_{\max}=30$--$48$ are shown}. 
    The finite-size scaling ansatz \rev{$G_{AB}(R_{\mathrm{CDW}},L)=L^{-\eta_{\psi}}f_{0}^{G}(R_{\mathrm{CDW}})$} is employed to extract the fermionic anomalous dimension $\eta_\psi$. 
    Panels show the collapsed curves and the extracted parameter $\eta_\psi$ for each window.
    }
    \label{fig:data_collapse_Gab_tV}
\end{figure*}

The sliding-window data-collapse results are shown for the correlation ratio (Fig.~\ref{fig:data_collapse_CR_tV}), the squared charge-density-wave order parameter (Fig.~\ref{fig:data_collapse_m2_tV}), and the off-diagonal Green's function (Fig.~\ref{fig:data_collapse_Gab_tV}). 

\begin{figure*}[htbp]
    \centering
    \includegraphics[width=0.85\textwidth]{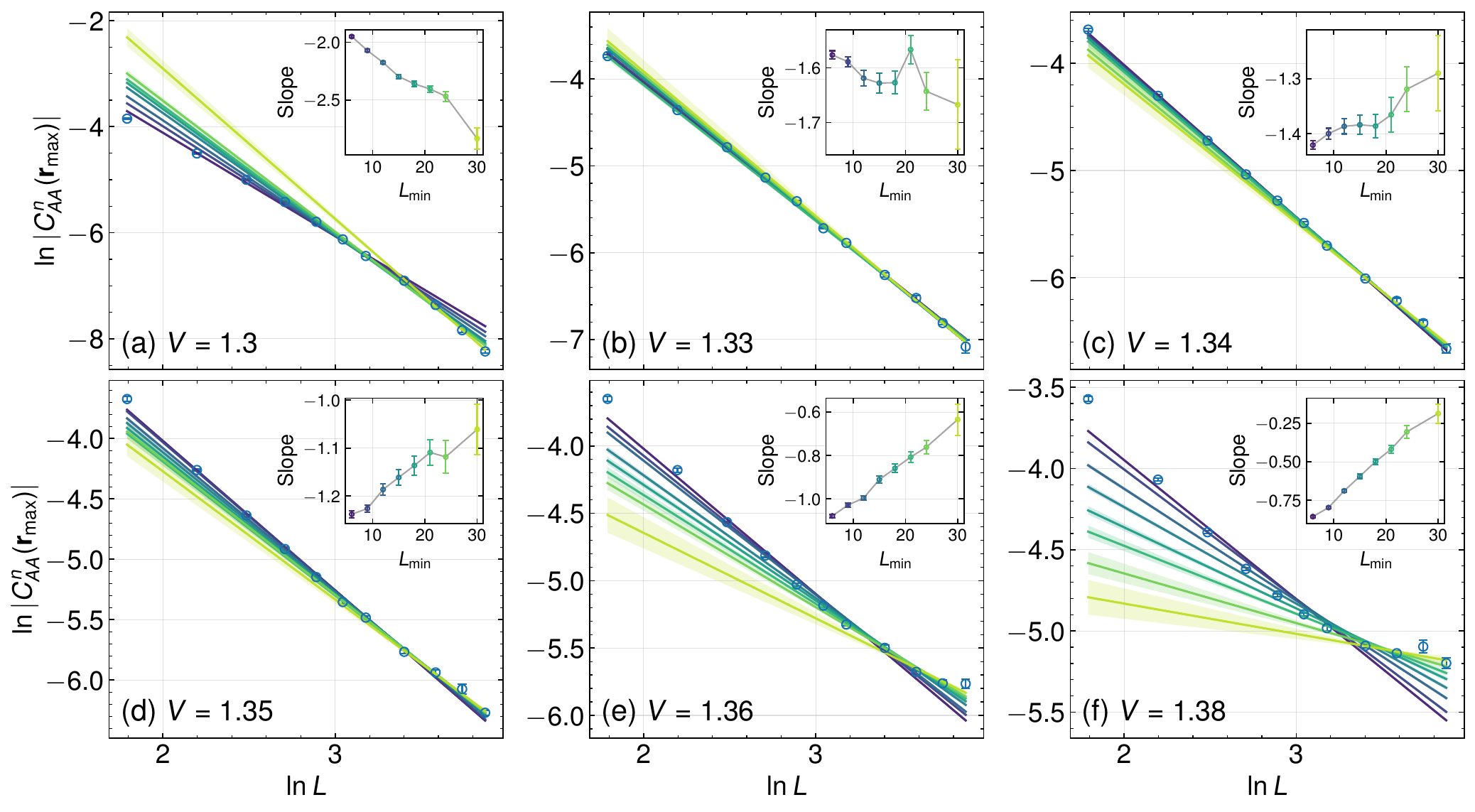}
    \caption{Finite-size scaling of the density-density correlation function \rev{$C^{n}_{AA}(\mathbf{r}_{\max})$ [Eq.~\eqref{equ:C_rmax_tV_def}]} at the maximum distance $\mathbf{r}_{\max} = \lfloor L/2 \rfloor (\mathbf{a}_2-\mathbf{a}_1)$ for the $t$-$V$ model; at the critical point, we expect a scaling behavior of $\sim 1/L^{1+\eta_\phi}$. 
    The plots show the double-logarithmic scaling of the correlation magnitude versus system size $L$ for various interaction strengths $V$. 
    Solid lines represent linear fits performed over the window $[L_{\min}, 48]$. 
    Insets display the fitted slope as a function of the fitting window lower bound $L_{\min}$.
    }
    \label{fig:decay_tV}
\end{figure*}

\begin{figure*}[htbp]
    \centering
    \includegraphics[width=0.82\textwidth]{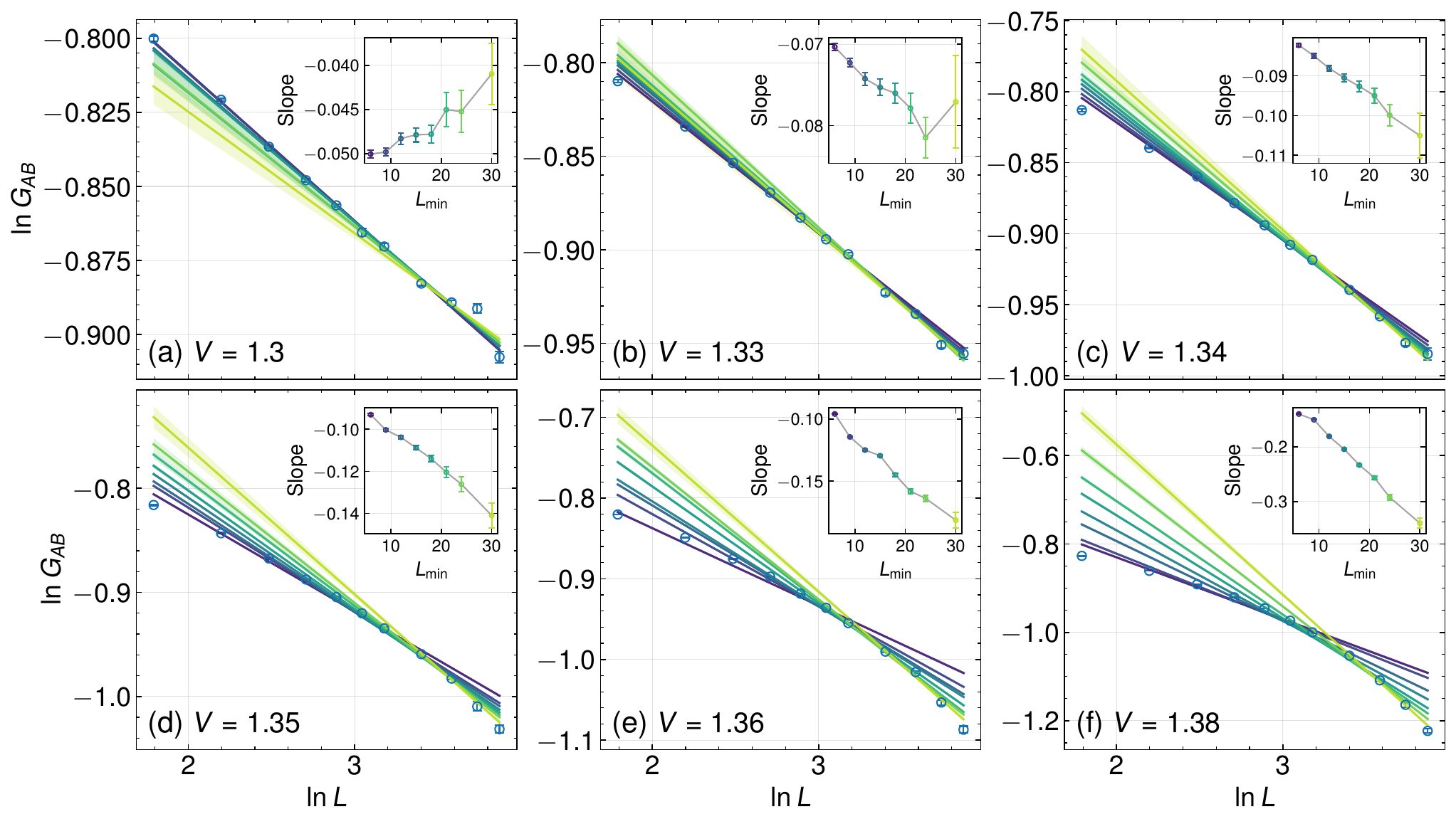}
    \caption{\rev{Finite-size scaling of the averaged off-diagonal Green's function modulus $G_{AB}$ at fixed interaction strengths $V$ for the $t$-$V$ model, in the same layout as Fig.~\ref{fig:decay_tV}; at the critical point, $G_{AB}\sim L^{-\eta_\psi}$.
    Solid lines and shaded bands show linear fits of $\ln G_{AB}$ versus $\ln L$ over $[L_{\min}, 48]$ and their $1\sigma$ uncertainties, and insets display the fitted slope as a function of $L_{\min}$.}}
    \label{fig:gab_decay_tV}

    \vspace{1.0em}
    \includegraphics[width=0.84\textwidth]{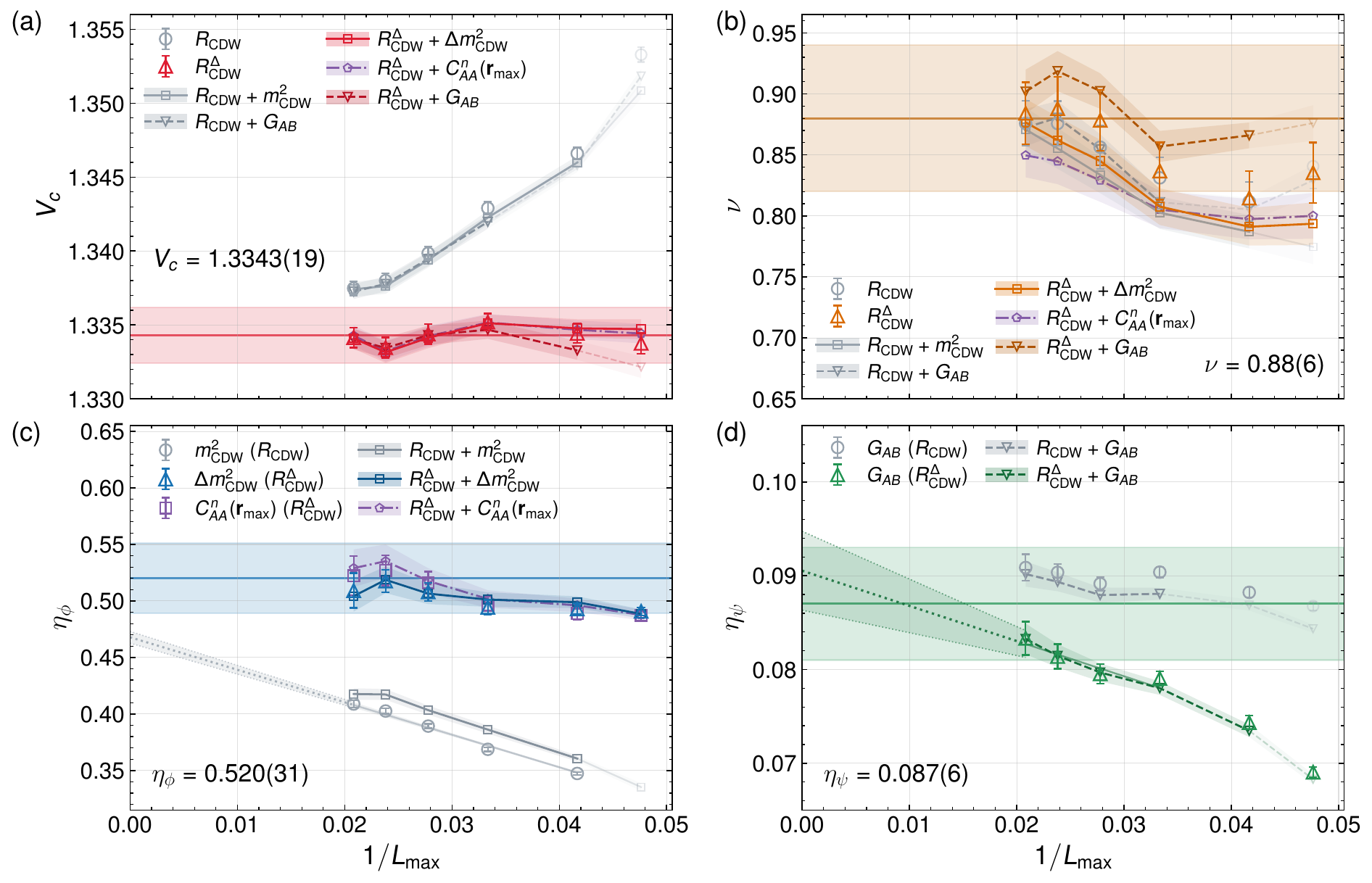}
    \caption{\rev{Critical quantities of the spinless $t$-$V$ model across equal-time observable families and data-collapse schemes, in the layout and conventions of Fig.~\ref{fig:fit_extrapolation_hubbard_comparison}, showing (a) $V_c$, (b) $\nu$, (c) $\eta_\phi$, and (d) $\eta_\psi$ versus $1/L_{\max}$ across conventional (gray) and background-suppressed (colored) separate and joint collapses.
    Dotted lines with shaded $1\sigma$ fans show the linear extrapolation of conventional $m^2_{\mathrm{CDW}}(R_{\mathrm{CDW}})$ in (c) and an illustrative linear extrapolation of $R^{\Delta}_{\mathrm{CDW}}+G_{AB}$ [$L_{\max}\ge 30$, giving $0.091(4)$] in (d); faded symbols and curves mark excluded small-size windows (Appendix~\ref{app:data_collapse_analysis}), and horizontal bands mark the combined final $t$-$V$ estimates in Eq.~\eqref{equ:final_estimates}.}}
    \label{fig:fit_extrapolation_tv_comparison}
\end{figure*}

\rev{Analogous to Eqs.~\eqref{equ:C_rmax_def}--\eqref{equ:R_Delta_def} in Sec.~\ref{sec:bg_suppressed}, the background-suppressed observables for the $t$-$V$ model are obtained by replacing $C_{\mathrm{AFM}}(\mathbf{k})$ with $C_{\mathrm{CDW}}(\mathbf{k})$ to define $\Delta m^2_{\mathrm{CDW}}\equiv C_{\mathrm{CDW}}(0)-C_{\mathrm{CDW}}(\delta\mathbf{k}_{1})$ and $R^{\Delta}_{\mathrm{CDW}}\equiv [C_{\mathrm{CDW}}(0)-C_{\mathrm{CDW}}(\delta\mathbf{k}_{1})]/[C_{\mathrm{CDW}}(0)-C_{\mathrm{CDW}}(\delta\mathbf{k}_{2})]$, together with the maximum-distance density-density correlation function}
\begin{equation}\label{equ:C_rmax_tV_def}
    \rev{C^{n}_{AA}(\mathbf{r}_{\max})\equiv\frac{1}{L^{2}}\sum_{i,j\in A}\delta_{\mathbf{r}_{i}-\mathbf{r}_{j},\,\mathbf{r}_{\max}}\left\langle\left(\hat{n}_{i}-\tfrac{1}{2}\right)\left(\hat{n}_{j}-\tfrac{1}{2}\right)\right\rangle.}
\end{equation}
\rev{Figures~\ref{fig:decay_tV} and~\ref{fig:gab_decay_tV} present the analogous fixed-$V$ analysis for $C^{n}_{AA}(\mathbf{r}_{\max})$ and $G_{AB}$, fitting linear slopes from $L_{\min}$ to $L=48$ and tracking them against $L_{\min}$ in the insets.
Inside the gapped CDW phase at $V\ge 1.36$ in panels~(e) and~(f), exponential decay on a nonzero CDW order floor flattens $\ln |C^{n}_{AA}(\mathbf{r}_{\max})|$ and drives its inset slope above $-1$, whereas without a constant floor $\ln G_{AB}$ bends downward and its inset slope steepens continuously with $L_{\min}$.
Since $V_c$ lies between the simulated couplings $V=1.33$ and $1.34$ across Figs.~\ref{fig:fit_extrapolation_bothmodels}(e),~\ref{fig:fit_extrapolation_bgsup}(e), and~\ref{fig:crossing_tV}(b), panels~(b) and~(c) of Figs.~\ref{fig:decay_tV} and~\ref{fig:gab_decay_tV} bracket criticality from the semimetallic and CDW sides.
Consistently, between panels~(b) and~(c), the large-size slopes from $L_{\min}=24$ span $-1.64(3)$ to $-1.32(4)$ in Fig.~\ref{fig:decay_tV} and $-0.082(3)$ to $-0.100(3)$ in Fig.~\ref{fig:gab_decay_tV}, whose implied exponents enclose the data-collapse estimates of $\eta_\phi$ in Figs.~\ref{fig:fit_extrapolation_bothmodels}(g) and~\ref{fig:fit_extrapolation_bgsup}(g) and of $\eta_\psi$ in Figs.~\ref{fig:fit_extrapolation_bothmodels}(h) and~\ref{fig:fit_extrapolation_bgsup}(h) within uncertainties.}

\rev{Figure~\ref{fig:fit_extrapolation_tv_comparison} overlays the conventional and background-suppressed $t$-$V$ sliding-window trajectories (both separate and joint collapses) in the same format as Fig.~\ref{fig:fit_extrapolation_hubbard_comparison}.
In panels~(a)--(c), suppressing the short-range background flattens $V_c$ across all windows and brings $\nu$ and $\eta_\phi$ onto consistent large-size plateaus across separate and joint fits, yielding the conservatively rounded $1\sigma$-union intervals $V_c=1.3343(19)$, $\nu=0.88(6)$, and $\eta_\phi=0.520(31)$ in Eq.~\eqref{equ:final_estimates}.}

\rev{In panel~(d), the conventional and background-suppressed trajectories provide complementary upper and lower bounds on $\eta_\psi$.
Unlike in the Hubbard model, where an anomalous $\nu$ in $R_{\mathrm{AFM}}+G_{AB}$ rules out the conventional $\eta_\psi\approx 0.19$ plateau [Fig.~\ref{fig:fit_extrapolation_hubbard_comparison}(b)], in the $t$-$V$ model $R_{\mathrm{CDW}}+G_{AB}$ shows no anomaly in either parameter, as its $V_c$ tracks $R_{\mathrm{CDW}}$ and $R_{\mathrm{CDW}}+m^2_{\mathrm{CDW}}$ within the conventional family [panel~(a)] while its $\nu$ agrees across both families [panel~(b)].
We therefore retain the conventional plateaus [$0.0898(25)$ from $G_{AB}(R_{\mathrm{CDW}})$ and $0.0898(18)$ from $R_{\mathrm{CDW}}+G_{AB}$], which serve as an upper bound because the conventional $V_c$ still lies slightly on the CDW side where $G_{AB}$ decays faster [Figs.~\ref{fig:gab_decay_tV}(b) and~\ref{fig:gab_decay_tV}(c)].
Conversely, both $G_{AB}(R^{\Delta}_{\mathrm{CDW}})$ and $R^{\Delta}_{\mathrm{CDW}}+G_{AB}$ drift upward across all accessible windows, a nonuniversal, observable-dependent effect seen only in $G_{AB}$ that may reflect slight over-subtraction of a $\mathbf{k}$-dependent background via the farther shell $\delta\mathbf{k}_{2}$ in $R^{\Delta}_{\mathrm{CDW}}$ alongside subleading corrections in $G_{AB}$ itself.
Their largest-window ($[21,48]$) frontier values, $0.0833(18)$ and $0.0833(17)$, thus serve only as a lower bound, with larger system sizes needed for a tighter background-suppressed determination.
Taking the conservatively rounded $1\sigma$ union of the two bounds yields $\eta_\psi=0.087(6)$ in Eq.~\eqref{equ:final_estimates}, and an illustrative linear extrapolation of $R^{\Delta}_{\mathrm{CDW}}+G_{AB}$ ($L_{\max}\ge 30$) gives $0.091(4)$, consistent with this interval within uncertainties.}

\FloatBarrier
%%%%%%%%%%%%%%%%%%%%%%%%%%%%%%%%%%%%%%%%%%%%%%%%%%%%%
\section{\rev{Covariance propagation in the joint bootstrap}}\label{app:covariance_extrapolation}
%%%%%%%%%%%%%%%%%%%%%%%%%%%%%%%%%%%%%%%%%%%%%%%%%%%%%
\begingroup
\revcolor

The joint bootstrap of Appendix~\ref{app:data_collapse_analysis} propagates two layers of statistical correlation that a conventional window-by-window analysis discards (Fig.~\ref{fig:covariance_layers}), namely the data-sharing-induced inter-window covariance $\Sigma$ [Eq.~\eqref{equ:window_covariance}] across overlapping fitting windows and the intra-run covariance $\Lambda$ among quantities measured in the same simulation run.
Here we show that both are present in our data and switch them on one at a time, starting from a decoupled bootstrap that carries neither.
We follow their effect on the three width-six trajectories that undergo linear $1/L_{\max}$ extrapolation in Fig.~\ref{fig:fit_extrapolation_bothmodels}, namely Hubbard $U_c$ and $\eta_\phi$ ($L_{\max}=33$--$72$) and $t$-$V$ $\eta_\phi$ ($L_{\max}=24$--$48$).
Table~\ref{tab:covariance_stages} collects the extrapolations at each stage.

\begin{figure}[tp]
    \centering
    \includegraphics[width=\columnwidth]{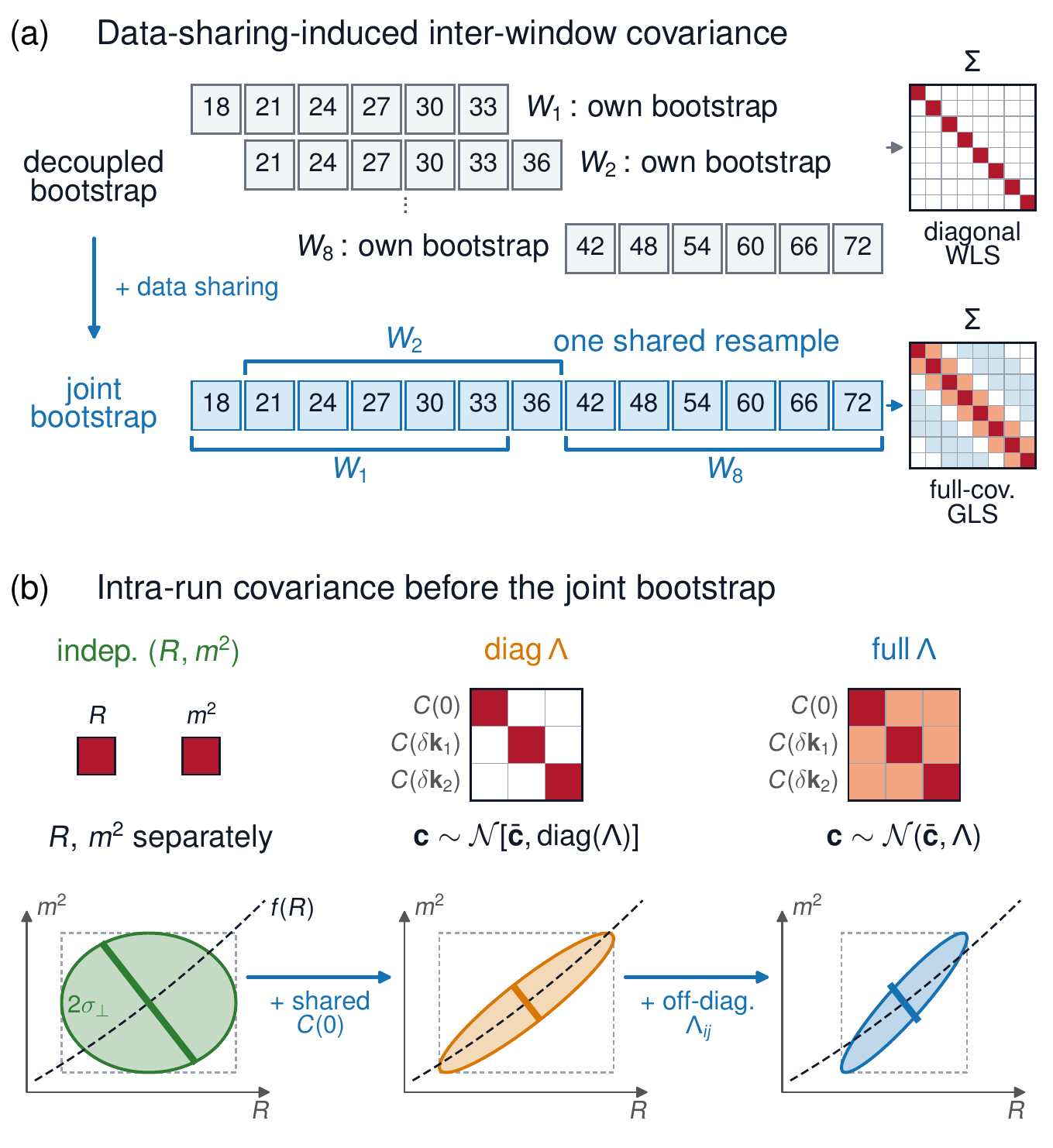}
    \caption{\rev{Two layers of covariance examined in this appendix.
    (a) In the decoupled bootstrap, each fitting window (the Hubbard windows $W_1$, $W_2$, and $W_8$ are shown) has its own bootstrap, and only $\mathrm{diag}(\Sigma)$ is available.
    In the joint bootstrap, all windows are fitted to the same resample, which yields the full matrix $\Sigma$.
    The patterns of $\Sigma$ are schematic.
    (b) Three nested ways of drawing the data at one simulation point, with the resulting noise ellipse of $(R,m^2)$ inside its marginal bounding box (dashed rectangle).
    The dashed line is the collapse curve $m^2=f(R)$, and the thick bar marks the spread $2\sigma_\perp$ normal to it, which alone perturbs the collapse.
    The ellipses are schematic, and the narrowing of $\sigma_R$ under $\mathrm{full}\,\Lambda$ occurs in the Hubbard model only.}}
    \label{fig:covariance_layers}
\end{figure}

\begin{table}[tp]
    \centering
    \caption{\rev{Linear $1/L_{\max}$ extrapolations at each stage of covariance propagation, with $\chi^2_{\mathrm{red}}$ of the fit in brackets ($\mathrm{dof}=6$ for the Hubbard model and $3$ for the $t$-$V$ model).
    The first row uses diagonal WLS and the others full-covariance GLS.
    The last row is the joint bootstrap of Appendix~\ref{app:data_collapse_analysis}.}}
    \label{tab:covariance_stages}
    \begingroup
    \revcolor
    \footnotesize
    \setlength{\tabcolsep}{1pt}
    \begin{tabular}{lccc}
    \toprule
    Stage & Hubbard $U_c$ & Hubbard $\eta_\phi$ & $t$-$V$ $\eta_\phi$ \\
    \midrule
    decoupled & $3.6731(34)$ [$0.72$] & $0.766(8)$ [$0.31$] & $0.473(15)$ [$0.11$] \\
    $\mathrm{indep.}\,(R,m^2)$ & $3.671(5)$ [$1.64$] & $0.765(11)$ [$0.35$] & $0.469(19)$ [$0.23$] \\
    $\mathrm{diag}\,\Lambda$ & $3.6708(45)$ [$1.70$] & $0.776(5)$ [$1.25$] & $0.469(6)$ [$3.03$] \\
    $\mathrm{full}\,\Lambda$ & $3.6702(35)$ [$4.03$] & $0.7790(31)$ [$2.95$] & $0.468(6)$ [$2.97$] \\
    \bottomrule
    \end{tabular}
    \endgroup
\end{table}

\subsection{Data sharing and inter-window covariance}\label{app:covariance_joint}

The decoupled bootstrap [Fig.~\ref{fig:covariance_layers}(a)] bootstraps and collapses each fitting window on its own and then extrapolates the window estimates by diagonal weighted least squares (diagonal WLS), which assumes zero covariance between windows.
The joint bootstrap instead fits every window to the same resample, so the replicates of windows that share system sizes fluctuate together and yield the full matrix $\Sigma$ for the full-covariance GLS extrapolation [Eq.~\eqref{equ:window_gls}].
To isolate this step, both workflows here perturb $R$ and $m^2$ independently at each simulation point [$\mathrm{indep.}\,(R,m^2)$ in Fig.~\ref{fig:covariance_layers}(b)].
Each window then has the same marginal distribution in both, so we evaluate the decoupled extrapolation on the joint replicates by retaining only $\mathrm{diag}(\Sigma)$, and the two extrapolations differ solely in the off-diagonal entries of $\Sigma$.

Figures~\ref{fig:covariance_extrapolation}(a)--(c) show that these entries are far from negligible.
Adjacent width-six windows, which share five of their six system sizes, have Pearson correlations $r_{ij}=\Sigma_{ij}/\sqrt{\Sigma_{ii}\Sigma_{jj}}$ of $0.52$--$0.58$ on average.
At larger separations, $r_{ij}$ changes sign and develops a weak negative tail, more pronounced in the Hubbard model, before vanishing once windows no longer overlap.
The sign change is a leverage effect, because each estimate is set by the size trend across its window and a fluctuation at one size therefore shifts it in opposite directions at the upper-$L$ and lower-$L$ edges.
Distant windows share only sizes that lie at the upper edge of the earlier window and the lower edge of the later one.
This pattern is set by the window overlap and changes little when the intra-run covariance is included in Appendix~\ref{app:covariance_paired}.

\begin{figure*}[tp]
    \centering
    \includegraphics[width=\textwidth]{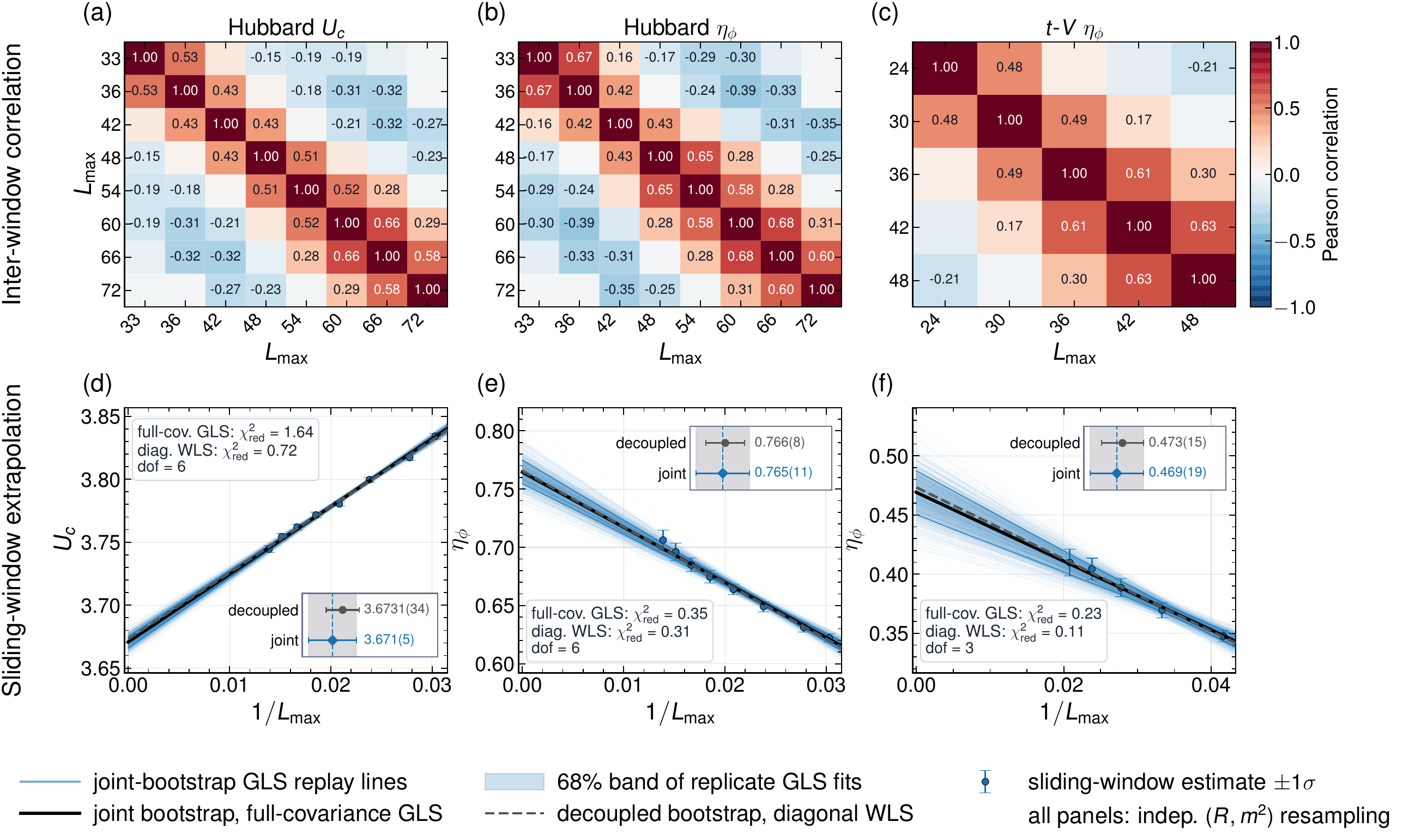}
    \caption{\rev{Data-sharing-induced inter-window correlations and their role in the $1/L_{\max}$ extrapolation, with $R$ and $m^2$ resampled independently at each simulation point.
    (a)--(c) Pearson correlation matrices $r_{ij}$ of the window replicates.
    (d)--(f) Sliding-window estimates versus $1/L_{\max}$ with the diagonal WLS fit of the decoupled bootstrap and the full-covariance GLS fit of the joint bootstrap.
    Thin blue lines and the blue band show GLS fits to individual replicate vectors $\mathbf{p}^{(b)}$ and their central $68\%$ interval.
    Insets compare the intercepts, with gray shading marking the $1\sigma$ range of the joint-bootstrap intercept, and boxes report $\chi^2_{\mathrm{red}}$.
    All error bars denote $1\sigma$ uncertainties.}}
    \label{fig:covariance_extrapolation}
\end{figure*}

Restoring them has two consequences for the extrapolation [Figs.~\ref{fig:covariance_extrapolation}(d)--(f) and Table~\ref{tab:covariance_stages}].
First, the intercepts shift by at most $0.5$ standard errors, while their standard errors grow by $27\%$--$38\%$, because diagonal WLS counts positively correlated neighboring windows as independent constraints.
Second, $\chi^2_{\mathrm{red}}$ rises by a factor of $1.1$--$2.3$.
Positive correlation between adjacent windows suppresses the variance of their difference, so $\Sigma^{-1}$ penalizes smooth departures from linearity more strongly than $\mathrm{diag}(\Sigma)^{-1}$ does.
For Hubbard $U_c$, this already signals a departure from a linear $1/L_{\max}$ drift.
For $\eta_\phi$, $\chi^2_{\mathrm{red}}$ remains far below unity, because the independent resampling of $R$ and $m^2$ inflates the window uncertainties.

\subsection{Intra-run covariance}\label{app:covariance_paired}

At each simulation point, $(L,U)$ or $(L,V)$, $\Lambda$ contains the covariances among the momentum shells and, through the shared $C(0)$, a covariance between $R=1-C(\delta\mathbf{k}_1)/C(0)$ and $m^2=C(0)$.
Figure~\ref{fig:covariance_paired}(a) shows the corresponding intra-run Pearson correlations over all $L\ge 6$ simulation points, whose distributions average out the finite-bin fluctuations of the individual jackknife estimates.
In both models, $R$ and $m^2$ are almost perfectly correlated, with medians of $+0.93$ in the Hubbard model and $+0.96$ in the $t$-$V$ model.
The momentum-shell correlations are strong and positive in the Hubbard model (medians $+0.71$ between $C(0)$ and $C(\delta\mathbf{k}_1)$ and $+0.95$ between $C(\delta\mathbf{k}_1)$ and $C(\delta\mathbf{k}_2)$) but weak in the $t$-$V$ model ($-0.28$ and $+0.23$).

\begin{figure*}[tp]
    \centering
    \includegraphics[width=\textwidth]{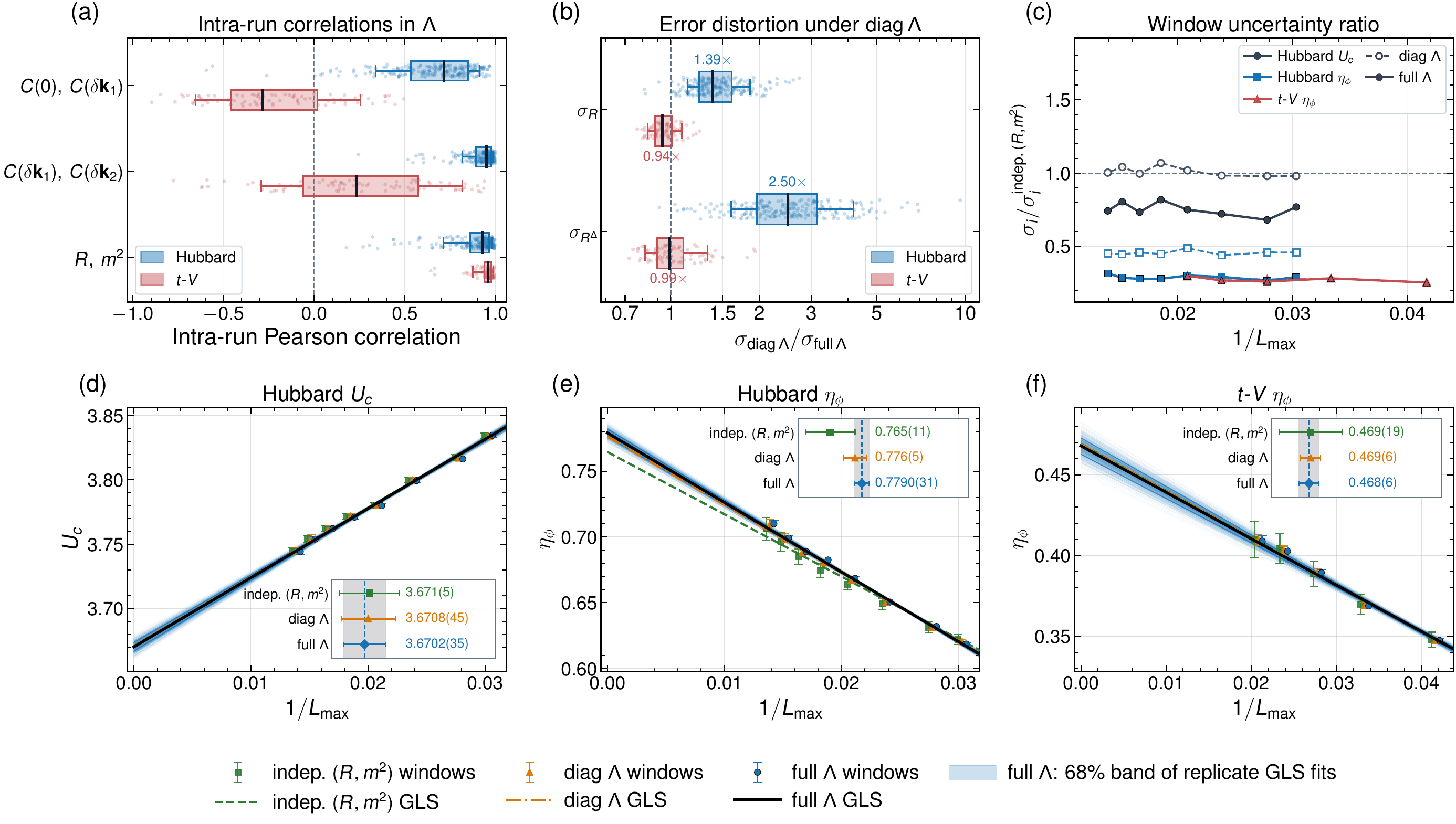}
    \caption{\rev{Intra-run covariance and its effect on the sliding-window analysis.
    (a) Distributions over all $L\ge 6$ simulation points entering the collapses ($218$ for the Hubbard model and $82$ for the $t$-$V$ model) of the intra-run Pearson correlations, with one dot per simulation point. Boxes span the interquartile range, with median bars and $10\%$--$90\%$ whiskers.
    (b) Pointwise uncertainty ratio $\sigma_{\mathrm{diag}\,\Lambda}/\sigma_{\mathrm{full}\,\Lambda}$ of $R$ and $R^\Delta$, annotated with medians.
    (c) Per-window standard-error ratio $\sigma_i/\sigma_i^{\mathrm{indep.}\,(R,m^2)}$ ($\sigma_i\equiv\sqrt{\Sigma_{ii}}$) under $\mathrm{diag}\,\Lambda$ (open symbols) and $\mathrm{full}\,\Lambda$ (filled symbols).
    (d)--(f) Sliding-window estimates (offset horizontally for visibility) and full-covariance GLS extrapolations under the three resampling schemes defined in the text. Blue bands mark the central $68\%$ interval of replicate GLS fits under $\mathrm{full}\,\Lambda$. Insets compare the intercepts, with gray shading marking the $1\sigma$ range of the $\mathrm{full}\,\Lambda$ intercept.}}
    \label{fig:covariance_paired}
\end{figure*}

The intra-run covariance has two sources, which we add one at a time while keeping the full-covariance GLS extrapolation fixed [Fig.~\ref{fig:covariance_layers}(b)].
The first is that $R$ and $m^2$ are computed from the same $C(0)$.
\textit{Diagonal-$\Lambda$ resampling} ($\mathrm{diag}\,\Lambda$) draws $\mathbf{c}\sim\mathcal{N}[\bar{\mathbf{c}},\mathrm{diag}(\Lambda)]$, evaluates both $R$ and $m^2$ from the drawn $\mathbf{c}$, and propagates $E_i$ from $\mathrm{diag}(\Lambda)$.
They then become correlated through the shared $C(0)$, while each keeps its marginal error bar from $\mathrm{indep.}\,(R,m^2)$.
The second is the covariance among different correlation functions, namely the off-diagonal entries of $\Lambda$.
\textit{Full-$\Lambda$ resampling} ($\mathrm{full}\,\Lambda$) draws $\mathbf{c}\sim\mathcal{N}(\bar{\mathbf{c}},\Lambda)$ as in Appendix~\ref{app:data_collapse_analysis} and includes both.
For the pair $(R,m^2)$, the two sources are the two terms of the covariance obtained by error propagation (Appendix~\ref{app:data_collapse_analysis}),
\begin{equation}\label{equ:R_m2_covariance}
    \sigma_{R,m^2}=\frac{1-R}{m^2}\,\sigma_{m^2}^2-\frac{\mathrm{cov}[C(0),C(\delta\mathbf{k}_1)]}{m^2},
\end{equation}
of which $\mathrm{diag}\,\Lambda$ retains only the first.

First, $\mathrm{diag}\,\Lambda$ compresses the window uncertainties of $\eta_\phi$ by more than half in both models [Fig.~\ref{fig:covariance_paired}(c)] and reduces the intercept uncertainty more than twofold in the Hubbard model and more than threefold in the $t$-$V$ model (Table~\ref{tab:covariance_stages}).
As a control, collapsing $R$ alone with no second observable fluctuating with it leaves Hubbard $U_c$ essentially unchanged, confirming that the $\eta_\phi$ compression comes from the $R$--$m^2$ covariance.
The origin of this compression is geometric [Fig.~\ref{fig:covariance_layers}(b)].
The data collapse absorbs this $R$--$m^2$ covariance because the two fluctuate together along the curve $m^2=f(R)$, and tangential fluctuations only slide data points along it without affecting the fit.
Since $m^2$ grows with $R$, their positive correlation aligns the noise ellipse with the curve and shrinks the normal component $\sigma_\perp$ at fixed marginal variances.
For fluctuations $(\delta X_i,\delta Y_i)$ around the rescaled curve $Y=F(X)$, Eq.~\eqref{equ:effective_variance} directly reflects this projection through $E_i^2=\mathrm{Var}[\delta Y_i-F^{\prime}(X_i)\,\delta X_i]=[1+F^{\prime}(X_i)^2]\,\sigma_\perp^2$.
Its cross term $-2F^{\prime}(X_i)\sigma_{X_iY_i}$ removes the fluctuations that run along the curve.
The same holds whenever the fluctuations of two collapsed observables follow their mutual trend.
For example, $G_{AB}$ decreases with $R$ and is anticorrelated with it within a run (median correlation about $-0.3$ in both models), so its collapse against $R$ benefits in the same way, although more weakly.

Second, $\mathrm{full}\,\Lambda$ matters where the momentum-shell correlations are strong.
In the Hubbard model, neglecting them overestimates $\sigma_R$ by a median factor of $1.39$ and $\sigma_{R^\Delta}$ [Eq.~\eqref{equ:R_Delta_def}] by $2.50$ [Fig.~\ref{fig:covariance_paired}(b)].
Restoring them lowers the $U_c$ window uncertainties to $0.68$--$0.82$ of their $\mathrm{indep.}\,(R,m^2)$ values and reduces the $\eta_\phi$ intercept uncertainty by a further third.
In the $t$-$V$ model, $\sigma_R$ and $\sigma_{R^\Delta}$ are nearly unchanged, and $\mathrm{diag}\,\Lambda$ already captures the gain.
Compared with $\mathrm{indep.}\,(R,m^2)$, $\mathrm{full}\,\Lambda$ shifts the intercepts by only $0.1$ standard errors for both Hubbard $U_c$ and $t$-$V$ $\eta_\phi$, and by $1.4$ standard errors for Hubbard $\eta_\phi$ once $\sigma_{R,m^2}$ is restored in the point weights $E_i$ [Figs.~\ref{fig:covariance_paired}(d)--(f)].
Between $\mathrm{diag}\,\Lambda$ and $\mathrm{full}\,\Lambda$, the intercepts agree within $0.6$ standard errors throughout.

The compressed window uncertainties also sharpen the goodness-of-fit test of the linear extrapolation (Table~\ref{tab:covariance_stages}).
As the covariance layers are added, $\chi^2_{\mathrm{red}}$ increases and exceeds unity for all three trajectories under $\mathrm{full}\,\Lambda$, most clearly in the Hubbard model, exposing a residual curvature in $1/L_{\max}$ that the inflated uncertainties of the earlier stages concealed.

\clearpage
\endgroup

%%%%%%%%%%%%%%%%%%%%%%%%%%%%%%%%%%%%%%%%%%%%%%%%%%%%%
\section{Fitting-window scan in data-collapse analysis}\label{app:fitting_window_scan}
%%%%%%%%%%%%%%%%%%%%%%%%%%%%%%%%%%%%%%%%%%%%%%%%%%%%%

\begin{figure*}[htbp]
    \centering
    \includegraphics[width=\textwidth]{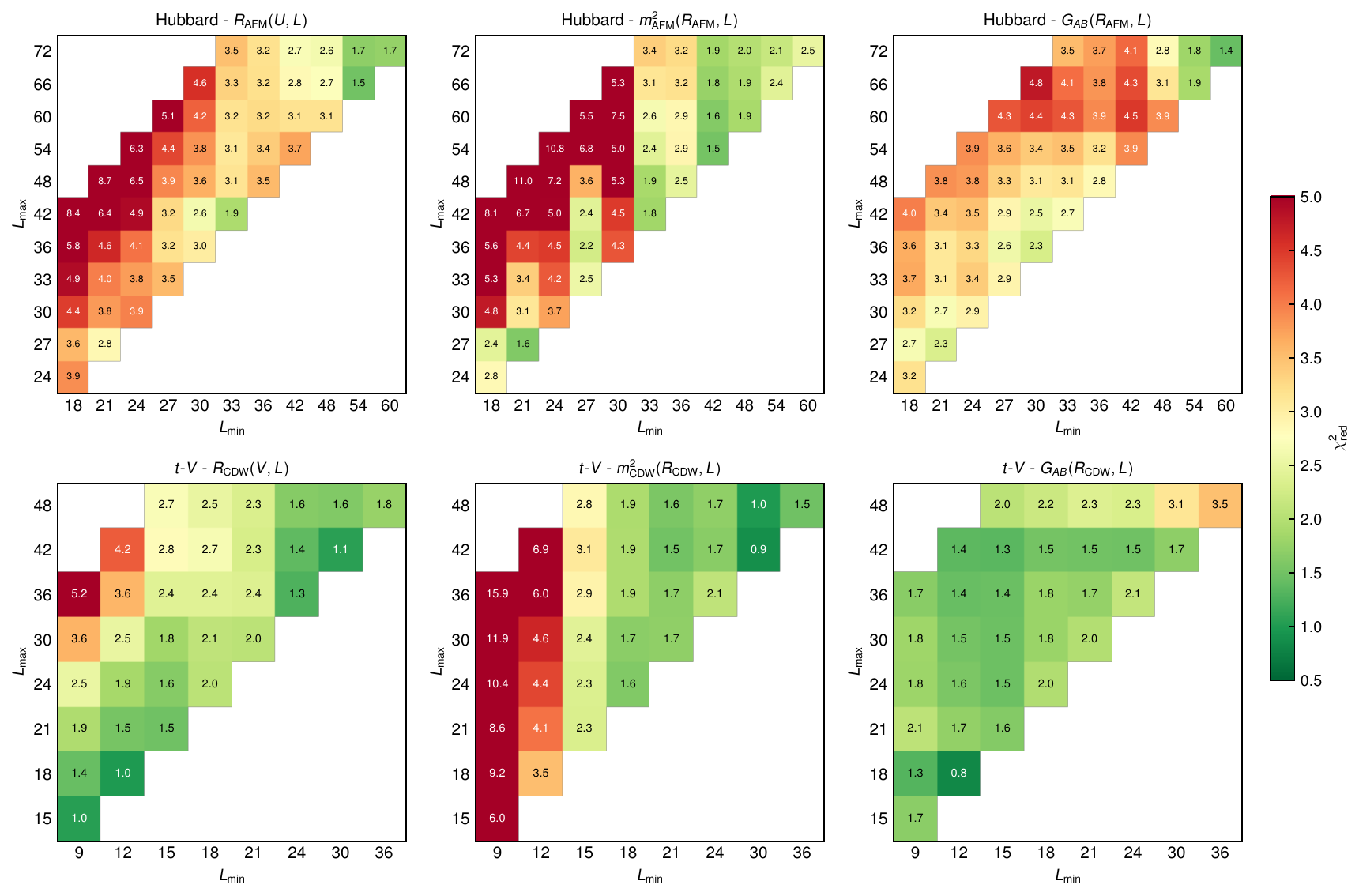}
    \caption{Heatmaps of the reduced chi-squared $\chi^2_{\mathrm{red}}$ for the data-collapse fits across different size windows $[L_{\text{min}},L_{\text{max}}]$.
    The top (bottom) row shows results for the Hubbard ($t$-$V$) model, and the three columns correspond to the correlation-ratio collapse $R(U,L)$ [or $R(V,L)$], the \rev{order-parameter} collapse $m^2(R,L)$, and the \rev{averaged-modulus} collapse \rev{$G_{AB}(R,L)$}, respectively.
    \rev{Each $\chi^2_{\mathrm{red}}$ is evaluated with all fitting parameters fixed at their bootstrap means.}
    From left to right and top to bottom, the window width decreases (i.e., fewer system sizes are included in each fit), and one sees that $\chi^2_{\mathrm{red}}$ also decreases in most cases.
    Colors are clipped at $\chi^2_{\mathrm{red}}=5$ for visualization, while the numbers in each cell indicate the actual values.}
    \label{fig:chi2red_heatmap}
\end{figure*}

\begin{figure*}[htbp]
    \centering
    \includegraphics[width=\textwidth]{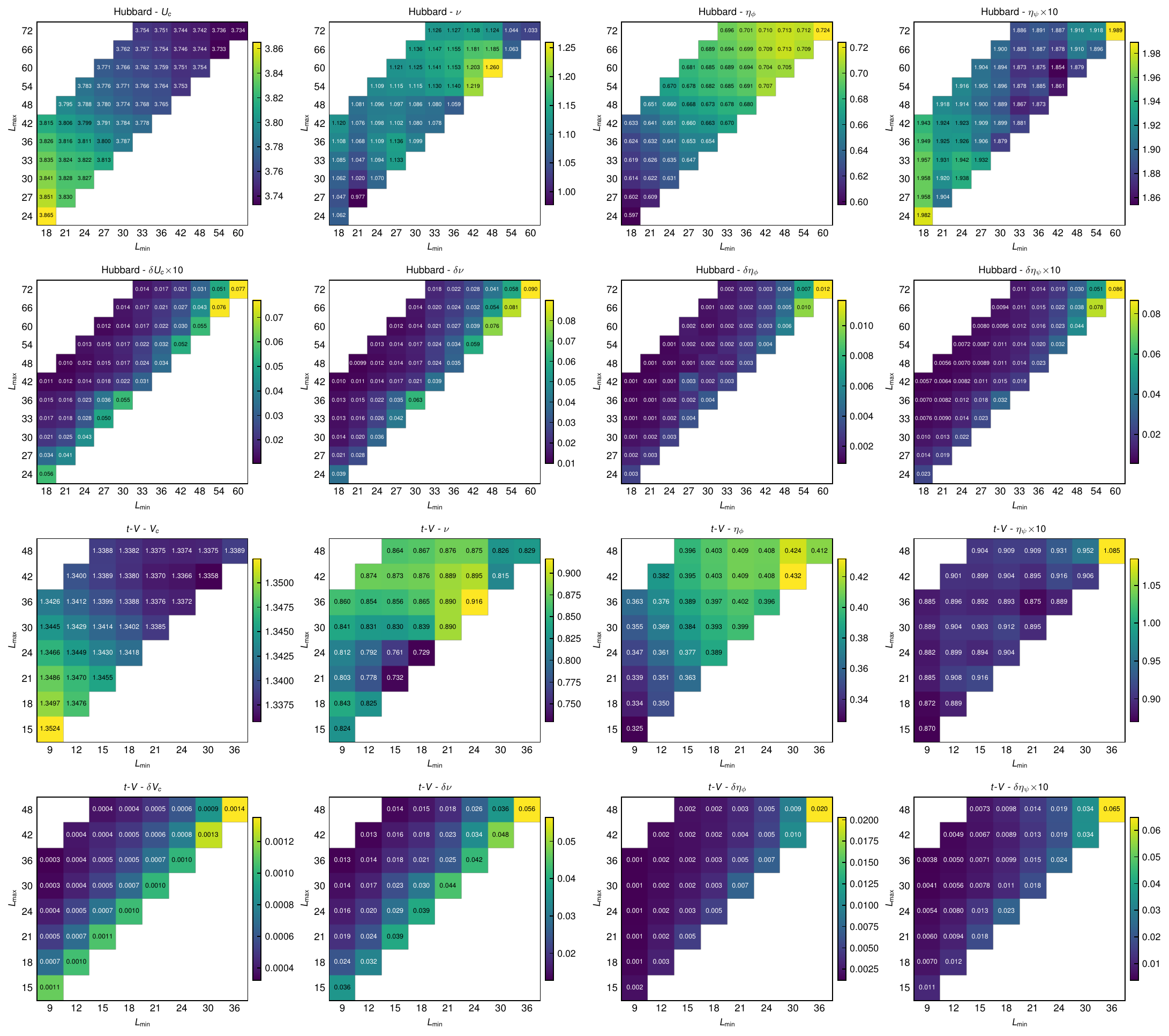}
    \caption{Heatmaps of the fitted critical parameters and their bootstrap uncertainties across different size windows $[L_{\text{min}},L_{\text{max}}]$.
    For each model, the upper row shows the fitted central values of $(X_c,\nu,\eta_\phi,\eta_\psi)$, and the lower row shows the corresponding bootstrap standard deviations.
    Here $X_c=U_c$ for the Hubbard model (top two rows) and $X_c=V_c$ for the $t$-$V$ model (bottom two rows).
    From left to right and top to bottom, the window width decreases (i.e., fewer system sizes are included in each fit), and one sees that the uncertainties increase in most cases.
    \rev{Reading a horizontal row from left to right, i.e., increasing $L_{\min}$ at fixed $L_{\max}$, traces the shrinking-window procedure of Appendix~\ref{app:shrinking_window}, while reading a constant-width diagonal from lower left to upper right traces the sliding-window trajectory of Fig.~\ref{fig:fit_extrapolation_bothmodels}.}}
    \label{fig:critical_exponents_heatmap}
\end{figure*}

In this section, we summarize a comprehensive scan of the fitting window in linear system size, parameterized by its lower and upper cutoffs $L_{\text{min}}$ and $L_{\text{max}}$.
We vary the window width from three to \rev{eight} consecutive system sizes.
For each window, we perform the same data-collapse fits as in Appendix~\ref{app:data_collapse_analysis} and record the final reduced chi-squared $\chi^2_{\mathrm{red}}$, the fitted critical parameters, and their bootstrap uncertainties.
\rev{Windows of all widths are fitted on the same resamples of the covariance-aware joint bootstrap, and each window estimate and its uncertainty are the mean and standard deviation over the resulting replicates.}
The results are shown in Figs.~\ref{fig:chi2red_heatmap} ($\chi^2_{\mathrm{red}}$) and \ref{fig:critical_exponents_heatmap} (critical point and exponents).

\begin{figure*}[htbp]
    \centering
    \includegraphics[width=\textwidth]{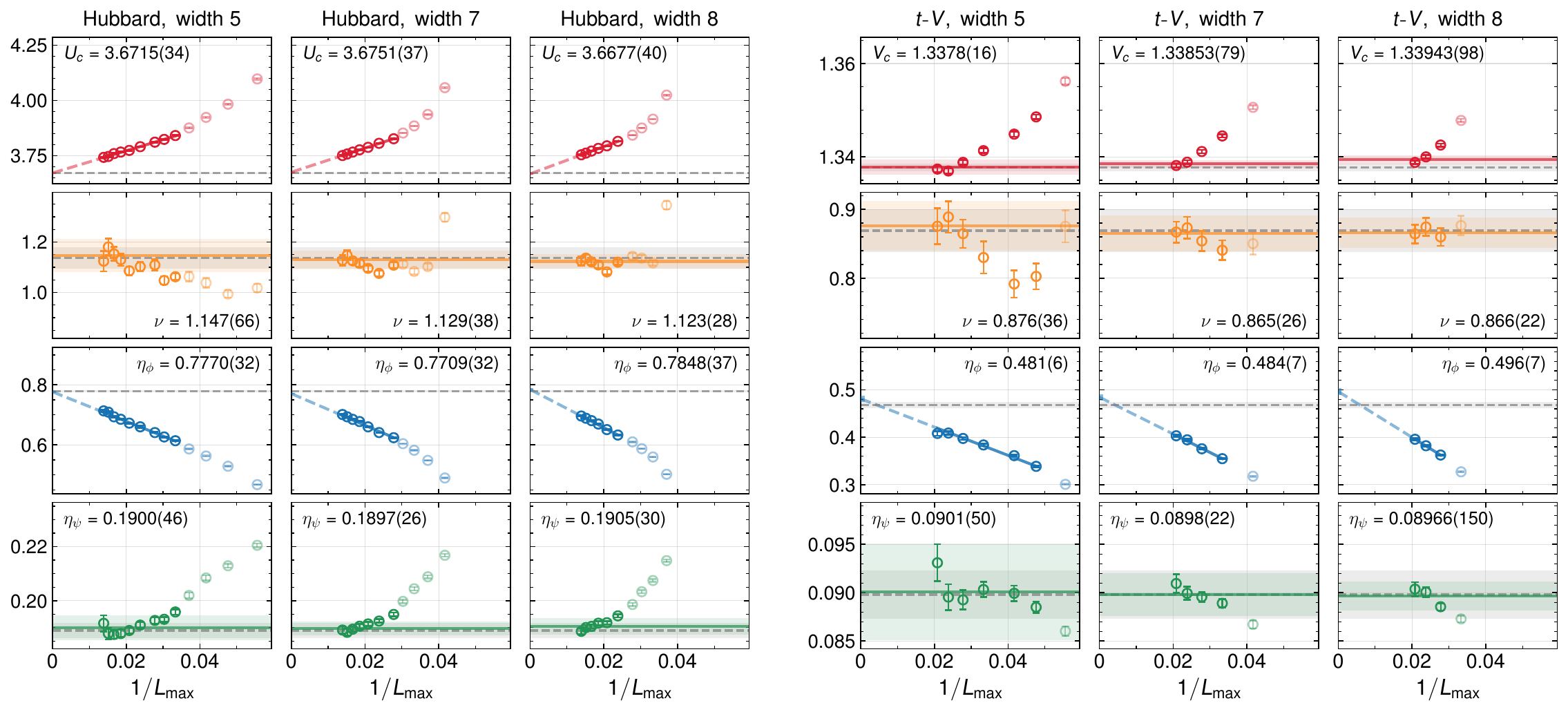}
    \caption{\rev{Sliding-window analysis of Fig.~\ref{fig:fit_extrapolation_bothmodels} repeated with windows of five, seven, and eight consecutive sizes.
    The three left (right) columns show the Hubbard ($t$-$V$) model, with the window width given above each column.
    The four rows show $X_c$, $\nu$, $\eta_\phi$, and $\eta_\psi$, and panels of the same model in a given row share the same vertical scale.
    Faded symbols mark excluded windows ($L_{\min}<18$ for Hubbard, $L_{\min}<9$ for $t$-$V$, as in Fig.~\ref{fig:fit_extrapolation_bothmodels}).
    Colored lines, bands, and intercept bars follow the conventions of Fig.~\ref{fig:fit_extrapolation_bothmodels}, and gray dashed horizontal lines and shaded bands mark the width-six conventional baseline estimates from Fig.~\ref{fig:fit_extrapolation_bothmodels}.}}
    \label{fig:sliding_window_width}
\end{figure*}

\rev{The sliding-window analysis in Fig.~\ref{fig:fit_extrapolation_bothmodels} uses windows of six consecutive sizes, which lie on one constant-width diagonal of Fig.~\ref{fig:critical_exponents_heatmap}.
To test this choice, Fig.~\ref{fig:sliding_window_width} repeats the analysis with windows of five, seven, and eight consecutive sizes.
We keep the lower cutoff of the width-six windows, $L_{\min}\geq18$ for the Hubbard model and $L_{\min}\geq9$ for the $t$-$V$ model, and extract the thermodynamic-limit estimates following Appendix~\ref{app:data_collapse_analysis}.
Most resulting estimates agree with the width-six values within their uncertainties.
The largest shifts occur for the linearly extrapolated $\eta_\phi$, which for the $t$-$V$ model gives $0.481(6)$, $0.484(7)$, and $0.496(7)$ for widths five, seven, and eight compared with $0.468(6)$ for width six, and for the Hubbard model gives $0.7770(32)$, $0.7709(32)$, and $0.7848(37)$ compared with $0.7790(31)$, or up to $3.3\sigma$ (Table~\ref{tab:fss_summary}).
Narrower windows give noisier trajectories, whereas wider windows leave fewer windows above the cutoff, only three for the $t$-$V$ model at width eight.}

\begin{figure*}[htbp]
    \centering
    \includegraphics[width=\textwidth]{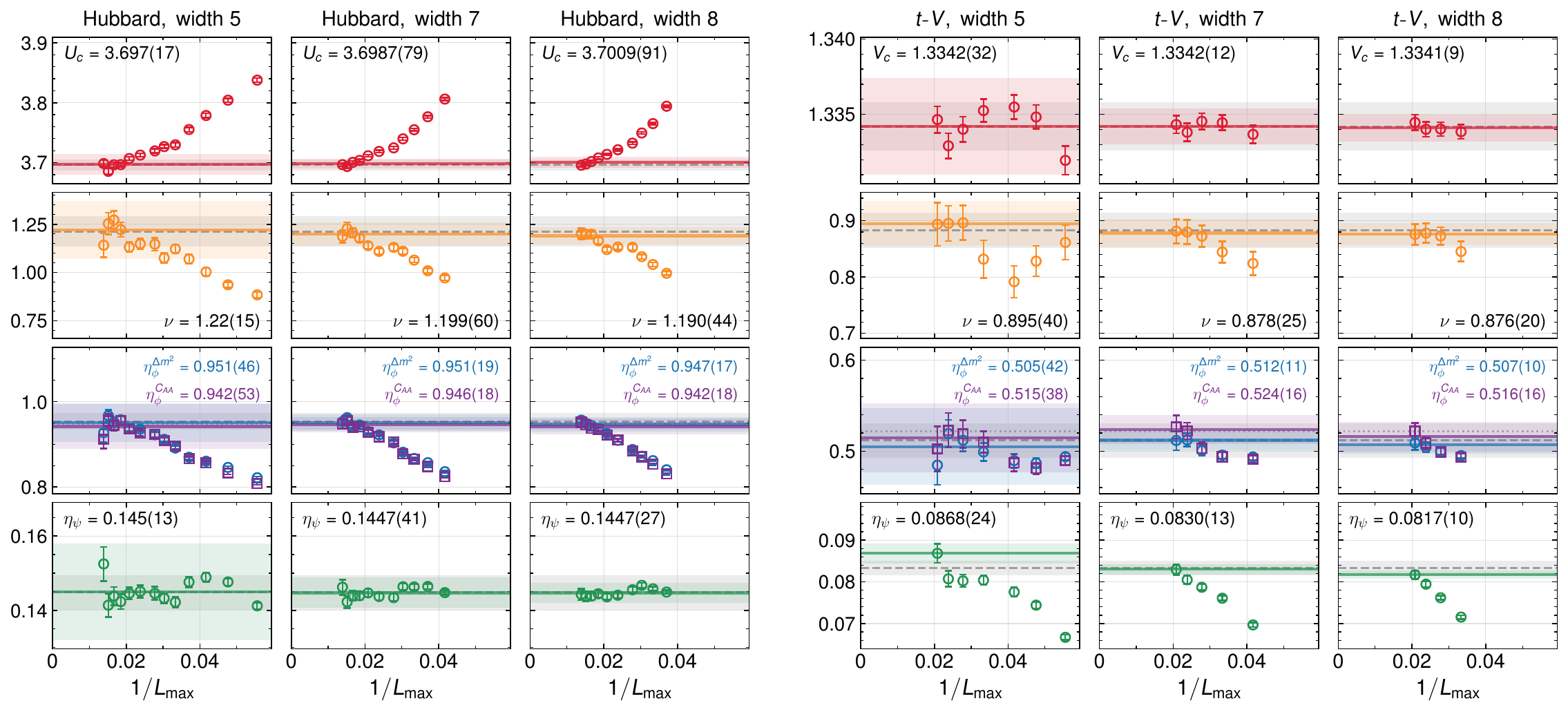}
    \caption{\rev{Sliding-window analysis of the background-suppressed observables (Fig.~\ref{fig:fit_extrapolation_bgsup}) repeated with windows of five, seven, and eight consecutive sizes, in the layout of Fig.~\ref{fig:sliding_window_width}.
    Window-selection cutoffs follow Sec.~\ref{sec:bg_suppressed} ($L_{\max}\ge 48$ for Hubbard $U_c$ at width five; $L_{\max}\ge 42$ for $t$-$V$ $\eta_\phi$ at widths seven and eight).
    In the $\eta_\phi$ row, blue circles and purple squares show the estimates from $\Delta m^2$ and $C_{AA}(\mathbf{r}_{\max})$, respectively, with their values given in the same colors.
    Colored lines, bands, and intercept bars follow the conventions of Fig.~\ref{fig:fit_extrapolation_bgsup}, and gray dashed horizontal lines and shaded bands mark the width-six estimates from Fig.~\ref{fig:fit_extrapolation_bgsup}.}}
    \label{fig:sliding_window_width_bgsup}
\end{figure*}

\rev{Figure~\ref{fig:sliding_window_width_bgsup} repeats the same test for the background-suppressed observables of Sec.~\ref{sec:bg_suppressed}, keeping the window admission and plateau cutoffs of Fig.~\ref{fig:fit_extrapolation_bgsup}.
All their plateau estimates agree with the width-six values within $0.5\sigma$ for both models.
In particular, the Hubbard plateaus give $U_c=3.697(17)$, $3.6987(79)$, and $3.7009(91)$ and $\eta_\phi=0.951(46)$, $0.951(19)$, and $0.947(17)$ from $\Delta m^2_{\mathrm{AFM}}$ for widths five, seven, and eight, compared with $3.6963(91)$ and $0.954(19)$ for width six.}

\begin{figure}[t]
    \centering
    \includegraphics[width=0.97\columnwidth]{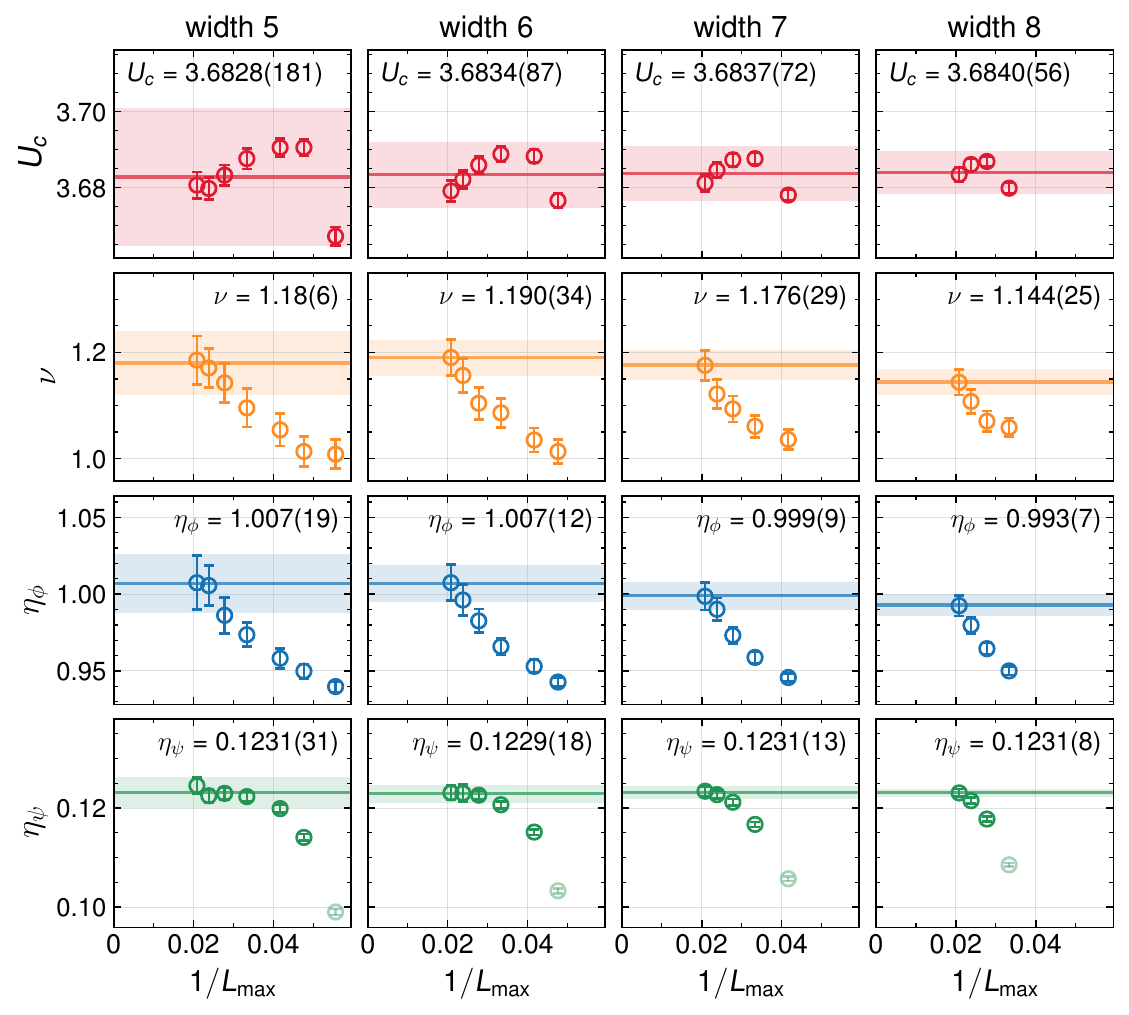}
    \caption{\rev{Sliding-window width scan (widths 5--8) of the Hubbard $\beta=L$ spin susceptibility ($L\le 48$), with shared row scales.
    Faded symbols mark excluded windows ($L_{\min}=6$ for $\eta_\psi$, $\chi^2_{\mathrm{red}}\approx 9$--$14$).
    Colored bands mark the reported values, namely envelopes over all windows for $U_c$ and over $L_{\min}\ge 15$ for $\eta_\psi$, and largest-window frontier values for the still-climbing $\nu$ and $\eta_\phi$ (two largest windows at width five).}}
    \label{fig:sliding_window_width_susc}

    \vspace{0.7em}
    \includegraphics[width=0.97\columnwidth]{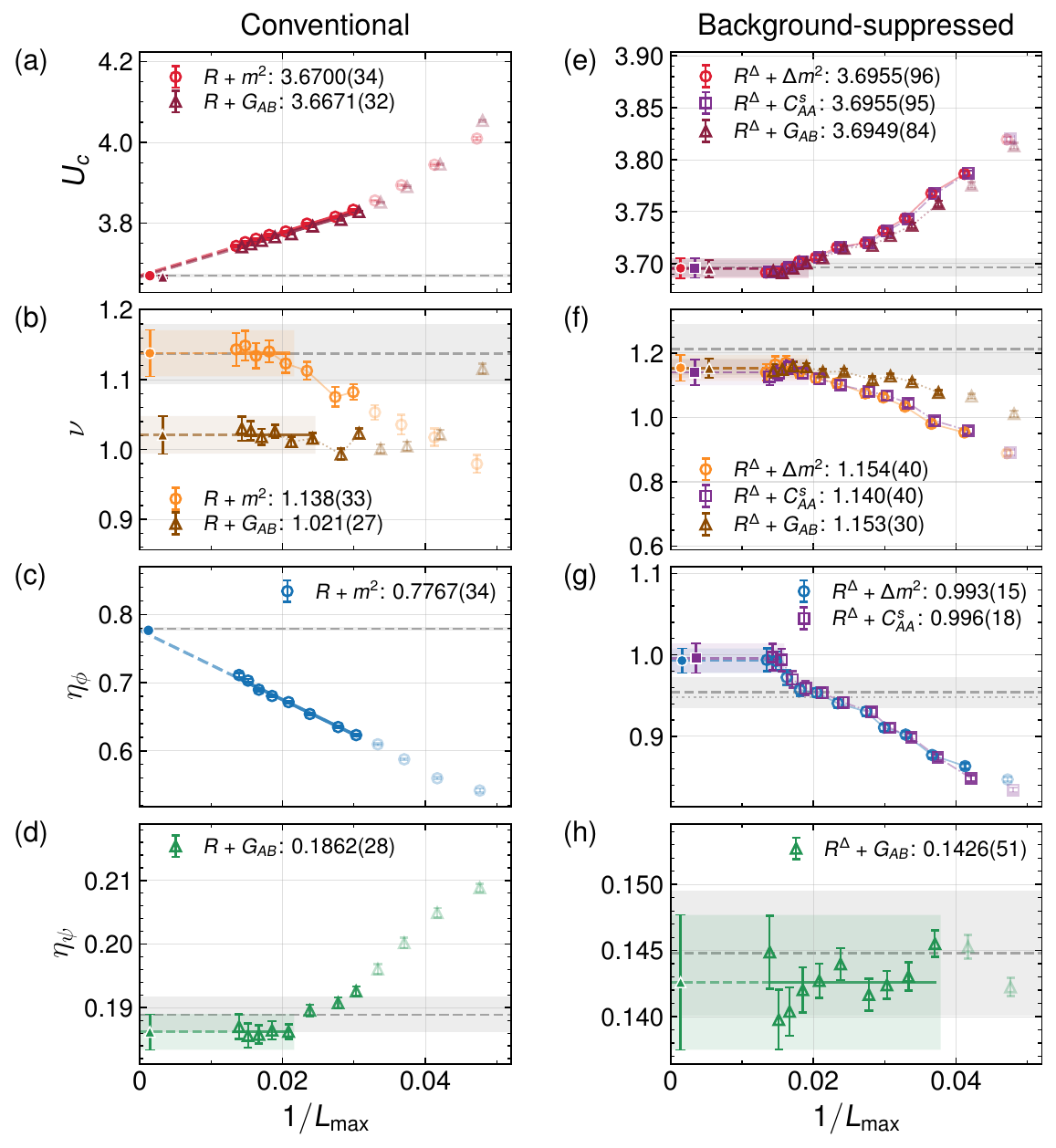}
    \caption{\rev{Width-six joint collapses of the Hubbard model for conventional [(a)--(d)] and background-suppressed [(e)--(h)] equal-time observables.
    Faded symbols mark excluded windows ($L_{\min}<18$ conventional; $L_{\min}=6$ for $R^{\Delta}+\Delta m^2$ and $R^{\Delta}+C^s_{AA}$; $L_{\min}\le 9$ with $\chi^2_{\mathrm{red}}=9$--$33$ for $R^{\Delta}+G_{AB}$).
    Colored bands and lines follow Secs.~\ref{sec:gnh_criticality} and~\ref{sec:bg_suppressed} [with GLS extrapolations in (a) and (c), $\chi^2_{\mathrm{red}}=3.04,5.07$ for $U_c$ and $1.71$ for $\eta_\phi$, and left-edge bars for the thermodynamic-limit estimates], and gray bands mark the separate-collapse references (Figs.~\ref{fig:fit_extrapolation_bothmodels} and~\ref{fig:fit_extrapolation_bgsup}).}}
    \label{fig:hubbard_joint_collapse}
\end{figure}

\rev{Figure~\ref{fig:sliding_window_width_susc} shows the corresponding window-width scan for the Hubbard $\beta=L$ spin susceptibility across widths five to eight.
For widths five, seven, and eight, the envelopes $U_c=3.6828(181)$, $3.6837(72)$, and $3.6840(56)$ agree with the width-six estimate $3.6834(87)$ within $0.1\sigma$, and $\eta_\psi$ from $G_{AB}\,(R^{\chi}_{\mathrm{AFM}})$ levels off for $L_{\min}\ge 15$ at $0.1231(31)$, $0.1231(13)$, and $0.1231(8)$, compared with $0.1229(18)$ for width six.
Since $\nu$ and $\eta_\phi$ still climb at every width, we quote their largest-window values as frontier lower bounds [e.g., $\nu=1.190(34)$ and $\eta_\phi=1.007(12)$ at width six], averaging the two largest windows at width five (Table~\ref{tab:fss_summary}).}

\begin{figure}[t]
    \centering
    \includegraphics[width=0.97\columnwidth]{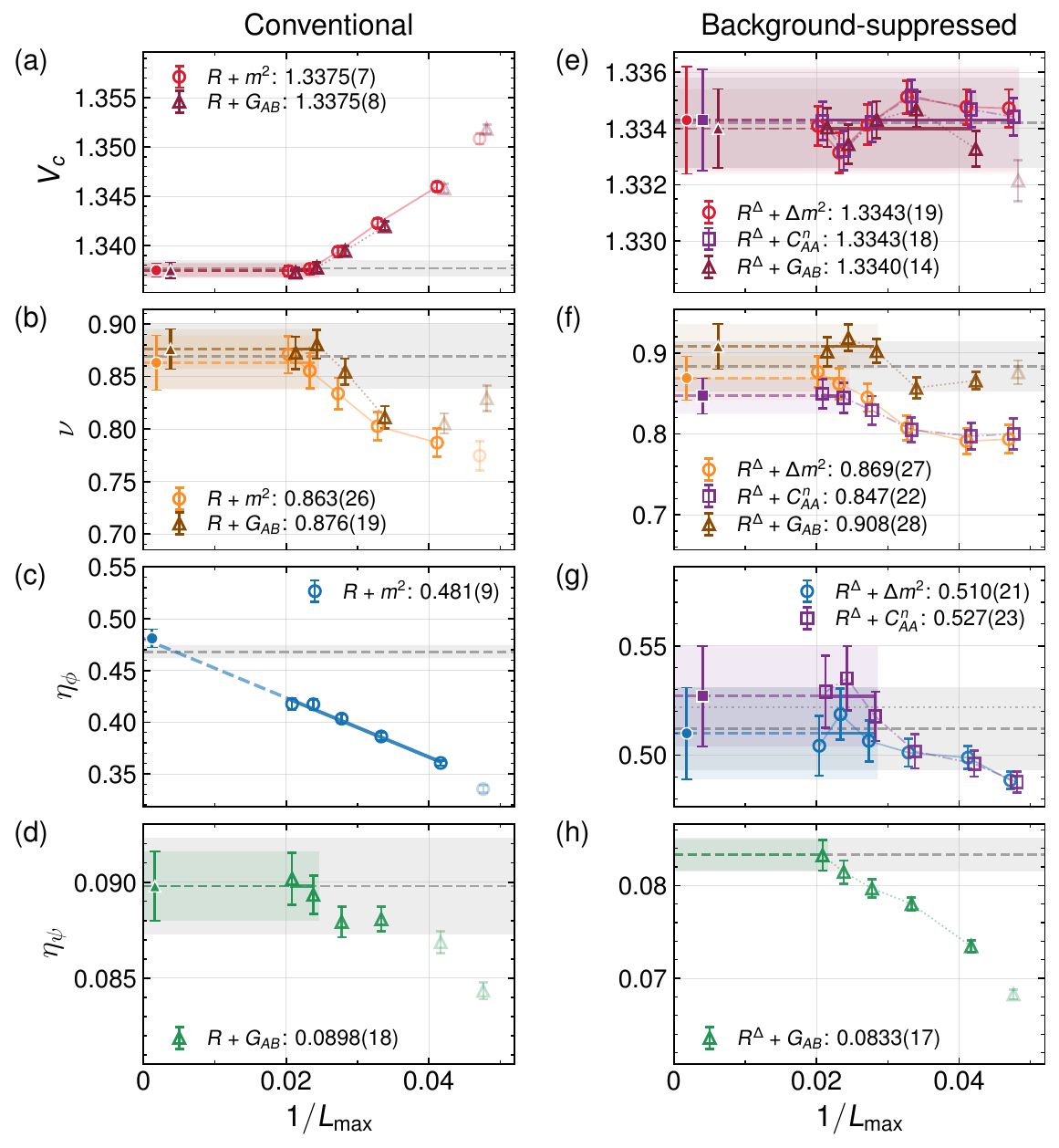}
    \caption{\rev{Width-six joint collapses of the $t$-$V$ model for conventional [(a)--(d)] and background-suppressed [(e)--(h)] equal-time observables, in the layout and conventions of Fig.~\ref{fig:hubbard_joint_collapse}.
    Faded symbols mark excluded windows ($L_{\min}=6$ for $R+m^2$ and $R^{\Delta}+G_{AB}$, $L_{\min}\le 9$ for $R+G_{AB}$; $\chi^2_{\mathrm{red}}=4.3$--$19$ versus $\lesssim 3$ in the other windows), the diagonal line in (c) shows a GLS extrapolation ($\chi^2_{\mathrm{red}}=1.03$), and gray bands mark the separate-collapse references (Figs.~\ref{fig:fit_extrapolation_bothmodels} and~\ref{fig:fit_extrapolation_bgsup}).}}
    \label{fig:tv_joint_collapse}
\end{figure}

\rev{Figures~\ref{fig:hubbard_joint_collapse} and~\ref{fig:tv_joint_collapse} show the width-six joint collapses across both equal-time families for the Hubbard and $t$-$V$ models, respectively, excluding high-$\chi^2_{\mathrm{red}}$ small-size windows.
All trajectories and extracted parameters agree with their separate-collapse counterparts (Table~\ref{tab:fss_summary}), apart from the conventional Hubbard $R_{\mathrm{AFM}}+G_{AB}$ $\nu$ shift discussed in Sec.~\ref{sec:bg_suppressed}.}

\clearpage
%%%%%%%%%%%%%%%%%%%%%%%%%%%%%%%%%%%%%%%%%%%%%%%%%%%%%
\section{\rev{Comparison with other finite-size analysis strategies}}\label{app:conventional_analysis}
%%%%%%%%%%%%%%%%%%%%%%%%%%%%%%%%%%%%%%%%%%%%%%%%%%%%%

\rev{In this section, we compare the sliding-window data collapse of the main text with two conventional analysis strategies---crossing-point and shrinking-window---applied to the same data.
Table~\ref{tab:analysis_strategies} compares the input data for each finite-size estimate and the route from the resulting sequence to the TDL.
The subsections below present this comparison on the conventional observables and then apply both alternative strategies to the background-suppressed observables (Appendix~\ref{app:crossing_bgsup}).}

\begin{table*}[t]
    \centering
    \caption{\rev{Comparison of three finite-size analysis strategies applied to the same data.}}
    \label{tab:analysis_strategies}
    {\small\revcolor
    \begin{tabular*}{\textwidth}{@{\extracolsep{\fill}}lC{4.9cm}C{3.5cm}C{3.8cm}@{}}
    \toprule
    & Sliding-window (this work) & Crossing-point & Shrinking-window \\
    \midrule
    \begin{tabular}[t]{@{}l@{}}Input data \\(per finite-size estimate)\end{tabular} & one fixed-width fitting window $[L_{\min},L_{\max}]$ & one size pair $(L,rL)$ & one fitting window $[L_{\min},L_{\max}]$ at fixed $L_{\max}$ \\ \addlinespace[5pt]
    Route to the TDL & $\checkmark$ & $\checkmark$ & $\times$ \\ \addlinespace[3pt]
    Computational cost & high & negligible & medium \\ \addlinespace[3pt]
    Key vulnerability & many tunable choices (window width and scaling corrections) & statistical noise in individual sizes & loss of constraint in narrow windows \\ \addlinespace[3pt]
    Most suitable when & extended size sets, noisier data at large $L$ & uniformly high quality at all sizes & moderate size sets, for convergence checks \\
    \bottomrule
    \end{tabular*}}
\end{table*}

\subsection{\rev{Crossing-point analysis}}\label{app:crossing_point}

\rev{Following Ref.~\cite{Phys.Rev.Lett.2022Liu}, we define the crossing $g^\times(L,rL)$ of an RG-invariant, dimensionless quantity $R(g,L)$ between sizes $L$ and $rL$ ($r>1$) by $R(g^\times,L)=R(g^\times,rL)$; it provides a finite-size pseudocritical estimate of the critical coupling.
At criticality, $s(g,L)\equiv\partial R(g,L)/\partial g\sim L^{1/\nu}$, $m^2(g_c,L)\sim L^{-(1+\eta_\phi)}$, and $Z(g_c,L)\sim L^{-\eta_\psi}$ ($z=1$). Evaluating the size ratios at the crossing $g^\times$ then gives
\begin{align}
\frac{1}{\nu}(L) &= \frac{1}{\ln r}\ln\frac{s(g^\times,rL)}{s(g^\times,L)},\label{equ:crossing_nu}\\
\eta_\phi(L) &= -1-\frac{1}{\ln r}\ln\frac{m^2(g^\times,rL)}{m^2(g^\times,L)},\label{equ:crossing_etaphi}\\
\eta_\psi(L) &= -\frac{1}{\ln r}\ln\frac{Z(g^\times,rL)}{Z(g^\times,L)},\label{equ:crossing_etapsi}
\end{align}
In the present models, $g=U$ or $V$, $R$ is the corresponding correlation ratio, $m^2$ is the squared AFM or CDW order parameter, and $Z$ is estimated by the averaged modulus $G_{AB}$.
We take $r=2$ in the analysis. In each of $10^4$ bootstrap resamples, all system sizes are resampled jointly before the cubic-polynomial interpolations are evaluated, so crossings that share a size retain their covariance.
Each sequence $X(L)$ at fixed $r$ is then extrapolated as $X(L)=X_\infty+aL^{-b}$ by a full-covariance GLS fit with the bootstrap covariance matrix, where the drift exponent $b$ is fitted separately for each quantity and is not identified with a unique microscopic correction exponent.}

\begin{figure*}[htbp]
    \centering
    \includegraphics[width=\textwidth]{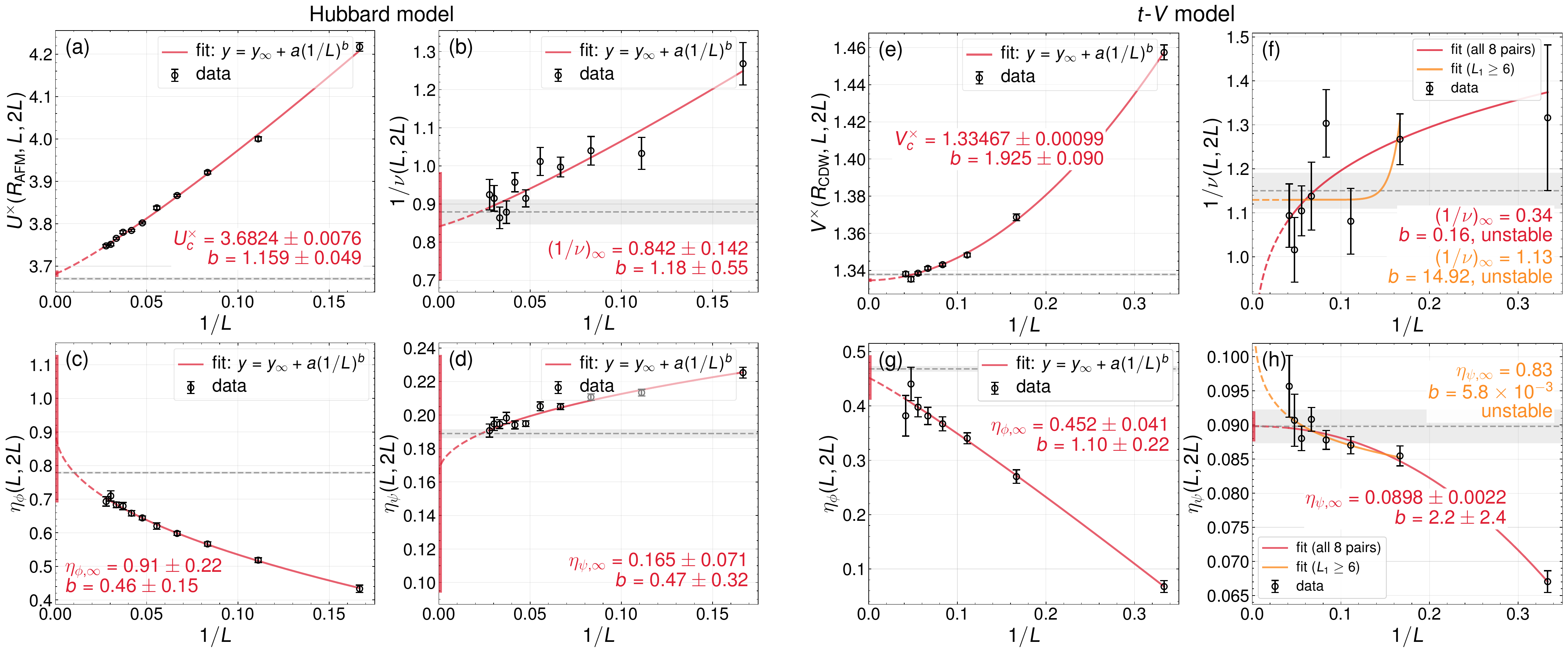}
    \caption{\rev{Crossing-point analysis for the Hubbard model [(a)--(d)] and the $t$-$V$ model [(e)--(h)].
    For each model, the four panels show the finite-size estimates from $(L,2L)$ pairs versus $1/L$ ($L$ being the smaller size of the pair): the crossing coupling, $1/\nu$, $\eta_\phi$, and $\eta_\psi$, together with power-law extrapolations whose intercepts and fitted exponents are labeled.
    Panels (a) and (e) present the same crossing-point data and power-law fits as Figs.~\ref{fig:correlation_ratio_crossing} and \ref{fig:crossing_tV}, respectively.
    Panels (f) and (h) show both the all-pair fit (red) and the $L\geq6$ fit excluding $(3,6)$ (yellow).
    Fits labeled ``unstable'' are unconstrained and yield no estimate.
    Gray dashed lines and shaded bands mark the sliding-window estimates and their uncertainty intervals from Fig.~\ref{fig:fit_extrapolation_bothmodels}, with $\nu$ inverted to $1/\nu$ in panels~(b) and (f).}}
    \label{fig:crossing_bothmodels}
\end{figure*}

\rev{Figure~\ref{fig:crossing_bothmodels} compares the two analyses directly. Although the individual crossing-point values are visibly noisier and their power-law extrapolations less well constrained than the sliding-window results, the two analyses agree within their uncertainty intervals for the five critical exponents with stable crossing-point extrapolations, and within about $1.5\sigma$ and $2.5\sigma$ for $U_c$ and $V_c$, respectively.
For $1/\nu$, inverting the extrapolated Hubbard intercept $0.842(142)$ [panel~(b)] gives $\nu=1.19(20)$.
Two $t$-$V$ panels illustrate how fragile the extrapolation can be.
For $1/\nu$ [panel~(f)], the crossing-point values scatter beyond their error bars and the power-law fit is unconstrained with or without the smallest size pair $(3,6)$, so we quote no crossing-point estimate of the $t$-$V$ exponent $\nu$.
For $\eta_\psi$ [panel~(h)], the fit is stable only when $(3,6)$ is included; without it, the fit is dominated by the noisiest large-size point.
Both cases illustrate the difficulty of crossing-point analysis when data quality varies across sizes. 
Repeating the analysis with the alternative fixed-offset $(L,L+6)$ pairing yields a still noisier crossing sequence whose extrapolation is correspondingly less reliable, since each crossing at large sizes combines two noisy data sets.}

\rev{The comparatively weaker agreement for the critical couplings reflects the different extrapolation strategies: the crossing-point analysis extrapolates a power law over all $(L,2L)$ pairs, whereas the sliding-window analysis uses only large-size windows.
For $U_c$, linear extrapolations of the last five and six crossing points give $3.6717(92)$ and $3.6739(61)$, consistent with the sliding-window value within $0.6\sigma$. For $V_c$, the largest-size crossing points fluctuate around the sliding-window value without a resolved drift.}

\rev{Assuming the analytic background controls the asymptotic drift~\cite{Phys.Rev.B2014Campostrini,Phys.Rev.B2015ParisenToldin}, the crossing-point estimates of $1/\nu$ and $\eta_\phi$ [Fig.~\ref{fig:crossing_bothmodels}(b), (c)] give $1/\nu+1-\eta_\phi=0.93(26)$ for the Hubbard model.
This value agrees within error bars with the crossing-coupling drift exponent $1.159(49)$ [panel~(a)], providing an internal consistency check of the crossing-point analysis.
The $t$-$V$ model lacks a stable crossing-point estimate of $1/\nu$, so the corresponding check is not available.}

\subsection{\rev{Shrinking-window data collapse}}\label{app:shrinking_window}

\begin{figure*}
    \centering
    \includegraphics[width=\textwidth]{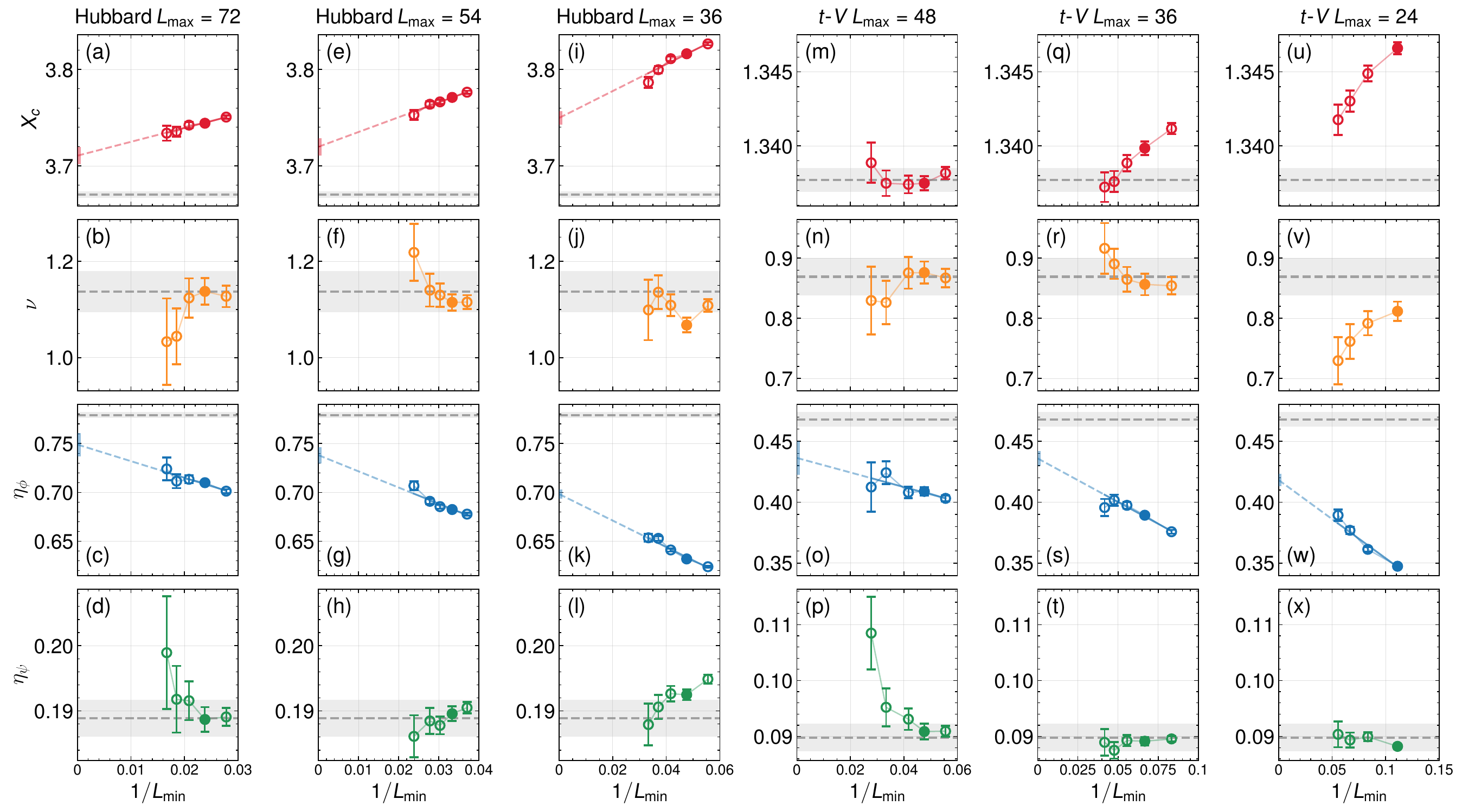}
    \caption{\rev{Shrinking-window analysis in a format analogous to Fig.~\ref{fig:fit_extrapolation_bothmodels}, providing a selective view of the fitting-window scan in Fig.~\ref{fig:critical_exponents_heatmap}.
    The three left (right) columns show the Hubbard ($t$-$V$) model at fixed terminal sizes $L_{\max}=72,54,36$ ($48,36,24$), and the four rows show $X_c$, $\nu$, $\eta_\phi$, and $\eta_\psi$.
    Within each model, panels in a given row share the same vertical scale.
    Each panel plots fitted parameters from windows $[L_{\min},L_{\max}]$ against $1/L_{\min}$, with the axis extended to zero.
    As $L_{\min}$ increases, the window width decreases from seven to three consecutive sizes, except in the $L_{\max}=24$ $t$-$V$ column, where it decreases from six to three.
    Filled markers denote the width-six windows used in Fig.~\ref{fig:fit_extrapolation_bothmodels}, and gray dashed horizontal lines and shaded bands mark the conventional baseline estimates from Fig.~\ref{fig:fit_extrapolation_bothmodels}.
    Thin lines are guides to the eye, whereas the colored solid and dashed lines shown in the Hubbard $X_c$ row and the $\eta_\phi$ rows are heuristic diagonal WLS linear fits in $1/L_{\min}$ and their continuations to $1/L_{\min}=0$, with short bars marking the resulting intercepts and nominal uncertainties, which ignore the data-sharing-induced inter-window covariance.}}
    \label{fig:shrinking_window}
\end{figure*}

\rev{The shrinking-window strategy increases $L_{\min}$ while holding $L_{\max}$ fixed, so $L_{\min}$ remains bounded from above and the resulting sequence cannot approach the TDL.
Figure~\ref{fig:shrinking_window} illustrates this limitation in practice, displaying the fixed-$L_{\max}$ sequences contained in the two-dimensional window scan of Fig.~\ref{fig:critical_exponents_heatmap} in a format analogous to Fig.~\ref{fig:fit_extrapolation_bothmodels}: each column fixes a different terminal size $L_{\max}$, so that the terminal-size dependence can be assessed directly.
Although extrapolating in $L_{\min}$ is thus meaningless in principle, we include heuristic linear extrapolations for the visibly drifting $U_c$ and $\eta_\phi$ to make the drift quantitative.}

\rev{The shortcomings of the shrinking-window strategy are twofold.
First, within each panel of Fig.~\ref{fig:shrinking_window}, the statistical cost of ``shrinking'' the window shows up directly in the error bars, which grow visibly from right to left (increasing $L_{\min}$ at fixed $L_{\max}$).
With only a few sizes inside the fit window, the collapse is less constrained, so statistical fluctuations in individual sizes carry greater leverage, particularly at the largest sizes where the data are noisiest---the same weakness as in the crossing-point analysis.
In the largest-$L_{\max}$ columns, shrinking leaves only these relatively noisy sizes, making any direct extrapolation to $1/L_{\min}\rightarrow0$ unreliable, quite apart from its legitimacy.}

\rev{Second, and more fundamental, comparing across the $L_{\max}$ columns, the fitted $U_c$ and $\eta_\phi$ drift systematically, and even the heuristic-fit intercepts drift with $L_{\max}$.
For the Hubbard $U_c$, the heuristic intercepts move from $3.750(8)$ at $L_{\max}=36$ to $3.711(10)$ at $L_{\max}=72$, approaching the conventional sliding-window estimate (gray band) without reaching it.
This drift directly contradicts the working assumption of the shrinking-window analysis with the leading-order ansatz used here: that the largest accessible sizes are already sufficient for the fit to return thermodynamic-limit parameters.
With this form, the omitted finite-size corrections are absorbed into the fitted parameters rather than removed, so approaching the TDL requires advancing $L_{\max}$, as the sliding-window sequence does, rather than shrinking the window at fixed $L_{\max}$.}

\subsection{\rev{Summary and trade-offs}}\label{app:strategy_tradeoffs}

\rev{Crossing-point and leading-order sliding-window analyses both account for scaling corrections through the size dependence of finite-size estimates.
For the crossing-point analysis, the leading scaling relations define the crossing coupling and effective critical exponents at finite $L$, whose residual drift is then extrapolated toward the thermodynamic limit.
Likewise, the leading-order data-collapse fits used in the main text absorb the omitted corrections into window-dependent effective parameters.
The sliding-window analysis tracks their evolution as the whole fitting window moves to larger sizes and extrapolates the remaining drift.}

\rev{This shared logic does not imply identical drift laws.
For the crossing coupling, finite-size scaling theory predicts a power-law drift with exponent $1/\nu+\min(\omega_{\mathrm{irr}},2-z-\eta_\phi)$, with $z=1$ here~\cite{Phys.Rev.B2014Campostrini,Phys.Rev.B2015ParisenToldin}.
For the sliding-window analysis, the drift law remains unclear, and no comparable theoretical prediction is available.
Power-law fits to the sliding-window trajectory also depend on the extrapolation variable: whereas $(2L)^{-b}=2^{-b}L^{-b}$ for a crossing pair, $L_{\max}/L_{\min}$ varies from $1.7$ to $3.5$ across our windows, and fits in $L_{\min}$ and $L_{\max}$ yield different $b$.}

\rev{Taken together, the three analyses illustrate the complementary trade-offs summarized in Table~\ref{tab:analysis_strategies}.
For the present data, whose largest sizes carry substantial statistical noise, the sliding-window analysis is more robust than the crossing-point analysis for thermodynamic-limit inference.
When each estimate is constrained by only a single size pair or by one of the narrowest shrinking windows, statistical fluctuations acquire greater leverage, as discussed above.
The sliding-window analysis reduces this leverage by pooling more sizes per estimate and uses the fixed-width points across successive $L_{\max}$ values to form a systematic trajectory in $1/L_{\max}$.
Varying the window width from six to five, seven, or eight sizes shifts all estimates by at most $1.4\sigma$, except for the linearly extrapolated $\eta_\phi$, which shifts by up to $3.3\sigma$ (Appendix~\ref{app:fitting_window_scan}).}

\subsection{\rev{Background-suppressed observables}}\label{app:crossing_bgsup}

\begin{figure*}[htbp]
    \centering
    \includegraphics[width=\textwidth]{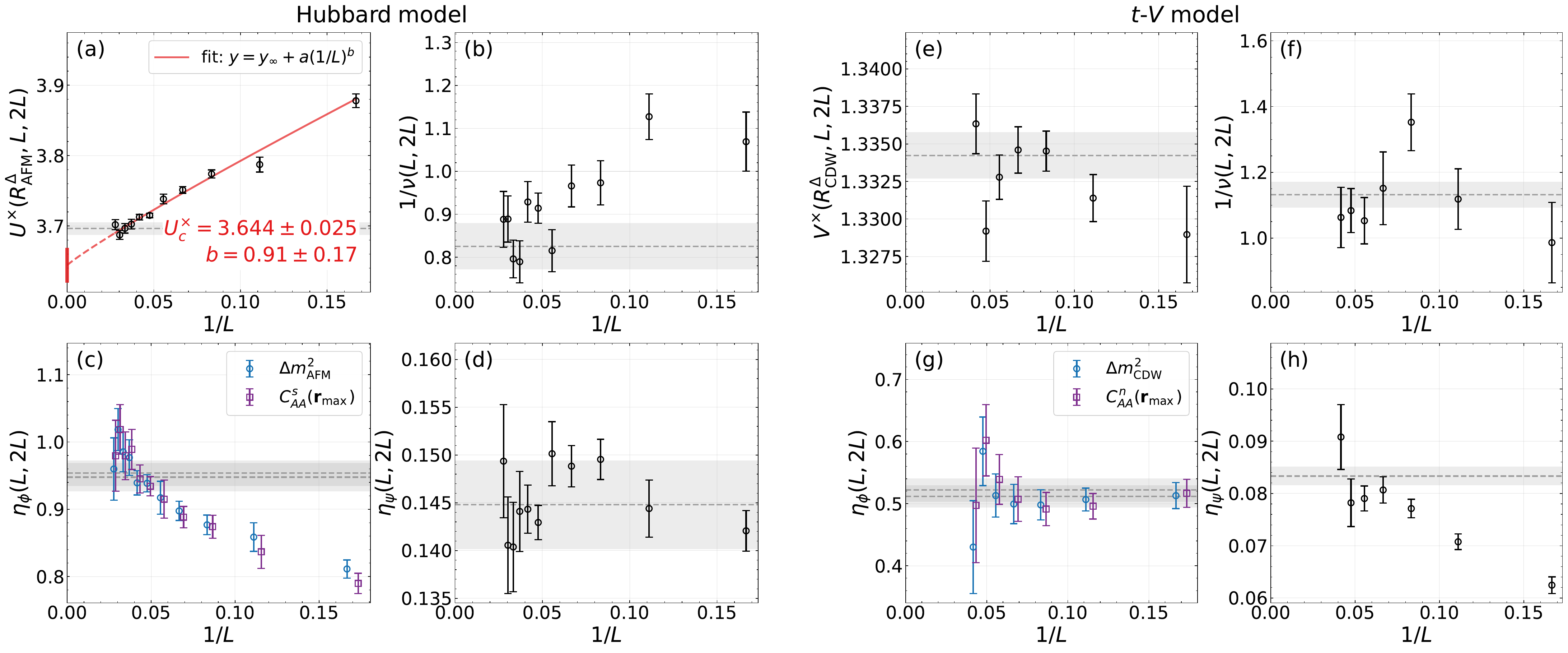}
    \caption{\rev{Crossing-point analysis of the background-suppressed equal-time observables for the Hubbard model [(a)--(d)] and the $t$-$V$ model [(e)--(h)], in the layout of Fig.~\ref{fig:crossing_bothmodels}.
    The crossings are those of $R^{\Delta}$, and $\eta_\phi$ is obtained from $\Delta m^2$ (circles) and $C_{AA}(\mathbf{r}_{\max})$ (squares, slightly displaced horizontally for clarity).
    A three-parameter power-law extrapolation is shown only where its drift exponent $b$ is constrained [panel~(a)].
    Gray dashed lines and shaded bands mark the sliding-window estimates and their uncertainty intervals from Fig.~\ref{fig:fit_extrapolation_bgsup}, with $\nu$ inverted to $1/\nu$ in panels~(b) and~(f).}}
    \label{fig:crossing_bgsup}
\end{figure*}

\begin{figure}[htbp]
    \centering
    \includegraphics[width=\columnwidth]{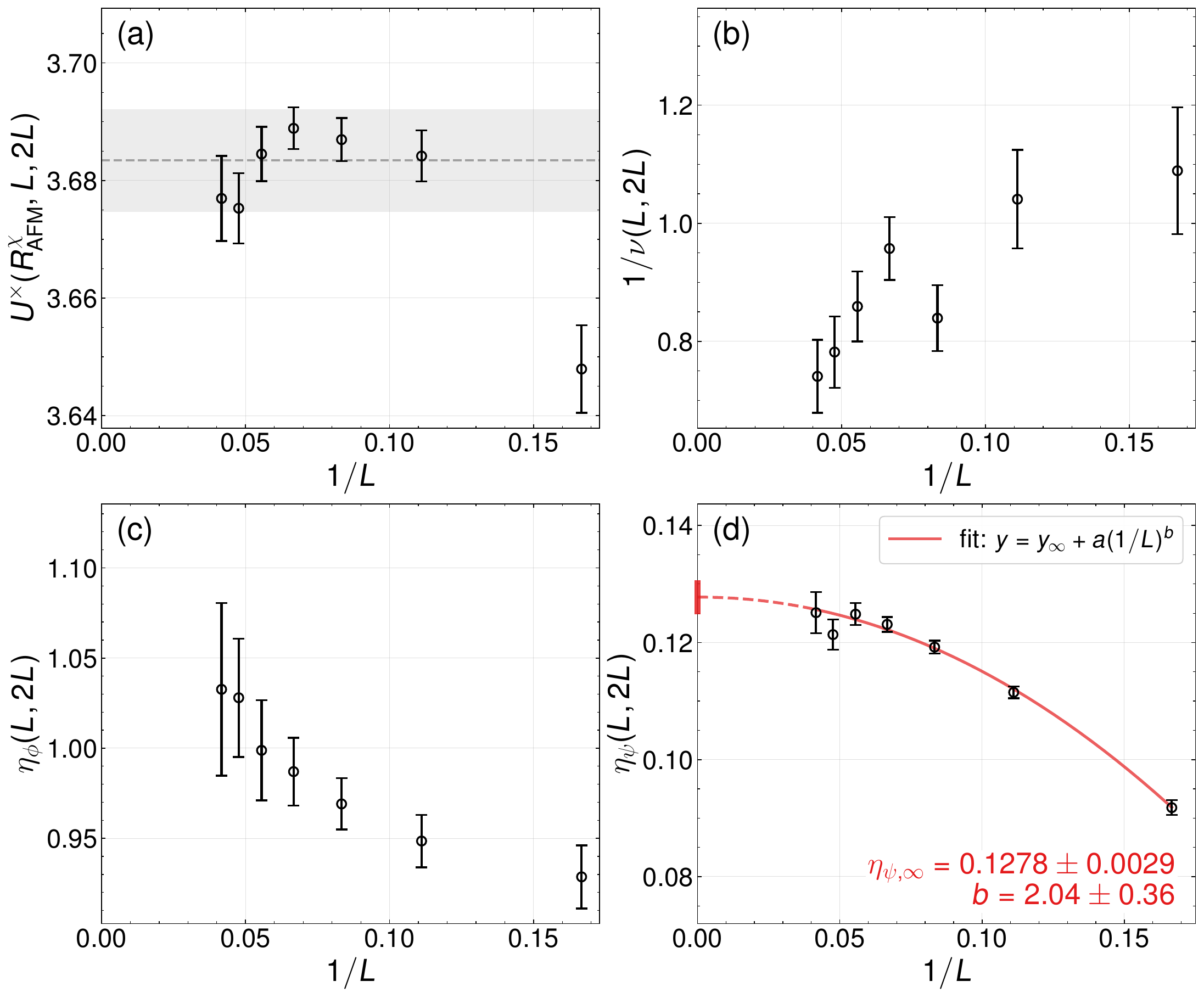}
    \caption{\rev{Crossing-point analysis of the spin susceptibility for the Hubbard model at $\beta=L$.
    Panels~(a)--(d) show the crossing coupling and $1/\nu$ from $R^{\chi}_{\mathrm{AFM}}$, $\eta_\phi$ from $L^{-3}\chi_{\mathrm{AFM}}(0)$, and $\eta_\psi$ from $G_{AB}$ at the crossings, across the seven $(L,2L)$ pairs with $L\le 24$ ($2L\le 48$).
    A three-parameter power-law extrapolation is constrained only in panel~(d).
    The gray dashed line and shaded band in panel~(a) mark the sliding-window $U_c$ estimate and its uncertainty interval from Fig.~\ref{fig:fit_extrapolation_hubbard_comparison}.}}
    \label{fig:crossing_susc}
\end{figure}

\rev{Finally, we apply both alternative strategies to the background-suppressed observables of Sec.~\ref{sec:bg_suppressed}, beginning with the $(L,2L)$ crossing-point analysis of Appendix~\ref{app:crossing_point}.
Replacing $R$ with $R^{\Delta}$ and $m^2$ with $\Delta m^2$ or $C_{AA}(\mathbf{r}_{\max})$ in Eqs.~\eqref{equ:crossing_nu}--\eqref{equ:crossing_etapsi} yields Fig.~\ref{fig:crossing_bgsup}, while using $R^{\chi}_{\mathrm{AFM}}$, $L^{-3}\chi_{\mathrm{AFM}}(0)$, and $G_{AB}$ at $\beta=L$ yields Fig.~\ref{fig:crossing_susc}.}

\begin{figure*}[tp]
    \centering
    \includegraphics[width=\textwidth]{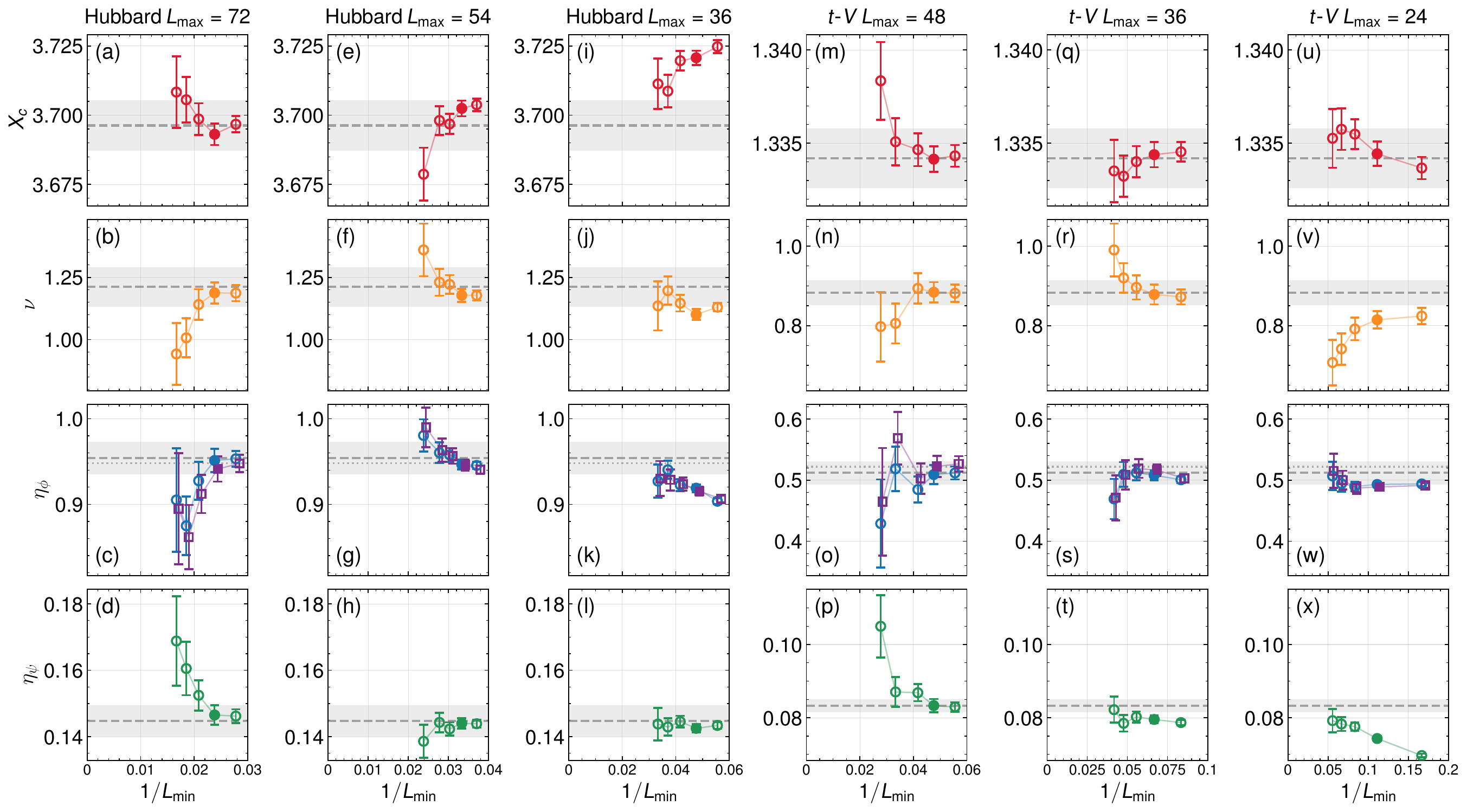}
    \caption{\rev{Shrinking-window analysis of the background-suppressed equal-time observables (Fig.~\ref{fig:fit_extrapolation_bgsup}), in the layout of Fig.~\ref{fig:shrinking_window}.
    The three left (right) columns show the Hubbard ($t$-$V$) model at fixed terminal sizes $L_{\max}=72,54,36$ ($48,36,24$), and the four rows show $X_c$, $\nu$, $\eta_\phi$, and $\eta_\psi$, with panels of the same model in a given row sharing the same vertical scale.
    As $L_{\min}$ increases, the window width shrinks from seven to three consecutive sizes in every column.
    Filled markers denote the width-six windows used in Fig.~\ref{fig:fit_extrapolation_bgsup}, thin lines are guides to the eye, and gray dashed horizontal lines and shaded bands mark the width-six estimates from Fig.~\ref{fig:fit_extrapolation_bgsup}.
    In the $\eta_\phi$ row, blue circles and purple squares (slightly displaced horizontally for clarity) denote $\Delta m^2$ and $C_{AA}(\mathbf{r}_{\max})$, respectively, with the dotted gray line marking the width-six plateau estimate from $C_{AA}(\mathbf{r}_{\max})$.}}
    \label{fig:shrinking_window_bgsup}
\end{figure*}

\rev{Under the three-parameter power-law fit $X(L)=X_{\infty}+aL^{-b}$ of Appendix~\ref{app:crossing_point}, suppressing the short-range background largely removes the drift needed to constrain the free exponent $b$, leaving the fit undetermined in ten of the twelve panels, where we therefore report no extrapolated values.
In the Hubbard crossing coupling of $R^{\Delta}_{\mathrm{AFM}}$ [Fig.~\ref{fig:crossing_bgsup}(a)], fitting all eleven pairs down to $(6,12)$ gives $U_c^{\times}=3.644(25)$ with $b=0.91(17)$, $2.0\sigma$ below the sliding-window plateau because a single power law is pulled downward by the steep small-$L$ drop rather than the flat large-$L$ crossings.
By contrast, in Fig.~\ref{fig:crossing_susc}(d), evaluating $G_{AB}$ at the $R^{\chi}_{\mathrm{AFM}}$ crossing exhibits a smooth upward drift across the seven pairs with $2L\le 48$, constraining $b=2.04(36)$ and extrapolating to $\eta_{\psi,\infty}=0.1278(29)$, consistent with the sliding-window susceptibility plateau $0.1229(18)$ and $3.1\sigma$ below the equal-time background-suppressed plateau $0.1448(47)$.}

\rev{Nevertheless, the finite-size $(L,2L)$ sequences themselves corroborate Sec.~\ref{sec:bg_suppressed} without invoking a data-collapse ansatz.
In Fig.~\ref{fig:crossing_bgsup}, all four Hubbard quantities settle onto the sliding-window bands at large $L$ (with $\eta_\psi$ flat across all pairs and both $\eta_\phi$ proxies agreeing pair by pair), while $t$-$V$ $V_c$, $1/\nu$, and $\eta_\phi$ fluctuate directly around their sliding-window bands and $t$-$V$ $\eta_\psi$ climbs monotonically toward the conventional range.
Likewise, in Figs.~\ref{fig:crossing_susc}(a)--\ref{fig:crossing_susc}(c), the susceptibility crossings above $(6,12)$ sit on the $U_c$ band and closely track the upward window trends in both $1/\nu$ and $\eta_\phi$, reaching $\eta_\phi=1.03(5)$ at $(24,48)$.
Although one could extract plateau envelopes directly from these converged $(L,2L)$ crossings, each point relies on a single size pair and carries larger point-to-point fluctuations and error bars, resulting in much wider confidence intervals than the six-size sliding windows (gray bands).}

\rev{Figure~\ref{fig:shrinking_window_bgsup} repeats the shrinking-window analysis of Appendix~\ref{app:shrinking_window} for the background-suppressed equal-time observables.
Compared with Fig.~\ref{fig:shrinking_window}, the clearest change appears in Hubbard $U_c$ and $\eta_\phi$ (and in $t$-$V$ $\eta_\phi$), where the wider windows at $L_{\max}=54$ and $72$ already lie on the gray sliding-window bands rather than drifting systematically across $L_{\max}$, leaving pre-plateau offsets only at smaller terminal sizes such as Hubbard $L_{\max}=36$ [panels~(i) and~(k)] and $t$-$V$ $\eta_\psi$ at $L_{\max}=24$ [panel~(x)].
Within a fixed $L_{\max}$ column, however, the sequence stays flat down to width three only where the statistics remain clean, notably $t$-$V$ at $L_{\max}=36$ [panels~(q)--(t)].
At the largest terminal sizes ($L_{\max}=54,72$ for Hubbard and $L_{\max}=48$ for $t$-$V$), narrowing the window to three or four noisy sizes instead inflates the error bars and pulls the central values off the band, showing that narrower windows bring no advantage when large-size noise dominates.}

\clearpage
%%%%%%%%%%%%%%%%%%%%%%%%%%%%%%%%%%%%%%%%%%%%%%%%%%%%%
\section{\rev{Sliding-window data collapse with leading corrections to scaling}}\label{app:correction_initialization}
%%%%%%%%%%%%%%%%%%%%%%%%%%%%%%%%%%%%%%%%%%%%%%%%%%%%%
\begingroup
\revcolor

In Sec.~\ref{sec:gnh_criticality}, the sliding-window data-collapse analysis of the conventional observables $R$, $m^2$, and $G_{AB}$ uses the leading scaling form, with finite-size corrections reflected in the size dependence of the fitted parameters.
Because the strong analytic background in $R$ and $m^2$ decays slowly with system size, these leading-form collapses yield $\chi^2_{\mathrm{red}}\sim10$--$100$ in the excluded small-size windows (Appendix~\ref{app:data_collapse_analysis}) and leave a pronounced drift in $U_c$ and $\eta_\phi$ up to the largest accessible sizes.
Rather than suppressing the background at the observable level (Sec.~\ref{sec:bg_suppressed}), here we test whether explicitly including a leading correction to scaling in the collapse ansatz can absorb this background and resolve the drift while keeping the fitted parameters well constrained.

We follow the data-collapse and bootstrap workflow of Appendix~\ref{app:data_collapse_analysis}, replacing the leading scaling form with the corrected ansatz in each window of six available sizes:
\begin{equation}\label{equ:corrected_ansatz}
    \begin{aligned}
        R(g,L)&=f_0^R(x)+L^{-\omega}f_1^R(x),\\
        m^2(R,L)&=L^{-1-\eta_\phi}\left[f_0^m(R)+L^{-\omega}f_1^m(R)\right],\\
        G_{AB}(R,L)&=L^{-\eta_\psi}\left[f_0^G(R)+L^{-\omega}f_1^G(R)\right].
    \end{aligned}
\end{equation}
Here, $x=(g-g_c)L^{1/\nu}$, with $g=U$ and $g_c=U_c$ for Hubbard, and $g=V$ and $g_c=V_c$ for the $t$-$V$ model.
The correlation ratio $R$ and squared order parameter $m^2$ refer to AFM order in Hubbard and CDW order in the $t$-$V$ model; $G_{AB}$ is the averaged modulus of Eq.~\eqref{equ:fss_G_AB}.
The functions $f_0$ and $f_1$ describe the leading and correction contributions, respectively.
Here, $\omega$ is the fitted leading correction-to-scaling exponent, which can effectively describe several correction terms over the available size range without isolating their physical origins.
We infer both functions using Harada's Bayesian scaling analysis~\cite{Phys.Rev.E2011Harada} with its composite-kernel extension for corrections to scaling~\cite{Phys.Rev.E2015Harada}.
To explore both rapidly and slowly decaying corrections, we initialize $\omega$ at $\omega_{\mathrm{init}}=10$ and $0.3$, respectively, with the latter motivated by $\epsilon$-expansion calculations~\cite{Phys.Rev.B2023Ladovrechis}.
In both cases, $\omega$ is fitted freely with all other settings unchanged, so the two corrected fits differ only in $\omega_{\mathrm{init}}$.
In Appendix~\ref{app:correction_bgsup}, we apply the same test to the background-suppressed equal-time observables of Sec.~\ref{sec:bg_suppressed}.

\subsection{Window fits and parameter trajectories}\label{app:correction_windows}

Figures~\ref{fig:correction_collapse_R} and \ref{fig:correction_collapse_m2_G} compare the leading form with corrected fits from both initial values in representative windows covering smaller and larger system sizes.
Figure~\ref{fig:correction_trajectory} shows the fit quality ($\chi^2_{\mathrm{red}}$), fitted correction exponents ($\omega$), and physical parameter estimates for the leading and corrected fits as $L_{\max}$ increases in the Hubbard model.
We compute $\chi^2_{\mathrm{red}}$ using the effective variance in Eq.~\eqref{equ:effective_variance}, accounting for uncertainty in the horizontal variable and its correlation with the fitted quantity.

\begin{figure*}[tp]
    \centering
    \includegraphics[width=0.85\textwidth]{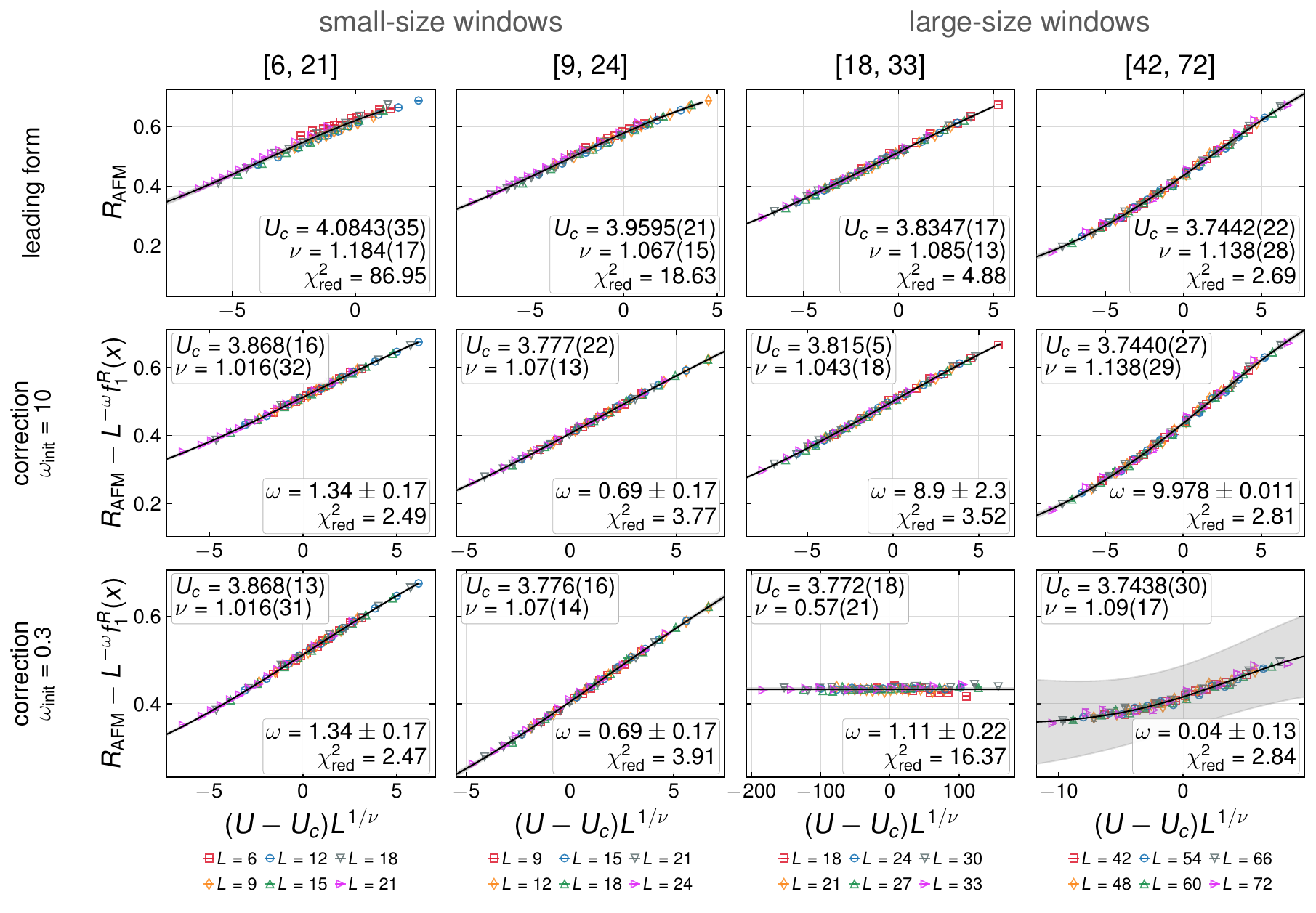}
    \caption{\rev{Hubbard correlation-ratio data collapses in four representative windows, $[6,21]$, $[9,24]$, $[18,33]$, and $[42,72]$, from left to right.
    Rows show the leading form and corrected fits initialized at $\omega_{\mathrm{init}}=10$ and $0.3$, respectively, with $\omega$ fitted freely.
    Each column uses identical input data; the corrected rows subtract the fitted correction term of Eq.~\eqref{equ:corrected_ansatz}.
    Each panel is drawn with all fitting parameters fixed at their bootstrap means (Appendix~\ref{app:data_collapse_analysis}), taken within the IQR fence of Fig.~\ref{fig:correction_trajectory} for the corrected rows, which set the horizontal scaling and the subtracted correction; axis ranges are chosen independently.
    Annotations give bootstrap parameter estimates and $\chi^2_{\mathrm{red}}$ evaluated at the bootstrap mean parameters.
    Gray bands show one posterior standard deviation of the inferred leading scaling function $f_0$ at fixed bootstrap mean parameters and are often too narrow to be visible.}}
    \label{fig:correction_collapse_R}
\end{figure*}

\begin{figure*}[tp]
    \centering
    \includegraphics[width=\textwidth,height=0.70\textheight,keepaspectratio]{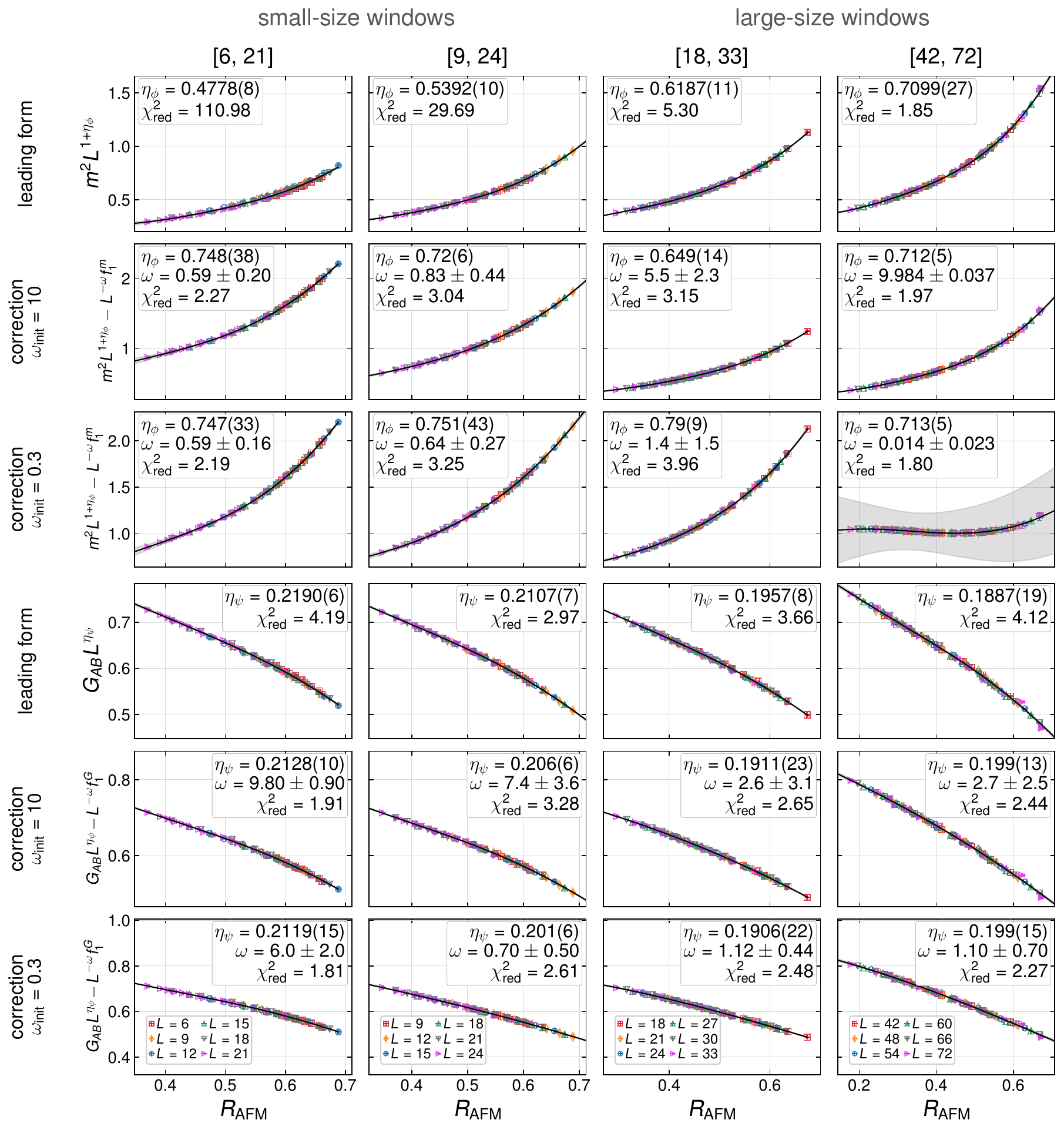}
    \caption{\rev{Hubbard $m^2_{\mathrm{AFM}}$ (upper three rows) and $G_{AB}$ (lower three rows) data collapses in the same four windows as Fig.~\ref{fig:correction_collapse_R}.
    Within each group, rows show the leading form and corrected fits initialized at $\omega_{\mathrm{init}}=10$ and $0.3$, respectively, with $\omega$ fitted freely.
    Fits for each observable and window use identical input data; corrected rows subtract the fitted correction from the rescaled observable.
    Corresponding columns share the same $R_{\mathrm{AFM}}$ range.
    As in Fig.~\ref{fig:correction_collapse_R}, panels are drawn with all fitting parameters fixed at their bootstrap means.
    Annotations give bootstrap parameter estimates and $\chi^2_{\mathrm{red}}$ evaluated at the bootstrap mean parameters.
    Gray bands show one posterior standard deviation of the inferred leading scaling function $f_0$ at fixed bootstrap mean parameters and are often too narrow to be visible.}}
    \label{fig:correction_collapse_m2_G}
\end{figure*}

\begin{figure*}[tp]
    \centering
    \includegraphics[width=0.9\textwidth]{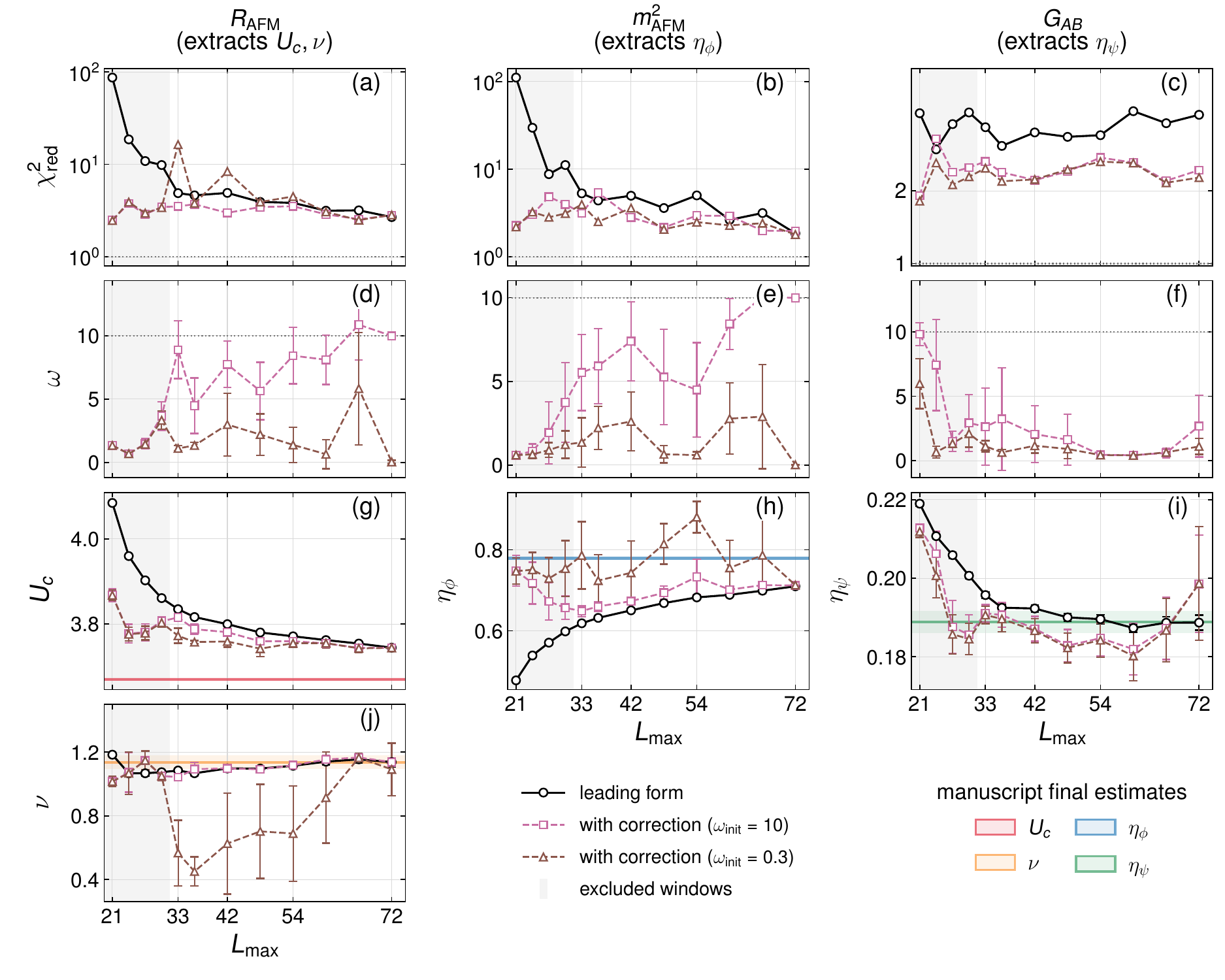}
    \caption{\rev{Leading and corrected fits for the Hubbard model as a function of $L_{\max}$, using windows of six sizes.
    Black open circles denote the leading form; purple open squares and brown open triangles denote corrected fits initialized at $\omega_{\mathrm{init}}=10$ and $0.3$, respectively, with $\omega$ fitted freely.
    Panels (a)--(c) show $\chi^2_{\mathrm{red}}$ evaluated at the bootstrap mean parameters, (d)--(f) the fitted $\omega$, and (g)--(j) the physical parameter estimates with bootstrap uncertainties.
    Window estimates and error bars are bootstrap means and standard deviations.
    For the corrected fits, they are taken after applying a permissive interquartile-range (IQR) fence separately to each fitted parameter, while the leading-form values are those of Fig.~\ref{fig:fit_extrapolation_bothmodels}.
    Values within $[Q_1-k\,\mathrm{IQR},Q_3+k\,\mathrm{IQR}]$ are retained, with $k=10$, where $Q_1$ and $Q_3$ are the first and third quartiles and $\mathrm{IQR}=Q_3-Q_1$.
    For $\nu$, the fence acts on $1/\nu$, and the window estimate averages $\nu$ over the retained replicates.
    Shading marks windows ($L_{\min}<18$) excluded from the main-text analysis (Appendix~\ref{app:data_collapse_analysis}); horizontal lines and bands indicate the final leading-form estimates and uncertainties reported there.}}
    \label{fig:correction_trajectory}
\end{figure*}

Adding the correction can substantially improve fits in the small-size windows excluded from the main-text analysis [e.g., the left two columns of Figs.~\ref{fig:correction_collapse_R} and \ref{fig:correction_collapse_m2_G}; the full excluded range is shaded gray in Fig.~\ref{fig:correction_trajectory}(a)--(c)].
For the correlation ratio and $m^2$, both initializations visibly improve the collapse in the smallest window, $[6,21]$, reducing $\chi^2_{\mathrm{red}}$ from $87.0$ and $111.0$ to below $3$.
Across the small-size windows, the corrected fits keep $\chi^2_{\mathrm{red}}$ between $2$ and $5$ for both observables and flatten its variation across windows.
For $G_{AB}$ and in the $L_{\max}\geq 33$ windows, the corrected fits bring at most a modest reduction in $\chi^2_{\mathrm{red}}$, typically from $3$--$5$ to $2$--$4$, and none of the three observables reaches $\chi^2_{\mathrm{red}}\approx1$ [Fig.~\ref{fig:correction_trajectory}(a)--(c)].
The corrected fits generally yield larger error bars on $U_c$, $\eta_\phi$, $\eta_\psi$, and $\nu$ than the leading form across the $L_{\max}\geq33$ windows [Fig.~\ref{fig:correction_trajectory}(g)--(j)].
With $\omega_{\mathrm{init}}=10$, the fitted $\omega$ of $R_{\mathrm{AFM}}$ and $m^2_{\mathrm{AFM}}$ increasingly stays near its starting value or spreads over a broad range around a large value ($\geq 5$) as the window sizes increase [Fig.~\ref{fig:correction_trajectory}(d), (e)].
These large exponents suppress the correction through $L^{-\omega}$, making $\omega$ weakly constrained while bringing the estimates of $U_c$, $\nu$, and $\eta_\phi$ close to the leading-form results.
With $\omega_{\mathrm{init}}=0.3$, the fitted $\omega$ often takes values of order unity or converges close to zero, while anomalous estimates of $\nu$, down to $0.45$, appear in the windows with $33\leq L_{\max}\leq54$ [Fig.~\ref{fig:correction_trajectory}(d)--(f), (j)].
Thus, improved fit quality can coexist with larger fluctuations and uncertainties in the window estimates.

\subsection{Bootstrap distributions and competing fit branches}\label{app:correction_bootstrap}

To understand the increased parameter fluctuations and the anomalous drop in $\nu$ at intermediate sizes [Fig.~\ref{fig:correction_trajectory}(j)], we examine the bootstrap distributions for a representative Hubbard window with $L_{\max}=54$ [Fig.~\ref{fig:correction_distributions}].
Compared with the leading form, both corrected fits yield much broader parameter distributions.
For $\eta_\phi$, the corrected distributions extend from $0.69$ to above $0.95$, far beyond the narrow leading-form peak at $0.683$, and the one for $\omega_{\mathrm{init}}=0.3$ is centered near $0.88$ [panel (c)].
For $\eta_\psi$, both corrected distributions are centered near $0.184$, below the leading-form value $0.190$, with long tails [panel (d)].
Notably, starting from $\omega_{\mathrm{init}}=0.3$ produces double peaks in $U_c$ and $\nu$, with $\nu$ branches near $0.48$ and $1.12$ [panels (a), (b)].
The high-$\nu$ branch lies close to the leading-form results, while the low-$\nu$ branch accounts for $68\%$ of the replicates.
The low-$\nu$ branch recurs throughout the $L_{\max}=33$--$54$ windows and is most populated at $L_{\max}=36$ ($98\%$ of the replicates), where the window estimate reaches its minimum [Fig.~\ref{fig:correction_trajectory}(j)].
The bimodal bootstrap distributions of Hubbard $U_c$ and $\nu$ reveal a deeper instability: a single central value and error bar cannot adequately represent the competing fit branches.

\begin{figure*}[tp]
    \centering
    \includegraphics[width=0.9\textwidth]{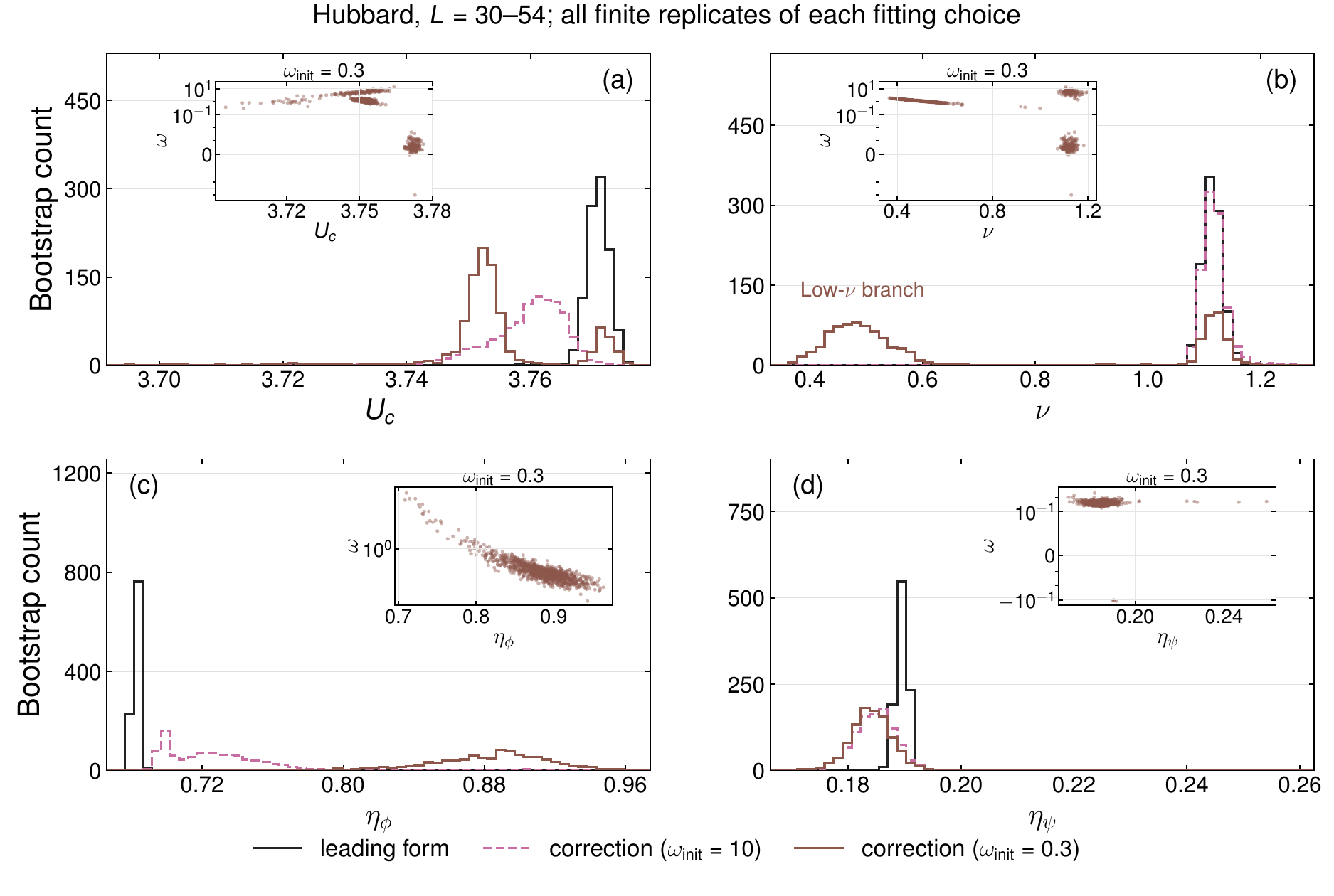}
    \caption{\rev{Bootstrap parameter distributions for the Hubbard window with $L_{\max}=54$, showing all finite replicates (at least 999 of 1000) for each fitting choice and parameter.
    Black, pink, and brown histograms denote the leading form and corrected fits initialized at $\omega_{\mathrm{init}}=10$ and $0.3$, respectively.
    Each inset pairs the displayed parameter with $\omega$ from the same fit initialized at $0.3$; the $\omega$ axis uses a symmetric logarithmic scale, linear for $|\omega|<10^{-3}$.}}
    \label{fig:correction_distributions}
\end{figure*}

\begin{figure*}[tp]
    \centering
    \includegraphics[width=0.9\textwidth]{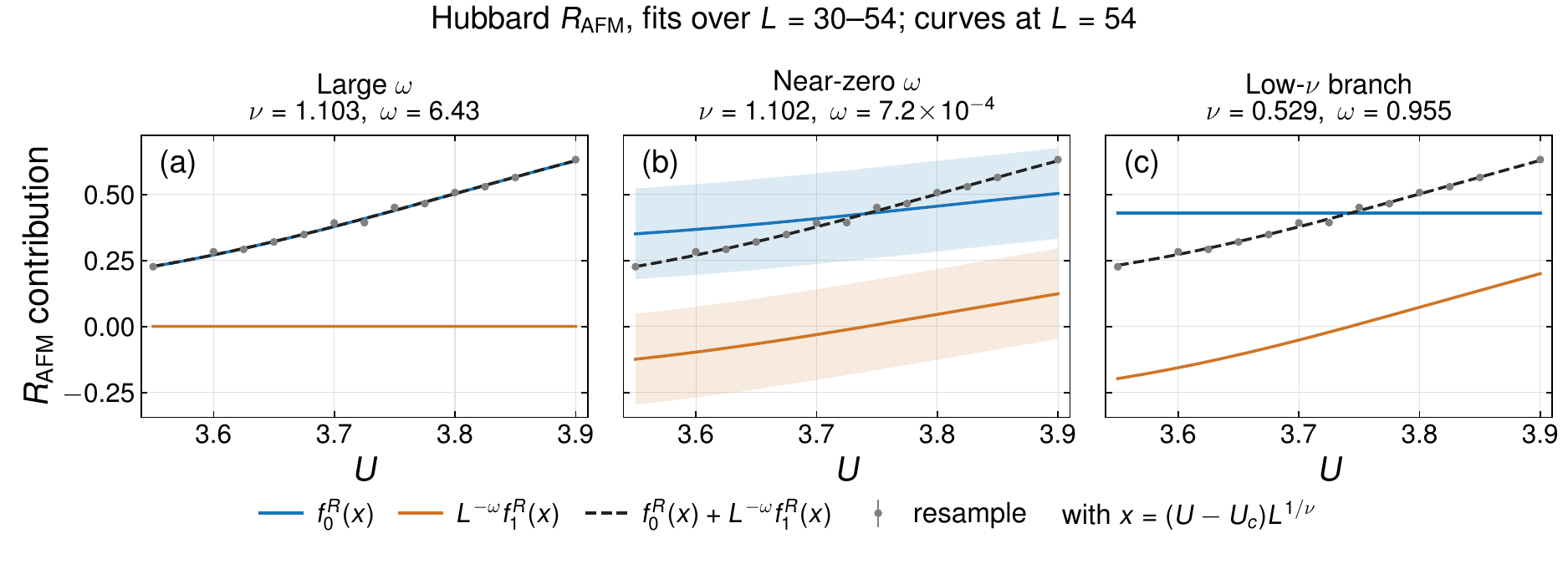}
    \caption{\rev{Decomposition of the Hubbard correlation-ratio fits into leading and correction contributions.
    The three fits use the same new resample of $R_{\mathrm{AFM}}(U,L)$ over the full $L_{\max}=54$ window, obtained by perturbing each point with Gaussian noise of its standard error.
    Initial values of $(U_c,\nu,\omega)$ and the kernel parameters are taken from representative replicates of the three groups in Fig.~\ref{fig:correction_distributions}(a,b).
    Panel titles report the final fitted results of $\nu$ and $\omega$.
    For clarity, only the $L=54$ data (gray points) and fitted contributions are shown as functions of $U$.
    Blue and orange curves show the leading and correction contributions, and black dashed curves their sum.
    Bands show one posterior standard deviation of each contribution at fixed fitted hyperparameters and are mostly too narrow to be visible in panels (a) and (c).}}
    \label{fig:correction_branches}
\end{figure*}

The $\omega$ inset in Fig.~\ref{fig:correction_distributions}(b) further separates the high-$\nu$ branch into large-$\omega$ ($\omega\gtrsim5$) and near-zero-$\omega$ groups, giving three representative fit outcomes together with the low-$\nu$ branch.
Using representative parameters from each group as initial values, we refit the same new resample of $R_{\mathrm{AFM}}(U,L)$ over the $L_{\max}=54$ window.
Figure~\ref{fig:correction_branches} decomposes each fitted curve at $L=54$ into the leading term $f_0^R(x)$ and correction $L^{-\omega}f_1^R(x)$ of Eq.~\eqref{equ:corrected_ansatz}.
All three fits describe the data well, but assign the variation with $U$ differently to the leading and correction terms.
\begin{itemize}
    \item \textbf{High $\nu$, large $\omega$ [Fig.~\ref{fig:correction_branches}(a)].}
    The large $\omega$ and small fitted correction amplitude make $L^{-\omega}f_1^R(x)$ negligible, giving $R_{\mathrm{AFM}}(U,L)\simeq f_0^R(x)$.
    This explains its close agreement with the leading-form fit.
    \item \textbf{High $\nu$, near-zero $\omega$ [Fig.~\ref{fig:correction_branches}(b)].}
    With $L^{-\omega}\simeq1$ across the fitting window, $R_{\mathrm{AFM}}(U,L)\simeq f_0^R(x)+f_1^R(x)\equiv F^R(x)$.
    The two components can vary widely in opposite directions while their sum remains tightly constrained.
    Their strong negative posterior covariance keeps the uncertainty in the sum small despite the broad individual bands in panel (b).
    \item \textbf{Low $\nu$, finite $\omega$ [Fig.~\ref{fig:correction_branches}(c)].}
    The fitted $f_0^R(x)$ turns out to be nearly constant and nonzero, leaving almost all of the $U$ dependence to $L^{-\omega}f_1^R(x)$.
    The slope at the fitted $U_c$ therefore follows $\partial R_{\mathrm{AFM}}/\partial U\propto L^{1/\nu-\omega}$.
    The finite $\omega$ thus compensates for the smaller $\nu$, bringing $1/\nu-\omega\simeq0.937$ close to $1/\nu\simeq0.907$ in panel (a).
\end{itemize}
Good fits to the $R_{\mathrm{AFM}}(U,L)$ data therefore do not ensure stable parameter estimates.
The fitted $\omega$ remains sensitive to the window and initialization for each observable [Fig.~\ref{fig:correction_trajectory}(d)--(f)].
The low-$\nu$ fit, in particular, is inconsistent with the assumed leading--subleading hierarchy.

\subsection{Thermodynamic-limit estimates}\label{app:correction_limits}

Using the window-selection and estimation procedures of the main text, the corrected fits with $\omega_{\mathrm{init}}=10$ yield thermodynamic-limit estimates consistent with the leading-form results within $2.2\sigma$ for Hubbard $U_c$ and $t$-$V$ $\eta_\phi$, and within $1\sigma$ for all other parameters of both models [Fig.~\ref{fig:correction_extrapolation}].
With $\omega_{\mathrm{init}}=0.3$, competing fit branches in the Hubbard $U_c$ and the broadened, shifted distributions of $\eta_\phi$ in both models obscure any monotonic drift in $1/L_{\max}$.
We therefore take envelope averages over all retained windows for these three quantities [Fig.~\ref{fig:correction_extrapolation}(a), (e), (f)] instead of linear extrapolations.
All $\omega_{\mathrm{init}}=0.3$ estimates likewise agree with the leading-form results within $2.3\sigma$ for Hubbard $U_c$ and within $1\sigma$ for all other parameters (Table~\ref{tab:fss_summary}).
However, because the freely fitted $\omega$ introduces additional freedom to compensate changes in the other fit parameters, the corrected fits carry larger error bars, especially for $\omega_{\mathrm{init}}=0.3$.
Even with $\omega_{\mathrm{init}}=10$, they produce noisier window trajectories, making the approach to the thermodynamic limit harder to discern than with the leading form.
Explicitly fitting $L^{-\omega}$ therefore cannot resolve the background-induced drift while keeping the parameters well constrained, motivating our retention of the leading scaling form in Sec.~\ref{sec:gnh_criticality} and the observable-level background suppression in Sec.~\ref{sec:bg_suppressed}.

\begin{figure*}[tp]
    \centering
    \includegraphics[width=\textwidth,height=0.66\textheight,keepaspectratio]{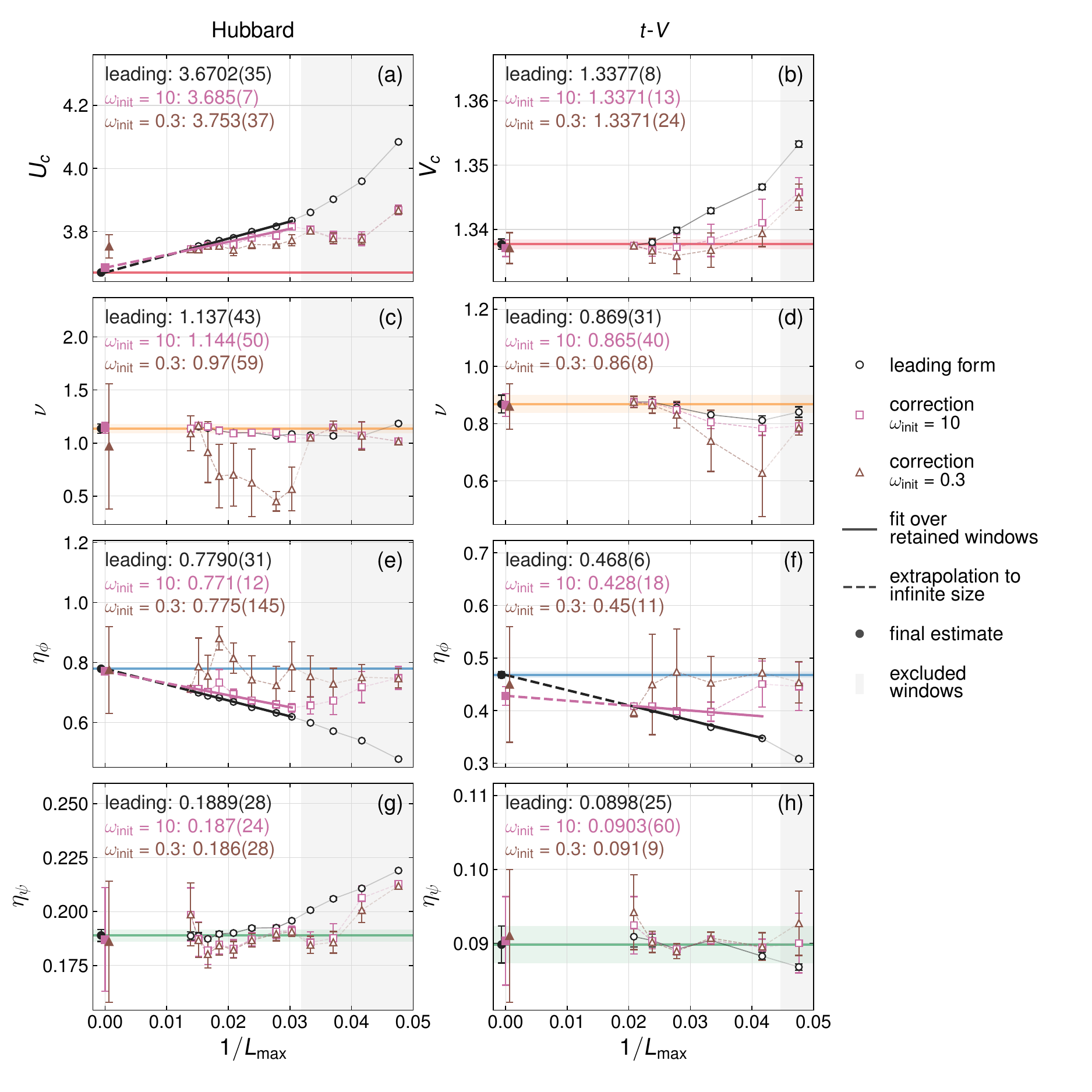}
    \caption{\rev{Thermodynamic-limit estimates from leading and corrected scaling fits for the Hubbard model (left) and the $t$-$V$ model (right).
    Black open circles, purple open squares, and brown open triangles show window estimates for the leading form and corrected forms initialized at $\omega_{\mathrm{init}}=10$ and $0.3$, respectively; $\omega$ is fitted freely in both corrected forms.
    Window estimates and error bars are defined as in Fig.~\ref{fig:correction_trajectory}; thin connecting lines are guides to the eye.
    Thermodynamic-limit estimates use the window selections of the main text.
    In panels (a), (e), and (f), thick solid lines show the full-covariance GLS fits linear in $1/L_{\max}$ for the leading form and $\omega_{\mathrm{init}}=10$, and thick dashed lines extend them to $1/L_{\max}=0$.
    All other estimates, including those for $\omega_{\mathrm{init}}=0.3$ in these panels, are envelope averages.
    Filled markers show the resulting intercept or envelope estimates and their uncertainties, offset horizontally for clarity.
    Shaded regions mark excluded windows ($L_{\min}<18$ for Hubbard, $L_{\min}<9$ for $t$-$V$; Appendix~\ref{app:data_collapse_analysis}).
    Horizontal lines and bands indicate the leading-form estimates and their uncertainties.}}
    \label{fig:correction_extrapolation}
\end{figure*}

\subsection{Background-suppressed observables}\label{app:correction_bgsup}

Finally, we verify that the leading-form collapses of the background-suppressed equal-time observables (Sec.~\ref{sec:bg_suppressed}) are robust against an explicit correction to scaling.
For the Hubbard model, we apply Eq.~\eqref{equ:corrected_ansatz} with $R^{\Delta}_{\mathrm{AFM}}$ replacing $R$ throughout and $C^{s}_{AA}(\mathbf{r}_{\max})$ or $\Delta m^2_{\mathrm{AFM}}$ replacing $m^2$, using the same windows and initial values of $\omega$ (Fig.~\ref{fig:correction_bgsup_trajectory}).

\begin{figure*}[tp]
    \centering
    \includegraphics[width=\textwidth]{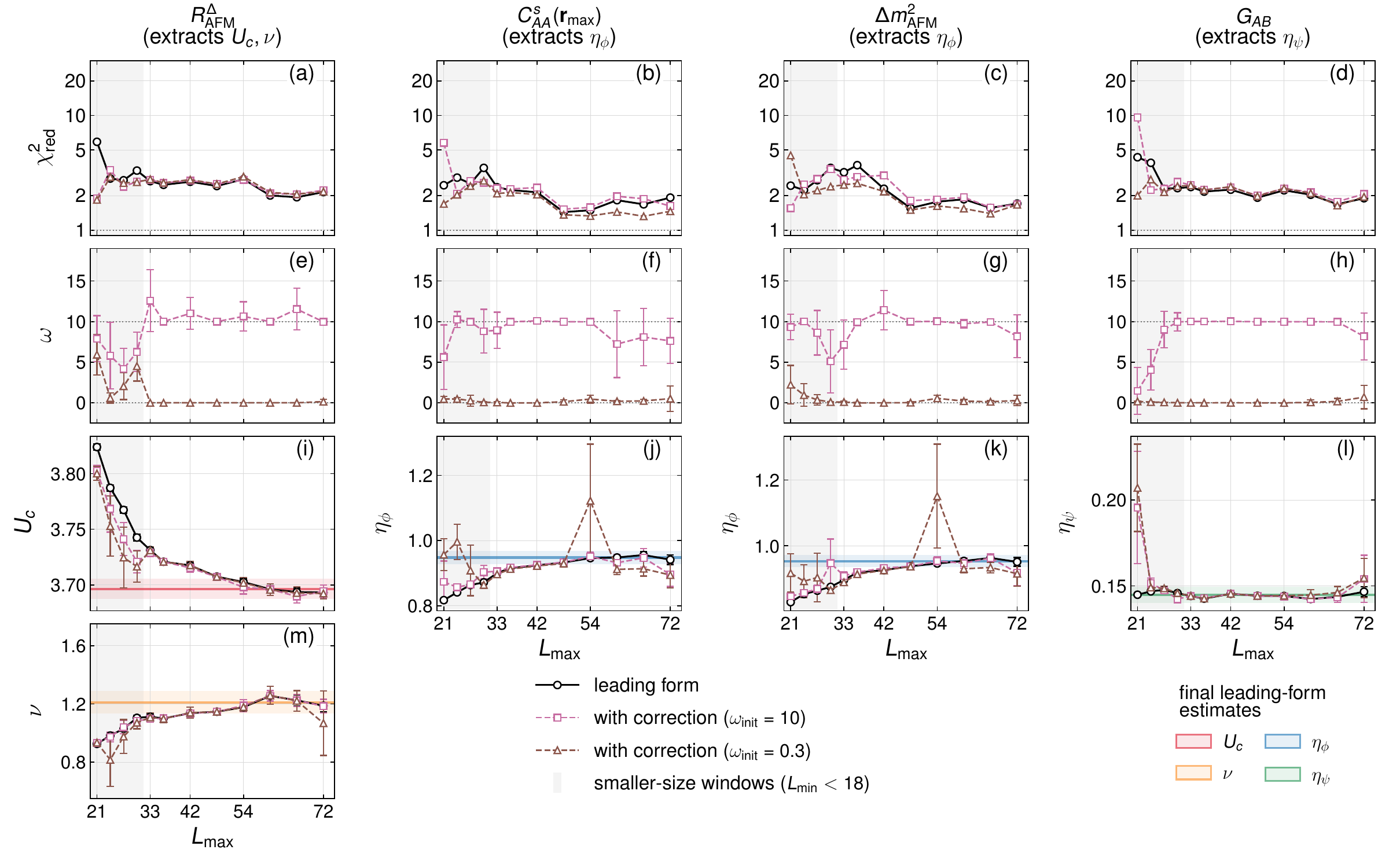}
    \caption{\rev{Leading and corrected fits of the background-suppressed equal-time observables for the Hubbard model as a function of $L_{\max}$, in the same layout and conventions as Fig.~\ref{fig:correction_trajectory}.
    Columns correspond to $R^{\Delta}_{\mathrm{AFM}}$, $C^{s}_{AA}(\mathbf{r}_{\max})$, $\Delta m^2_{\mathrm{AFM}}$, and $G_{AB}$, with the latter three collapsed against $R^{\Delta}_{\mathrm{AFM}}$.
    Panels (a)--(d) show $\chi^2_{\mathrm{red}}$, (e)--(h) the fitted $\omega$, and (i)--(m) the physical parameter estimates with bootstrap uncertainties.
    Horizontal lines and bands indicate the respective final leading-form estimates and uncertainties from Sec.~\ref{sec:bg_suppressed}.}}
    \label{fig:correction_bgsup_trajectory}
\end{figure*}

Unlike the conventional observables, the leading form already describes all windows well ($\chi^2_{\mathrm{red}}\approx 1.4$--$6$), and adding $L^{-\omega}$ brings no systematic improvement [Fig.~\ref{fig:correction_bgsup_trajectory}(a)--(d)].
For $L_{\max}\geq33$, the fitted $\omega$ remains controlled by $\omega_{\mathrm{init}}$, staying near $10$ for $\omega_{\mathrm{init}}=10$ and collapsing toward zero for $\omega_{\mathrm{init}}=0.3$ [Fig.~\ref{fig:correction_bgsup_trajectory}(e)--(h)].
These two regimes correspond to the unconstrained outcomes in Fig.~\ref{fig:correction_branches}(a) and (b), where the correction is either suppressed by $L^{-\omega}$ or degenerate with the leading term.
Across the $L_{\max}\geq33$ windows, the corrected estimates of $U_c$, $\nu$, $\eta_\phi$, and $\eta_\psi$ agree with the leading-form ones within $2\sigma$, but with larger uncertainties, especially for $\omega_{\mathrm{init}}=0.3$ [Fig.~\ref{fig:correction_bgsup_trajectory}(i)--(m)].
Once the background is suppressed, an explicit correction term is thus neither resolved by the data nor needed, confirming our use of the leading scaling form in Sec.~\ref{sec:bg_suppressed}.
\clearpage
\endgroup

%%%%%%%%%%%%%%%%%%%%%%%%%%%%%%%%%%%%%%%%%%%%%%%%%%%%%
\section{\rev{Summary of critical parameters across analysis procedures}}\label{app:fss_summary}
\begin{figure}[h!]\hsize=\textwidth\linewidth=\textwidth\columnwidth=\textwidth
    \centering
    \includegraphics[width=\textwidth]{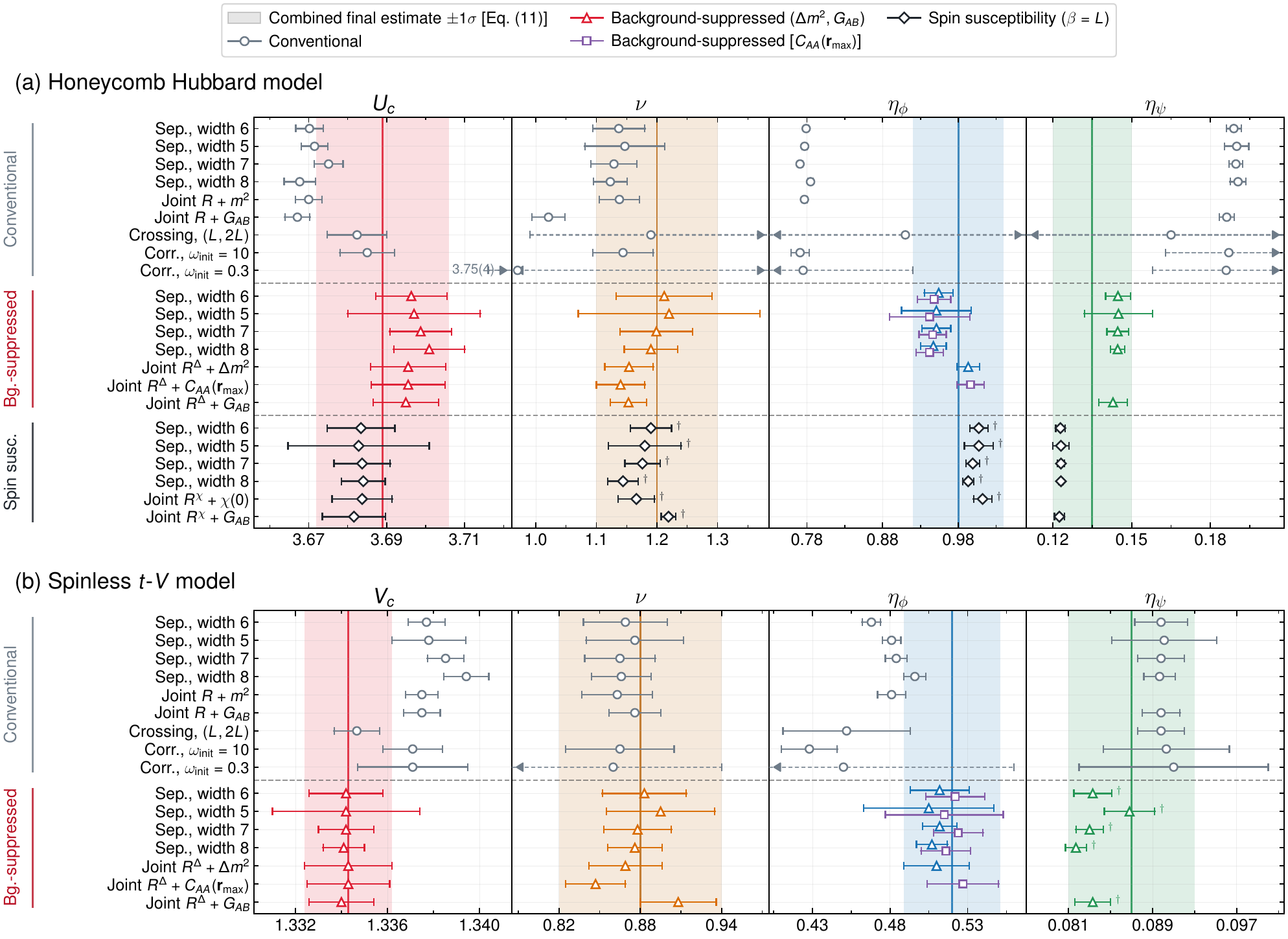}
    \caption{\rev{Comparison of thermodynamic-limit critical couplings and exponents across analysis procedures for (a) the honeycomb Hubbard model and (b) the spinless $t$-$V$ model (numerical values in Table~\ref{tab:fss_summary}).
    Horizontal error bars denote $1\sigma$ statistical uncertainties within each procedure, with dashed lines and arrowheads indicating wide uncertainties extending beyond the panel range (and the off-scale Hubbard $\omega_{\mathrm{init}}=0.3$ $U_c$ annotated).
    Vertical lines and shaded bands mark our combined final estimates and $1\sigma$ intervals in Eq.~\eqref{equ:final_estimates}; daggers denote largest-window frontier estimates that serve as lower bounds.}}
    \label{fig:fss_summary_forest}
\end{figure}

%%%%%%%%%%%%%%%%%%%%%%%%%%%%%%%%%%%%%%%%%%%%%%%%%%%%%
\begingroup
\revcolor

Figure~\ref{fig:fss_summary_forest} and Table~\ref{tab:fss_summary} compare the thermodynamic-limit critical couplings and exponents obtained across Secs.~\ref{sec:gnh_criticality}--\ref{sec:bg_suppressed} and Appendices~\ref{app:FSS_tV} and~\ref{app:fitting_window_scan}--\ref{app:correction_initialization}.
For the background-suppressed observables and spin susceptibility, the plateau values at window widths five, seven, and eight shift by at most $0.5\sigma$ from the width-six values and stay within the combined final $1\sigma$ intervals built from them [Eq.~\eqref{equ:final_estimates}].
Within the conventional family, leading-order separate and joint collapses also cluster tightly (apart from the distorted $\nu$ in the Hubbard $R_{\mathrm{AFM}}+G_{AB}$ fit discussed in Sec.~\ref{sec:bg_suppressed}), though their small statistical error bars expose slight residual window drift, whereas crossing-point extrapolations and explicit $L^{-\omega}$ fits remain broadly compatible at the cost of much larger uncertainties.
Across observable families, all procedures overlap around a common parameter region in the $t$-$V$ model, whereas in the Hubbard model the conventional estimates of $U_c$, $\eta_\phi$, and $\eta_\psi$ form a distinct cluster systematically displaced from the background-suppressed and susceptibility results (Sec.~\ref{sec:bg_suppressed}).

\endgroup

\onecolumngrid

\begin{table}[h!]\hsize=\textwidth\linewidth=\textwidth\columnwidth=\textwidth
    \caption{\rev{Thermodynamic-limit estimates for the honeycomb Hubbard and spinless $t$-$V$ models from the analysis procedures in Secs.~\ref{sec:gnh_criticality}--\ref{sec:bg_suppressed} and Appendices~\ref{app:FSS_tV} and~\ref{app:fitting_window_scan}--\ref{app:correction_initialization}.
    Error bars are statistical uncertainties within each procedure.
    Crossing-point entries are quoted only for the conventional observables from the power-law extrapolations in Fig.~\ref{fig:crossing_bothmodels}, with $\nu$ inverted from $1/\nu$ and the unstable $t$-$V$ $\nu$ fit omitted (Appendix~\ref{app:crossing_point}); once background drift is suppressed, power-law crossing fits are largely unconstrained (Appendix~\ref{app:crossing_bgsup}).
    For the background-suppressed observables, the two $\eta_\phi$ entries at each window width are obtained from $\Delta m^2$ and, in brackets, $C_{AA}(\mathbf{r}_{\max})$; joint collapses at width six are listed under each family following Figs.~\ref{fig:hubbard_joint_collapse} and~\ref{fig:tv_joint_collapse} (and the separate-collapse rules for susceptibility).
    The susceptibility data are available only for the Hubbard model ($L\le 48$), where $U_c$ and $\eta_\psi$ reach plateaus; daggered entries denote largest-window ($L_{\max}=48$, or $L_{\max}\ge 42$ for susceptibility width five) frontier estimates that serve as lower bounds (Sec.~\ref{sec:bg_suppressed}).
    Boldface entries mark the combined final estimates adopted in Eq.~\eqref{equ:final_estimates}.}}
    \label{tab:fss_summary}
    {\small\revcolor
    \begin{tabular*}{\textwidth}{@{\extracolsep{\fill}}lcccccccc@{}}
    \toprule
    & \multicolumn{4}{c}{Honeycomb Hubbard model} & \multicolumn{4}{c}{Spinless $t$-$V$ model} \\
    \cmidrule(lr){2-5}\cmidrule(l){6-9}
    Procedure & $U_c$ & $\nu$ & $\eta_\phi$ & $\eta_\psi$ & $V_c$ & $\nu$ & $\eta_\phi$ & $\eta_\psi$ \\
    \midrule
    \multicolumn{9}{@{}l}{\textbf{Conventional observables}} \\
    \addlinespace[0.08em]
    \multicolumn{9}{@{}l}{Sliding window, separate collapse (leading form)} \\
    \quad width 6 (Sec.~\ref{sec:gnh_criticality}) & $3.6702(35)$ & $1.137(43)$ & $0.7790(31)$ & $0.1889(28)$ & $1.3377(8)$ & $0.869(31)$ & $0.468(6)$ & $0.0898(25)$ \\
    \quad width 5 & $3.6715(34)$ & $1.147(66)$ & $0.7770(32)$ & $0.1900(46)$ & $1.3378(16)$ & $0.876(36)$ & $0.481(6)$ & $0.0901(50)$ \\
    \quad width 7 & $3.6751(37)$ & $1.129(38)$ & $0.7709(32)$ & $0.1897(26)$ & $1.33853(79)$ & $0.865(26)$ & $0.484(7)$ & $0.0898(22)$ \\
    \quad width 8 & $3.6677(40)$ & $1.123(28)$ & $0.7848(37)$ & $0.1905(30)$ & $1.33943(98)$ & $0.866(22)$ & $0.496(7)$ & $0.08966(150)$ \\
    \addlinespace[0.06em]
    \multicolumn{9}{@{}l}{Sliding window, joint collapse (width 6)} \\
    \quad $R+m^2$ & $3.6700(34)$ & $1.138(33)$ & $0.7767(34)$ & -- & $1.3375(7)$ & $0.863(26)$ & $0.481(9)$ & -- \\
    \quad $R+G_{AB}$ & $3.6671(32)$ & $1.021(27)$ & -- & $0.1862(28)$ & $1.3375(8)$ & $0.876(19)$ & -- & $0.0898(18)$ \\
    \addlinespace[0.08em]
    Crossing point, $(L,2L)$ & $3.6824(76)$ & $1.19(20)$ & $0.91(22)$ & $0.165(71)$ & $1.33467(99)$ & -- & $0.452(41)$ & $0.0898(22)$ \\
    \addlinespace[0.08em]
    \multicolumn{9}{@{}l}{Sliding window, separate collapse (corrected form)} \\
    \quad $\omega_{\mathrm{init}}=10$ & $3.685(7)$ & $1.144(50)$ & $0.771(12)$ & $0.187(24)$ & $1.3371(13)$ & $0.865(40)$ & $0.428(18)$ & $0.0903(60)$ \\
    \quad $\omega_{\mathrm{init}}=0.3$ & $3.753(37)$ & $0.97(59)$ & $0.775(145)$ & $0.186(28)$ & $1.3371(24)$ & $0.86(8)$ & $0.45(11)$ & $0.091(9)$ \\
    \addlinespace[0.08em]
    \midrule
    \addlinespace[0.08em]
    \multicolumn{9}{@{}l}{\textbf{Background-suppressed observables}} \\
    \addlinespace[0.06em]
    \multicolumn{9}{@{}l}{Sliding window, separate collapse (leading form)} \\
    \quad width 6 (Sec.~\ref{sec:bg_suppressed}) & $3.6963(91)$ & $1.212(79)$ & $0.954(19)$ & $0.1448(47)$ & $1.3342(16)$ & $0.883(31)$ & $0.512(19)$ & $0.0833(18)^{\dagger}$ \\[-1.4pt]
     & & & $[0.948(22)]$ & & & & $[0.522(19)]$ & \\[1.0pt]
    \quad width 5 & $3.697(17)$ & $1.22(15)$ & $0.951(46)$ & $0.145(13)$ & $1.3342(32)$ & $0.895(40)$ & $0.505(42)$ & $0.0868(24)^{\dagger}$ \\[-1.4pt]
     & & & $[0.942(53)]$ & & & & $[0.515(38)]$ & \\[1.0pt]
    \quad width 7 & $3.6987(79)$ & $1.199(60)$ & $0.951(19)$ & $0.1447(41)$ & $1.3342(12)$ & $0.878(25)$ & $0.512(11)$ & $0.0830(13)^{\dagger}$ \\[-1.4pt]
     & & & $[0.946(18)]$ & & & & $[0.524(16)]$ & \\[1.0pt]
    \quad width 8 & $3.7009(91)$ & $1.190(44)$ & $0.947(17)$ & $0.1447(27)$ & $1.3341(9)$ & $0.876(20)$ & $0.507(10)$ & $0.0817(10)^{\dagger}$ \\[-1.4pt]
     & & & $[0.942(18)]$ & & & & $[0.516(16)]$ & \\
    \addlinespace[0.06em]
    \multicolumn{9}{@{}l}{Sliding window, joint collapse (width 6)} \\
    \quad $R^{\Delta}+\Delta m^2$ & $3.6955(96)$ & $1.154(40)$ & $0.993(15)$ & -- & $1.3343(19)$ & $0.869(27)$ & $0.510(21)$ & -- \\
    \quad $R^{\Delta}+C_{AA}(\mathbf{r}_{\max})$ & $3.6955(95)$ & $1.140(40)$ & $[0.996(18)]$ & -- & $1.3343(18)$ & $0.847(22)$ & $[0.527(23)]$ & -- \\
    \quad $R^{\Delta}+G_{AB}$ & $3.6949(84)$ & $1.153(30)$ & -- & $0.1429(54)$ & $1.3340(14)$ & $0.908(28)$ & -- & $0.0833(17)^{\dagger}$ \\
    \addlinespace[0.08em]
    \midrule
    \addlinespace[0.08em]
    \multicolumn{9}{@{}l}{\textbf{Spin susceptibility}} \\
    \addlinespace[0.06em]
    \multicolumn{9}{@{}l}{Sliding window, separate collapse (leading form)} \\
    \quad width 6 (Sec.~\ref{sec:bg_suppressed}) & $3.6834(87)$ & $1.190(34)^{\dagger}$ & $1.007(12)^{\dagger}$ & $0.1229(18)$ & -- & -- & -- & -- \\
    \quad width 5 & $3.6828(181)$ & $1.18(6)^{\dagger}$ & $1.007(19)^{\dagger}$ & $0.1231(31)$ & -- & -- & -- & -- \\
    \quad width 7 & $3.6837(72)$ & $1.176(29)^{\dagger}$ & $0.999(9)^{\dagger}$ & $0.1231(13)$ & -- & -- & -- & -- \\
    \quad width 8 & $3.6840(56)$ & $1.144(25)^{\dagger}$ & $0.993(7)^{\dagger}$ & $0.1231(8)$ & -- & -- & -- & -- \\
    \addlinespace[0.06em]
    \multicolumn{9}{@{}l}{Sliding window, joint collapse (width 6)} \\
    \quad $R^{\chi}_{\mathrm{AFM}}+\chi_{\mathrm{AFM}}(0)$ & $3.6837(77)$ & $1.166(30)^{\dagger}$ & $1.012(12)^{\dagger}$ & -- & -- & -- & -- & -- \\
    \quad $R^{\chi}_{\mathrm{AFM}}+G_{AB}$ & $3.6816(81)$ & $1.219(12)^{\dagger}$ & -- & $0.1225(19)$ & -- & -- & -- & -- \\
    \addlinespace[0.08em]
    \midrule
    \addlinespace[0.08em]
    \textbf{Combined final (Sec.~\ref{sec:discussion})} & $\textbf{3.689(17)}$ & $\textbf{1.20(10)}$ & $\textbf{0.98(6)}$ & $\textbf{0.135(15)}$ & $\textbf{1.3343(19)}$ & $\textbf{0.88(6)}$ & $\textbf{0.520(31)}$ & $\textbf{0.087(6)}$ \\
    \bottomrule
    \end{tabular*}}
\end{table}
\FloatBarrier
\twocolumngrid

%%%%%%%%%%%%%%%%%%%%%%%%%%%%%%%%%%%%%%%%%%%%%%%%%%%%%
\section{Gross-Neveu universality classes}\label{app:Gross-Neveu}
%%%%%%%%%%%%%%%%%%%%%%%%%%%%%%%%%%%%%%%%%%%%%%%%%%%%%

The previously reported critical exponents for the $N=8$ Gross-Neveu-Heisenberg and $N=4$ Gross-Neveu-Ising universality classes are summarized in Tables~\ref{tab:gn_heis} and~\ref{tab:gn_is}, respectively.

\begin{table*}[tp]
    \centering
    \small
    \caption{Estimated critical exponents in the $N=8$ Gross-Neveu-Heisenberg (GNH) universality class reported in previous studies. For quantum Monte Carlo studies of the Hubbard model, the critical coupling $U_c$ is listed when available; $N_{\max}$ denotes the largest number of lattice sites used in the simulations.
    *\rev{Inverted from the $1/\nu$ values in Refs.~\cite{Phys.Rev.D2017Zerf,Phys.Rev.B2026Hands}.}
    $^\dagger$These values are estimated by collapsing $m^2$ without the correction term, as recommended by the authors of Ref.~\cite{Phys.Rev.X2016Otsuka} in their later work~\cite{Phys.Rev.B2020Otsuka}.}
    \label{tab:gn_heis}
    \begin{tabular}{@{} >{\raggedright\arraybackslash}p{10.0cm} C{1.65cm} C{1.65cm} C{1.65cm} C{1.65cm} r @{}}
        \toprule
        Method & $\nu$ & $\eta_\phi$ & $\eta_\psi$ & $U_c$ & $N_{\max}$ \\
        \midrule
        4-loop $4-\epsilon$ expansion (Pad\'e [2/2])~\cite{Phys.Rev.D2017Zerf} & 1.5562* & 0.9985 & 0.1833 &  &  \\
        4-loop $4-\epsilon$ expansion (Pad\'e [3/1])~\cite{Phys.Rev.D2017Zerf} & 1.2352 & 0.9563 & 0.1560 &  &  \\
        leading-loop $2+\epsilon$ expansion interpolated with 4-loop $4-\epsilon$~\cite{Phys.Rev.B2023Ladovrechis} & 1.20(18) & 1.01(6) & 0.13(3) &  &  \\
        Large-$N$ ($\eta_\phi$ and $\nu$ are $O(1/N^2)$; $\eta_\psi$ is $O(1/N^3)$)~\cite{Phys.Rev.D2018Gracey} & 1.1823 & 1.1849 & 0.1051 &  &  \\
        FRG~\cite{PRB2014L.Janssen} & 1.295 & 1.015 & 0.084 &  &  \\
        FRG~\cite{Phys.Rev.B2018Knorr} & 1.258 & 1.032 & 0.071(2) &  &  \\
        \addlinespace[0.25em]
        \midrule
        \addlinespace[0.25em]
        \multicolumn{6}{@{}l}{\textbf{Honeycomb Hubbard model}}\\
        \addlinespace[0.15em]
        PQMC (This work) & \rev{1.20(10)} & \rev{0.98(6)} & \rev{0.135(15)} & \rev{3.689(17)} & 10368 \\
        \rev{FT-DQMC ($\beta t=L$)~\cite{ArXiv2609H.Du}} & \rev{1.097(7)} & \rev{0.72(2)} & \rev{0.155(6)} & \rev{3.731(4)} & \rev{3528} \\
        PQMC (Finite-time scaling)~\cite{PRL2026YK.Yu} & 1.17(7) & 0.60(6) & 0.15(4) & 3.91(3) & 3042 \\
        PQMC (Finite-time scaling; average over three results)~\cite{Nat.Commun.2025Zeng} & 1.02(3) & 0.50(8) &  &  & 1152 \\
        HMC~\cite{Phys.Rev.B2021Ostmeyer} & 1.181(43) & 0.52(1) &  & 3.835(14) & 20808 \\
        FT-DQMC (fixed to a low temperature)~\cite{Phys.Rev.B2019Buividovich} & 0.928 & 0.624 &  & 3.942 & 1152 \\
        PQMC~\cite{Phys.Rev.X2016Otsuka} & 1.02(1) & 0.65(3)$^\dagger$ & 0.20(2) & 3.85(1) & 2592 \\
        PQMC~\cite{Phys.Rev.B2015ParisenToldin} & 0.84(4) & 0.70(15) &  & 3.80(1) & 648 \\
        \addlinespace[0.25em]
        \midrule
        \addlinespace[0.25em]
        \multicolumn{6}{@{}l}{\textbf{Other lattice models}}\\
        \addlinespace[0.15em]
        \rev{FT-DQMC~\cite{ArXiv2611J.Vaeth} ($\pi$-flux Hubbard)} & \rev{1.10(3)} & \rev{1.09(4)} &  &  & \rev{1024} \\
        \rev{HMC~\cite{Phys.Rev.B2026Hands} (Domain wall fermion lattice field theory)} & \rev{1.58(6)*} & \rev{1.42(8)} &  &  & \rev{576} \\
        PQMC~\cite{Phys.Rev.B2025Lang} (SLAC fermion square Hubbard) & 1.02(3) & 0.73(1) & 0.09(1) &  & 361 \\
        FT-DQMC~\cite{Phys.Rev.Lett.2021Xu} ($d$-wave-rotor square Hubbard) & 0.99(8) & 0.55(2) &  &  & 400 \\
        FT-DQMC~\cite{Phys.Rev.B2021Liu} (Honeycomb plaquette interaction) & 0.90(3) & 0.80(9) & 0.29(2) &  & 1152 \\
        PQMC~\cite{Phys.Rev.B2020Otsuka} ($d$-wave square Hubbard) & 1.05(5) & 0.75(4) & 0.23(4) &  & 1600 \\
        PQMC~\cite{Science2018Tang} (Honeycomb long-range interaction) & 0.88(5) & 0.80(16) &  &  & 450 \\
        HMC~\cite{Phys.Rev.B2018Buividovich} (Honeycomb Extended Hubbard) & 1.162 & 0.872(44) &  &  & 648 \\
        PQMC~\cite{Phys.Rev.X2016Otsuka} ($\pi$-flux Hubbard) & 1.02(1) & 0.64(6)$^\dagger$ & 0.23(2) &  & 1600 \\
        \bottomrule
    \end{tabular}
\end{table*}

\begin{table*}[tp]
    \centering
    \small
    \caption{Estimated critical exponents in the $N=4$ Gross-Neveu-Ising (GNI) universality class reported in previous studies. For quantum Monte Carlo studies of the $t$-$V$ model, the critical coupling $V_c$ is listed when available; $N_{\max}$ denotes the largest number of lattice sites used in the simulations. 
    For the large-$N$ results, the Pad\'e-Borel resummation is applied following~\cite{Phys.Rev.B2018Ihrig}.
    Note that the conformal bootstrap result~\cite{J.HighEnerg.Phys.2023Erramilli} is obtained for the $O(N)$-GNY model rather than the chiral-Ising GNY (GNI) model with $O(N/2)^2\rtimes Z_2$ symmetry; however, these two models are nearly degenerate for the leading low-energy operators and become distinguishable only at higher perturbative orders~\cite{J.HighEnerg.Phys.2023Erramilli,Phys.Rev.B2023Wanga}.}
    \label{tab:gn_is}
    \begin{tabular}{@{} >{\raggedright\arraybackslash}p{10.0cm} C{1.65cm} C{1.65cm} C{1.65cm} C{1.65cm} r @{}}
        \toprule
        Method & $\nu$ & $\eta_\phi$ & $\eta_\psi$ & $V_c$ & $N_{\max}$ \\
        \midrule
        4-loop $4-\epsilon$ expansion (Borel resummation)~\cite{Phys.Rev.B2018Ihrig} & 0.898(30) & 0.487(12) & 0.102(12) &  &  \\
        5-loop $4-\epsilon$ expansion interpolated with 4-loop $2+\epsilon$ (Pad\'e [5/5])~\cite{PRD2025JA.Gracey} & 0.8925 & 0.5070 &  &  &  \\
        Large-$N$ ($\eta_\phi$ and $\nu$ are $O(1/N^2)$; $\eta_\psi$ is $O(1/N^3)$)~\cite{Phys.Lett.B1992Gracey,Int.J.Mod.Phys.A1994Graceya,Int.J.Mod.Phys.A1994Gracey} & 1.066 & 0.509 & 0.106 &  &  \\
        FRG~\cite{Phys.Rev.B2016Knorr} & 0.93(1) & 0.5506 & 0.0654 &  &  \\
        Conformal bootstrap~\cite{J.HighEnerg.Phys.2023Erramilli} & 0.9083(83) & 0.5155(30) & 0.08712(32) &  &  \\
        \addlinespace[0.25em]
        \midrule
        \addlinespace[0.25em]
        \multicolumn{6}{@{}l}{\textbf{Honeycomb $t$-$V$ model}}\\
        \addlinespace[0.15em]
        PQMC (This work) & \rev{0.88(6)} & \rev{0.520(31)} & \rev{0.087(6)} & \rev{1.3343(19)} & 4608 \\
        PQMC (Finite-time scaling)~\cite{Sci.Adv.2026YK.Yu} & 0.77(12) & 0.49(5) & 0.073(4) & 1.35(1) & 1458 \\
        PQMC (Finite-time scaling; average over three results)~\cite{Nat.Commun.2025Zeng} & 0.74(1) & 0.19(4) &  &  & 1152 \\
        CTQMC~\cite{Phys.Rev.B2016Hesselmann} & 0.74(4) & 0.275(25) &  & 1.359 & 882 \\
        SSEQMC~\cite{Phys.Rev.B2016Wang} & 0.72(9) &  &  & 1.36 & 1152 \\
        PQMC~\cite{NewJ.Phys.2015Li} & 0.77(3) & 0.45(2) &  & 1.355 & 1152 \\
        CTQMC~\cite{NewJ.Phys.2014Wang} & 0.80(3) & 0.302(7) &  & 1.356 & 450 \\
        \addlinespace[0.25em]
        \midrule
        \addlinespace[0.25em]
        \multicolumn{6}{@{}l}{\textbf{Other lattice models ($\pi$-flux $t$-$V$ model)}}\\
        \addlinespace[0.15em]
        Fermion-bag QMC (continuous time)~\cite{Phys.Rev.D2020Huffman} & 0.89(1) & 0.51(3) &  &  & 4096 \\
        Fermion-bag QMC (discrete time, $\Delta_\tau=1$)~\cite{Phys.Rev.D2020Huffman} & 0.94(3) & 0.49(4) &  &  & 10000 \\
        PQMC~\cite{NewJ.Phys.2015Li} & 0.79(4) & 0.43(2) &  &  & 484 \\
        CTQMC~\cite{NewJ.Phys.2014Wang} & 0.80(6) & 0.318(8) &  &  & 400 \\
        \bottomrule
    \end{tabular}
\end{table*}

\clearpage

%%%%%%%%%%%%%%%%%%%%%%%%%%%%%%%%%%%%%%%%%%%%%%%%%%%%%
\bibliography{reference_submatrixLR}
%%%%%%%%%%%%%%%%%%%%%%%%%%%%%%%%%%%%%%%%%%%%%%%%%%%%%

\end{document}